%% file: luca_ghislotti_phd_thesis.tex
\pdfoutput=1
\documentclass{settings}

\usepackage[english]{babel} 
\usepackage[utf8]{inputenc} 
\usepackage[T1]{fontenc}   
\usepackage{graphicx}  
\usepackage{eso-pic}    
\usepackage{setspace}  
\usepackage{longtable}  
\usepackage{pdfpages}   
\usepackage{fancyhdr}   
\usepackage{hyperref}
\usepackage{afterpage} 
\usepackage{datetime}
\usepackage{bm}
\usepackage{ifthen}
\usepackage[detect-all=true]{siunitx}
\usepackage{newtxtext,newtxmath}
\usepackage{pdfpages}
\usepackage{dirtree}
\usepackage{cancel}
\usepackage{floatrow}
\usepackage{comment}
\usepackage{multirow}
\usepackage[colorinlistoftodos, textsize=small]{todonotes}
\usepackage{textcomp}
\usepackage{gensymb}
\usepackage{makecell}
\usepackage[verbose]{placeins}
\usepackage{ragged2e}
\usepackage{nccmath}
\usepackage[absolute,overlay]{textpos}
\usepackage{titlesec}
\usepackage[version=4]{mhchem}
\usepackage[notransparent]{svg}
\usepackage{csquotes}
\usepackage{xcolor}
\usepackage{booktabs}
\usepackage[capitalise,nameinlink]{cleveref}
\usepackage{pifont}
\usepackage{textcomp}

\hypersetup{
    pdfauthor={Luca Ghislotti},
    pdftitle={Design, Testing and Calibration of the GAPS Experiment Si(Li) Tracker Readout ASIC: From the First Flight Campaign Toward the Second Mission Upgrade},
    pdfsubject={Design, Testing and Calibration of the GAPS Experiment Si(Li) Tracker Readout ASIC: From the First Flight Campaign Toward the Second Mission Upgrade},
    pdfkeywords={ASIC, Silicon Detectors, Readout Electronics, CMOS Technology, Dark Matter},
    pdfcreator={LaTeX Workshop for VSCode},
    pdfproducer={pdfTeX-1.40.24}
}

\newcommand{\cmark}{\textcolor{green!60!black}{\ding{51}}} % ✓ green
\newcommand{\xmark}{\textcolor{red}{\ding{55}}}            % ✗ red

\newdateformat{monthyeardate}{%
  \monthname[\THEMONTH] \THEYEAR}

\newcommand\blankpage{
    \null
    \thispagestyle{empty}
    \addtocounter{page}{0}
    \newpage}

    \fancypagestyle{plain}{
    \fancyhf{}
    \fancyfoot[C]{\thepage}

}

\hypersetup{
    colorlinks,
    linkcolor={black!100!black},
    citecolor={black!100!black},
    urlcolor={black!100!black}
}

\makeatletter
\def\smallunderbrace#1{\mathop{\vtop{\m@th\ialign{##\crcr
   $\hfil\displaystyle{#1}\hfil$\crcr
   \noalign{\kern3\p@\nointerlineskip}%
   \tiny\upbracefill\crcr\noalign{\kern3\p@}}}}\limits}
\makeatother

\newfloatcommand{capbtabbox}{table}[][\FBwidth]

\ifthenelse{\boolean{final}}{
    \setcounter{tocdepth}{2}
}{}

\begin{document}
\pagenumbering{gobble}
\input{preamble/titlepage/titlepage_EN_logo.tex}
\blankpage

%-------------------------------------------------------------------------------
%	PREAMBLE
%-------------------------------------------------------------------------------

\startpreamble
\frontmatter

%-------------------------------------------------------------------------------
%	ABSTRACT
%-------------------------------------------------------------------------------

\input{preamble/abstract}

%-------------------------------------------------------------------------------
%	TABLE OF CONTENTS
%-------------------------------------------------------------------------------

\tableofcontents

%-------------------------------------------------------------------------------
%	THESIS MAIN TEXT
%-------------------------------------------------------------------------------

\addtocontents{toc}{\vspace{2em}}
\mainmatter

% Introduction
\unnumberedchapterintroduction{Introduction}
\input{maintext/introduction}

% Chapter 1: The GAPS experiment
\numberedchapter
\input{maintext/chapters/chapter1}

% Chapter 2: Characterization of a flight-ready GAPS Si(Li) tracker module
\numberedchapter
\input{maintext/chapters/chapter2.tex}

% Chapter 3: Calibration of the GAPS Si(Li) tracker for in-flight operation
\numberedchapter
\input{maintext/chapters/chapter3.tex}

% Chapter 4: Design of the ANTARES4 readout ASIC
\numberedchapter
\input{maintext/chapters/chapter4.tex}

% Chapter 5: Characterization of the ANTARES4 readout ASIC
\numberedchapter
\input{maintext/chapters/chapter5.tex}

% Conclusions
\unnumberedchapterconclusions{Conclusions}
\input{maintext/conclusions}

%-------------------------------------------------------------------------------
%	APPENDICES
%-------------------------------------------------------------------------------

\appendix
\renewcommand{\chaptername}{Appendix}
\addtocontents{toc}{\vspace{1em}}

% Appendix A: Contribution to the GAPS integration and first Antarctic campaign
\numberedappendix
\input{maintext/appendix/appendixA}

%-------------------------------------------------------------------------------
%	BIBLIOGRAPHY
%-------------------------------------------------------------------------------

\ifthenelse{\boolean{final}}{
    \hypersetup{urlcolor=ieeeblue}
    \bibliographychapter
    % Pre-generated bibliography (IEEEtran style), no BibTeX run needed
    \input{thesis_bibliography.tex}

}{}

\end{document}

%% file: preamble/titlepage/titlepage_EN_logo.tex
\thispagestyle{empty}
{\fontsize{13}{15}\selectfont

    \begin{textblock*}{\textwidth}(2.98cm,1.5cm)
        \begin{center}
            \vskip0.5cm
            \includegraphics[scale=0.75]{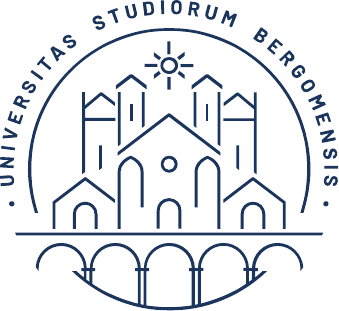}
        \end{center}
    \end{textblock*}

    \begin{textblock*}{\textwidth}(2.98cm,7cm)
        \begin{center}
            {
                \fontsize{16}{19}\selectfont
                \textsc{Università degli Studi di Bergamo}
            }
        \end{center}
        \begin{center}
            % Department of Engineering and Applied Sciences
            School of Doctoral Studies
        \end{center}
        \vspace{-0.65cm}
        \begin{center}
            Ph.D. Program in Engineering and Applied Sciences
        \end{center}
        \vspace{-0.65cm}
        \begin{center}
            XXXVIII Cycle - SSD: IINF-01/A - Electronics
        \end{center}
        \vspace{1cm}
        \begin{center}
            {
                \fontsize{20}{22}\selectfont
                \bfseries
                Design, Testing and Calibration of the\\
                \vspace{0.16em}
                GAPS Experiment Si(Li) Tracker Readout ASIC:\\
                \vspace{0.16em}
                From the First Flight Campaign Toward\\
                \vspace{0.35em}
                the Second Mission Upgrade
            }
        \end{center}
        \vspace{9.5cm}
        \begin{center}
            Academic Year 2024 / 2025
        \end{center}
    \end{textblock*}

    \begin{textblock*}{\textwidth}(3.05cm,18.4cm)
        \begin{longtable}{r l}
            \renewcommand{\arraystretch}{1.5}
            Advisor:           & Prof. Massimo Manghisoni \\
            Co-Advisor:        & Elisa Riceputi, Ph.D.    \\\\
            Ph.D. Coordinator: & Prof. Alessandra Marini
        \end{longtable}
    \end{textblock*}

    \begin{textblock*}{\textwidth}(13.1cm,22.4cm)
        Ph.D. Thesis
    \end{textblock*}
    \begin{textblock*}{\textwidth}(13.1cm,23.1cm)
        Luca GHISLOTTI
    \end{textblock*}
    \begin{textblock*}{\textwidth}(13.1cm,23.8cm)
        Student ID: 1052975
    \end{textblock*}
}

%% file: preamble/abstract.tex
\chapter*{Abstract}

\par
The General Antiparticle Spectrometer is a \hbox{balloon-borne} experiment designed to perform three \hbox{long-duration} flights from Antarctica to search for \hbox{low-energy} cosmic antinuclei, in particular antideuterons, as an indirect evidence of dark matter.
Its detection method is based on the formation and decay of exotic atoms, identified through the emission of characteristic \hbox{x-rays} and the subsequent annihilation into secondary particles.
At the core of the instrument is the silicon tracker, composed of \hbox{lithium-drifted} silicon detectors operated at \SI{-40}{\celsius} and read out by a custom \hbox{application-specific} integrated circuit, SLIDER32, implemented in \SI{180}{\nano\meter} CMOS technology.
This thesis reports on the integration, calibration, and validation of the SLIDER32-\hbox{based} tracker readout electronics.
Laboratory characterization of a single tracker module with an \hbox{Americium-241} source and cosmic muons was complemented by calibration and testing of the full tracker in its flight configuration, leading to the first \hbox{x-ray} spectrum acquisition from a \hbox{Cadmium-109} source.
These activities formed part of the integration and testing phase of the experiment for the first flight, which culminated in the complete instrument assembly and validation during the Antarctic campaign of 2024.
Building on this experience, the work introduces the design and characterization of a new prototype integrated circuit conceived as an upgrade to SLIDER32: the ANTARES4 chip, fabricated in \SI{65}{\nano\meter} CMOS technology.
The device integrates eight analog channels up to the shaping stage and implements innovative solutions for \hbox{dynamic-signal} compression exploiting \hbox{dynamic-threshold} MOS transistors, together with an improved detector leakage current compensation circuit at the preamplifier input.
Its laboratory characterization provides a first validation of these design choices, laying the foundations for the \hbox{next-generation} tracker readout system to be adopted in the second flight of the experiment.

\vspace{1.5em}
\noindent\textbf{Keywords:} ASIC, Silicon Detectors, Readout Electronics, CMOS Technology, Dark Matter

%% file: maintext/introduction.tex
\chapter*{Introduction}

\par
The nature of dark matter remains one of the most compelling open questions in contemporary physics.
While gravitational evidence for its existence is well established across galactic and cosmological scales, its underlying nature is still unknown.
Among indirect detection strategies, the observation of low-energy cosmic ray antinuclei, such as antideuterons, offers a particularly clean probe owing to the extremely low expected background from conventional astrophysical processes in this energy range.
Although such events are exceedingly rare, their detection could provide critical clues to the nature of dark matter.

\par
The General AntiParticle Spectrometer (GAPS) experiment is designed to explore this detection channel through long-duration balloon flights in the stratosphere.
It is optimized for the identification of low-energy cosmic antideuterons through a novel detection technique based on the formation and decay of exotic atoms.
As incoming antideuterons slow down and stop in the detector material, they are captured by silicon atoms, forming excited exotic atomic states.
These atoms de-excite by emitting characteristic \hbox{x-rays}, followed by annihilation into secondary particles such as pions and protons.
This distinctive sequence provides a powerful means of discriminating antideuterons from the astrophysical background in a previously unexplored energy range.

\par
At the core of the GAPS detection system is the silicon tracker, composed of over one thousand lithium-drifted silicon detectors arranged in a modular structure.
The detectors, segmented into eight strips and operated at \SI{-40}{\celsius}, enable simultaneous \hbox{x-ray} spectroscopy and charged-particle tracking.  
To read out charge deposits with sufficient resolution across the required input dynamic range, each module is instrumented with a dedicated front-end board hosting a custom application-specific integrated circuit (ASIC).
The SLIDER32 ASIC, developed in \SI{180}{\nano\meter} CMOS technology, is the current flight \hbox{front-end} for the first mission.
It integrates 32 analog channels, each consisting of a \hbox{low-noise} charge-sensitive amplifier (CSA) with dynamic signal compression, a shaping stage, a self-trigger circuit, and a shared differential 11-bit analog-to-digital converter (ADC), followed by a digital back-end for configuration and data readout.

\par
This thesis presents the work carried out over the past three years on the development, integration, and validation of the GAPS tracker readout electronics.
The first part focuses on the laboratory characterization of a single tracker module using a dedicated setup, with the aim of verifying that the readout chain based on the SLIDER32 ASIC met all design specifications.
This included spectral measurements with an Americium-241 \hbox{x-ray} source and cosmic muons.
The validation activity was then extended to the full tracker system in its flight configuration, involving the development and deployment of custom calibration and test procedures implemented within the flight software.
This effort enabled the first \hbox{x-ray} spectrum acquisition from a Cadmium-109 source using the fully assembled tracker, and continued through the Antarctic integration campaign of December 2024, where the complete instrument was assembled and validated ahead of its first flight scheduled for December 2025.

\par
The extensive testing and calibration activities carried out during the integration phase of the tracker provided a detailed understanding of the performance of the \hbox{SLIDER32-based} readout system under flight-like conditions, highlighting both its strengths and limitations.
Building upon this experience, the second part of this thesis is devoted to the design of a prototype ASIC upgrade in \SI{65}{\nano\meter} CMOS technology, intended to serve as the basis for the tracker readout system for the second GAPS flight, planned from 2026 onward.
The \antareslogo (ANTiparticle Asic REadout for Si(Li) detectors with four CSA feedback configurations) chip is an R\&D prototype integrating eight analog channels up to the shaping stage.
It explores new circuit solutions for dynamic signal compression, introducing a novel approach based on the dynamic-threshold MOSFET (DTMOS) topology.
The ASIC also features an improved implementation of the leakage current compensation circuit at the preamplifier input, designed to increase the leakage current compensation capabilities of the previous version.
The characterization of the prototype ASIC was carried out using a dedicated laboratory test setup, aimed at evaluating the performance of the analog front-end up to the shaper output.
These initial measurements represent a first step toward the validation of the proposed circuit implementation, guiding the development of the next-generation readout channel for the GAPS tracker.

\par
The thesis is structured as follows.
\Cref{chapter1} provides an overview of the GAPS experiment and the role of the Si(Li) tracker within its detection strategy.
\Cref{chapter2} presents the laboratory characterization of a single tracker module, including measurements with \hbox{x-rays} from an Americium-241 source and with cosmic muons.
\Cref{chapter3} describes the integration and validation of the full tracker system in flight configuration, detailing the procedures developed for test and calibration, and reporting the first measurements obtained with a Cadmium-109 x-ray source and with cosmic muons using the fully assembled instrument.
\Cref{chapter4} introduces the design of the new prototype ASIC developed in \SI{65}{\nano\meter} CMOS technology, discussing the key innovations for dynamic signal compression and leakage current compensation.
Finally, \Cref{chapter5} presents the test methodology and the results of the characterization of the prototype chip.

\par
The appendix collects supplementary material relevant to the topics discussed in the main chapters.  
Appendix~\ref{appendix_gaps_collaboration} documents the author's direct involvement and participation in the experiment, detailing the contribution to the GAPS integration and first Antarctic campaign, including the instrument integration and testing (I\&T) phases and the activities carried out during the Antarctic deployment.

% % --- PDF version ---
% \ifthenelse{\boolean{final}}{
% \par
% The appendix collects supplementary material relevant to the topics discussed in the main chapters.  
% Appendix~\ref{appendix_gaps_collaboration} documents the author's direct involvement and participation in the experiment, detailing the contribution to the GAPS integration and first Antarctic campaign, including the instrument integration and testing (I\&T) phases and the activities carried out during the Antarctic deployment.
% }{}

% % --- Print version ---
% \ifthenelse{\boolean{print}}{
% \par
% The appendices collect supplementary material relevant to the topics discussed in the main chapters.  
% Appendix~\ref{appendix_gaps_collaboration} documents the author's direct involvement and participation in the experiment, detailing the contribution to the GAPS integration and first Antarctic campaign, including the instrument integration and testing (I\&T) phases and the activities carried out during the Antarctic deployment.
% Appendix~\ref{appendix_sili_model} presents the Si(Li) strip detector simulation model and its equivalent model for charge calculation.
% Appendix~\ref{appendix_schematics} describes the \antareslogo ASIC test board and setup, with schematics, BOMs, and PCB layouts.
% }{}

%% file: maintext/chapters/chapter1.tex
\chapter{The GAPS experiment}\label{chapter1}

%-------------------------------------------------------------------------------
%   The dark matter unsolved mystery
%-------------------------------------------------------------------------------
\section{The unsolved dark matter mystery}

\par
According to the most recent results of the Planck experiment~\cite{collaborationPlanck2013Results2014}, our universe is composed of \SI{68}{\percent} dark energy, \SI{27}{\percent} dark matter, and \SI{5}{\percent} baryonic matter.
The latter refers to the form of matter primarily composed of baryons---subatomic particles such as protons and neutrons, each consisting of three quarks.
This category includes all ``ordinary'' matter, such as stars, planets, interstellar gas and all living organisms.
It is the component of the universe that interacts through the electromagnetic force, making it detectable via electromagnetic radiation.
Unlike baryonic matter, the origins of dark energy and dark matter remain unknown, representing one of the major unresolved problems in modern cosmology.

%   Dark matter candidates and detection techniques
\subsection{Dark matter candidates and detection techniques}

\sloppy
\par
The existence of dark matter was originally postulated by Fritz Zwicky in 1933 through his analysis of galaxy velocity dispersions in the Coma cluster, which revealed gravitational effects that could not be explained by visible matter alone~\cite{zwickyMassesNebulaeClusters1937}.
Subsequent observations have reinforced the need for a non-luminous, non-baryonic form of matter that interacts primarily through gravity.
WIMPs (weakly interacting massive particles) are among the most widely studied candidates for dark matter~\cite{roszkowskiWIMPDarkMatter2018}.
These hypothetical particles are electrically neutral, possess a mass ranging from a few GeV to several \si{\tera\electronvolt}, and interact via the weak nuclear force and gravity but not electromagnetically---rendering them invisible to conventional detectors.
Their theoretical grounding in extensions of the Standard Model have made WIMPs a central focus of dark matter research.

\par
Dark matter detection strategies are generally categorized into three types: particle colliders, direct searches, and indirect searches~\cite{lowellIndirectDarkMatter2018}.
Direct and indirect searches aim to study relic WIMPs already present in the universe~\cite{boezioPAMELAIndirectDark2009}, while collider experiments attempt to create WIMPs.
Although these detection methods differ operationally, they rely on complementary background models, contributing collectively to our understanding of dark matter.

\par
Indirect searches for dark matter aim to detect the byproducts of dark matter annihilation or decay, such as electrons~\cite{pinettiDarkMatterDecay2025}, positrons~\cite{adrianiObservationAnomalousPositron2009}, gamma-rays~\cite{hooperDarkMatterAnnihilation2011}, antiprotons~\cite{cuiPossibleDarkMatter2017}, and antideuterons~\cite{fornengoDarkMatterSearches2013,fukePresentStatusFuture2017}.
These particles can be observed in cosmic rays, which are \hbox{high-energy} particles originating from space that are composed mainly of protons and heavier nuclei, with a smaller fraction of electrons and antiparticles.
Cosmic rays thus serve as an indirect probe for dark matter, as rare components like antideuterons may result from exotic processes involving dark matter rather than from standard astrophysical sources.

\par
Satellite and balloon-based instruments allow for accurate measurement of cosmic rays near the top of the atmosphere, making it possible to identify these rare signatures from the more abundant background.
As discussed in~\cite{boezioCosmicRayDetection2020}, the study of cosmic rays also provides insight into their acceleration and propagation through the galaxy while offering sensitivity to physics beyond the Standard Model.

%   The role of antideuterons in indirect detection
\subsection{The role of antideuterons in indirect detection}

\par
Among the byproducts of dark matter annihilation or decay, antideuterons have emerged as particularly promising due to the extremely low background expected from conventional astrophysical processes~\cite{donatoAntimatterSupersymmetricDark2000}.
This makes them a potentially clean signature for new physics, including WIMP annihilation or decay.
Theoretical models predict that antideuterons produced through dark matter interactions will exhibit distinctive energy spectra, peaking at low kinetic energies below \SI{1}{\giga\electronvolt/$n$}.
In contrast, secondary antideuterons produced by cosmic-ray collisions with interstellar matter exhibit a steeply falling spectrum at low energies, providing a relatively ``background-free'' window for discovery~\cite{donatoAntideuteronsSignatureSupersymmetric2000}.

% --- PDF version ---
\ifthenelse{\boolean{final}}{
\par
In this context, experiments like the Alpha Magnetic Spectrometer (AMS-02)~\cite{battistonAntimatterSpectrometerAMS022008,IbarraWild2013} on the International Space Station (ISS), the Balloon-borne Experiment with a \hbox{Super-solenoidal} Spectrometer (BESS)~\cite{myersCosmicRay1H2005}, and the General AntiParticle Spectrometer (GAPS)~\cite{haileyIndirectSearchDark2009,ongGAPSExperimentSearch2018,vondoetinchemGAPSDarkMatter2016,vannucciniGAPSLowenergyAntimatter2018,osteriaGAPSBalloonborneCosmicray2020} are at the forefront of indirect dark matter searches.
GAPS, in particular, is optimized for low-energy antideuterons, offering high sensitivity in the low-background energy range~\cite{rogersSensitivityGAPSExperiment2023}.
The detection of even a few low-energy antideuterons could offer strong evidence for dark matter, making this channel an important component of broader indirect detection efforts~\cite{baerLowEnergyAntideuterons2005}.
}{}

% --- Print version ---
\ifthenelse{\boolean{print}}{
In this context, experiments like the Alpha Magnetic Spectrometer (AMS-02)~\cite{battistonAntimatterSpectrometerAMS022008,IbarraWild2013} on the International Space Station (ISS)~\cite{delucasInternationalSpaceStation1996}, the Balloon-borne Experiment with a Super-solenoidal Spectrometer (BESS)~\cite{myersCosmicRay1H2005}, and the General AntiParticle Spectrometer (GAPS)~\cite{haileyIndirectSearchDark2009,ongGAPSExperimentSearch2018,vondoetinchemGAPSDarkMatter2016,vannucciniGAPSLowenergyAntimatter2018,osteriaGAPSBalloonborneCosmicray2020} are at the forefront of indirect dark matter searches.
GAPS, in particular, is optimized for low-energy antideuterons, offering high sensitivity in the low-background energy range~\cite{rogersSensitivityGAPSExperiment2023}.
The detection of even a few low-energy antideuterons could offer strong evidence for dark matter, making this channel an important component of broader indirect detection efforts~\cite{baerLowEnergyAntideuterons2005}.
}{}

\par
In addition to its sensitivity to antideuterons, GAPS will offer unprecedented access to the low-energy regime of cosmic antiprotons, further enhancing its capability to probe dark matter interactions and other exotic phenomena~\cite{saffoldCosmicAntihelium3Nuclei2021a}.
Its novel particle identification approach based on exotic atom formation not only allows for clean separation from abundant background species but also enables high-statistics measurement of antiprotons in a kinetic energy range below \SI{0.25}{\giga\electronvolt/$n$}, never reached by previous experiments.
These complementary measurements will not only strengthen the discovery potential of the GAPS mission but also provide crucial validation of the detection technique for rare-event searches involving heavier antinuclei, such as antihelium.
As such, GAPS represents a key milestone in the exploration of antimatter in cosmic rays and the search for physics beyond the Standard Model.

\begin{figure}[t!]
    \centering
    \includegraphics[width=0.65\textwidth]{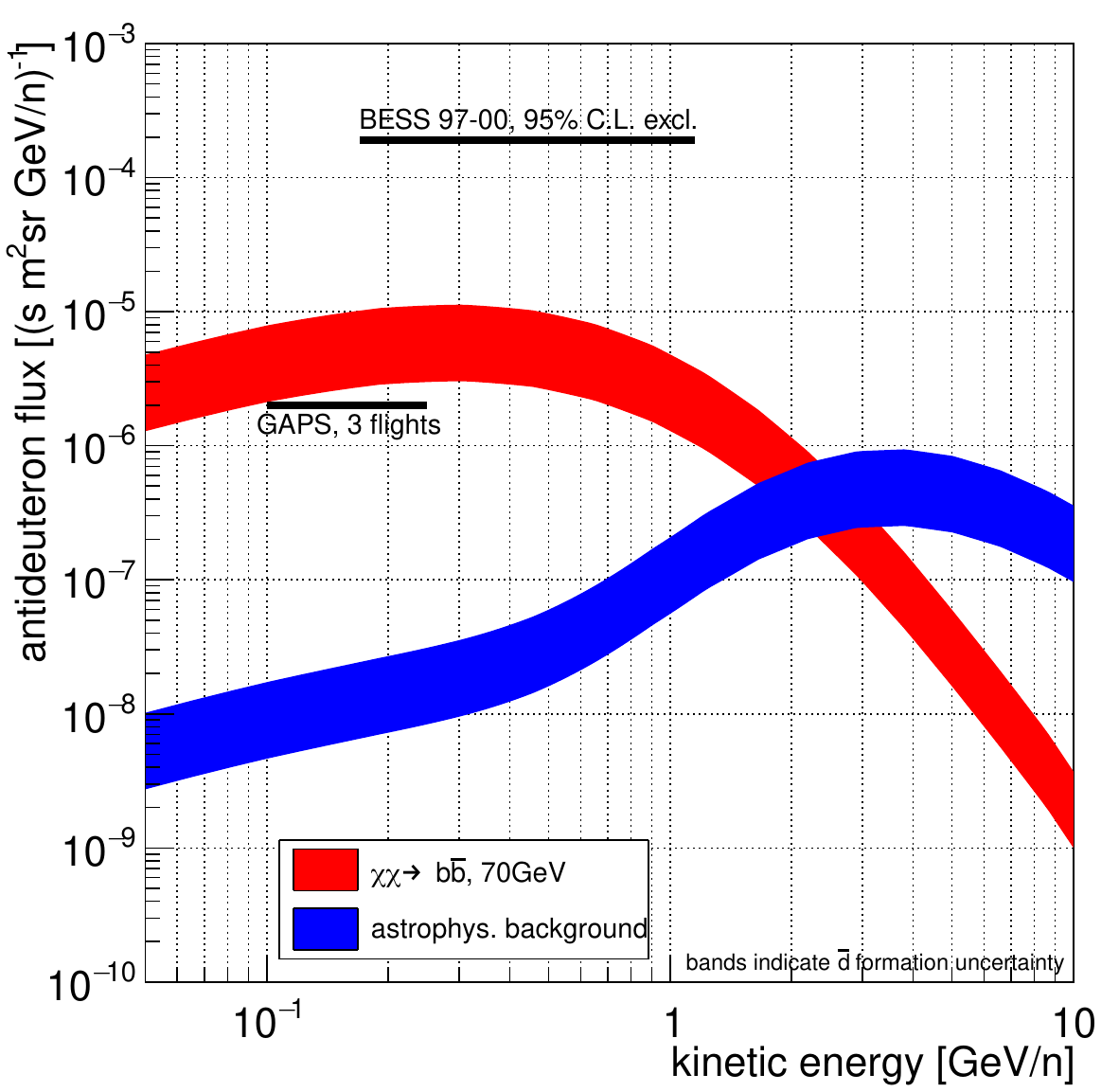}
    \caption{Projected sensitivity of GAPS to the antideuteron flux after three 35-day flights compared with the BESS antideuteron exclusion limit. Also shown are the predicted antideuteron fluxes from \SI{70}{\giga\electronvolt} WIMP annihilations into \( b\bar{b} \) final states and the expected astrophysical background. Data are taken from~\cite{aramakiAntideuteronSensitivityGAPS2016a,korsmeierProspectsVerifyPossible2018,fukeSearchCosmicRayAntideuterons2005}. Figure courtesy of the GAPS Collaboration.}
    \label{fig:ch1-antideuteron_sensitivity}
\end{figure}

\par
\Cref{fig:ch1-antideuteron_sensitivity} shows the projected sensitivity of the GAPS experiment to the cosmic antideuteron flux, assuming three long-duration balloon flights of 35 days each.
The GAPS sensitivity curve is displayed alongside the \SI{95}{\percent} confidence level exclusion limit established by BESS (1997--2000), highlighting the substantial improvement in sensitivity at low kinetic energies. 
Also shown are theoretical predictions for the antideuteron flux resulting from WIMP annihilations into \( b\bar{b} \) final states, assuming a dark matter particle mass of \SI{70}{\giga\electronvolt}.
The expected astrophysical background is also included, with the widths of the signal and background bands reflecting uncertainties in the antideuteron formation process.
The plot highlights the low-background regime accessible to GAPS, where even a few detected events could provide a potentially unambiguous signature of dark matter.

\par
This sensitivity projection illustrates the unique capability of GAPS to probe a region of parameter space that is largely inaccessible to existing instruments such as AMS-02 and BESS, particularly at kinetic energies below \SI{0.25}{\giga\electronvolt/$n$}. 
In this low-energy regime, the background from secondary astrophysical processes is expected to be minimal, such that even a single antideuteron detection would represent a major discovery, offering compelling evidence for new physics---such as dark matter annihilation or decay. 
By combining a large geometric acceptance with a novel detection technique based on exotic atom formation and decay, GAPS is positioned to set new constraints on dark matter models and open a previously unexplored observational window on low-energy cosmic antinuclei.

\begin{figure}[tbp]
    \centering
    \includegraphics[width=.85\textwidth]{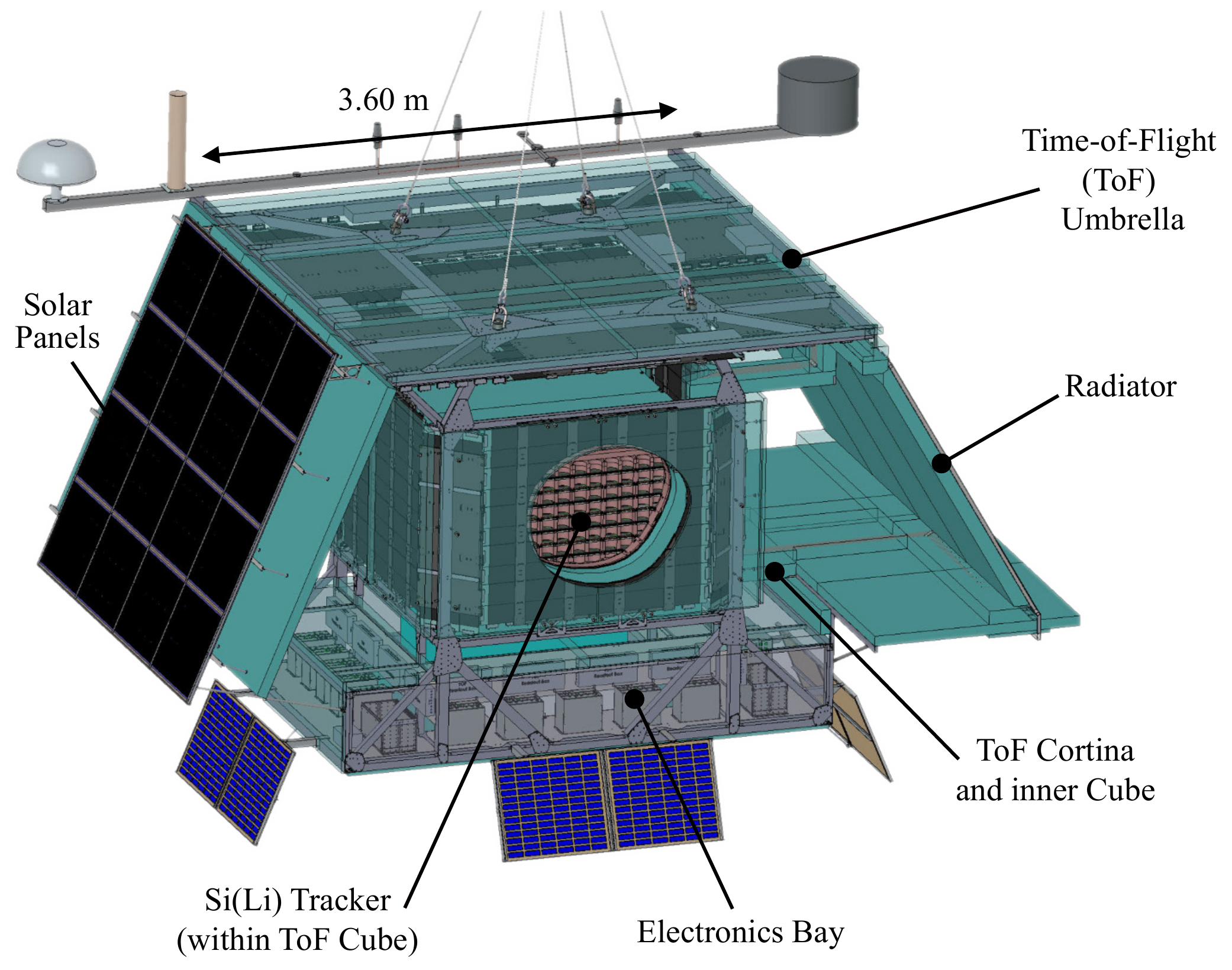}
    \caption{Schematic overview of the GAPS instrument, showing the three components of the time-of-flight (ToF) system as well as the Si(Li) tracker. Figure courtesy of the GAPS Collaboration.}
    \label{fig:ch1-gaps_instrument}
\end{figure}

%-------------------------------------------------------------------------------
%   The GAPS instrument
%-------------------------------------------------------------------------------
\section{The GAPS instrument}

%   Design of the detection apparatus
\subsection{Design of the detection apparatus}

\par
The GAPS science payload has a mass of approximately \SI{2300}{\kilogram} and is carried by a \hbox{helium-filled} balloon suitable for long-duration stratospheric flights.
The mission is scheduled to launch from the NASA Long Duration Balloon (LDB) Facility located at the National Science Foundation (NSF) McMurdo station in Antarctica.
Three flights of ${\sim}35$ days each are planned, for a total flight time of over 100 days at an altitude of approximately \SI{37}{\kilo\meter}, beginning with the first flight in the 2025 austral summer.

\par
Antarctica offers uniquely favorable conditions for long-duration balloon missions.
During the austral summer, a stable stratospheric wind pattern known as the polar vortex forms around the continent, allowing balloons to follow circular trajectories for several weeks.
The continuous daylight ensures uninterrupted solar power for the payload electronics and minimizes thermal fluctuations, supporting stable flight conditions.
Furthermore, the remote location reduces risk during termination and recovery, as the Antarctic airspace is largely unrestricted and free from commercial traffic---making it an ideal site for high-altitude, near-space scientific missions.

\sloppy
\par
\Cref{fig:ch1-gaps_instrument} illustrates the GAPS science payload.
It consists of a time-of-flight (ToF) system that provides both the particle velocity measurements and triggering capabilities.
The ToF surrounds the internal Si(Li) tracker which functions as the target and x-ray spectrometer.
Comprising 160 scintillator paddles ranging in length from \SIrange{1.5}{1.8}{\meter}, the ToF provides a timing resolution of $\SI{<400}{\pico\second}$, enabled by silicon-photomultiplier (SiPM) sensors and custom readout electronics~\cite{feldmanGAPSTimeofFlightDetector2024}.
This subsystem includes an inner Cube that encloses the tracker volume, an Umbrella panel located \SI{90}{\centi\meter} above the Cube, and a Cortina that surrounds the vertical sides of the Cube with a spacing of \SI{30}{\centi\meter}.

\begin{figure}[tbp]
    \centering
    \includegraphics[width=0.93\textwidth]{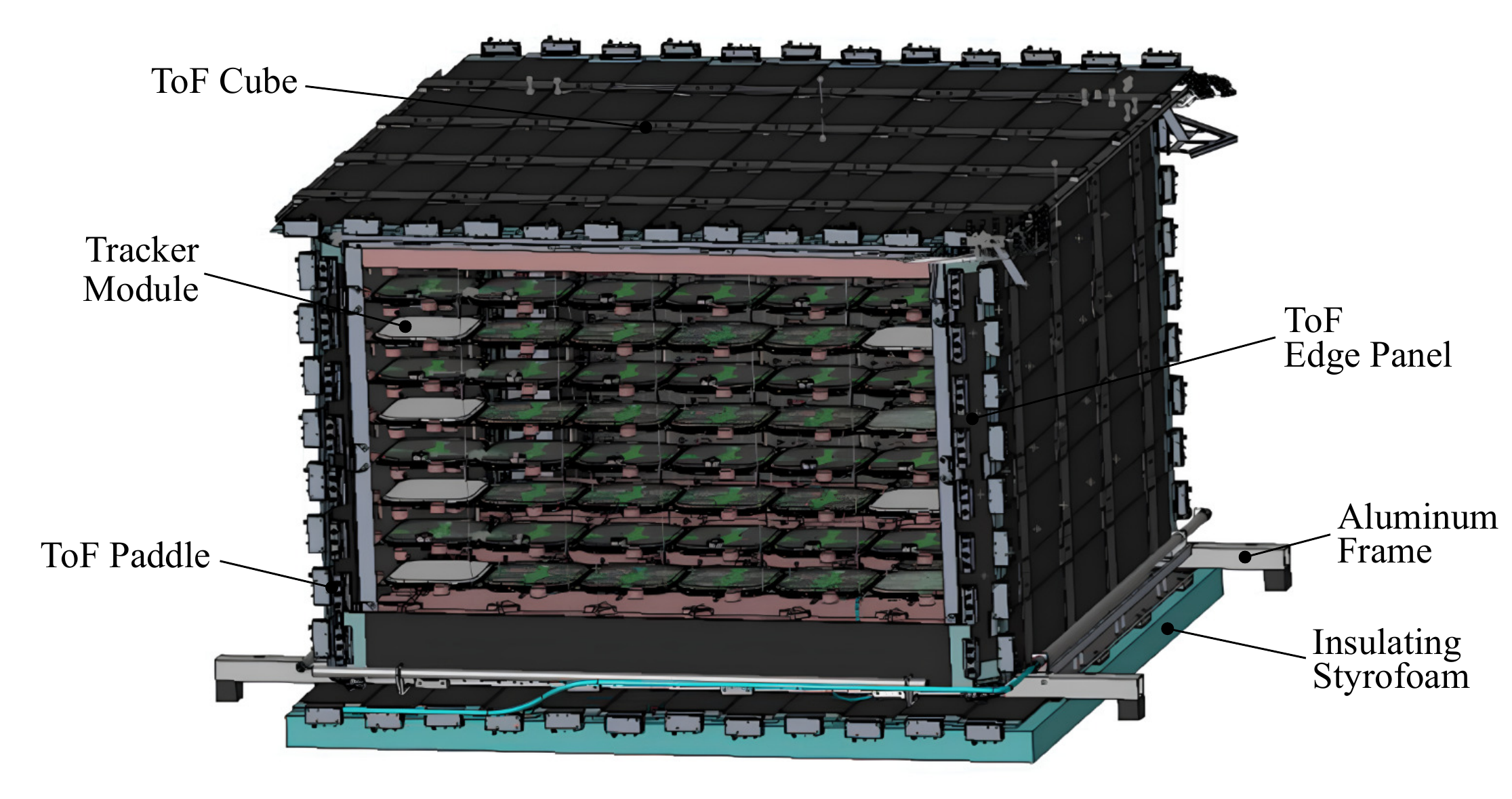}
    \caption{Detailed view of the GAPS Si(Li) Tracker assembly. The tracker is enclosed within the ToF Cube and comprises multiple layers of tracker modules, each containing four lithium-drifted silicon detectors. Figure courtesy of the GAPS Collaboration.}
    \label{fig:ch1-gaps_tracker}
\end{figure}

\par
The Si(Li) Tracker, centrally positioned within the instrument, consists of more than 1000 lithium-drifted silicon detectors, each having a diameter of \SI{10}{\centi\meter}, a thickness of \SI{2.5}{\milli\meter}, and segmented into eight strips.
The \SI{2.5}{\meter^3} tracker volume follows a modular design, with each module housing four detectors read out by a custom mixed-signal application-specific integrated circuit (ASIC) with 32 readout channels~\cite{riceputiExperimentalResultsCharacterization2023}.
For the first flight, the modules are arranged in seven layers with six rows of six modules each.  
Three additional layers are included on the bottom to provide further target mass and thermal balance.  
A detailed view of the tracker assembly is shown in \Cref{fig:ch1-gaps_tracker}.

\par
The tracker is cooled during flight to \SI{-40}{\celsius} using an innovative oscillating heat pipe (OHP) passive cooling system~\cite{okazakiDevelopmentMeterscaleOshaped2014, fukeDesignApplicationMultiloop2023} coupled to a radiator.
During on-ground calibrations, the ground cooling system (GCS) provides thermal coupling to a metal plate cooled by circulating methanol at \SI{-80}{\celsius}, while in flight the radiator points away from the Sun.

\begin{figure}[tbp]
    \centering
    \includegraphics[width=0.8\textwidth]{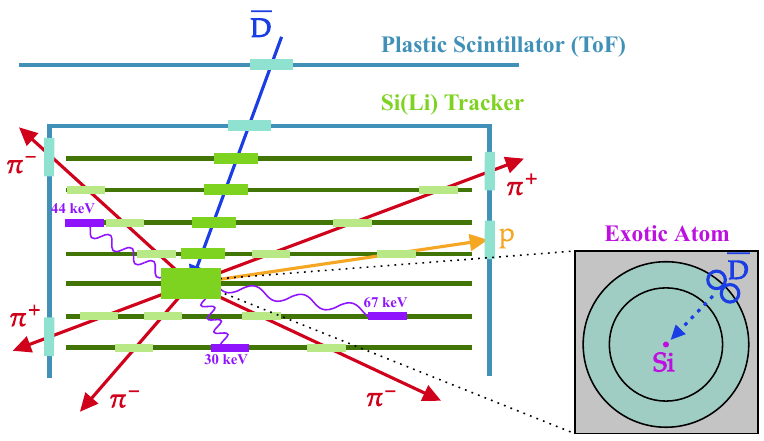}
    \caption{GAPS detection method: an antiparticle slows down and stops in the Si(Li) target, forming an exotic atom. The atomic x-rays are emitted as it de-excites, followed by charged particle emission from the nuclear annihilation. Figure courtesy of the GAPS Collaboration.}
    \label{fig:ch1-gaps_detection}
\end{figure}

%   The GAPS particle identification concept
\subsection{The GAPS particle identification concept}\label{sec:gaps_particle_identification}

\par
The GAPS experiment uses a novel particle identification method based on the formation, de-excitation, and annihilation of exotic atoms, schematically represented in \Cref{fig:ch1-gaps_detection}.
It involves capturing an antiparticle into a target material with subsequent formation of exotic atoms.
An exotic atom is formed when a negatively charged particle---other than an electron---is captured by an ordinary atom and replaces one of its electrons. 
This occurs when an incoming antiparticle, such as an antideuteron or antiproton, slows down and is captured by a target atom, forming an exotic atom in which the antiparticle orbits the nucleus.

\par
During operation, the ToF system measures the velocity and direction of the incoming antiparticle, providing the experiment's trigger.
The antiparticle loses energy as it traverses the detector and eventually stops in a strip of a Si(Li) detector, which serves as the target material. There, it forms an exotic atom in an excited state. 
This excited state is highly unstable; the exotic atom rapidly de-excites through the emission of Auger electrons, atomic x-rays, pions ($\pi^+$ and $\pi^-$), and other secondary particles (e.g., protons), all of which deposit energy in both the tracker and the ToF system~\cite{aramakiMeasurementAtomicXray2013}.
Since the atomic number of the target material is known, the Bohr formula for atomic x-ray energies can be used to uniquely determine the mass of the captured particle~\cite{moriNovelAntimatterDetector2002}:

\begin{equation}
    E_n = -\frac{Z^2 R_y}{n^2},
    \label{eq:bohr_basic}
\end{equation}

\noindent
where \( E_n \) is the energy of the level with principal quantum number \( n \), \( Z \) is the atomic number of the nucleus, and \( R_y \) is the Rydberg constant, which has a value of approximately \SI{13.6}{\electronvolt} for the hydrogen atom.
This formula describes the discrete energy levels of a negatively charged particle, such as an electron, bound to a nucleus by the Coulomb interaction.

\par
In the case of exotic atoms---where a negatively charged particle heavier than the electron, such as an antiproton or an antideuteron, is captured by an atomic nucleus---the standard Bohr formula must be modified to account for the reduced mass of the system~\cite{philConstitutionAtomsMolecules1913}.
The energy levels of such atoms are given by:

\begin{equation}
    E_n = -\frac{\mu Z^2 e^4}{2 (4 \pi \varepsilon_0)^2 \hbar^2 n^2},
    \label{eq:bohr_exotic}
\end{equation}

\noindent
where \( E_n \) is the binding energy of the level with principal quantum number \( n \), \( Z \) is the atomic number of the nucleus, \( e \) is the elementary charge, \( \varepsilon_0 \) is the vacuum permittivity, and \( \hbar \) is the reduced Planck constant.
The quantity \( \mu \) represents the reduced mass of the two-body system and is defined as follows:

\begin{equation}
\mu = \frac{m_1 m_2}{m_1 + m_2},
\end{equation}

\begin{figure}[tbp]
    \centering
    \includegraphics[width=0.8\textwidth]{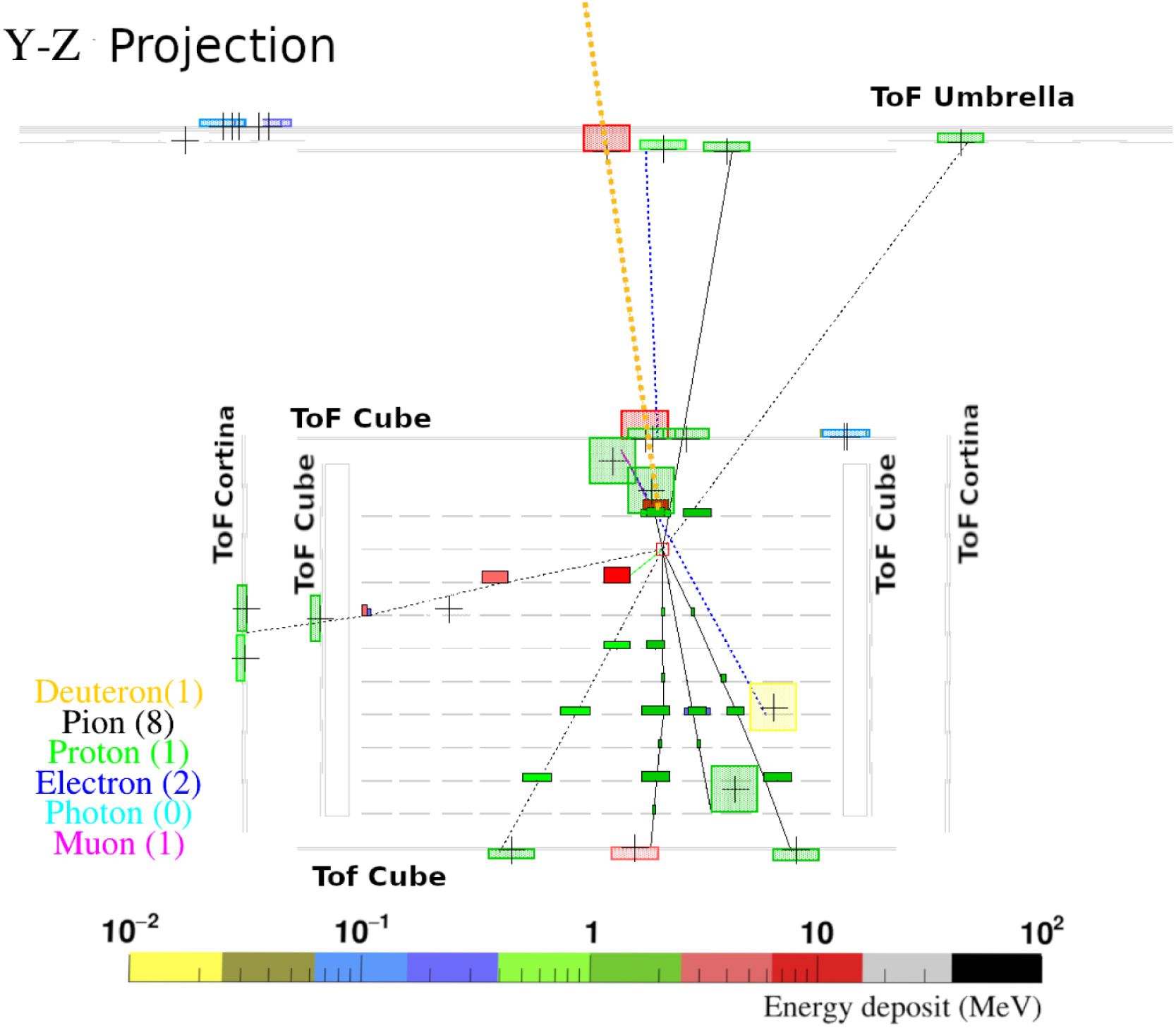}
    \caption{Simulated Y--Z projection of an antideuteron event with $v = 0.28\,c$, illustrating the full detector response including particle trajectories, energy deposits, and the reconstructed annihilation vertex~\cite{muniniAntinucleusAnnihilationReconstruction2021}. Figure courtesy of the GAPS Collaboration.}
    \label{fig:ch1-gaps_reconstructed_event}
\end{figure}

\noindent
where \( m_1 \) and \( m_2 \) are the masses of the captured antiparticle and the nucleus, respectively.
Since \( \mu \gg m_e \) for antiprotons and antideuterons, the binding energies in exotic atoms are significantly greater than those in ordinary atoms.
This leads to the emission of x-rays with characteristic energies during de-excitation, which depend on both the captured particle and the atomic number of the target.
These x-rays provide a unique and identifiable signature, enabling the instrument to discriminate between different antiparticle species.
In addition, the stopping vertex of the antiparticle can be reconstructed from the tracks of the resulting annihilation products, enabling a complete reconstruction of the event topology.

\par
\Cref{fig:ch1-gaps_reconstructed_event} shows a simulated event corresponding to an antideuteron with a velocity of $v = 0.28\,c$, displayed in the Y--Z projection of the detector~\cite{muniniIdentificationCosmicrayLight2024}.
The diagram presents a side view of the entire apparatus, where solid lines represent the Monte Carlo trajectories of particles and dotted lines represent antiparticles.
Each trajectory is color-coded according to particle species, as indicated in the legend. Energy deposits within active detector volumes are visualized using colored boxes, with the color scale representing the deposited energy in MeV.
The widths of the tracker hit boxes correspond to the strip dimensions, while the heights are proportional to the deposited energy.
ToF hit boxes are shown \hbox{edge-on} in this view, with heights also scaled by energy deposition.
Crosses inside the ToF boxes represent the one-sigma spatial resolution and are centered on the reconstructed hit positions.
The open red square marks the Monte Carlo location of the annihilation vertex.
This visualization highlights the characteristic topology of an antideuteron annihilation event, including secondary interactions and energy deposits across various subdetectors~\cite{tiberioReconstructionAntinucleusannihilationEvents2021}.

\par
The ability to reconstruct such event topologies is enabled by the combined use of the ToF system and the silicon tracker, which together provide sub-nanosecond timing, spatial resolution, and energy deposition information over a large sensitive volume.
This allows for the measurement of the incoming antiparticle velocity, its stopping point within the tracker, and the full annihilation star structure resulting from the exotic atom decay~\cite{muniniAntinucleusAnnihilationReconstruction2021}.
The emission of de-excitation x-rays, charged pions and protons produces a rich set of observables that can be combined to identify the antinucleus species with high confidence.

\par
The detection concept was validated through accelerator testing of a GAPS prototype at the KEK facility (High Energy Accelerator Research Organization)~\cite{iwashitaKEKDigitalAccelerator2011} in Tsukuba, Japan, in April 2004, which demonstrated the simultaneous observation of characteristic x-ray transitions and pion annihilation stars, confirming the feasibility of the method~\cite{Hailey_2006}.
Building on this validation, the novel particle identification approach was subsequently demonstrated in flight with the ``pGAPS'' prototype, launched from the Japanese Aerospace Exploration Agency (JAXA) Taiki Aerospace Research Field (TARF) balloon facility in Hokkaido, Japan, in June 2012~\cite{fukePGAPSExperimentEngineering2013,vondoetinchemFlightGAPSPrototype2014,mognetPrototypeGAPSPGAPS2014}.

%   Instrument integration and testing
\subsection{Instrument integration and testing}

\par
Since 2022, the GAPS instrument has been undergoing integration and testing in preparation for its first long-duration balloon flight, scheduled for December 2025~\cite{muniniIntegrationCalibrationGAPS2024a}.
The instrument has been built from the ground up on a custom-designed aluminum frame that houses all the components of the science payload, including the tracker, the ToF system, the cooling system, and the readout electronics.
GAPS integration started at Bates Massachusetts Institute of Technology (MIT) laboratories with the assembly of the mechanical structure.
Tracker integration also started at MIT, with a total of five Si(Li) tracker planes integrated, requiring the assembly and testing of 180 modules and 720 detectors.

\begin{figure}[tbp]
    \centering
    \includegraphics[width=0.8\textwidth]{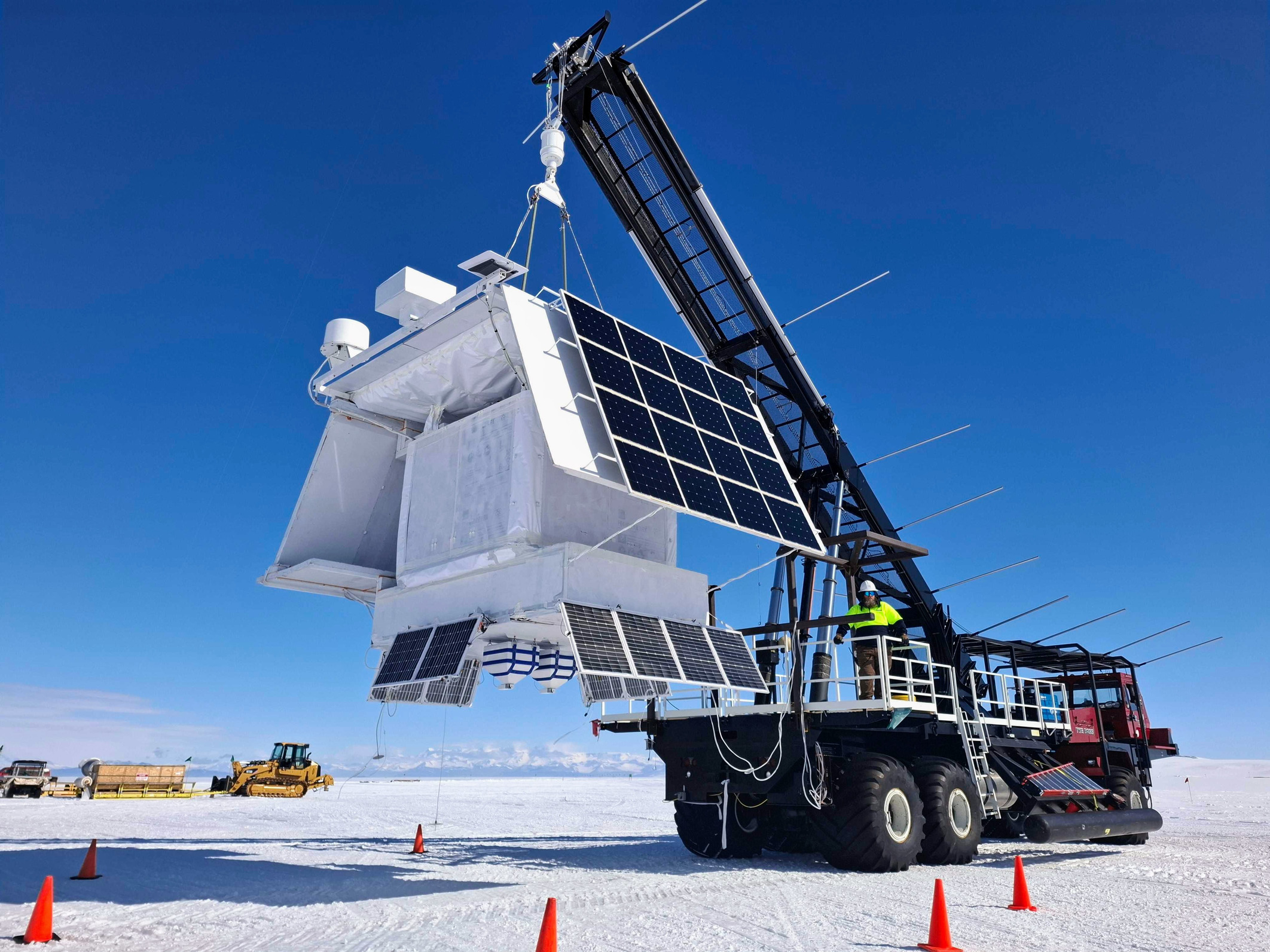}
    \caption{The GAPS instrument in its flight-ready configuration, suspended from the launch vehicle at the NASA Long Duration Balloon (LDB) facility at McMurdo Station, Antarctica. Picture courtesy of Philip von Doetinchem.}
    \label{fig:ch1-gaps_antarctica}
\end{figure}

\par
In September 2022, the instrument was moved to the Space Sciences Laboratory (SSL) at the University of California, Berkeley, where the tracker integration was completed.
During this phase, the tracker was calibrated using cosmic muons, radioactive sources, and exploiting the self-calibration modes of the Si(Li) detector readout electronics~\cite{riceputi32AnalogChannels2023}.
The cooling system was also integrated and tested to ensure the required thermal performance for tracker operation during on-ground calibration.
Throughout the integration process, tracker temperatures were monitored, and the cooling system was validated to maintain the tracker at its designated operating temperature~\cite{fukeThermalControlSystem2022}.
Following tracker integration, parts of the ToF Cube and Umbrella were integrated and tested at SSL, including the scintillator paddles, SiPMs and readout electronics.
The ToF system was calibrated for timing response, energy scale and velocity estimation using cosmic muons, and its trigger functionality was successfully validated.

\par
In June 2023, the GAPS payload was transferred to the National Technical Systems (NTS) facility in El Segundo, Los Angeles, for an instrument-level thermal vacuum (TVAC) test. 
This test assessed the thermal performance of the instrument in a vacuum environment, simulating flight conditions and validating the thermal model.
Tracker and electronics temperatures were monitored throughout the test, confirming consistent operation in accordance with thermal predictions. The instrument remained fully operational during the multi-day test, with the tracker successfully collecting cosmic muon data and demonstrating the functionality of the complete system.

\par
In July 2023, the payload was moved to the Nevis Laboratories at Columbia University in Irvington, New York, for integration of the remaining components of the ToF system, including the Umbrella and its readout and trigger electronics. 
An extensive data-taking campaign with cosmic muons and x-ray sources was conducted to calibrate both the Si(Li) tracker and the ToF system.

\par
As the final step before deployment to Antarctica, the GAPS instrument was moved to the NASA Columbia Scientific Balloon Facility (CSBF) in Palestine, Texas, for final integration and validation before flight.
All subsystems---including the power system with solar panels, the data acquisition system, and the NASA-provided telemetry---were integrated and tested.
A hang test was conducted to verify the mechanical integrity of the fully assembled instrument.
During this test, the instrument was suspended from the launch vehicle and powered on, confirming that all components were functioning as expected~\cite{tiberioforthegapscollaborationIndirectSearchDark2025}.

\begin{figure}[tbp]
    \centering
    \includegraphics[width=0.75\textwidth]{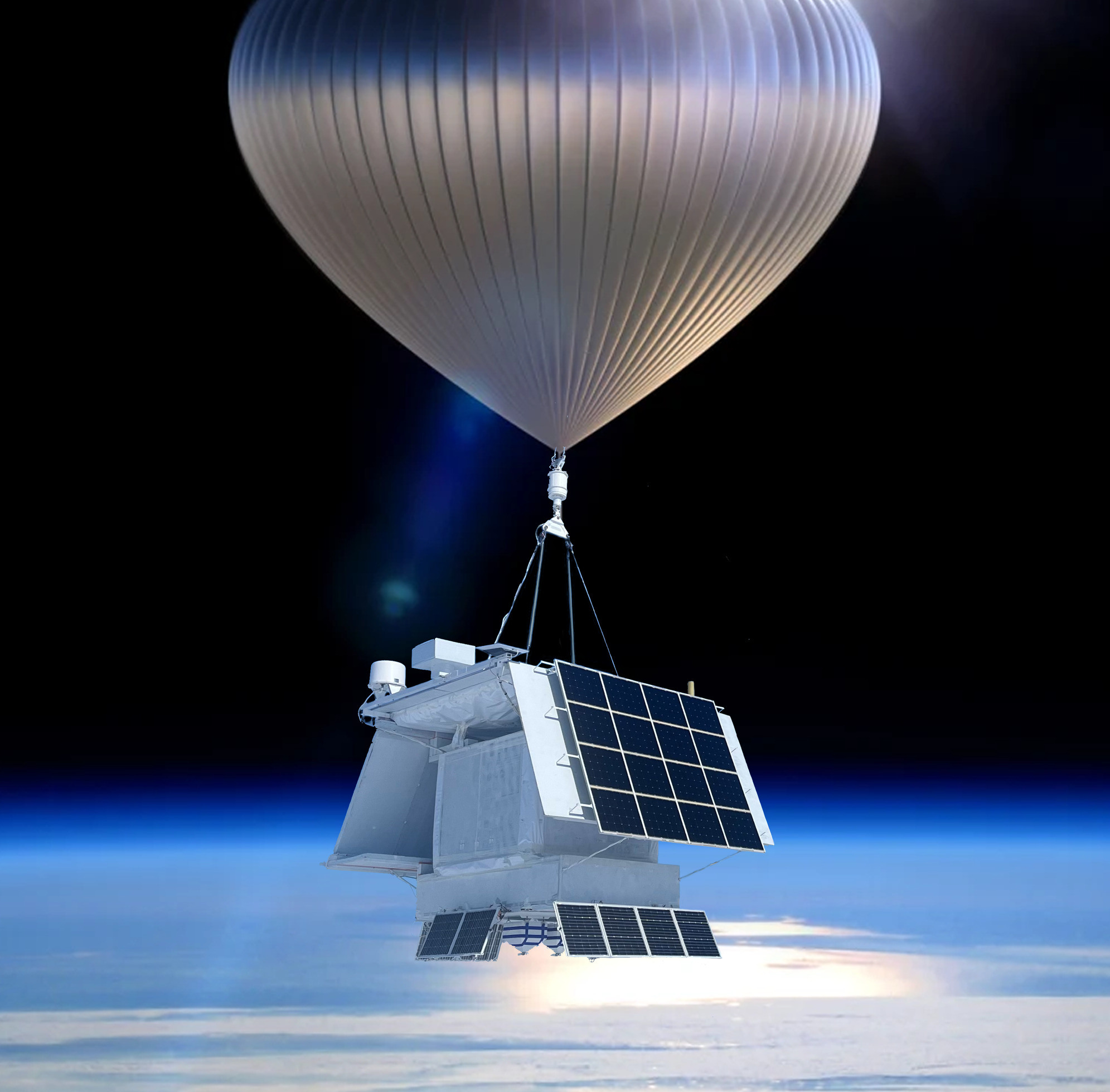}
    \caption{Rendering of the GAPS instrument in flight, suspended from a high-altitude scientific balloon during Antarctic operations. The balloon is not to scale with respect to the instrument. The image illustrates the instrument's configuration at float altitude. Picture courtesy of Kelsey Lee.}
    \label{fig:ch1-gaps_flight}
\end{figure}

\par
The first GAPS campaign in Antarctica was conducted from October 2024 to January 2025, with the instrument assembled and tested at the NASA LDB facility at McMurdo Station, Antarctica. 
\Cref{fig:ch1-gaps_antarctica} shows the GAPS instrument in its flight-ready configuration, suspended from the launch vehicle at the NASA LDB facility.
The picture was taken during the commissioning deployment in December 2024, which served to prepare and validate the instrument in anticipation of its first scientific flight scheduled for December 2025.
A constructed rendering of the GAPS instrument in flight at float altitude is shown in \Cref{fig:ch1-gaps_flight}, illustrating its configuration when suspended from a high-altitude balloon during the flight.

%% file: maintext/chapters/chapter2.tex
\chapter{Characterization of a flight-ready GAPS Si(Li) tracker module}\label{chapter2}

\par
This chapter reports the detailed laboratory characterization of a complete flight-ready GAPS Si(Li) tracker module~\cite{manghisoniXrayParticleDetection2025}.
The first part focuses on the validation of the \hbox{front-end} electronics, including transient response and noise performance.
In the second part, the detector response to radiation is evaluated through dedicated acquisitions using an Americium-241 source and atmospheric muons at ground level, enabling the extraction of energy resolution and effective thresholds, as well as the evaluation of the detection capability under different triggering configurations.

%-------------------------------------------------------------------------------
%   The GAPS Si(Li) tracker module
%-------------------------------------------------------------------------------
\section{The GAPS Si(Li) tracker module}

\par
The GAPS Si(Li) tracker is constructed using a modular design and consists of ten planes, seven of which are equipped with silicon detectors.
The detectors are grouped into modules of four, with the modules organized in rows of six, each row being read out by a dedicated field-programmable gate array (FPGA) board.
These boards are centrally controlled by the flight computer.

%   Detector module description
\subsection{Detector module description}
\begin{figure}[t!]
    \centering
    \includegraphics[width=0.84\textwidth]{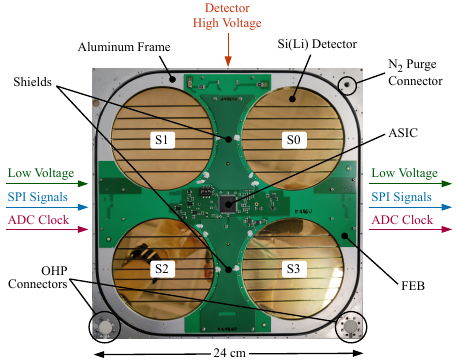}
    \caption{GAPS tracker module. The module consists of four Si(Li) detectors mounted in an aluminum frame and connected to an FEB that hosts the readout ASIC. The top and bottom aluminized polypropylene windows are not shown.}
    \label{fig:ch2-GAPS_module_description}
\end{figure}

The tracker module, shown in~\Cref{fig:ch2-GAPS_module_description}, covers an area of $24\times\SI{24}{\centi\meter\squared}$.
Each module contains four Si(Li) detectors, individually segmented into eight parallel strips, yielding a total of 32 readout channels per module.
The detectors are mounted in a machined aluminum frame that provides mechanical support and electrical grounding for the Si(Li) detectors through their guard rings.
A dedicated front-end board is integrated into the module and hosts a custom front-end ASIC responsible for processing the signals from the detector strips.
Two holes on the bottom of the frame serve as connectors for the OHP thermal control system (not shown in the figure).
These connectors are thermally coupled to the aluminum frame, allowing heat to be extracted from the detectors.
The cooling system maintains the detector temperature between \SI{-35}{\celsius} and \SI{-40}{\celsius} to ensure a nominal strip leakage current of \SIrange{2.5}{5}{\nano\ampere} during operation.

\begin{figure}[t!]
    \centering
    \includegraphics[width=1\textwidth]{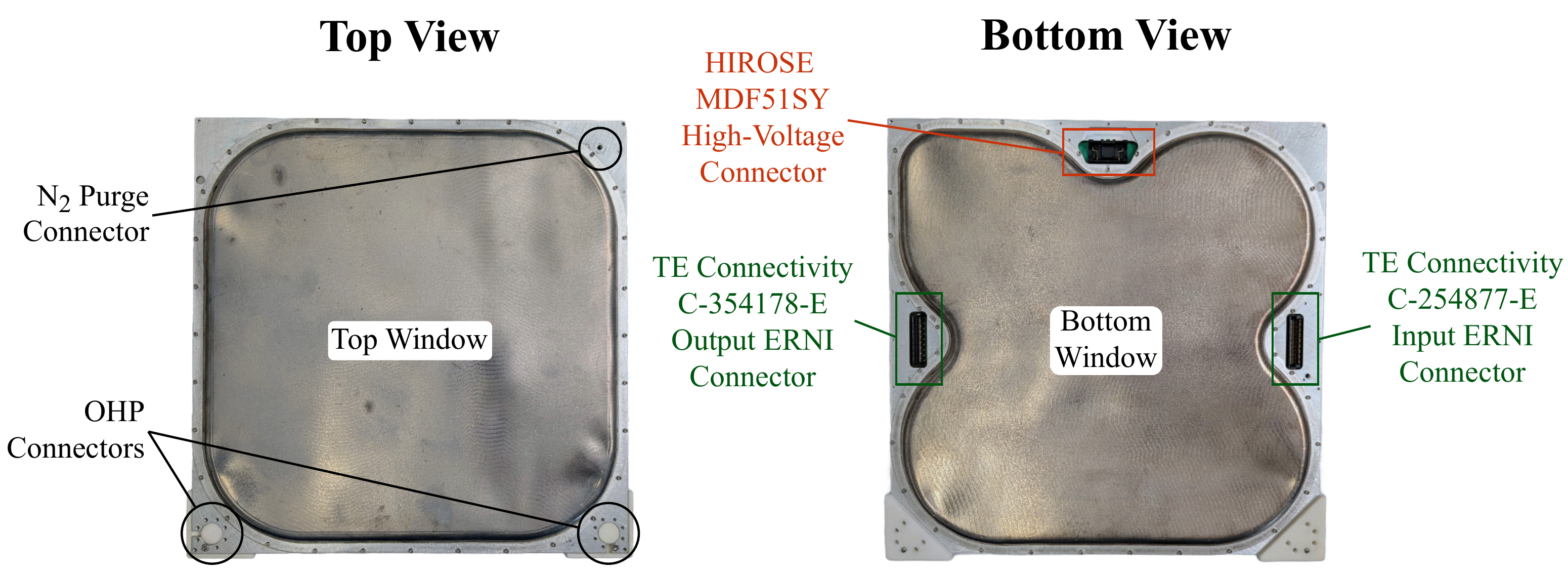}
    \caption{Top view (left) and bottom view (right) of the GAPS tracker module with the protective aluminized polypropylene windows mounted. In the bottom view, the HIROSE MDF51SY high-voltage connector for Si(Li) detector bias,   together with the TE Connectivity C-254877-E (input) and C-354178-E (output) ERNI connectors, are visible.}
    \label{fig:ch2-module_top_bottom}
\end{figure}

\par
Each module is enclosed between two aluminized polypropylene windows (not shown in~\Cref{fig:ch2-GAPS_module_description}) to protect the detectors during assembly and operation.
These windows form a sealed volume that is purged with dry nitrogen gas via a dedicated connector to suppress humidity and prevent condensation.
They also shield the detectors from stray light, airborne contaminants, and electromagnetic interference, as illustrated in~\Cref{fig:ch2-module_top_bottom}.

\par
Modules are assembled into structures known as tracker rows.  
Each row consists of six aligned modules and spans \SI{1.44}{\meter} in length.
Detector modules are connected in a row via custom flex-rigid printed circuit boards (PCBs) that distribute the ASIC low-voltage power and control signals along the row.
The \SI{-250}{\volt} high-voltage detector bias is supplied via a dedicated connector located at the bottom of each module.
The complete tracker is constructed by combining these rows into multiple planes, enabling the measurement of x-rays from exotic atom \hbox{de-excitation} as well as the three-dimensional reconstruction of charged particle trajectories through the detection volume.

%   The Si(Li) detector
\subsection{The Si(Li) detector}\label{sec:ch2-sili_detector}

\par
The Si(Li) detector array inside the GAPS tracker plays a crucial role in the experiment's detection scheme, providing the absorption depth, active area, tracking efficiency, and \hbox{x-ray} energy resolution necessary for exotic atom particle identification.
Unlike conventional silicon detectors, which are typically constrained to small active areas and require cryogenic operation, the GAPS Si(Li) detector features an overall diameter of \SI{10}{\centi\meter}, of which about \SI{9}{\centi\meter} in diameter constitute the active area segmented into eight readout strips, and an overall thickness of \SI{2.5}{\milli\meter}, with \SI{2.29}{\milli\meter} ($\SI{>90}{\percent}$) of sensitive thickness.
It is biased from the bottom with a voltage between \SI{-200}{\volt} and \SI{-300}{\volt}, achieving full depletion at \SI{-100}{\volt}.
The strip capacitance is \SI{38}{\pico\farad} at an operating bias voltage of \SI{-250}{\volt}.  
A leakage current of \SI{2.5}{\nano\ampere} is achieved at \SI{-40}{\celsius} and \SI{-250}{\volt}, with an expected maximum of \SI{5}{\nano\ampere} at the same temperature and bias voltage~\cite{kozaiDevelopingMassproductionModel2019}.

\par
Lithium drifting is the defining process that distinguishes a Si(Li) detector from a conventional silicon one, and is the key to obtaining a very thick and fully depleted sensitive volume.
In a standard $p$-$n$ silicon detector, the depletion depth is limited by the substrate resistivity and the maximum reverse bias that the junction can withstand.
As a consequence, typical silicon detectors have thicknesses of a few hundred micrometers.
Si(Li) detectors overcome this limitation by compensating the acceptor concentration in $p$-type silicon through the controlled drift of lithium ions during the fabbrication process.
The resulting compensated bulk behaves like high-resistivity intrinsic silicon, enabling full depletion over millimeter-scale thicknesses, far beyond what is achievable with ordinary silicon. 

\par
To meet the stringent requirements of the GAPS experiment, a dedicated production process was developed for large-area, segmented Si(Li) detectors capable of operating at relatively high temperatures ($\SI{\sim-40}{\celsius}$), without cryogenic cooling~\cite{perezFabricationLowcostLargearea2018a}.
This design provides the necessary stopping power for low-energy antiparticles while allowing characteristic x-rays to escape for identification.
The process uses high-purity, floating-zone, $p$-type silicon.
It relies on precision lithium diffusion and drift control, and the intentional retention of thin undrifted layers to reduce leakage current and optimize passivation.

\begin{figure}[tbp]
    \centering
    \includegraphics[width=0.9\textwidth]{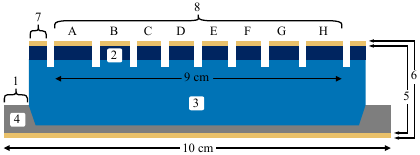}
    \caption{Cross-sectional diagram of an eight-strip Si(Li) detector (not to scale). Key features include the top-hat brim~(1), $n$\textsuperscript{+} Li-diffused layer~(2), compensated active volume~(3), undrifted $p$-type regions~(4), Ni~(5) and Au~(6) contacts, grooves defining the guard ring~(7), and active strips~(8), labeled A to H. Diagram by Field Rogers, adapted from~\cite{rogersLargeareaSiLiDetectors2019}.}
    \label{fig:ch2-SiLi_diagram}
\end{figure}

\par
Each detector begins as a high-resistivity, boron-doped, $p$-type silicon wafer.
A thin $n$\textsuperscript{+} layer ($\SI{\sim0.1}{\milli\meter}$) is formed on the top surface via thermal diffusion.
Lithium ions are then drifted downward through the bulk of the wafer under an applied electric field, forming a compensated, intrinsic-like volume extending through \SI{2.29}{\milli\meter} of the \SI{2.5}{\milli\meter} thickness.
This region constitutes the sensitive volume of the detector.
The remaining undrifted $p$-type silicon forms a thin layer at the bottom and a top peripheral region known as the “top-hat brim,” where a $\SI{\sim3}{\milli\meter}$-wide, $\SI{\sim1}{\milli\meter}$-deep trench prevents lithium penetration.
This structural feature, along with a $\SI{\sim1}{\milli\meter}$-wide, $\SI{\sim0.3}{\milli\meter}$-deep groove that defines a grounded guard ring, suppresses surface leakage currents and ensures reliable operation at elevated temperatures.
The guard ring electrically isolates the active area, which is segmented into eight equal-area strips by additional parallel grooves.

\par
Both detector surfaces are metallized with approximately \SI{20}{\nano\meter} of nickel followed by around \SI{100}{\nano\meter} of gold, ensuring reliable ohmic contacts.
Signal is read from the $n$\textsuperscript{+}-side, while bias is applied to the bottom $p$-side.
This configuration, combined with the compact segmentation and optimized geometry, yields excellent spatial resolution and enables x-ray spectroscopy with energy resolution better than \SI{4}{\kilo\electronvolt} FWHM (full width at half maximum) in the \SIrange{10}{100}{\kilo\electronvolt} range at temperatures between \SI{-35}{\celsius} and \SI{-45}{\celsius}, without the need for cryogenic cooling systems~\cite{rogersLargeareaSiLiDetectors2019}.

\par
\begin{figure}[tbp]
    \centering
    \includegraphics[width=0.75\textwidth]{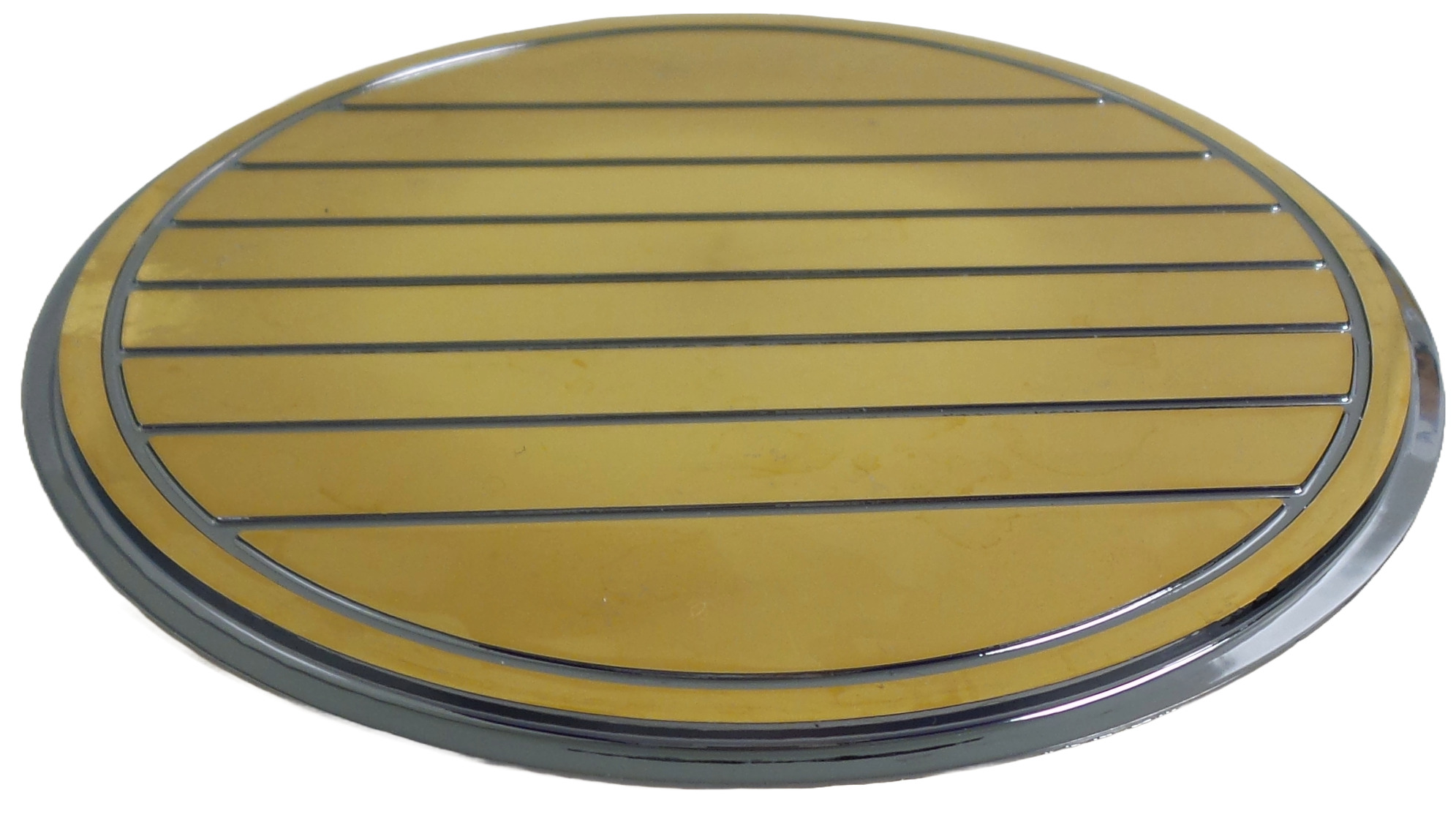}
    \caption{Photograph of a fabricated eight-strip GAPS Si(Li) detector. The image shows the top ($n$-side) surface with etched grooves that electrically isolate the strips and define the outer guard ring. These circular detectors have a diameter of \SI{10}{\centi\meter} and a typical thickness of $\SI{\sim2.5}{\milli\meter}$, optimized to provide sufficient stopping power for low-energy cosmic-ray antiparticles while allowing efficient escape and detection of de-excitation x-rays. Picture courtesy of the GAPS collaboration.}
    \label{fig:ch2-SiLi_picture}
\end{figure}

A schematic cross-section of the detector structure is shown in~\Cref{fig:ch2-SiLi_diagram}, highlighting all relevant features such as the lithium-drifted active region, metallic contacts, guard ring, and segmentation grooves.
The eight readout strips are labeled from A to H.
A photo of a fabricated eight-strip flight detector, illustrating the actual implementation of this geometry, is presented in~\Cref{fig:ch2-SiLi_picture}.

\par
Following fabrication, each of the more than 1100 detectors underwent extensive \hbox{low-temperature} testing at \SI{-40}{\celsius} to verify performance prior to integration.
Detectors were assembled into modules, instrumented with discrete low-noise preamplifiers, and irradiated with \hbox{Cadmium-109} sources to characterize the energy resolution of the silicon strips.
Leakage current measurements were also performed at varying temperature conditions.
This campaign, carried out at the Massachusetts Institute of Technology (MIT) and the University of Hawai\textquoteleft i at M\={a}noa (UHM), demonstrated that over \SI{90}{\percent} of the detectors were operable within the payload and that more than \SI{80}{\percent} of the strips achieved energy resolutions below \SI{4}{\kilo\electronvolt} FWHM with leakage currents under \SI{50}{\nano\ampere} at \SI{-37}{\celsius}, suitable for x-ray spectroscopy.
Detectors not meeting these criteria were classified as ``particle tracking'' quality, as they are still able to provide position and energy-deposition information for charged particles, but with reduced energy resolution.
The results validated a pre-selection model based on room-temperature leakage current measurement at \SI{-250}{\volt} bias and enabled the optimized, quality-based placement of detectors within the tracker during the subsequent integration phase~\cite{xiaoLargescaleDetectorTesting2023}. 

%   The front-end board
\subsection{The front-end board}

\par
The front-end board (FEB), positioned centrally inside the module as shown in~\Cref{fig:ch2-GAPS_module_description}, performs several essential functions within the module.
It hosts the SLIDER32 ASIC, housed in a 128-pin plastic thin-quad flat package (TQFP128) with a footprint of \hbox{$14\times\SI{14}{\milli\metre\squared}$}.
The board also includes all components required for the proper operation of the ASIC, such as three low-dropout (LDO) regulators that receive AVDD, DVDD, and CAL biases from the low-voltage power supply (LVPS) and generate the \SI{1.8}{\volt} and \SI{3.3}{\volt} supplies needed for the ASIC and its internal blocks.
Additionally, the board features a three-bit switch to assign the serial peripheral interface (SPI) address, a 16-bit DAC for calibration, a temperature sensor, and pull-down resistors for the control signal lines.

\par
Another key function of the FEB is to provide and filter the \SI{-250}{\volt} high voltage supply needed to reverse-bias the four Si(Li) detectors.
This high voltage is supplied by the high-voltage power supply (HVPS) and enters the board from the top via a connector (HIROSE MDF51SY) mounted on the bottom side.
The FEB also establishes both power (AVDD, DVDD, CAL) and communication (SPI control signals and ADC clock) links between the flight computer and the ASIC and enables connectivity between ASICs across six modules connected in series.

\par
Bias and signal connections are routed to the board through a miniaturized \hbox{50-pin} ERNI connector (TE Connectivity C-254877-E) on the bottom left side and are extracted through a matching ERNI connector of opposite polarity (TE Connectivity C-354178-E) on the right side. The connectors are accommodated in cutouts of the aluminum frame, which are sealed internally with gaskets to guarantee an airtight seal.
The board has a cross-shaped geometry and measures $22.7\times\SI{22.7}{\centi\metre\squared}$, leaving room for the four Si(Li) detectors, which are wire-bonded to the ASIC via vertical tracks in the central region.
The tracks are covered by custom shields---\hbox{single-plane} ground layers placed directly above the tracks---to suppress noise pickup and external electromagnetic interference.
The FEB is implemented as a six-layer, \SI{1.6}{\milli\metre}-thick printed circuit board (PCB) using Rogers RO4350B for the core and RO4450F for the prepreg layers.
These materials were chosen to minimize dielectric losses that could affect the noise performance of the detector readout~\cite{yanSimpleMiniatureUltrawideband2016,liDesignCompactUltraWideband2023}.

%   The SLIDER32 ASIC
\subsection{The SLIDER32 ASIC}

\par
Signal readout from the Si(Li) detector strips is performed by a custom-designed ASIC named SLIDER32 (32 channels Si-LI DEtector Readout), fabricated using a \SI{180}{\nano\meter} CMOS technology~\cite{manghisoni32channelsReadoutASIC2023}.
Each chip comprises 32 low-noise analog channels designed to read out the strips from four detectors as well as an 11-bit successive approximation register (SAR) analog-to-digital converter (ADC) along with a digital control and communication back-end.
The analog front-end is based on a charge-sensitive amplifier (CSA) featuring a non-linear MOS capacitor in the feedback loop of the amplifier, which introduces a dynamic signal compression mechanism~\cite{manghisoniDynamicCompressionSignal2015,manghisoniDynamicCompressionSignal2018}.
This enables the ASIC to maintain high resolution at low energies while supporting a wide input dynamic range extending from \SI{10}{\kilo\eV} to \SI{100}{\mega\electronvolt}~\cite{ratti2DImagerXray2016a}.

\par
The CSA incorporates a Krummenacher feedback network that performs both charge restoration and leakage current compensation.
A dedicated circuit allows direct readout of the strip leakage current with a resolution of \SI{200}{\pico\ampere}---essential for detector diagnostics during on-ground calibration activities. 
The voltage signal at the output of the CSA is processed by a unipolar, second-order, semi-Gaussian, time-invariant $\text{CR-RC}^2$ filter with eight selectable peaking times between \SI{0.25}{\micro\second} and \SI{1.66}{\micro\second}.
This shaped signal is sampled either via an external trigger signal or using the dedicated self-trigger circuit.
The self-trigger is implemented using an active differentiator followed by a zero-crossing detector, which generates a trigger synchronized with the shaper's peaking time.
This trigger initiates a single-ended to differential sample \& hold (S\&H) block, which outputs a signal proportional to the peak of the shaped pulse.

\par
The stored differential signal is then digitized by a shared 11-bit successive approximation register (SAR) ADC, realized in a hybrid capacitive--resistive architecture. 
The ADC provides an effective resolution close to 10 bits with a conversion time of about \SI{1}{\micro\second}, corresponding to a quantization step of $\SI{\sim1}{\kilo\electronvolt}$ in the x-ray energy range, while keeping the average power consumption below \SI{1}{\milli\watt}. 

\begin{figure}[t!]
    \centering
    \includegraphics[width=0.9\textwidth]{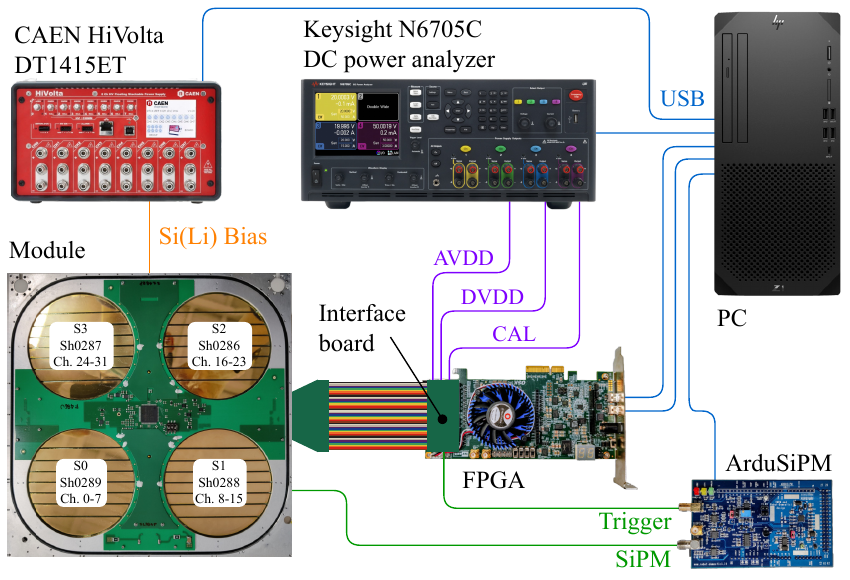}
    \caption{Block diagram of the test setup for evaluating the Si(Li) tracker module. The diagram illustrates the connections between the power supplies, the FPGA control unit, the host PC, the external scintillator and trigger system, and the module under test. The module is positioned rotated \SI{180}{\degree} with respect to the view in \Cref{fig:ch2-GAPS_module_description}. The $5\times\SI{5}{\centi\meter\squared}$ scintillator with SiPMs connected to the ArduSiPM readout board is centrally placed $\SI{\sim5}{\centi\meter}$ below detector S0 (\texttt{Sh0289}, channels 0--7) and is not visible.}
    \label{fig:ch2-module_setup_diagram}
\end{figure}

%-------------------------------------------------------------------------------
%   Test setup
%-------------------------------------------------------------------------------
\section{Test setup}\label{sec:ch2-module_test_setup}

\par
The experimental setup used to characterize the module is shown in~\Cref{fig:ch2-module_setup_diagram}.
It includes a Keysight N6705C DC Power Analyzer, which supplies the analog (AVDD) and digital (DVDD) voltages as well as the calibration bias (CAL) to the FEB.
A CAEN HiVolta DT1415ET high-voltage power supply was used to provide the \SI{-250}{\volt} reverse bias required for the four Si(Li) detectors.
Digital communication with the FEB was managed by an ALTERA Cyclone V FPGA, which transmits SPI commands according to the protocol expected by the ASIC's digital back-end.

\par
To interface power and signal connections between the instruments and the FEB, a custom-designed interface board was employed.
This board routes the supply voltages from the power analyzer and digital control signals from the FPGA to the FEB through a \SI{50}{\centi\meter}-long, 40-wire flexible ribbon cable.
A host PC, connected to the FPGA via two USB cables, runs a Python-based testing program specifically developed for performing automated evaluations of the SLIDER32 ASIC.

\par
In addition to this core test setup, two auxiliary systems were used to assess the module's detection capabilities.
For x-ray testing, an \hbox{Americium-241} source was positioned beneath Si(Li) detector \#0 (corresponding to ASIC channels 0--7).
For cosmic muon detection, a scintillator coupled to Hamamatsu S13360-1325CS silicon photomultipliers (SiPMs) was placed \SI{5}{\centi\metre} below the same detector.
The SiPMs were read out using an ArduSiPM board, a fully integrated ionizing particle detection and acquisition unit based on an open-source Arduino Due microcontroller board and a custom shield~\cite{bocciArduSiPMCompactTrasportable2014,agostiniMeasurementCosmicRay2023}.
All measurements were conducted with the detector module housed in an environmental chamber (ACS DY110) which maintained the target operating temperature of \SI{-40}{\celsius} expected during the flight.

\begin{table}[tbp]
    \centering
    {\renewcommand{\arraystretch}{1.2}
    \setlength{\tabcolsep}{4.5pt}
    \begin{tabular}{c*{4}{ccc}}
        \toprule
        \multirow{3}{*}{Strip} &
        \multicolumn{3}{c}{\texttt{Sh0289} (S0)} &
        \multicolumn{3}{c}{\texttt{Sh0288} (S1)} &
        \multicolumn{3}{c}{\texttt{Sh0286} (S2)} &
        \multicolumn{3}{c}{\texttt{Sh0287} (S3)} \\
        \cmidrule(lr){2-4} \cmidrule(lr){5-7} \cmidrule(lr){8-10} \cmidrule(lr){11-13}
        & $I_{\text{leak}}$ & $E_{\text{res}}$ & $C_{d}$ 
        & $I_{\text{leak}}$ & $E_{\text{res}}$ & $C_{d}$ 
        & $I_{\text{leak}}$ & $E_{\text{res}}$ & $C_{d}$ 
        & $I_{\text{leak}}$ & $E_{\text{res}}$ & $C_{d}$ \\
        & (\si{\nano\ampere}) & (\si{\kilo\electronvolt}) & (\si{\pico\farad})
        & (\si{\nano\ampere}) & (\si{\kilo\electronvolt}) & (\si{\pico\farad})
        & (\si{\nano\ampere}) & (\si{\kilo\electronvolt}) & (\si{\pico\farad})
        & (\si{\nano\ampere}) & (\si{\kilo\electronvolt}) & (\si{\pico\farad}) \\
        \midrule
        A & 2.14 & 3.44 & 39.1 & 2.77 & 3.37 & 38.7 & 2.80 & 3.95 & 39.1 & 2.80 & 3.38 & 38.5 \\
        B & 2.14 & 3.40 & 39.0 & 2.22 & 3.54 & 38.4 & 3.00 & 3.87 & 38.9 & 3.11 & 3.46 & 38.5 \\
        C & 1.88 & 3.74 & 38.7 & 2.00 & 3.50 & 38.6 & 2.03 & 4.02 & 38.7 & 2.38 & 3.61 & 38.3 \\
        D & 2.16 & 3.77 & 38.7 & 2.47 & 3.70 & 38.3 & 2.28 & 4.18 & 38.9 & 1.93 & 3.52 & 38.5 \\
        E & 2.38 & 3.69 & 38.3 & 2.42 & 3.58 & 38.4 & 1.87 & 4.31 & 39.0 & 1.62 & 3.43 & 38.3 \\
        F & 2.33 & 3.59 & 38.6 & 2.35 & 3.31 & 38.5 & 2.38 & 4.21 & 39.4 & 1.87 & 3.49 & 38.8 \\
        G & 2.54 & 3.53 & 38.2 & 2.07 & 3.19 & 38.5 & 2.09 & 3.97 & 39.1 & 1.58 & 3.37 & 38.5 \\
        H & 2.86 & 3.51 & 38.3 & 2.77 & 3.35 & 38.6 & 1.20 & 4.34 & 39.3 & 3.20 & 3.41 & 38.7 \\
        \bottomrule
    \end{tabular}
    }
    \caption{Summary of Si(Li) strip parameters measured at \SI{-37}{\celsius} with \SI{-250}{\volt} bias during the detector testing campaign~\cite{xiaoLargescaleDetectorTesting2023} for the four sensors used in the module under test. The table reports the fitted leakage current $I_{\text{leak}}$ obtained from the noise model based on measured x-ray energy resolutions, the FWHM energy resolution $E_{\text{res}}$ at a peaking time of \SI{1}{\micro\second}, and the measured strip capacitance $C_d$.}
    \label{tab:ch2-sili_parameters}
\end{table}

%   Overview of the module under test
\subsection{Si(Li) detector characterization for the tested module}

\begin{figure}[tbp]
    \centering
    \includegraphics[width=0.85\textwidth]{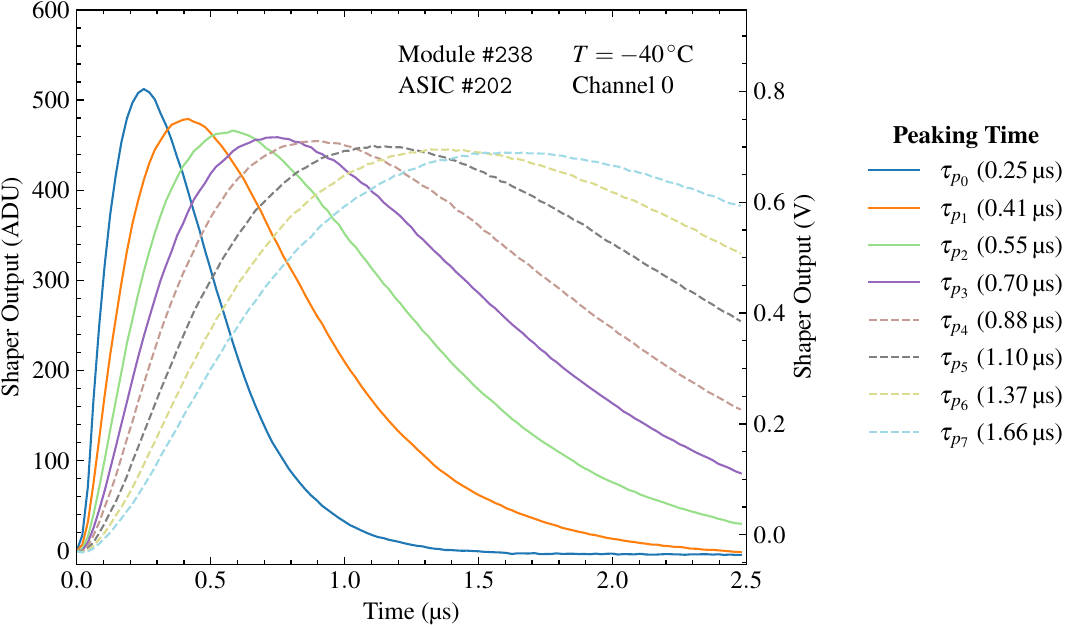}
    \caption{Digitized waveform reconstructed at the output of the analog channel for the eight selectable peaking times from \SI{0.25}{\micro\second} to \SI{1.66}{\micro\second}. An equivalent input energy of \SI{841}{\kilo\electronvolt} is injected at the channel input. ADU (analog-to-digital unit) is the width of the ADC bin: $\SI{1}{ADU}=\SI{1.76}{\milli\volt}$.}
    \label{fig:ch2-waveform_scan}
\end{figure}

\par
The module tested in this work was \texttt{\#238}, instrumented with FEB \texttt{\#202I} and ASIC \texttt{\#202}.
The four Si(Li) detectors---\texttt{Sh0289} (S0), \texttt{Sh0288} (S1), \texttt{Sh0286} (S2), and \texttt{Sh0287} (S3)---were characterized prior to module assembly, with the results reported in~\Cref{tab:ch2-sili_parameters}.
These parameters were measured at \SI{-37}{\celsius} with a \SI{-250}{\volt} detector bias using a dedicated test setup and a discrete preamplifier during the GAPS Si(Li) detector test campaign at MIT and UHM~\cite{rogersLargeareaSiLiDetectors2019}.
The leakage current was extracted by fitting the measured x-ray energy resolutions using the following noise model, valid for the discrete preamplifier employed in the test~\cite{xiaoLargescaleDetectorTesting2023}:

\begin{equation}
    \text{ENC}^2 = \left( 2qI_{\text{leak}} + \frac{4k_BT}{R_p} F_i \tau_p \right) + 4k_BTC^2_{\text{tot}} \left( R_s + \frac{\Gamma}{g_m} \right) \frac{F_\upsilon}{\tau_p} + 2\pi A_f C^2_{\text{tot}} F_{\upsilon f}.
\end{equation}

\noindent
The equivalent noise charge (ENC) is related to the detector FWHM energy resolution $E_{\text{res}}$ via

\begin{equation}
    E_{\text{res}} = 2.35 \varepsilon \frac{\text{ENC}}{q},
    \label{eq:ch2-eres_enc}
\end{equation}

\noindent
where $\varepsilon\approx\SI{3.6}{\electronvolt}$ is the ionization energy per electron-hole pair in silicon~\cite{kleinBandgapDependenceRelated1968}, and $q$ is the elementary charge. 
The noise model contains three free parameters: the strip leakage current $I_{\text{leak}}$, the coefficient of $1/f$ noise $A_f$, and the series resistance $R_s$ arising from the readout connection in the test setup, as described in~\cite{xiaoLargescaleDetectorTesting2023}.
The other terms are as follows: $C_{\text{tot}} = C_{\text{strip}} + C_{\text{FET}} + C_{\text{int}} + C_{\text{stray}}$ is the total strip capacitance obtained from the sum of the Si(Li) strip capacitance ($C_{\text{strip}}$), the preamplifier FET capacitance ($C_{\text{FET}}$), and any additional inter-electrode ($C_{\text{int}}$) and stray ($C_{\text{stray}}$) capacitances.
$k_B$ is the Boltzmann constant, $T$ is the absolute temperature, $g_m$ is the transconductance of the input JFET device, and $R_p = \SI{100}{\mega\ohm}$ is the preamplifier parallel resistance.
The coefficient $\Gamma = 1$ corresponds to the unity excess-noise factor assigned to the JFET transconductance contribution in the noise model adopted in~\cite{xiaoLargescaleDetectorTesting2023}; it is not derived from device geometry nor related to MOSFET short-channel effects.
$F_i$, $F_\upsilon$, and $F_{\upsilon f}$ are the shaping coefficients for the fourth-order Gaussian shaper employed in the preamplifier signal processing chain.
The energy resolution ($E_{\text{res}}$, \si{\kilo\electronvolt}~FWHM) was obtained using the same fourth-order Gaussian shaper with a \SI{1}{\micro\second} peaking time.

\begin{figure}[tbp]
    \centering
    \includegraphics[width=0.75\textwidth]{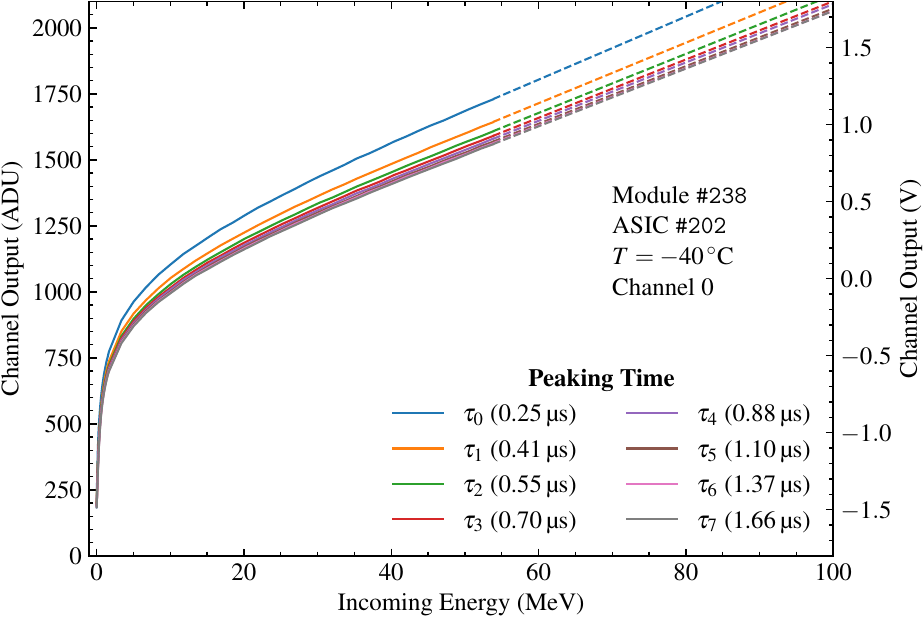}
    \caption{Amplitude of the differential signal at the output of the S\&H for channel~0 measured as a function of the energy released in the Si(Li) strip detector from \SI{10}{\kilo\electronvolt} to \SI{100}{\mega\electronvolt}, varying the peaking time from $\tau_0$ (\SI{0.25}{\micro\second}) to $\tau_7$ (\SI{1.66}{\micro\second}). Dashed lines provide the extrapolation from $\SI{\sim55}{\mega\electronvolt}$ to the input dynamic range limit.}
    \label{fig:ch2-fdt_allpts}
\end{figure}

%-------------------------------------------------------------------------------
%   Experimental results
%-------------------------------------------------------------------------------
\section{Experimental results}\label{ch2-module_testing}

\par
This section presents the experimental results from the characterization of the Si(Li) tracker module detailed in the previous sections.
First, the readout ASIC was validated by performing a series of automated tests to verify its functionality and confirm it met the performance requirements for flight-qualified electronics.
The characterization of the readout electronics was also carried out to ensure that the performance of the ASIC was consistent with the results obtained at ASIC level using a dedicated test board, as reported in~\cite{manghisoni32channelsReadoutASIC2023,ghislottiCharacterisationReadoutElectronics2022}.

\par
The module was then operated in its \hbox{flight-ready} configuration to read out energy deposits from the Si(Li) detectors, both from x-rays emitted by the Americium-241 source and from cosmic-ray muons at ground level, using the dedicated scintillator and SiPM readout system.

%   Front-end electronics characterization
\subsection{Front-end electronics characterization}

\subsubsection{Channel transient response}

\par
The first step in validating the module was to evaluate the dynamic response of the ASIC mounted on the FEB.
The analysis focused on characterizing the \hbox{time-domain} behavior of the shaper output for each channel, with emphasis on its compatibility with the expected shaping function and timing performance.
The transient response was measured at the output of the shaper by digitizing the signal with the integrated S\&H stage followed by the 11-bit SAR ADC.
The test was conducted by injecting an equivalent signal corresponding to a particle energy deposition of \SI{841}{\kilo\electronvolt} into the channel.
For each of the eight selectable shaping configurations---from $\tau_0$ (\SI{0.25}{\micro\second}) to $\tau_7$ (\SI{1.66}{\micro\second})---the shaped waveform was acquired and analyzed.

\par
The acquired waveforms, plotted in~\Cref{fig:ch2-waveform_scan} for channel~0, are expressed in \hbox{analog-to-digital} units (ADUs) as measured at the ADC output. 
The plot also includes the corresponding voltage value at the output of the differential S\&H, considering a conversion factor of \SI{1}{ADU} = \SI{1.76}{\milli\volt}.
The pedestal component was subtracted to isolate the shaped signal and the resulting waveforms exhibit a time profile consistent with the expected \hbox{$\text{CR}$--$\text{RC}^2$} response for all peaking time settings.
In addition, the effective peaking times were extracted from the waveforms by measuring the delay between the injection trigger and the signal peak.
The measured values are in good agreement with the nominal design specifications for each shaping configuration, confirming the correct timing behavior of the analog processing chain.

\begin{figure}[tbp]
    \centering
    \includegraphics[width=0.85\textwidth]{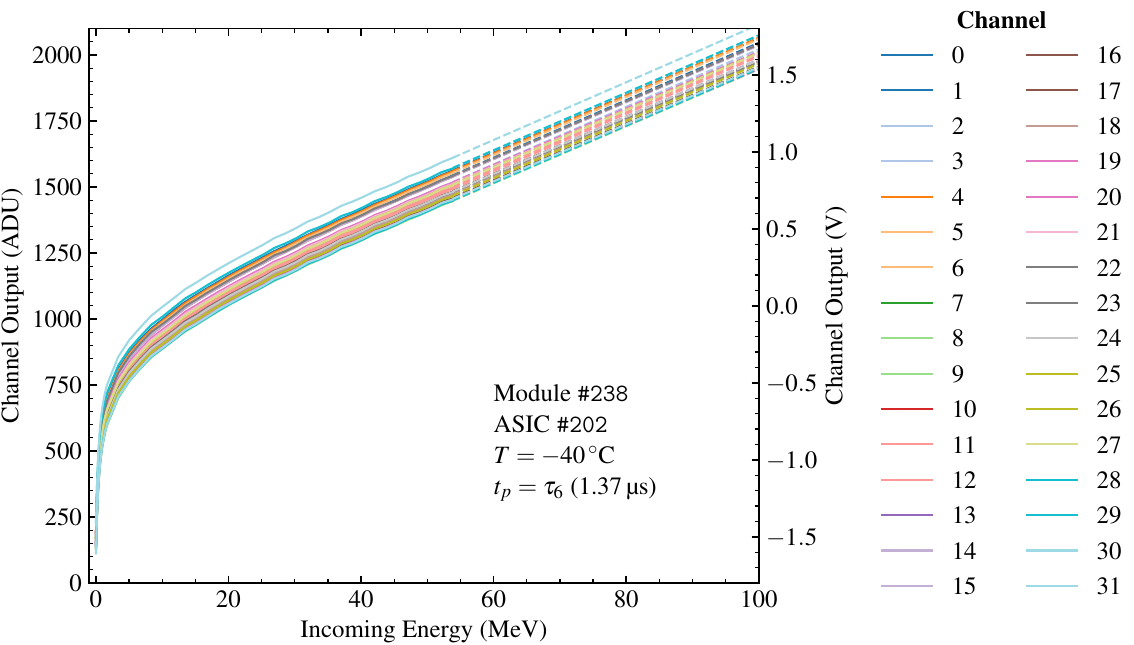}
    \caption{Amplitude of the differential signal at the output of the S\&H for all 32 ASIC channels at a fixed peaking time of $\tau_6 = \SI{1.37}{\micro\second}$. The signal is shown as a function of the energy released in the Si(Li) strip detector from \SI{10}{\kilo\electronvolt} to \SI{100}{\mega\electronvolt}. Dashed lines provide the extrapolation from $\SI{\sim55}{\mega\electronvolt}$ to the input dynamic range upper limit of \SI{100}{\mega\electronvolt}.}
    \label{fig:ch2-fdt_allch}
\end{figure}

%   Channel transcharacteristic and gain
\subsubsection{Channel input--output transcharacteristic and gain}\label{sec:ch2-ch_fdt}

\par
For each peaking time setting, the input--output transcharacteristic of the full analog readout channel was obtained by measuring the peak value of the differential signal digitized at the output of the S\&H, as a function of the energy released in the detector, as shown in~\Cref{fig:ch2-fdt_allpts}.
As expected, the response exhibits a nearly bilinear shape, consistent with the performance of the ASIC tested in a standalone configuration using a dedicated test bench setup~\cite{manghisoni32channelsReadoutASIC2023}.
It must be noted that due to a limitation in the calibration circuit, the measurement of the input--output transcharacteristic of the readout channel was restricted to a maximum value of \SI{55}{\mega\electronvolt}. 
However, as shown by the extrapolated dashed lines in~\Cref{fig:ch2-fdt_allpts}, the compression implemented at the CSA level allows the channel to accommodate an input dynamic range extending up to \SI{100}{\mega\electronvolt} within the ADC's \SIrange{-1.8}{1.8}{\volt} rail-to-rail input range.

\par
\Cref{fig:ch2-fdt_allch} displays the input--output transcharacteristic of all 32 channels at a fixed peaking time of $\tau_6 = \SI{1.48}{\micro\second}$, under the same conditions as the previous plot.
The figure highlights channel-to-channel variation, showing a spread of \SI{26.5}{ADU} at \SI{84.1}{\kilo\electronvolt} (corresponding to \SI{9.1}{\percent}) and \SI{40.8}{ADU} at \SI{25.2}{\mega\electronvolt} (\SI{3.2}{\percent}).
Despite this intrinsic mismatch, all channels exhibit a correct and consistent response across the input energy range.
Moreover, this dispersion does not limit the instrument performance, as channel-by-channel gain corrections are applied during post-processing and are included in the dedicated calibration procedure presented in~\Cref{chapter3}.

\begin{table}[tbp]
    \centering
    \renewcommand{\arraystretch}{1.2}
    \setlength{\tabcolsep}{6pt}
    \begin{tabular}{c c cc cc}
        \toprule
        \multicolumn{2}{c}{Peaking Time} & 
        \multicolumn{4}{c}{Gain} \\
        \cmidrule(lr){1-2} \cmidrule(lr){3-6}
        \multirow{2}{*}{Setting} & \multirow{2}{*}{$\tau_p$ (\si{\micro\second})} & 
        \multicolumn{2}{c}{x-ray} & \multicolumn{2}{c}{Particle} \\
        \cmidrule(lr){3-4} \cmidrule(lr){5-6}
        & & (\si{ADU/\kilo\electronvolt}) & (\si{\milli\volt/\kilo\electronvolt}) & 
        (\si{ADU/\mega\electronvolt}) & (\si{\milli\volt/\mega\electronvolt}) \\
        \midrule
        0 & $0.25 \pm 0.00$ & $1.51 \pm 0.09$ & $2.66 \pm 0.16$ & $12.00 \pm 0.11$ & $21.12 \pm 0.19$ \\
        1 & $0.40 \pm 0.01$ & $1.16 \pm 0.07$ & $2.04 \pm 0.13$ & $11.49 \pm 0.08$ & $20.22 \pm 0.14$ \\
        2 & $0.56 \pm 0.02$ & $1.05 \pm 0.05$ & $1.85 \pm 0.09$ & $11.19 \pm 0.09$ & $19.70 \pm 0.16$ \\
        3 & $0.70 \pm 0.04$ & $1.03 \pm 0.04$ & $1.82 \pm 0.08$ & $11.12 \pm 0.06$ & $19.57 \pm 0.11$ \\
        4 & $0.89 \pm 0.02$ & $1.00 \pm 0.04$ & $1.76 \pm 0.07$ & $11.05 \pm 0.09$ & $19.44 \pm 0.15$ \\
        5 & $1.12 \pm 0.03$ & $0.98 \pm 0.03$ & $1.72 \pm 0.06$ & $11.02 \pm 0.08$ & $19.39 \pm 0.14$ \\
        6 & $1.35 \pm 0.01$ & $0.96 \pm 0.03$ & $1.69 \pm 0.05$ & $10.93 \pm 0.06$ & $19.24 \pm 0.11$ \\
        7 & $1.59 \pm 0.03$ & $0.95 \pm 0.03$ & $1.66 \pm 0.05$ & $10.93 \pm 0.07$ & $19.23 \pm 0.12$ \\
        \bottomrule
    \end{tabular}
    \caption{Gain and peaking time measurements for the eight selectable peaking times across all 32 ASIC channels. The \textit{x-ray} gain figure is measured in the \SIrange{10}{100}{\kilo\electronvolt} energy range, and the \textit{Particle} gain figure is measured in the \SIrange{25}{100}{\mega\electronvolt} energy range. Gains are reported in both ADU and volts, assuming a conversion factor of $\SI{1}{ADU}=\SI{1.76}{\milli\volt}$.}
    \label{tab:ch2-channel_gain_pts}
\end{table}

\par
\Cref{tab:ch2-channel_gain_pts} summarizes the gain figures obtained from linear interpolation of the \hbox{bi-linear} input-out transcharacteristic in the two relevant energy regions.
The \textit{x-ray} gain figure was obtained interpolating the \SIrange{10}{100}{\kilo\electronvolt} energy region, while the \textit{Particle} gain figure was obtained from interpolation between \SI{25}{\mega\electronvolt} and \SI{100}{\mega\electronvolt}.
Gains are reported in both \si{\milli\volt/\kilo\electronvolt} and \si{\milli\volt/\mega\electronvolt}.
The results are consistent with those obtained during the characterization of the ASIC in the dedicated standalone test setup.

%   Si(Li) strip leakage current
\subsubsection{Si(Li) strip leakage current}

\begin{figure}[tbp]
    \centering
    \includegraphics[width=0.75\textwidth]{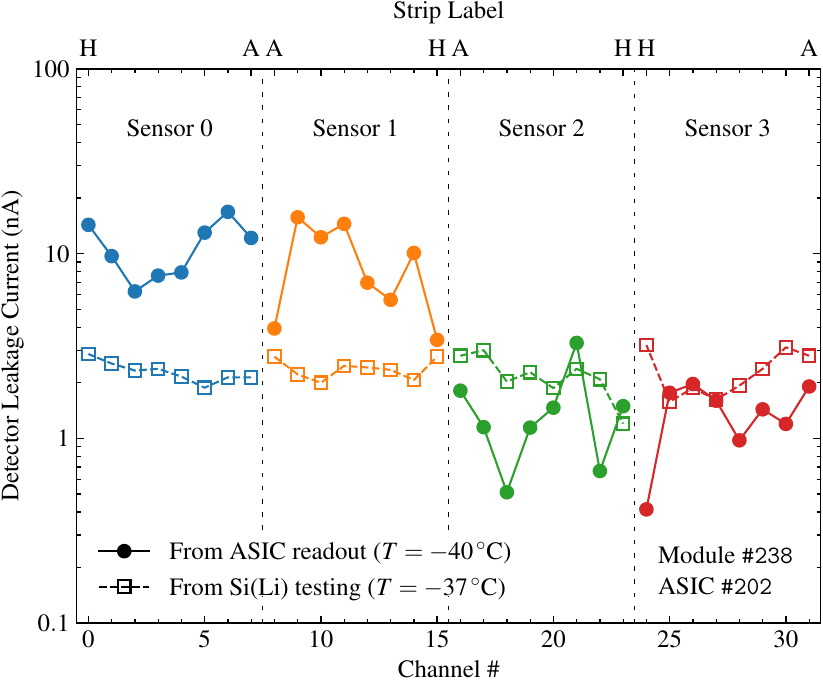}
    \caption{Strip leakage currents measured on the sensors mounted on the module, using a custom circuit integrated into the feedback network of the ASIC CSA (solid circles). For comparison, leakage current values from the GAPS Si(Li) testing campaign (see~\Cref{tab:ch2-sili_parameters}) are also shown (open squares).}
    \label{fig:ch2-sili_leakage_comparison}
\end{figure}

\par
One of the features integrated into the ASIC readout chain is a dedicated on-chip circuit that enables direct measurement of the strip leakage current through the feedback network of the CSA.
Figure~\ref{fig:ch2-sili_leakage_comparison} shows a comparison between the leakage currents measured using this internal ASIC feature and those obtained during the GAPS Si(Li) sensor testing campaign (\Cref{tab:ch2-sili_parameters}). The leakage current values extracted using the ASIC are summarized in~\Cref{tab:ch2-chan-ileak-eres}. These data correspond to the four sensors installed on the module under test.

\par
Overall, there is good agreement between the two measurement sets, indicating that the ASIC-based readout circuit reproduces the expected leakage current behavior with satisfactory accuracy.
However, a significant deviation is observed for Sensor~0 and 1, where the ASIC reports a systematic increase in leakage current of approximately \SI{10}{\nano\ampere} compared to the values measured during the sensor qualification phase.
This discrepancy is noteworthy, as the nominal leakage current for the strips is typically in the nanoampere range.
Given the magnitude of the difference, it is unlikely that the origin lies in the sensors themselves.

\par
A plausible explanation is the influence of environmental humidity, to which Si(Li) sensors are particularly sensitive.
While the module is normally stored at room temperature in a dry box with relative humidity maintained below \SI{15}{\percent}, the climate chamber used for testing was set to \SI{10}{\percent} relative humidity.
Nevertheless, humidity control cannot be guaranteed during temperature transitions, particularly when cycling between ambient conditions and \SI{-40}{\celsius}.
Transient exposure to moisture during these phases may have contributed to the increased leakage current observed by the ASIC.

\par
A closer inspection of the mechanical layout of the module provides an explanation for why this effect is observed only in Sensors~0 and~1.
As shown in \Cref{fig:ch2-GAPS_module_description,fig:ch2-module_top_bottom}, the aluminum frame of the module includes a hole located adjacent to these two sensors, used for flowing dry $\mathrm{N_2}$ gas to suppress humidity.
In the present measurement campaign, no dry-gas flushing was used, and this opening was left exposed.
As a consequence, Sensors~0 and~1 were more susceptible to moisture ingress during temperature cycling, whereas Sensors~2 and~3, located farther from this opening and partially shielded by the internal frame geometry, were significantly less affected.

\par
It should also be noted that the ASIC-based leakage current measurements were performed at \SI{-40}{\celsius}, while those from the Si(Li) sensor testing campaign were taken at \SI{-37}{\celsius}.
A temperature reduction of \SI{6}{\celsius} near the operating point is expected to reduce the leakage current by approximately a factor of 2.
Assuming the leakage current is dominated by the bulk silicon (rather than, e.g., surface effects), $I_{\text{leak}}$ is expected to scale with absolute temperature $T$ as follows~\cite{chilingarovTemperatureDependenceCurrent2013}:

\begin{equation}
    I_{\text{leak}} \propto T^2 \exp \left[ \frac{E_{\text{gap}}}{2k_B T} \right],
\end{equation}

\noindent
where $E_{\text{gap}}\approx\SI{1.14}{\electronvolt}$ is the silicon band-gap energy near \SI{-40}{\celsius}~\cite{bludauTemperatureDependenceBand1974} and $k_B$ is the Boltzmann constant.

\begin{table}[tbp]
    \centering
    {\renewcommand{\arraystretch}{1.3}
    \setlength{\tabcolsep}{4pt}
    \begin{tabular}{c*{4}{ccc}}
        \toprule
        \multirow{3}{*}{Strip} &
        \multicolumn{3}{c}{\texttt{Sh0289} (S0)} &
        \multicolumn{3}{c}{\texttt{Sh0288} (S1)} &
        \multicolumn{3}{c}{\texttt{Sh0286} (S2)} &
        \multicolumn{3}{c}{\texttt{Sh0287} (S3)} \\
        \cmidrule(lr){2-4} \cmidrule(lr){5-7} \cmidrule(lr){8-10} \cmidrule(lr){11-13}
        & Ch. & $I_{\text{leak}}$ & $E_{\text{res}}$ 
        & Ch. & $I_{\text{leak}}$ & $E_{\text{res}}$
        & Ch. & $I_{\text{leak}}$ & $E_{\text{res}}$
        & Ch. & $I_{\text{leak}}$ & $E_{\text{res}}$ \\
        & \# & (\si{\nano\ampere}) & (\si{\kilo\electronvolt})
        & \# & (\si{\nano\ampere}) & (\si{\kilo\electronvolt})
        & \# & (\si{\nano\ampere}) & (\si{\kilo\electronvolt})
        & \# & (\si{\nano\ampere}) & (\si{\kilo\electronvolt}) \\
        \midrule
        A & 7  & 12.16 & 4.56 & 8  & 3.93 & 4.69 & 16 & 1.81 & 5.61 & 31 & 1.91 & 4.98 \\
        B & 6  & 16.83 & 4.58 & 9  & 15.73 & 4.37 & 17 & 1.15 & 4.84 & 30 & 1.20 & 5.72 \\
        C & 5  & 12.99 & 4.40 & 10 & 12.25 & 4.25 & 18 & 0.51 & 4.91 & 29 & 1.44 & 5.43 \\
        D & 4  & 7.91  & 4.48 & 11 & 14.49 & 4.36 & 19 & 1.14 & 5.24 & 28 & 0.98 & 5.14 \\
        E & 3  & 7.60  & 4.27 & 12 & 6.96  & 4.56 & 20 & 1.47 & 5.26 & 27 & 1.60 & 4.84 \\
        F & 2  & 6.24  & 4.56 & 13 & 5.62  & 4.71 & 21 & 3.29 & 5.44 & 26 & 1.97 & 4.70 \\
        G & 1  & 9.69  & 4.48 & 14 & 10.08 & 4.62 & 22 & 0.67 & 5.14 & 25 & 1.77 & 5.39 \\
        H & 0  & 14.31 & 4.20 & 15 & 3.41  & 4.52 & 23 & 1.49 & 4.84 & 24 & 0.41 & 5.02 \\
        \bottomrule
    \end{tabular}
    }
    \caption{Leakage current $I_{\text{leak}}$ and energy resolution FWHM $E_{\text{res}}$ at $\tau_4=\SI{0.89}{\micro\second}$ peaking time measured on the module under test at \SI{-40}{\celsius} with \SI{-250}{\volt} detector bias. The corresponding ASIC channel number for each detector strip is also reported.}
    \label{tab:ch2-chan-ileak-eres}
\end{table}

%   Energy resolution and ENC
\subsubsection{Energy resolution and ENC}

\par
An essential test of the module involves evaluating how the electronic noise of the readout channel is affected when a detector is connected to it.
To distinguish between the characteristic x-rays emitted by de-exciting antiprotonic and antideuteronic exotic atoms, the Si(Li) detector's readout channel must achieve an energy resolution of approximately \SI{4}{\kilo\electronvolt} FWHM in the \SIrange{10}{100}{\kilo\electronvolt} range.  
The energy resolution is ultimately limited by electronic noise, which arises from two primary sources: the Si(Li) detector itself and the analog front-end electronics.

\par
The dominant source of detector noise is the leakage current, which gives rise to shot noise.  
The analog readout chain adds further noise contributions, generally categorized as series white noise, series $1/f$ noise, and parallel noise.
The total electronic noise can be analytically expressed in terms of ENC, which captures the combined effects of all noise sources referred to the input of the CSA.  
For a semi-Gaussian shaper, the squared ENC can be written as follows:

\begin{equation}
    \text{ENC}^2 = \left(C_d^*\right)^2 \left( A_1 \frac{S_w}{t_p} + A_2 2\pi A_f \right) + A_3 S_p t_p.
    \label{eq:ch2-enc}
\end{equation}

\begin{figure}[tbp]
    \centering
    \includegraphics[width=0.75\textwidth]{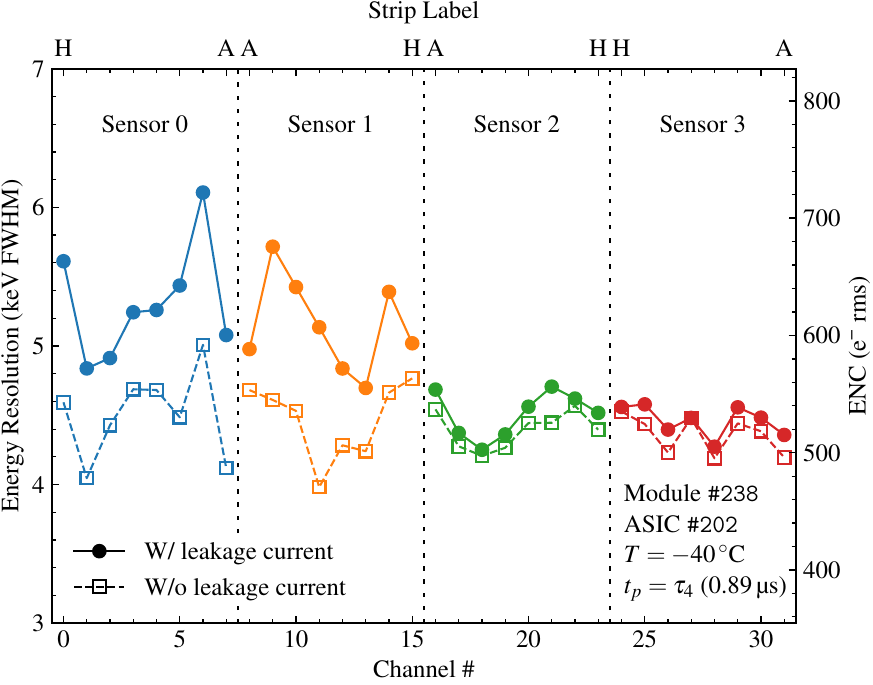}
    \caption{Measured FWHM energy resolution and ENC as a function of the ASIC readout channel number and corresponding sensor label. Measurements were taken at \SI{-40}{\celsius} for a peaking time $\tau_4 = \SI{0.89}{\micro\second}$ (solid circles). The plot also shows the corrected resolution obtained after subtracting the shot noise contribution attributed to the detector strip leakage current (open squares).}
    \label{fig:ch2-module_enc}
\end{figure}

\noindent
In this expression, $C_d^*$ denotes the total capacitance shunting the input node of the readout channel, including the detector capacitance $C_d$ and parasitic contributions; $S_w$ represents the frequency-independent (white) power spectral density of the input-referred series noise source and $A_f$ is the $1/f$ noise coefficient of the same series source. % $t_p$ is the peaking time;
The final term includes $S_p$, the white power spectral density of the input-referred parallel noise source, and is proportional to the peaking time $t_p$.  
The constants $A_1 = 0.85$, $A_2 = 0.54$, and $A_3 = 0.64$ are the shaping coefficients for a second-order unipolar semi-Gaussian shaper~\cite{degeronimoMOSFETOptimizationDeep2004}, and $t_p$ is the peaking time of the filter.
The parallel noise component arises primarily from shot noise generated by the detector leakage current $I_{\text{leak}}$ and by the current flowing in the Krummenacher feedback network $I_K$, with a total power spectral density given by:

\begin{equation}
    S_p = 2q (I_{\text{leak}} + I_K),
    \label{eq:ch2-shot_noise}
\end{equation}

\noindent
where $q$ is the elementary charge.  
This formulation highlights that the overall noise depends on both input capacitance and the peaking time, with the series terms dominating at short peaking times and the parallel contribution becoming dominant at longer peaking times~\cite{degeronimoMOSFETOptimizationDeep2004}.  

\begin{figure}[tbp]
    \centering
    \includegraphics[width=0.75\textwidth]{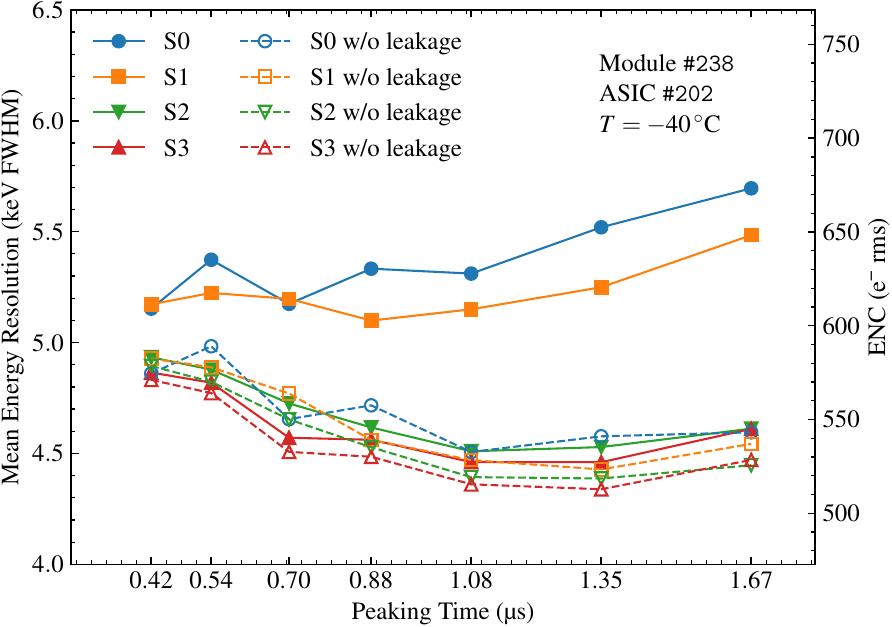}
    \caption{Mean FWHM energy resolution and ENC as a function of peaking time, ranging from \SIrange{0.42}{1.67}{\micro\second}, for the four detectors in the tested module (solid lines). Noise figures after subtraction of the shot noise contribution from strip leakage current are also shown (dashed lines).}
    \label{fig:ch2-module_enc_pt}
\end{figure}

\par
In practice, the ENC was determined from the standard deviation of the pedestal noise distribution measured at the output of each channel.
To ensure a reliable estimation of the intrinsic electronic noise and to mitigate the impact of deterministic interference from common mode noise---arising from both environmental disturbances and coupling effects intrinsic to the detector and its biasing configuration---a common mode noise subtraction was applied prior to the ENC calculation.
Based on the observed correlation patterns in the acquired data, the correction was performed locally within each detector group.
For each acquisition sample $x$, the mean pedestal across the eight channels connected to the same Si(Li) sensor \( p_{\text{cm}}[x] \) is defined as:

\begin{equation}
    p_{\text{cm}}[x] = \frac{1}{8} \sum_{y=y_0}^{y_0+7} p[x,y],
\end{equation}

\noindent
where $y_0$ denotes the first channel of each detector group, and $p[x,y]$ is the sampled pedestal for channel $y$.  
The corrected pedestal distribution was then obtained as:

\begin{equation}
    p'[x,y] = p[x,y] - p_{\text{cm}}[x],
\end{equation}

\noindent
This behavior is consistent with coupling effects associated with the detector's biasing and grounding configuration, which affects groups of channels connected to the same sensor.

\par
After subtracting the local common mode component, the ENC was calculated by dividing the standard deviation of the corrected pedestal distribution $\sigma_{\text{ped}}$ by the channel gain $\mu_{\text{ch}}$ extracted from a linear interpolation of the low-energy portion of the \hbox{input--output} channel transcharacteristic.
The result was then expressed in units of electrons \hbox{root-mean-square} (rms) and energy FWHM:

\begin{equation}
    \text{ENC FWHM} = 2.35 \frac{\sigma_{\text{ped}}}{\mu_{\text{ch}}} \SI{0.841}{\kilo\electronvolt/DAC_{inj}}.
    \label{eq:ch2-enc_pedestal}
\end{equation}

\noindent
The calibration factor of $\SI{0.841}{\kilo\electronvolt/DAC_{inj}}$ accounts for the injection circuit topology and was derived from the product of the injection capacitance (\SI{1.184}{\pico\farad}) and the DAC least significant bit (LSB) voltage step (\SI{31.25}{\micro\volt}), normalized to the energy-to-charge conversion factor for silicon (\SI{0.044}{\femto\coulomb/\kilo\electronvolt}, based on the average energy of $\SI{\sim3.6}{\electronvolt}$ per electron-hole pair).  
A 16-bit external DAC mounted on the FEB was used for charge injection.  
This resulted in an overall calibration factor of $\SI{0.841}{\kilo\electronvolt/DAC_{inj}}$ applied to all ENC calculations.

\begin{figure}[tbp]
    \centering
    \includegraphics[width=0.75\textwidth]{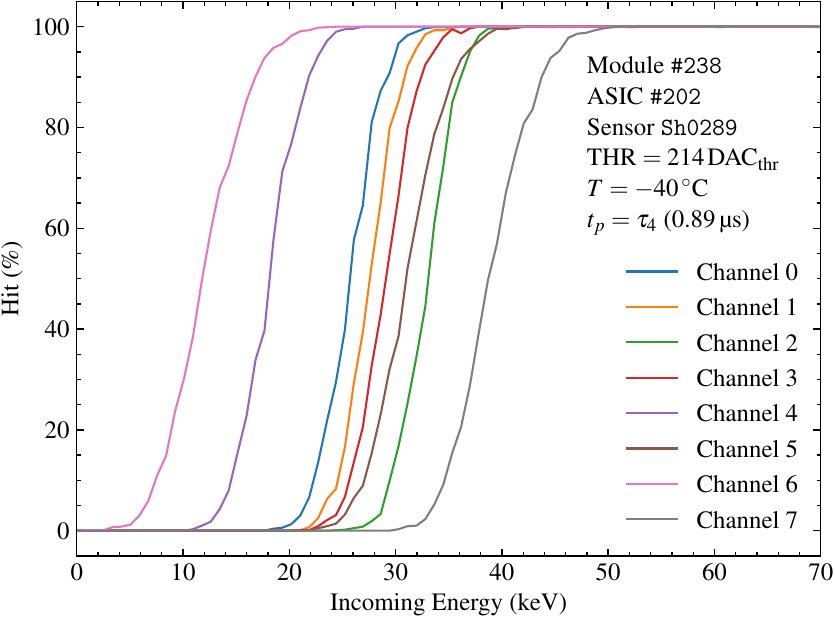}
    \caption{Trigger efficiency curves for channels 0 to 7 obtained from a charge scan performed in self-trigger mode at $\tau_4 = \SI{0.89}{\micro\second}$ peaking time with a global threshold setting of $\SI{214}{DAC_{thr}}$. For each channel, the injected energy was swept across a range while holding the threshold constant, allowing extraction of the effective threshold and energy resolution. The hit probability, defined as the fraction of events exceeding the threshold, is plotted against the incoming energy.}
    \label{fig:ch2-module_charge-scan}
\end{figure}

\par
The leakage currents measured by the ASIC for each channel, along with the corresponding energy resolutions FWHM, are reported in~\Cref{tab:ch2-chan-ileak-eres} to evaluate the noise performance across the module.
Figure~\ref{fig:ch2-module_enc} shows the ENC and corresponding energy resolution measured for all 32 analog readout channels of the module at a peaking time \hbox{$\tau_4 = \SI{0.89}{\micro\second}$}.  
These values, derived from calibration data, are plotted against the ASIC channel number, with sensor labels indicated on the top $x$-axis.  
Solid circles represent the measured energy resolution, expressed in \si{\kilo\electronvolt} FWHM, while a secondary vertical axis reports the ENC in electrons rms.

\par
The channels associated with Sensor~2 and~3 exhibit resolutions near the \SI{4}{\kilo\electronvolt} target, whereas those connected to Sensor~0 and~1 display degraded performance, with an energy resolution in the range of \SIrange{5}{6}{\kilo\electronvolt}.  
This degradation is consistent with the higher leakage currents observed in these sensors, as shown in Figure~\ref{fig:ch2-sili_leakage_comparison}, and reflects the expected contribution of shot noise to the total ENC.

\par
To better isolate the impact of leakage current on the energy resolution, the shot noise contribution was estimated using~\eqref{eq:ch2-shot_noise} and subtracted in quadrature from the measured ENC values.  
The resulting corrected resolutions are shown in the same plot as open squares.  
After correction, the energy resolution across all channels converges to the \SIrange{4}{5}{\kilo\electronvolt} range, demonstrating that the excess noise observed in Sensor~0 and~1 is primarily attributable to elevated strip leakage current.

\begin{table}[tbp]
    \centering
    \renewcommand{\arraystretch}{1.2}
    \setlength{\tabcolsep}{10pt}
    \begin{tabular}{ccc}
        \toprule
        Channel & THR & $E_{\text{res}}$ \\
                \# & (\si{\kilo\electronvolt}) & (\si{\kilo\electronvolt}) \\
        \midrule
        0 & 26.0 & 5.7 \\
        1 & 27.9 & 5.6 \\
        2 & 32.5 & 5.7 \\
        3 & 29.2 & 5.4 \\
        4 & 18.2 & 5.4 \\
        5 & 30.2 & 6.1 \\
        6 & 12.2 & 6.6 \\
        7 & 39.2 & 6.3 \\
        \midrule
        Mean $\pm$ std & $26.93 \pm 8.39$ & $5.85 \pm 0.44$ \\
        \bottomrule
    \end{tabular}
    \caption{Extracted channel thresholds (THR) and energy resolution ($E_{\text{res}}$) values from the charge scan performed in self-trigger mode with a peaking time of $\tau_4 = \SI{0.89}{\micro\second}$ for a global threshold setting of $\SI{214}{DAC_{thr}}$. The threshold corresponds to the energy at \SI{50}{\percent} trigger efficiency, while $E_{\text{res}}$ quantifies the FWHM energy resolution for each channel.}
    \label{tab:ch2_trigger_thresholds}
\end{table}

\par
The same procedure was applied to the seven peaking times for which data were available and are reported in~\Cref{fig:ch2-module_enc_pt}.
Results are reported as the mean energy resolution FWHM and ENC for the four detectors in the module, both before (solid lines) and after (dashed lines) subtracting the shot noise caused by leakage current. 
It is evident that for Sensor~0 and 1, where the effect of leakage current on ENC is stronger, the corrected resolution aligns with that of Sensor~2 and 3 especially for the longer peaking times.
This is consistent with the noise model presented in~\eqref{eq:ch2-enc}.

\par
These results validate the noise model presented earlier and emphasize the importance of selecting low-leakage Si(Li) sensors to meet resolution requirements in the energy range relevant to x-ray detection.  
Moreover, this work contributed to defining standard procedures for leakage current measurement and analysis during the on-ground calibration and validation of the GAPS instrument---both during hardware integration and pre-flight commissioning---thus demonstrating the impact of leakage current on the overall energy resolution performance of the experiment.

%   X-ray detection with an Americium-241 source
\subsection{X-ray detection with an Americium-241 source}\label{sec:ch2-americium}

\begin{figure}[tbp]
    \centering
    \includegraphics[width=0.8\textwidth]{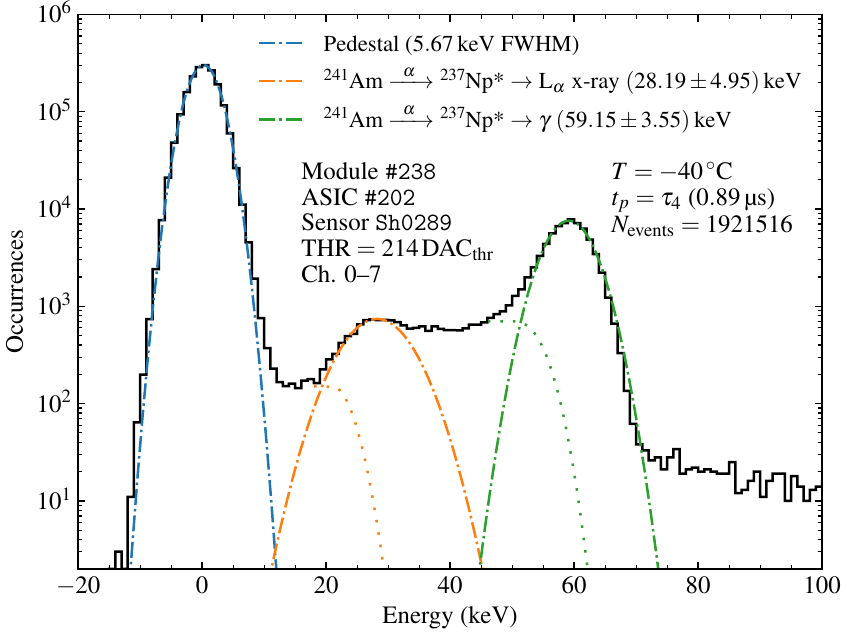}
    \caption{Energy spectrum acquired in self-trigger mode using a \ce{^{241}Am} source, with the global threshold set to 214 for channels 0--7 of sensor \texttt{Sh0289}. The spectrum shows the pedestal peak as well as two photopeaks resulting from the $\alpha$-decay of \ce{^{241}Am}: the L$_\alpha$ x-ray at $(28.19 \pm 4.95)\,\si{\kilo\electronvolt}$ and the $\gamma$ photopeak at $(59.15 \pm 3.55)\,\si{\kilo\electronvolt}$. The acquisition was performed at \SI{-40}{\celsius} over a one-hour period with a peaking time of $\tau_4 = \SI{0.89}{\micro\second}$ and a global threshold setting of $\SI{214}{DAC_{thr}}$. The histogram bin width is \SI{1}{\kilo\electronvolt}.}
    \label{fig:ch2-am241_spectrum}
\end{figure}

\par
To demonstrate the module's capability to detect x-rays, a test was performed using a sample of \ce{^{241}Am}.
This radionuclide undergoes $\alpha$-decay to form the daughter nucleus \ce{^{237}Np} in an excited state (\ce{^{237}Np^*}).
The de-excitation of \ce{^{237}Np^*} occurs via the emission of characteristic $L_\alpha$ x-rays and prompt $\gamma$-rays, which are observable as distinct features in the acquired energy spectrum at \SI{26.34}{\kilo\electronvolt} and \SI{59.54}{\kilo\electronvolt}, respectively~\cite{chechevDecayDataEvaluation2010a}.
The radioactive source was positioned beneath Sensor~0 (\texttt{Sh0289}), which is associated with channels from 0 to 7, as illustrated in~\Cref{fig:ch2-module_setup_diagram}.

\par
During this test, the ASIC was operated in self-trigger mode.
An initial charge scan was performed at various global thresholds to determine the optimal setting for detecting the most prominent photopeak.
In this procedure, the threshold was kept constant while the injected charge was swept across a range of values to evaluate the trigger response of each channel.
The resulting trigger profiles, expressed in terms of trigger probability (\SIrange{0}{100}{\percent}), enabled the extraction of the effective threshold for each channel.

\begin{figure}[tbp]
    \centering
    \includegraphics[width=0.75\textwidth]{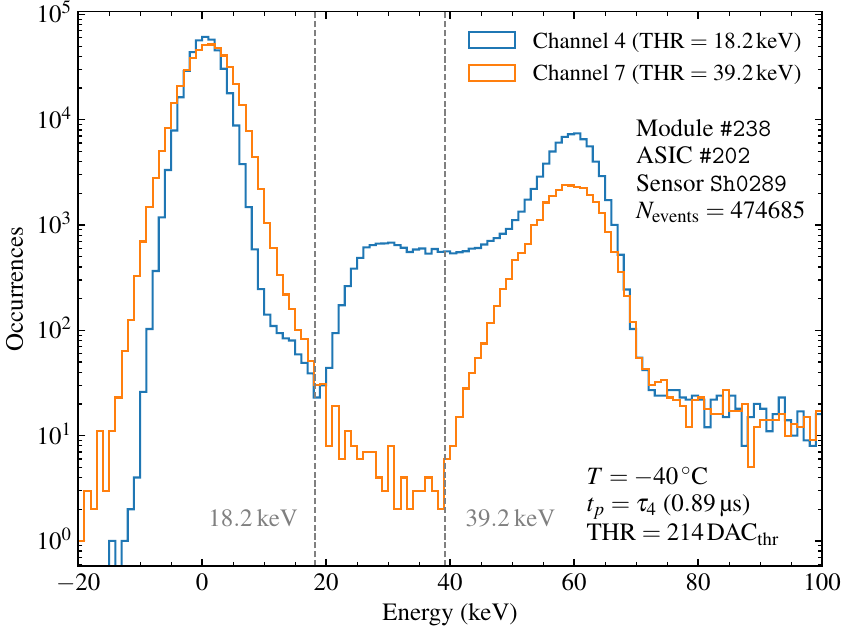}
    \caption{Comparison of energy spectra from channels~4 and~7 of sensor \texttt{Sh0289}, which exhibit significantly different effective thresholds: \SI{18.2}{\kilo\electronvolt} and \SI{39.2}{\kilo\electronvolt}, respectively. The acquisition was performed in self-trigger mode using a \ce{^{241}Am} source at \SI{-40}{\celsius}, with a peaking time of $\tau_4 = \SI{0.89}{\micro\second}$ and a global threshold setting of $\SI{214}{DAC_{thr}}$. Channel~4, with a lower threshold, records both the L$_\alpha$ x-ray and $\gamma$ emission peaks, while channel~7, limited by its higher threshold, does not detect the lower-energy component. The bin width is \SI{1}{\kilo\electronvolt}.}
    \label{fig:ch2-am241_ch4_7}
\end{figure}

\par
\Cref{fig:ch2-module_charge-scan} shows the results for the selected threshold setting of $\SI{214}{DAC_{thr}}$ and a peaking time of $\tau_4=\SI{0.89}{\micro\second}$, after fine threshold trimming was applied.
This trimming was implemented using a three-bit DAC that allows shifting the threshold of each channel over a total span of about \SI{40}{\kilo\electronvolt}, thereby minimizing dispersion among channels.
By interpolating the measured cumulative distribution functions (CDFs), the effective threshold and its standard deviation were extracted for each channel.
These values, when used in~\eqref{eq:ch2-eres_enc}, yield the corresponding energy resolution FWHM.
The results are summarized in~\Cref{tab:ch2_trigger_thresholds}.
The average energy resolution obtained with the charge scan was \SI{5.85}{\kilo\electronvolt} FWHM, in good agreement with the result obtained from pedestal sampling and reported in~\Cref{tab:ch2-chan-ileak-eres} for the same sensor and peaking time.

\par
A self-trigger acquisition using a \ce{^{241}Am} source was conducted over a one-hour period.
The ASIC was configured with a peaking time of $\tau_4 = \SI{0.89}{\micro\second}$, and only the discriminators for channels 0--7 were enabled.
The resulting pedestal-compensated cumulative energy spectrum, obtained by aggregating all triggered events across the eight active channels, is shown in~\Cref{fig:ch2-am241_spectrum} for an acquisition duration of one hour.
The spectrum exhibits the expected features from the $\alpha$-decay of \ce{^{241}Am}, including two photopeaks: an L$_\alpha$ x-ray at $(28.19 \pm 4.95)\,\si{\kilo\electronvolt}$ and a $\gamma$ photopeak at $(59.15 \pm 3.55)\,\si{\kilo\electronvolt}$.
These peaks were identified through interpolation of the spectrum using Gaussian functions.
A Compton shoulder is visible between the two peaks, corresponding to partial energy deposition events~\cite{als-nielsenXraysTheirInteraction2011}.
The histogram was constructed using a bin width of \SI{1}{\kilo\electronvolt}.
A Gaussian fit to the pedestal yielded an energy resolution of \SI{5.67}{\kilo\electronvolt}~FWHM.
This value is in agreement with the results from the pedestal noise analysis (\Cref{tab:ch2-chan-ileak-eres}) and the charge scan (\Cref{tab:ch2_trigger_thresholds}).

\par
The conversion from ADU to energy in \si{\kilo\electronvolt} was performed using a dedicated \hbox{input--output} transcharacteristic calibration procedure applied to each channel individually.
In this procedure, the gain was extracted through linear interpolation in the \SIrange{10}{100}{\kilo\electronvolt} region of the response curve, which was verified to exhibit linear behavior.
The resulting gain factors were then used to convert the recorded ADC values into physical energy units.
A detailed description of this calibration method is later presented in~\Cref{sec:ch3-fdt_calibration}.

\par
To illustrate the impact of channel-level threshold variation, the energy spectra acquired from channels~4 and~7 are shown in~\Cref{fig:ch2-am241_ch4_7}.
According to the charge scan results presented in~\Cref{tab:ch2_trigger_thresholds}, these two channels exhibit a significant difference in effective threshold, with values of \SI{18.2}{\kilo\electronvolt} for channel~4 and \SI{39.2}{\kilo\electronvolt} for channel~7.
While channel~4 does not represent the lowest threshold overall, it is sufficiently low to capture both the L$_\alpha$ x-ray and the $\gamma$ emission peaks from the \ce{^{241}Am} source.
In contrast, the higher threshold of channel~7 suppresses detection of the low-energy peak, resulting in a spectrum dominated by the \SI{59.2}{\kilo\electronvolt} $\gamma$ peak alone.
The acquisition for both channels was performed at \SI{-40}{\celsius}, in self-trigger mode, with a peaking time of $\tau_4 = \SI{0.89}{\micro\second}$, a global threshold of $\SI{214}{DAC_{thr}}$, and a bin width of \SI{1}{\kilo\electronvolt}.

%   Cosmic muon detection at ground level
\subsection{Cosmic muon detection at ground level}

\par
This section presents the results obtained from using the module to detect cosmic-ray muons. Data acquisition was carried out with the ASIC operating both in self-trigger mode---where the sampling signal is internally generated based on the output of the \hbox{zero-crossing} discriminator---and in external trigger mode, in which the trigger is provided by the ArduSiPM device.
For the external trigger setup, the scintillator used to detect coincident muon events was positioned approximately \SI{5}{\centi\meter} below Sensor~\texttt{Sh0289}.
Due to the limited area of the scintillator ($5\times\SI{5}{\centi\meter\squared}$), it was only possible to partially cover the sensor, targeting the region associated with channels~2, 3, 4, and~5.

\par
\begin{figure}[tbp]
    \centering
    \includegraphics[width=0.75\textwidth]{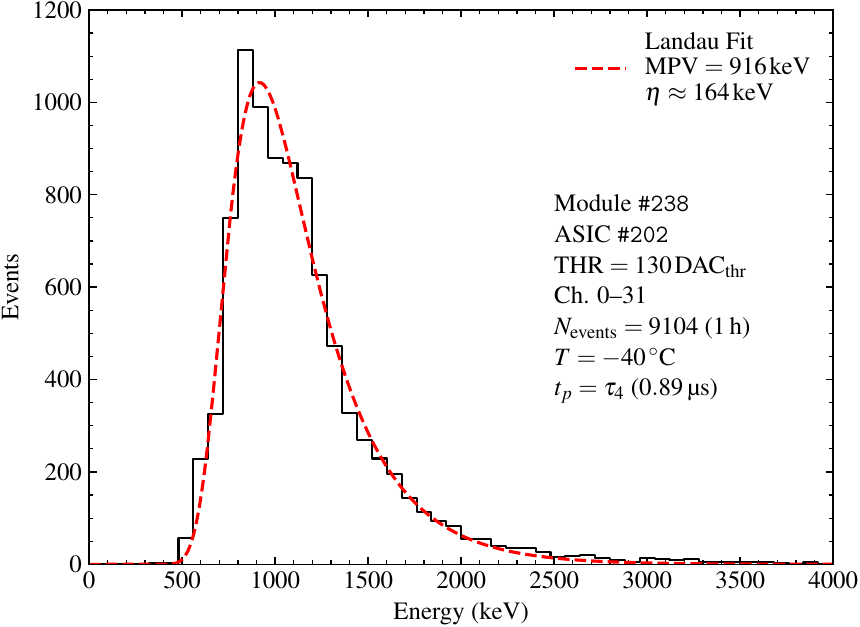}
    \caption{Muon detection spectrum acquired with the ASIC operating in self-trigger mode, with the global threshold set to $\SI{130}{DAC_{thr}}$ and zero suppression enabled. The data were collected over a period of one hour from channels 0 through 31 at a temperature of \SI{-40}{\celsius} and a peaking time of $\tau_4 = \SI{0.89}{\micro\second}$. A Landau fit yields an MPV of $\SI{916}{\kilo\electronvolt}$ and a scale parameter $\eta \approx \SI{164}{\kilo\electronvolt}$. In total, 9104 events were recorded. The histogram bin width is \SI{80}{\kilo\electronvolt}.}
    \label{fig:ch2-muons_self_trigger}
\end{figure}

\Cref{fig:ch2-muons_self_trigger} shows the energy spectrum recorded in self-trigger mode over a \hbox{one-hour} acquisition period.
For this configuration, all 32 channels of the ASIC were active and independently monitored for threshold crossings.
The ASIC was operated in zero suppression mode, whereby channel data are only read out if the shaper output exceeds the threshold and the internal comparator fires.
As a result, the pedestal peak is suppressed and does not appear in the spectrum.
The global threshold was set to $\SI{130}{DAC_{thr}}$, following a charge-scan procedure similar to that used for the \ce{^{241}Am} acquisition.
A higher threshold value was chosen than the one used for the americium measurements ($\SI{214}{DAC_{thr}}$) to match the increased energy deposition expected from cosmic muons and reduce the trigger rate associated to lower-energy events.

\par
The conversion from ADU to deposited energy in \si{\kilo\electronvolt} was performed using the same channel-by-channel input--output transcharacteristic calibration procedure described in~\Cref{sec:ch3-fdt_calibration}.
In this case, a cubic interpolation of the transcharacteristic was used to model the response and convert the data across the full energy range relevant for minimum ionizing particles (MIPs), approximately below \SI{5}{\mega\electronvolt}.

\par
The energy loss of a MIP, such as a muon traversing a thin silicon layer, is subject to statistical fluctuations that follow a Landau-Vavilov distribution~\cite{landauEnergyLossFast1944}.
Theoretical modeling predicts that the most probable energy loss per unit length ($\mathrm{d}E/\mathrm{d}x$) for a MIP in silicon is approximately \SI{368}{\electronvolt\per\micro\meter}.
For the \SI{2.29}{\milli\meter}-thick active volume of the Si(Li) sensor, this corresponds to a most probable energy deposition of $\SI{\sim842}{\kilo\electronvolt}$ for normally incident ($\theta = \SI{0}{\degree}$) particles~\cite{meroliEnergyLossMeasurement2011}.
The recorded spectrum was fitted with a Moyal function, which provides an analytical approximation to the Landau--Vavilov distribution~\cite{moyalTheoryIonizationFluctuations1955}:

\begin{equation}
    f(\lambda) = \frac{p_1}{\sqrt{2\pi}} \exp\left[-\frac{1}{2} \left( \lambda + e^{-\lambda} \right)\right], \quad \lambda = \frac{E - p_2}{p_3}.
    \label{eq:ch2-moyal}
\end{equation}

\noindent
Here, $p_1$ is a normalization constant, $p_2$ represents the most probable value (MPV) of the energy distribution, and $p_3$ is related to its width.
From the fit, an MPV of \SI{916}{\kilo\electronvolt} was obtained---approximately \SI{9}{\percent} higher than the theoretical value for perpendicular incidence.
This discrepancy is expected and can be explained by the angular dependence of the muon flux and the corresponding increase in path length through the detector.
While \SI{842}{\kilo\electronvolt} corresponds to a muon entering orthogonally to the sensor plane, the experimental setup with the ASIC operated in self-trigger configuration does not constrain the direction of arrival of cosmic muons.
In this case, all 32 channels of the ASIC are active, and the system is sensitive to any muon depositing enough energy to exceed the local threshold, regardless of its incidence angle.

\begin{figure}[tbp]
    \centering
    \includegraphics[width=0.75\textwidth]{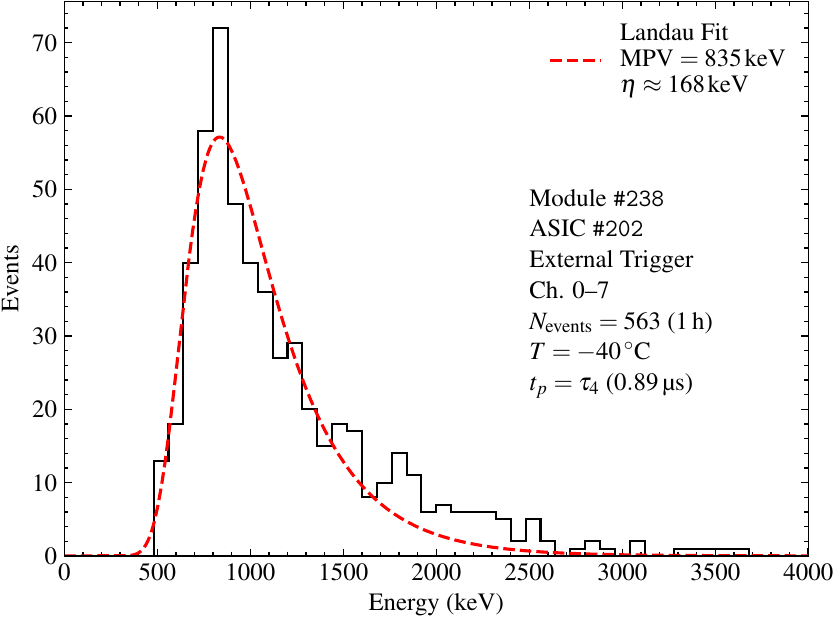}
    \caption{Muon detection spectrum acquired with the ASIC operating in external trigger mode. The trigger was provided by a $5\times\SI{5}{\centi\meter\squared}$ scintillator coupled to a SiPM and placed below Sensor~0. The spectrum is based on 563 events collected over one hour of acquisition. A Landau fit to the energy distribution yields an MPV of $\SI{835}{\kilo\electronvolt}$ and a scale parameter $\eta \approx \SI{168}{\kilo\electronvolt}$. The histogram bin width is \SI{80}{\kilo\electronvolt}.}
    \label{fig:ch2-muons_ext_trigger}
\end{figure}

\par
The angular distribution of atmospheric muons at sea level approximately follows a $\cos^2\theta$ dependence~\cite{danielWhyShouldWe2013}, favoring near-vertical directions but still allowing a significant fraction of inclined trajectories.
As the zenith angle $\theta$ increases, muons traverse a longer path through the detector, resulting in higher energy deposition proportional to $1/\cos\theta$.
Consequently, the MPV of the energy loss distribution shifts to higher values than predicted for normal incidence.
The measured spectrum thus reflects the convolution of energy losses over a wide range of zenith angles, leading to the observed MPV of \SI{916}{\kilo\electronvolt}---approximately \SI{9}{\percent} higher than the theoretical value for a muon passing perpendicularly through the \SI{2.29}{\milli\meter}-thick active volume.

\par
To complement the self-trigger measurements, data were also acquired using an external trigger configuration.
In this setup, a $5\times\SI{5}{\centi\meter\squared}$ plastic scintillator coupled to a SiPM was positioned $\SI{\sim5}{\centi\meter}$ beneath Sensor~0, whose eight strips are connected to channels~0--7 of the ASIC.
The system was configured to register an event only when a coincidence occurred between the ASIC and the external trigger signal generated by the scintillator.
This coincidence condition effectively reduces the accepted solid angle of the incoming muons, strongly favoring muons traveling close to normal incidence ($\theta \approx \SI{0}{\degree}$) owing to the limited separation between the scintillator and the silicon sensor.

\begin{figure}[t!]
    \centering
    \includegraphics[width=0.75\textwidth]{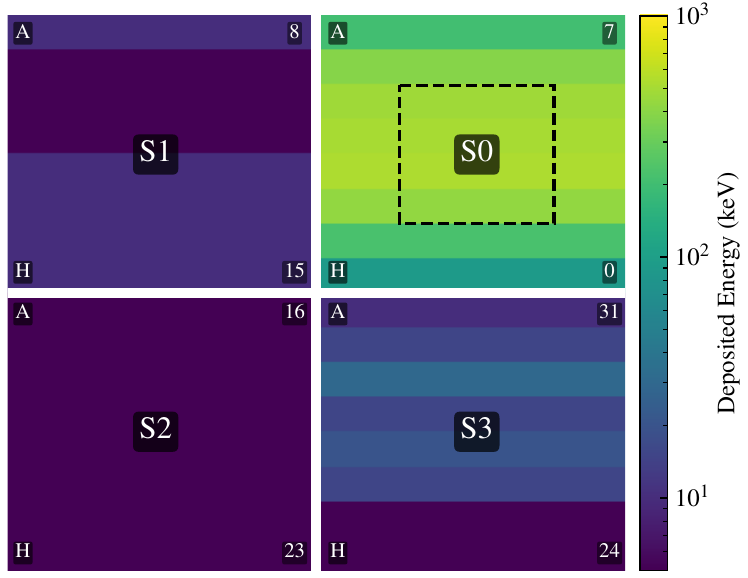}
    \caption{Map of total energy deposited on each strip of the module during the \hbox{one-hour} acquisition in external trigger mode. The layout reflects the physical arrangement of the four sensors (S0--S3), each composed of eight horizontal strips (A--H), for a total of 32 channels (0--31). Only Sensor~0 registers significant energy deposition, confirming that the majority of detected muons crossed the detector region aligned with the trigger scintillator (outlined by the black dashed box).}
    \label{fig:ch2-muons_ext_trigger_edep}
\end{figure}

\par
\Cref{fig:ch2-muons_ext_trigger} shows the energy spectrum acquired over a one-hour period under this configuration.
The histogram, constructed with a bin width of \SI{80}{\kilo\electronvolt}, is based on a total of 563 triggered events.
Compared to the self-trigger mode, the angular acceptance is more selective and leads to an energy deposition spectrum that is narrower and centered at a lower energy than that observed in the self-trigger mode.
This outcome is consistent with the expectation for near-vertical MIPs traversing the \SI{2.29}{\milli\meter}-thick silicon layer.
A Landau fit yields an MPV of \SI{835}{\kilo\electronvolt}, in good agreement with the theoretical value of \SI{842}{\kilo\electronvolt} calculated for orthogonal incidence.
The scale parameter of the distribution is $\eta \approx \SI{168}{\kilo\electronvolt}$ and is consistent with predictions for \SI{2.29}{\milli\meter}-thick silicon detectors~\cite{ferreroIntroductionUltraFastSilicon2021}.

\par
The effect of the trigger geometry is further illustrated by the spatial distribution of deposited energy across the detector channels.
As shown in \Cref{fig:ch2-muons_ext_trigger_edep}, only the channels connected to Sensor~0 register significant energy deposition, with the highest signals appearing on channels~2, 3, 4, and 5.
These channels correspond to the central strips directly aligned with the scintillator position, further confirming the spatial selectivity and directional sensitivity introduced by the external trigger configuration.

%-------------------------------------------------------------------------------
%   Summary and outlook
%-------------------------------------------------------------------------------
\section{Summary and outlook}

\par
The characterization campaign carried out on a flight-ready GAPS Si(Li) tracker module has demonstrated that the module is able to meet the stringent performance requirements of the experiment. 
The front-end electronics, based on the SLIDER32 ASIC, achieved an energy resolution close to \SI{4}{\kilo\electronvolt} FWHM in the \SIrange{10}{100}{\kilo\electronvolt} energy range, in line with the design target. 
The analysis of the Si(Li) detector leakage current highlighted its critical impact on noise performance and, consequently, on the overall energy resolution of the readout chain. 
On the basis of these results, standard procedures for leakage current measurement and analysis were defined, providing a reference for the on-ground calibration and validation of the GAPS instrument during hardware integration and \hbox{pre-flight} commissioning.

\par
These measurements confirm the results previously obtained on the ASIC alone using a dedicated test board, and represent the second step in the validation of the readout electronics. 
The results are in agreement with expectations and with the dedicated ASIC characterization~\cite{manghisoni32channelsReadoutASIC2023, ghislottiCharacterisationReadoutElectronics2022}, bridging the transition from single-chip testing to system-level operation. 
This activity completes the module-level validation, preceding the final step of the campaign on the fully integrated tracker, as discussed in~\Cref{chapter3}.

\par
As a final validation, dedicated measurements with an \ce{^{241}Am} source and with cosmic muons were used to verify the proper behavior of the entire readout chain, from the detector to the ADC. 
The use of a \ce{^{241}Am} source confirmed the module's capability to detect and resolve characteristic x-rays, while the cosmic muon tests at ground level validated its response to minimum-ionizing particles, demonstrating its suitability as both a spectrometer and a charged-particle tracker.

%% file: maintext/chapters/chapter3.tex
\chapter{Calibration of the GAPS Si(Li) tracker for in-flight operation}\label{chapter3}

\par
This chapter presents the calibration procedures developed for the GAPS Si(Li) tracker in preparation for the first balloon flight.  
The objective was to establish a robust methodology to determine the optimal energy threshold of each readout channel and to validate the performance of the full readout chain under flight-like conditions.  
To this end, a dedicated calibration campaign was conducted at the UC Berkeley Space Sciences Laboratory (SSL) in December 2022, during the tracker integration and on-ground validation phase.

\par
Beyond threshold determination, the calibration effort also encompassed the characterization of the input--output transcharacteristic of each channel, enabling the conversion from ADC units (ADU) to deposited energy in \si{\kilo\electronvolt}.  
Procedures were further developed to account for parasitic charge injection, to evaluate electronic noise, and to identify and mask noisy channels.  
The resulting workflow was subsequently implemented in the GAPS flight control software and adopted in the offline event reconstruction pipeline~\cite{ghislottiEnergyThresholdCalibration2024}.

\par
The methods described here provided the foundation for in-flight calibration, ensuring that the tracker can reliably detect signals across its full operating range.  
Their effectiveness was demonstrated through dedicated acquisitions with a Cadmium-109 radioactive source and atmospheric muons at ground level, culminating in the successful detection of the first x-rays with the flight-integrated instrument on December 13, 2022.

%-------------------------------------------------------------------------------
%   The GAPS Si(Li) tracker
%-------------------------------------------------------------------------------
\section{The GAPS Si(Li) tracker}

\par
Building on the laboratory characterization described in~\Cref{chapter2}, the calibration campaign presented in this chapter was conducted on multiple Si(Li) tracker modules integrated into the GAPS instrument during the final assembly phase.
While several validation procedures---such as the evaluation of the analog front-end response and noise performance---mirror those applied to individual modules under controlled laboratory conditions at \SI{-40}{\celsius}~\cite{reMixedsignalProcessorXray2023}, the opportunity to operate on the fully assembled tracker enabled the investigation of environmental and system-level effects that cannot be fully replicated in standalone setups.

\par
In the laboratory, temperature control was achieved using an environmental chamber capable of maintaining \SI{-40}{\celsius} with relative humidity below \SI{10}{\percent}.
In contrast, \hbox{on-ground} operation of the fully assembled tracker poses additional challenges.
The cooling system, based on the OHP principle, does not reach the \SI{-40}{\celsius} target during ground operation, resulting in elevated and spatially nonuniform module temperatures.
Due to the layered geometry of the tracker, module-to-module temperature variations are also observed.
Humidity control is maintained by continuously flushing dry nitrogen through the tracker volume; however, this solution provides a less stable and less tightly regulated environment than the controlled laboratory setup.
Higher operating temperatures lead to increased detector leakage currents, which in turn result in a greater parallel noise contribution from shot noise~\cite{spielerElectronicsNoise2005}.
This degrades the overall energy resolution of the readout channel, complicating both calibration and performance benchmarking. 
These factors must be considered when comparing data acquired under on-ground conditions to those obtained during controlled laboratory testing.

\begin{figure}[tbp]
    \centering
    \includegraphics[width=0.8\textwidth]{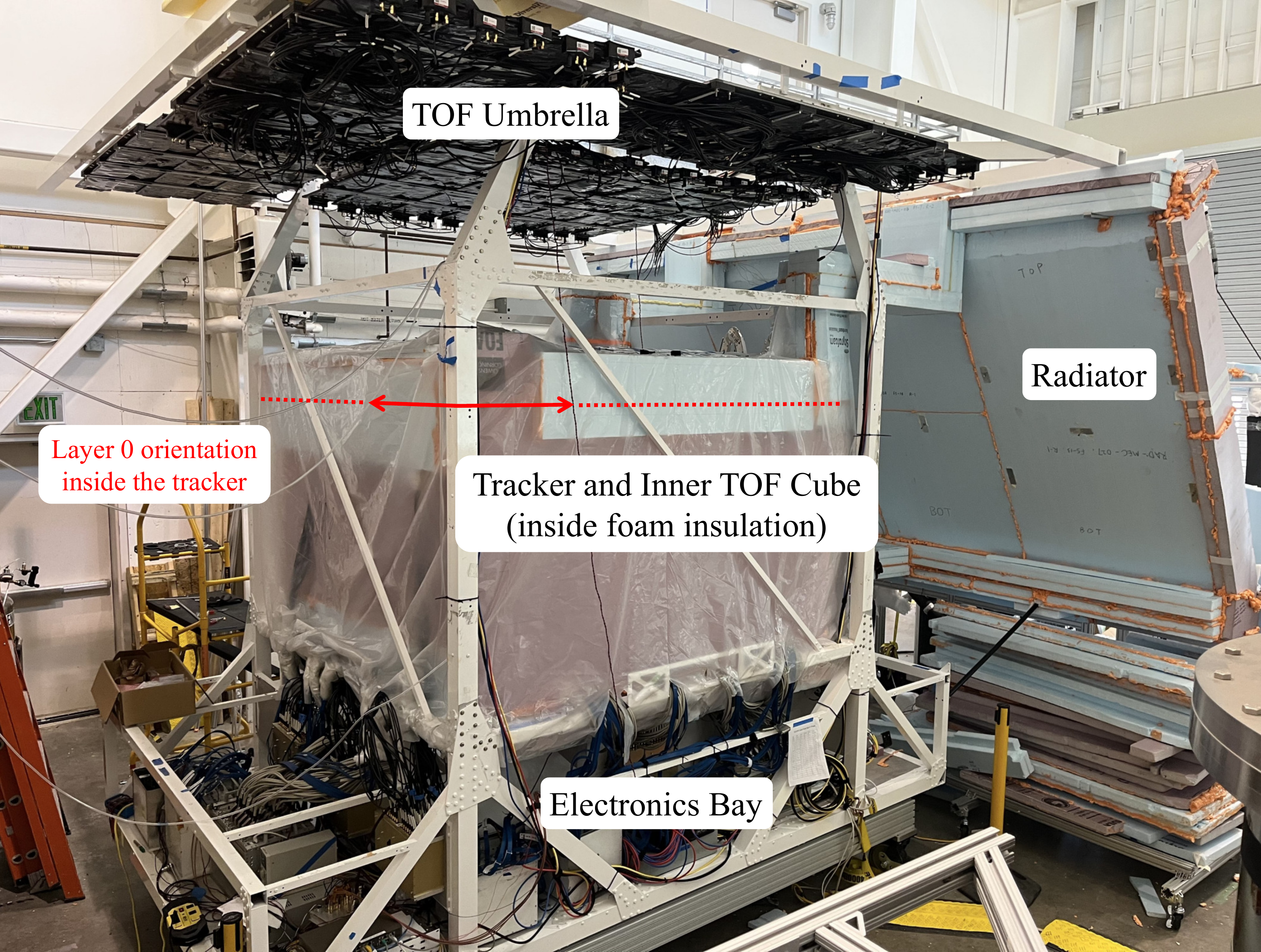}
    \caption{Payload integration in the high bay at UC Berkeley's Space Sciences Laboratory. The image shows the partially assembled gondola structure of the GAPS payload, including the foam-insulated tracker and inner ToF Cube, the electronics bay at the bottom, the radiator (connected to the cold plate of the OHP thermal control system), and the ToF Umbrella during installation. The red arrows indicate the viewing direction and orientation used for Layer~0 maps throughout this chapter. Picture courtesy of Field Rogers.}
    \label{fig:ch3-gaps_ssl}
\end{figure}

\par
An initial subset of tracker modules was read out using the same FPGA-based setup employed during the laboratory characterization campaign, in conjunction with the \hbox{Python-based} software framework developed for automated testing of the SLIDER32 ASIC.
After validating the consistency and reproducibility of the results, the calibration campaign was extended to the entire tracker volume.
In this phase, data acquisition was performed using the GAPS flight computer and custom-designed FPGA readout boards interfacing with the flight-qualified tracker front-end electronics.

\par
Throughout this work, the following labeling convention is adopted: the GAPS Si(Li) tracker is composed of ten horizontal layers, labeled from 0 to 9, of which seven are instrumented for the first flight.
Layer~0 corresponds to the topmost layer, while Layer~9 denotes the bottom layer of the tracker.
Each layer is subdivided into six rows, numbered from 0 to 5 (R0--R5).
Similarly, each row comprises six tracker modules, also numbered from 0 to 5 (M0--M5).
Within a row, Module~0 is located adjacent to the input of the signal chain that connects the modules in series, while Module~5 is placed at the opposite end.

\par
Figure~\ref{fig:ch3-gaps_ssl} shows the GAPS payload during the integration phase at the UC Berkeley Space Sciences Laboratory.
The photograph captures the gondola in a partially assembled state, with the foam-insulated Si(Li) tracker visible within the central ToF cube.
Also shown are the electronics bay at the bottom, which houses the tracker and ToF readout systems, and the radiator coupled to the cold plate of the OHP thermal control system.
At the top of the payload, the ToF Umbrella is shown during installation.
This configuration provided the environment for the first on-detector operation and threshold calibration of the flight hardware.

\begin{figure}[t!]
    \centering
    \includegraphics[width=.75\textwidth]{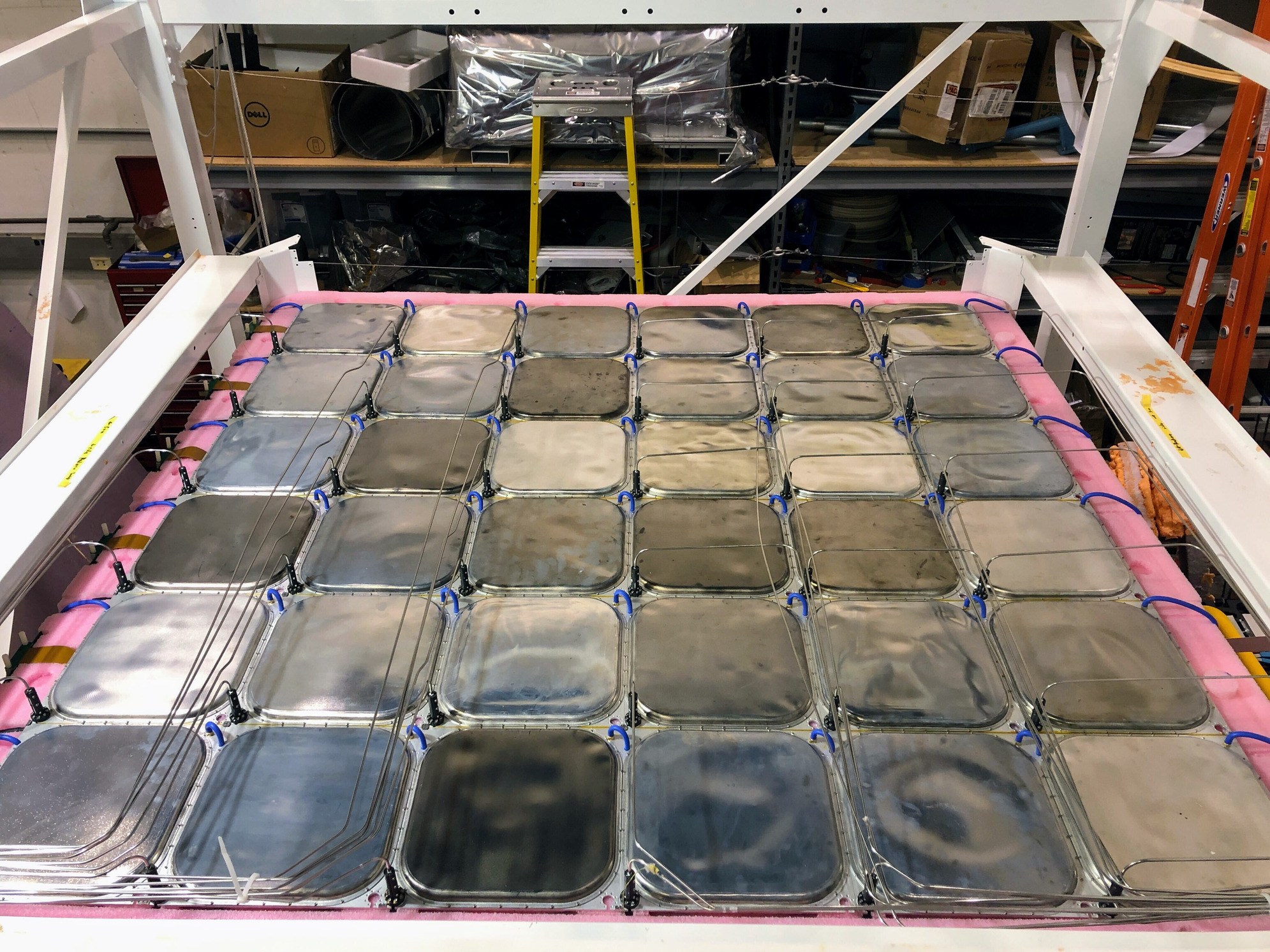}
    \caption{Layer~0 of the GAPS Si(Li) tracker during integration at SSL. The 6\,$\times$\,6 matrix of tracker modules (36 total) is visible with the protective aluminized polypropylene windows installed. Rows (R0--R5) follow the viewing direction and orientation indicated in~\Cref{fig:ch3-gaps_ssl}. Within each row the modules (M0--M5) are connected in series along the readout chain.}
    \label{fig:ch3-tracker_layer0}
\end{figure}

\par
\Cref{fig:ch3-tracker_layer0} shows Layer~0 of the GAPS Si(Li) tracker.
The 6\,$\times$\,6 matrix of modules is clearly visible, each covered by its aluminized polypropylene protective window.
The photograph also shows the stainless steel pipes of the OHP cooling system running across the module array, with collars interfacing to openings in the aluminum frames for thermal coupling.
Blue tubing is used to continuously flush dry nitrogen into the tracker volume, suppressing humidity and preventing condensation inside the sealed modules during \hbox{on-ground} cooling activities.

\par
\begin{figure}[t!]
    \centering
    \includegraphics[width=\textwidth]{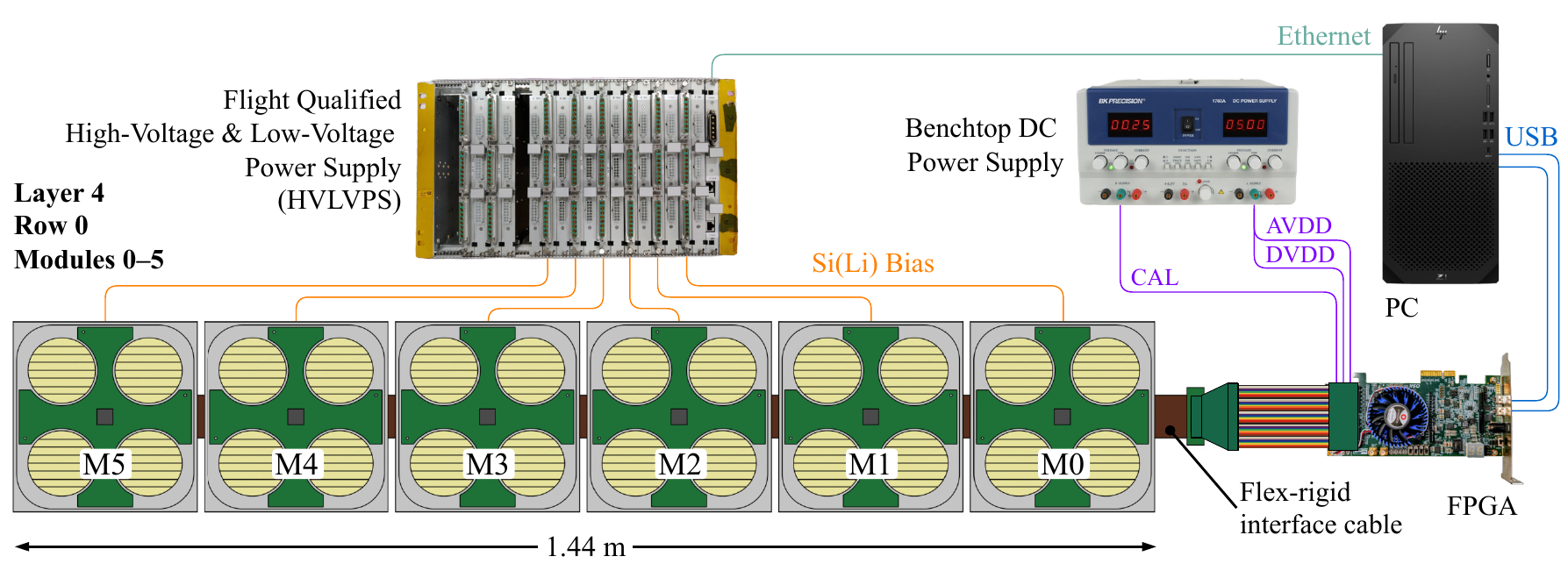}
    \caption{Test configuration used for the evaluation of Modules~0--5 in Row~0 of Layer~4 of the GAPS tracker. The setup extends the single-module infrastructure described in~\Cref{sec:ch2-module_test_setup} to operate on a tracker row. A ribbon cable provides temporary access through the tracker foam insulation, while the FPGA board interfaces with a host PC for control and data acquisition. Low-voltage power is supplied by a benchtop DC power supply, while the high-voltage bias is provided by the flight-qualified HVLVPS.}
    \label{fig:ch3-tracker_row_setup}
\end{figure}

The initial system-level calibration was performed on Module~0 in Row~0 of Layer~4, which was the first module read out using a test configuration similar to the one employed for individual module testing in the laboratory, as reported in~\Cref{sec:ch2-module_test_setup} (see~\Cref{fig:ch3-tracker_row_setup}).
Access to this specific row inside the tracker was temporarily achieved by routing a \SI{50}{\centi\meter}-long, 40-wire flexible ribbon cable through a dedicated cutout in the insulating styrofoam enclosure surrounding the tracker.
This cable directly connected the 50-pin TE Connectivity C-254877-E connector---integrated in the flex-rigid cable of the first module and normally interfaced with the row readout board (temporarily removed)---to the ribbon cable leading to the external FPGA.

\par
This setup enabled direct communication between the row of modules and an Altera Cyclone~V FPGA, which interfaced with a host PC via USB. Power was provided as follows: The \SI{1.8}{\volt} analog (AVDD) and digital (DVDD) low-voltage supplies, as well as the \SI{3.3}{\volt} calibration voltage (CAL), were sourced from a BK Precision 1760A benchtop DC power supply.
The \SI{-250}{\volt} high-voltage bias required for the Si(Li) detectors was supplied by the flight-qualified high-voltage and low-voltage power supply (HVLVPS)~\cite{scottiFrontendElectronicsGAPS2019}, as in the integrated tracker configuration.
In contrast, the low-voltage supply used in this test setup was external and temporary, required to power the row of modules when operated via the dedicated adapter board used for interfacing with the FPGA.

\begin{figure}[tbp]
    \centering
    \includegraphics[width=0.9\textwidth]{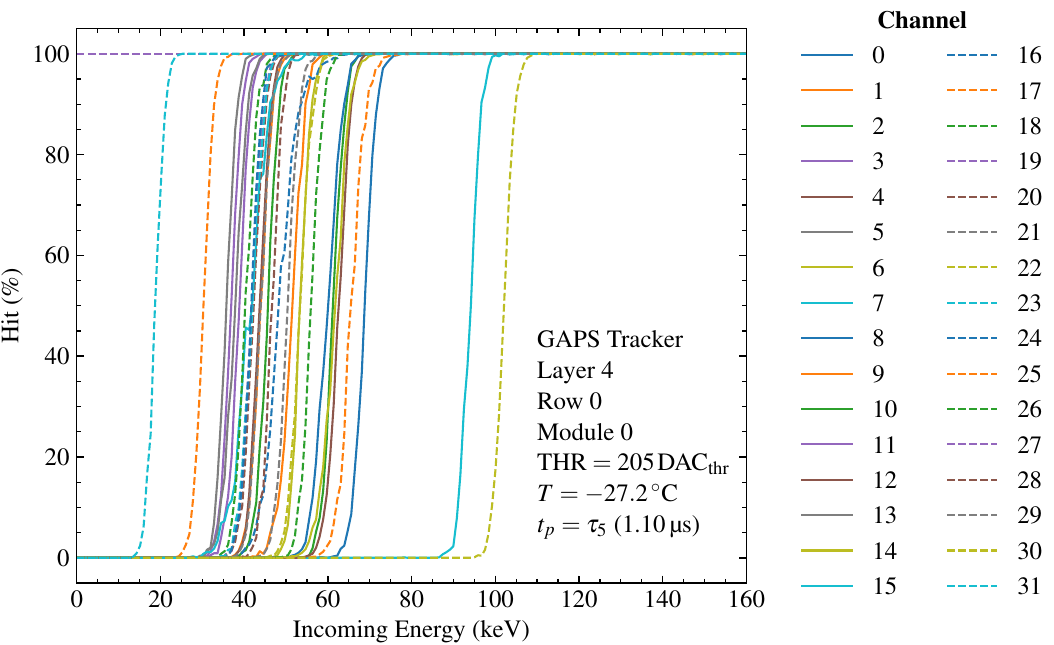}
    \caption{Charge scan results for all 32 analog channels of Layer~4, Row~0, Module~0, acquired at \SI{-27.2}{\celsius} with a global threshold setting of $\SI{205}{DAC_{thr}}$ and a peaking time $\tau_5 = \SI{1.10}{\micro\second}$. The \SI{50}{\percent} trigger probability point defines the uncorrected self-trigger energy threshold for each channel. Channel~27, which remained constantly above threshold, is excluded from the analysis.}
    \label{fig:ch3-charge_scan_raw}
\end{figure}

\par
Following this initial step, the calibration procedures were progressively extended to the entire tracker, which was operated using the full flight readout electronics and the dedicated flight computer.
For clarity and consistency, selected analyses are presented from representative subsets of the system, including modules located in Row~0 and Row~3 of Layer~0 as well as a module in Row~1 of Layer~2.
These regions were chosen to highlight both typical and edge-case behavior observed across the tracker.
Later in the chapter, broader results covering all modules in Layer~0 will be discussed.
The acquisition of cosmic muons, used to evaluate tracker performance with MIPs, is shown for Module~0 of Row~0 in Layer~0.

%-------------------------------------------------------------------------------
%   Channel energy threshold assessment
%-------------------------------------------------------------------------------
\section{Channel energy threshold assessment}

\par
This section presents the procedure developed to estimate the optimal self-trigger energy threshold of each analog channel of the GAPS tracker, considering systematic effects such as parasitic charge injection and channel-to-channel threshold dispersion.
These effects were mitigated through two key calibration steps: correction for parasitic injection induced by the charge injection circuitry and a per-channel fine threshold trimming using the integrated three-bit fine-threshold trimming DAC. 

\par
Combined with the evaluation of electronic noise and energy resolution, this procedure enabled both the precise configuration of the energy threshold for x-ray acquisition using a Cadmium-109 source and the identification of noisy channels.
Channels exhibiting abnormally low thresholds or high noise levels were identified and excluded from the final configuration, as they contributed to a high rate of spurious triggers that suppressed valid physics events.

\begin{figure}[t!]
    \centering
    \includegraphics[width=0.75\textwidth]{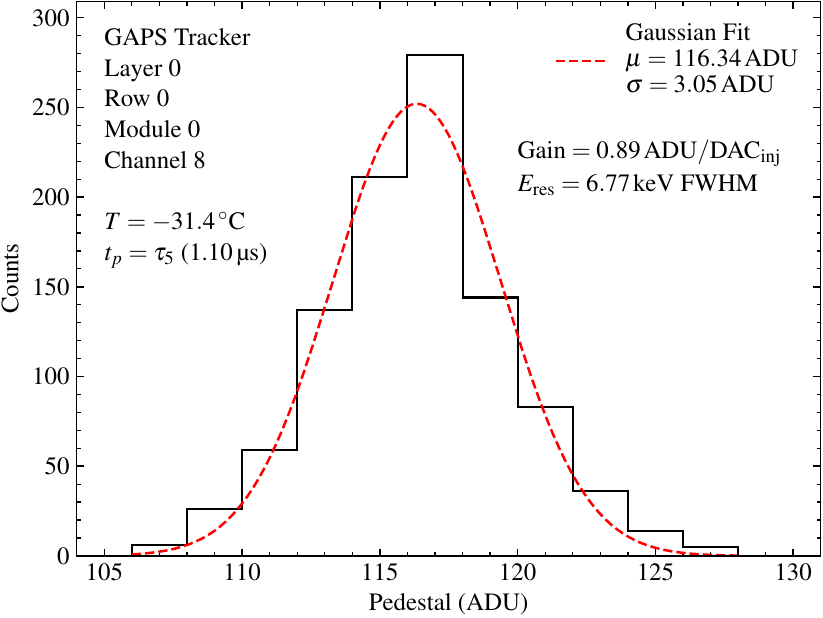}
    \caption{Gaussian fit of the pedestal distribution acquired from 1000 pedestal samples on Channel~8 in Layer~0, Row~0, Module~0, without charge injection. The extracted mean is used for parasitic injection compensation, while the standard deviation provides the pedestal noise estimate employed in ENC calculation. The histogram bin width is \SI{2}{ADU}.}
    \label{fig:ch3-pedestal_gaussian_fit}
\end{figure}

%   Charge scan
\subsection{Charge scan}

\par
To evaluate the effective energy threshold of each analog channel when operated in \hbox{self-trigger} mode, a dedicated charge scan procedure was performed.
The scan involved injecting known amounts of charge using the on-chip injection circuit, covering an energy range dependent on the threshold being characterized.
To account for threshold dispersion across all channels---both with and without fine trimming---the scan spanned approximately \SI{50}{\kilo\electronvolt} above and below the mean energy threshold associated with the selected \si{DAC_{inj}} setting.
For each injected energy value, the channel response was sampled 1000 times to account for spurious variations during charge injection and signal sampling.
The resulting trigger probability was then computed as the mean response and plotted as a function of the input energy.

\par
This procedure is particularly important when operating at low thresholds in the \hbox{low-energy} regime, where even small variations in channel thresholds can significantly impact performance.
Although the charge scan can, in principle, be carried out for all 256 threshold values settable via the eight-bit global threshold DAC, it becomes less critical at higher thresholds.
In those cases, the residual dispersion---on the order of a few tens of \si{\kilo\electronvolt} after fine threshold trimming---does not significantly degrade the energy resolution.

\par
The result of this measurement is shown in Figure~\ref{fig:ch3-charge_scan_raw}, which reports the hit probability curves obtained for all 32 channels of a single tracker module (Layer~4, Row~0, Module~0).
The energy threshold of each channel is defined as the energy value corresponding to a \SI{50}{\percent} hit probability, extracted by fitting the cumulative trigger response with a Gaussian error function model.
The following fitting function is used:

\begin{equation}
    \Phi(x,\mu,\sigma,k) = \frac{1}{2}k\left[1 + \mathrm{Erf}\left(\frac{x - \mu}{\sigma\sqrt{2}}\right)\right],
    \label{eq:charge_scan_fit}
\end{equation}

\noindent
where $x$ is the injected energy in \si{\kilo\electronvolt}, $\mu$ is the mean value representing the channel threshold, $\sigma$ is the standard deviation corresponding to the ENC, and $k$ is a normalization factor accounting for the total number of injections per energy step.
This approach enables a direct estimation of the ENC from the slope of the transition region in the trigger probability curve.

\begin{figure}[tbp]
    \centering
    \includegraphics[width=0.9\textwidth]{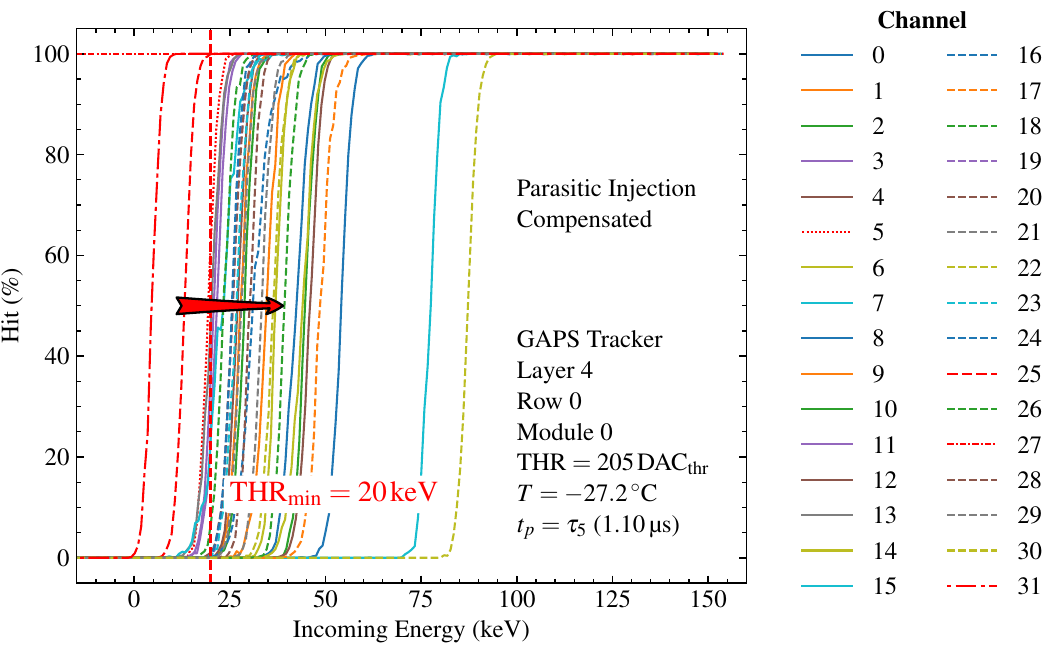}
    \caption{Charge scan results after parasitic injection compensation for all 32 channels of Layer~4, Row~0, Module~0. Channels with corrected thresholds below the minimum threshold limit of \SI{20}{\kilo\electronvolt} (indicated by the red dashed line) are flagged as hot channels. These channels are masked to suppress noise-induced events and preserve overall data quality. Channel~27, which remained constantly above threshold, is excluded from the analysis.}
    \label{fig:ch3-charge_scan_comp}
\end{figure}

%   Parasitic injection compensation
\subsection{Parasitic injection compensation}\label{sec:ch3-par_inj}

\par
In the presence of parasitic capacitance introduced by the calibration circuitry---primarily due to stray capacitance along the digital control lines operating between DGND (digital ground) and DVDD (digital $V_{\text{DD}}$)---the injected charge tends to be systematically overestimated.
This issue was first identified during the laboratory characterization of a tracker module, as presented in~\Cref{chapter2}, which involved energy deposition measurements using an Americium-241 source.
In that test, the global discriminator threshold was intentionally configured slightly above the expected \SI{59.54}{\kilo\electronvolt} photopeak, based on the assumption that no x-ray events should be detected.
Surprisingly, the photopeak remained visible in the acquired spectrum, prompting further investigation.

\begin{figure}[tbp]
    \centering
    \includegraphics[width=0.75\textwidth]{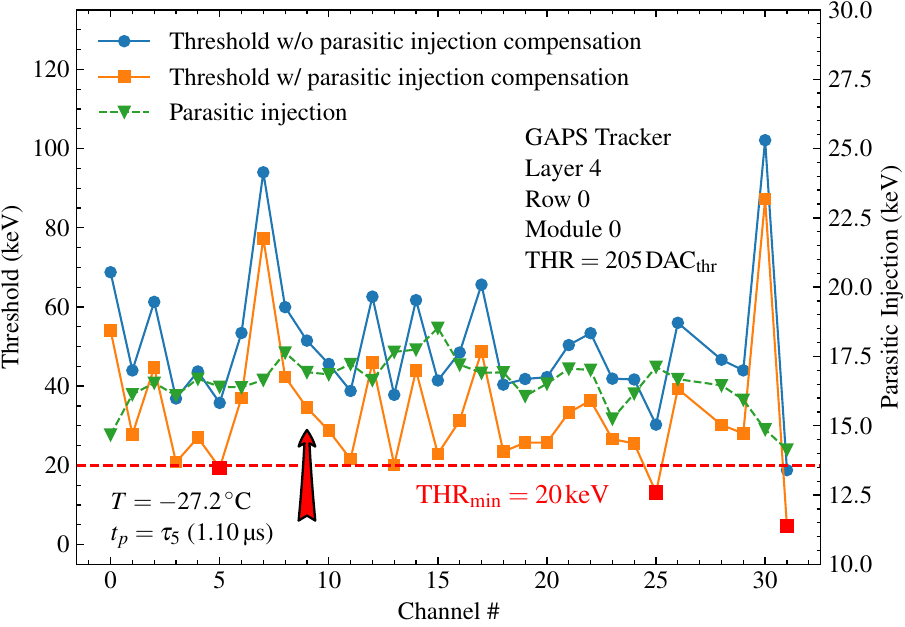}
    \caption{Extracted self-trigger thresholds for all channels of Layer~4, Row~0, Module~0 before and after parasitic injection compensation. The parasitic injection contribution, calculated as the offset between the channel pedestal and the intercept of the linear fit of the channel transfer function, is also shown. The horizontal dashed line indicates the minimum threshold limit of \SI{20}{\kilo\electronvolt} applied to suppress hot channels.}
    \label{fig:ch3-400_thresholds_parinj}
\end{figure}

\par
It was later determined that the apparent lowering of the energy threshold was due to parasitic injection, which effectively introduced an offset in the injected charge.
To correct for this effect, the parasitic contribution was subtracted from the measured threshold.
This compensation was achieved by estimating the intercept of the linear fit of the channel \hbox{input--output} transcharacteristic in the low-energy range (\SIrange{10}{100}{\kilo\electronvolt}) and comparing it to the measured pedestal position, obtained by sampling the channel baseline 1000 times without charge injection.
While the ENC was computed using the method described in~\eqref{eq:ch2-enc_pedestal}, the pedestal position was obtained by fitting the baseline distribution with a Gaussian function.
An example of such a fit is shown in Figure~\ref{fig:ch3-pedestal_gaussian_fit}, which illustrates the pedestal distribution for Channel~8 of Layer~0, Row~0, Module~0, as measured using the dedicated flight readout electronics.

\par
The corrected charge scan results are shown in Figure~\ref{fig:ch3-charge_scan_comp}.
In this version, parasitic injection has been compensated for all channels, revealing several with thresholds significantly below the imposed minimum threshold.
These channels exhibit abnormally high trigger rates during regular operation, often dominated by noise or spurious hits, and are classified as ``hot channels.''

\par
To mitigate the impact of hot channels on data acquisition, a minimum energy threshold of \SI{20}{\kilo\electronvolt} was imposed during configuration.
This value was chosen after evaluating multiple threshold settings and represents an optimal trade-off between suppressing \hbox{noise-dominated} channels and retaining sensitivity to relevant physical signals.
In particular, the \SI{20}{\kilo\electronvolt} limit ensures efficient detection of the \SI{88}{\kilo\electronvolt} photopeak from the Cadmium-109 source used during calibration while excluding channels with excessive trigger rates caused by low thresholds.
Channels with fitted thresholds below this limit were masked to suppress unwanted triggers.
This strategy preserves the effective bandwidth of the readout system and improves the overall signal-to-noise performance of the tracker by prioritizing real physics events over noise-induced triggers.

\begin{figure}[tbp]
    \centering
    \captionsetup[subfigure]{justification=justified,singlelinecheck=false}
    \begin{subfigure}{0.49\textwidth}
        \centering
        \includegraphics[width=\linewidth]{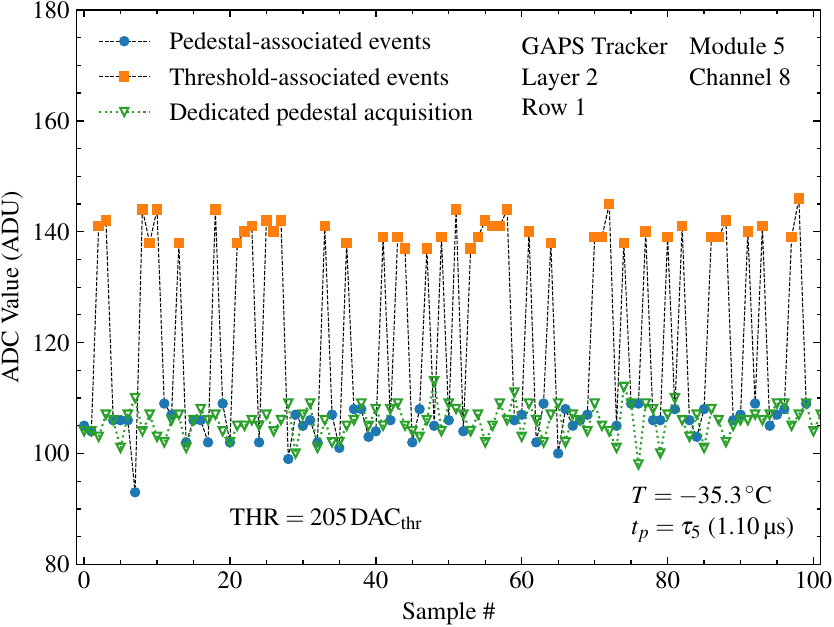}
        \caption{ADC values over time for Channel~8, showing pedestal events (blue circles) and threshold-induced events (orange squares), totaling 100 samples, alongside 100 events from dedicated pedestal sampling (green open triangles).}
        \label{fig:ch3-parinj_time}
    \end{subfigure}
    \hfill
    \begin{subfigure}{0.49\textwidth}
        \centering
        \includegraphics[width=\linewidth]{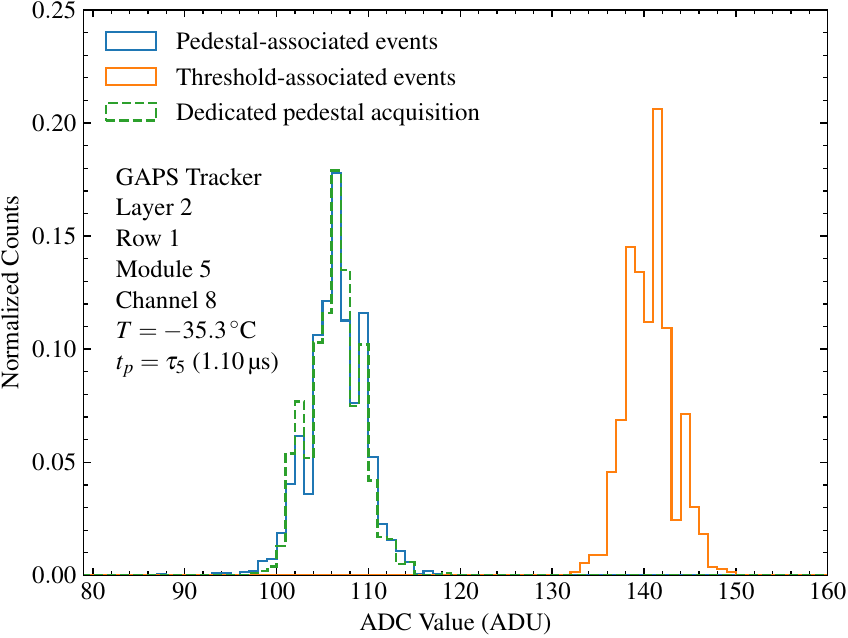}
        \caption{Histogram of ADC values, revealing a pedestal peak (blue) and a second peak (orange) due to the low effective threshold. The distribution from dedicated pedestal sampling (green, dashed) is overlaid for comparison. Bin width is \SI{1}{ADU}.}
        \label{fig:ch3-parinj_hist}
    \end{subfigure}
    \caption{Background data acquired on Channel~8, Layer~2, Row~1, Module~5, with $\SI{205}{DAC_{thr}}$ in self-trigger mode. The parasitic injection effect lowers the effective energy threshold to \SI{10.50}{\kilo\electronvolt}, resulting in a secondary peak in the spectrum. This behavior led to the masking of the channel, as its compensated threshold fell below the minimum acceptance limit of \SI{20}{\kilo\electronvolt}.}
    \label{fig:ch3-parinj_time_hist}
\end{figure}

\par
Figure~\ref{fig:ch3-400_thresholds_parinj} summarizes the threshold extraction results for each channel of Layer~4, Row~0, Module~0. Both uncorrected and parasitic injection-compensated thresholds are shown.
The difference between these sets of values reflects the impact of parasitic injection, which varies across channels due to layout-dependent channel-to-channel and process variations.
The parasitic injection component, extracted as the difference between the extrapolated pedestal and the linear fit intercept of the channel input--output transcharacteristic, is also plotted.
The minimum threshold limit of \SI{20}{\kilo\electronvolt} is indicated as a horizontal red dashed line. Channels falling below this threshold were classified as hot channels and masked to preserve data integrity.
In the specific case of Layer~4, Row~0, Module~0, Channels~5, 25, and 31 were identified as hot, while Channel~27 exhibited anomalous behavior, remaining constantly above threshold in the charge scans shown in~\Cref{fig:ch3-charge_scan_raw,fig:ch3-charge_scan_comp}.
The parasitic injection compensation procedure was applied to all 252 modules within the seven instrumented layers of the tracker.
For each of the 8064 channels, the parasitic contribution was estimated and subtracted from the raw threshold to obtain the effective energy threshold.

\par
A concrete manifestation of this effect was observed during background data acquisition on the GAPS tracker with a relatively low global discriminator threshold setting of $\SI{205}{DAC_{thr}}$.
In this evaluation, the tracker operated for several hours in self-trigger mode without any external x-ray source.
The objective was to investigate any spectral features beyond the pedestal peak, including potential energy deposits from MIPs produced by atmospheric muons.
However, the expected muon rate at ground level under this trigger configuration was too low to compensate for the significantly higher trigger rate from low-energy events, which were predominantly attributable to electronic noise.

\begin{figure}[tbp]
    \centering
    \includegraphics[width=\textwidth]{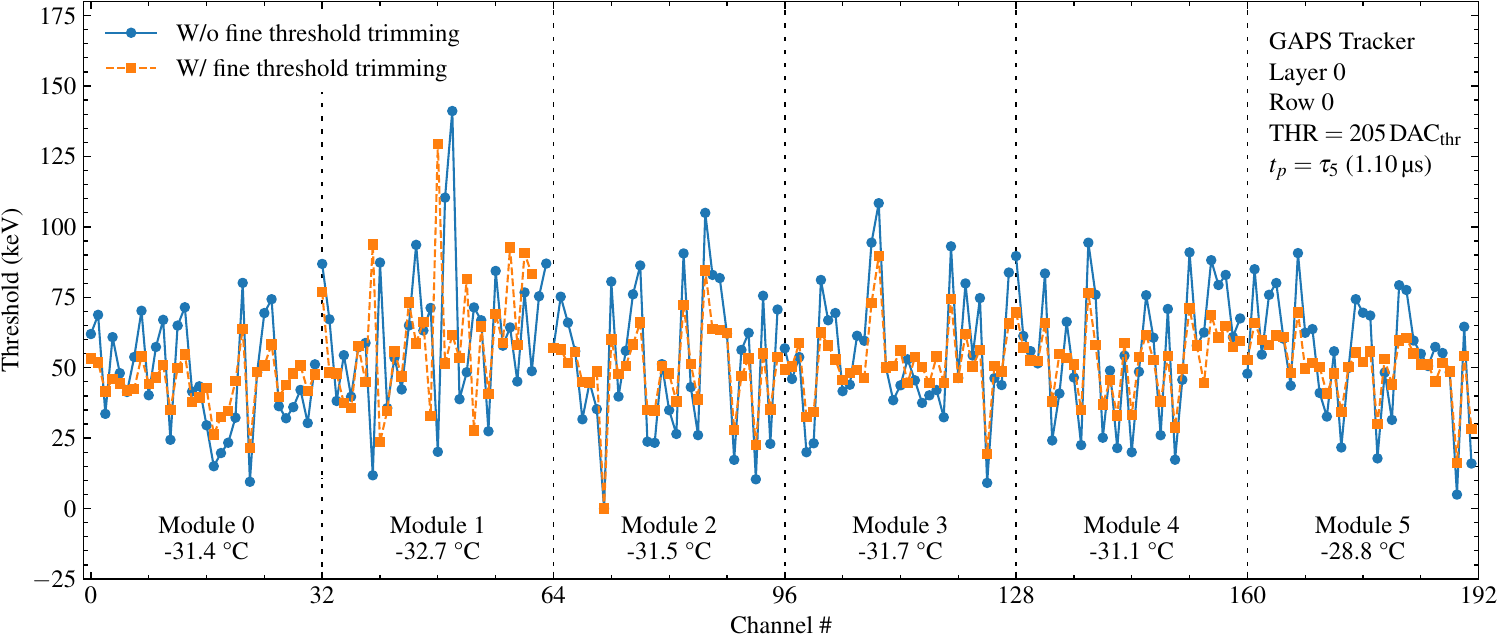}
    \caption{Self-trigger energy thresholds for all 192 channels of Layer~0, Row~0, Modules~0--5 before and after fine threshold trimming for a global threshold setting of $\SI{205}{DAC_{thr}}$. Each vertical band corresponds to one module, with its operating temperature reported in the legend. Fine threshold trimming significantly reduces the spread in thresholds across channels. The two curves represent data before and after trimming, respectively, and are reported without parasitic injection compensation.}
    \label{fig:ch3-L0R0M012345_thr_channels}
\end{figure}

\par
As shown in~\Cref{fig:ch3-parinj_time_hist}, the acquired background spectrum for Channel~8 of Layer~2, Row~1, Module~5 displays two distinct populations: a main cluster centered at approximately \SI{105.95}{ADU}, consistent with the pedestal mean obtained from a dedicated pedestal sampling run (\SI{105.85}{ADU}), and a second population around \SI{142.06}{ADU}, corresponding to repeated triggers caused by the low threshold of this specific channel.
From charge scan analysis, this channel had a nominal energy threshold of \SI{27.8}{\kilo\electronvolt} (with fine threshold trimming applied) and an energy resolution of \SI{4.55}{\kilo\electronvolt} FWHM.
However, after applying parasitic injection compensation, the threshold dropped to \SI{10.50}{\kilo\electronvolt}---well below the intended \SI{20}{\kilo\electronvolt} minimum threshold.
According to the hot channel suppression criteria described above, such channels were masked.
The behavior is clearly illustrated in both the time-domain (\Cref{fig:ch3-parinj_time}) and histogram (\Cref{fig:ch3-parinj_hist}) plots, which show the coexistence of legitimate pedestal events and spurious threshold-associated triggers.

%   Fine threshold trimming
\subsection{Fine threshold trimming}

\par
Since the global discriminator threshold is defined by a single eight-bit DAC shared across all channels within the ASIC, it cannot compensate for intrinsic channel-to-channel variations, such as those introduced by CMOS process mismatch~\cite{rattiOptimumDesignDACs2012}.  
This global DAC provides a resolution of \SI{1.6}{\milli\volt}, corresponding to an energy step of approximately \SI{8.83}{\kilo\electronvolt}.  
To overcome the limitation of a common threshold and reduce dispersion, each readout channel is equipped with a dedicated \hbox{3-bit} fine threshold trimming DAC, which allows local adjustment in steps of \SI{5}{\kilo\electronvolt} over a range of about \SI{40}{\kilo\electronvolt}~\cite{manghisoni32channelsReadoutASIC2023}.  
This \hbox{per-channel} trimming stage enables fine corrections around the global threshold, effectively minimizing dispersion relative to the common setting.

\par
\begin{figure}[tbp]
    \centering
    \includegraphics[width=0.75\textwidth]{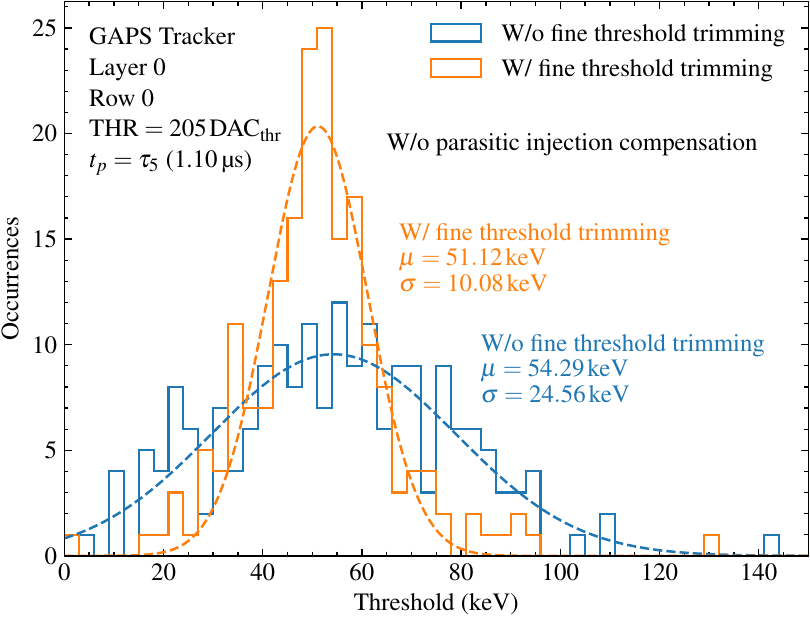}
    \caption{Histogram of the self-trigger energy thresholds for all modules of Layer~0, Row~0, Modules~0--5 (192 channels), before and after fine threshold trimming. Thresholds are extracted without applying parasitic injection compensation. The application of the three-bit fine threshold trimming reduces the standard deviation from \SI{24.56}{\kilo\electronvolt} to \SI{10.08}{\kilo\electronvolt}, confirming the effectiveness of the trimming procedure in reducing inter-channel threshold dispersion. The histogram bin width is \SI{3}{\kilo\electronvolt}.}
    \label{fig:ch3-400_thr_distribution_fthr}
\end{figure}

Figure~\ref{fig:ch3-400_thr_distribution_fthr} shows the distribution of self-trigger thresholds for all 192 channels of Layer~0, Row~0, Modules~0--5, both before and after applying fine threshold trimming.
The adjustment was performed without applying any parasitic injection compensation, in order to isolate the effect of the trimming procedure.
As expected, the trimming procedure significantly reduced threshold dispersion.
Prior to correction, the distribution followed a Gaussian shape with a standard deviation of \SI{24.56}{\kilo\electronvolt}, consistent with ASIC-level characterization.
After trimming, the spread was reduced to \SI{10.08}{\kilo\electronvolt}, confirming the effectiveness of the procedure in compensating for channel-level threshold variations.
The histogram bin width is \SI{3}{\kilo\electronvolt}.

\par
Figure~\ref{fig:ch3-L0R0M012345_thr_channels} provides a channel-by-channel visualization of the threshold values across all six modules of Layer~0, Row~0, before and after fine threshold trimming.
The plot highlights significant variation in untrimmed thresholds, both between modules and among channels within each module. 
The fine-trimming procedure reduces the threshold dispersion by approximately \SI{60}{\percent}, resulting in a narrower threshold distribution centered around \SI{\sim50}{\kilo\electronvolt} for a global threshold setting of $\SI{205}{DAC_{thr}}$, even without parasitic injection compensation.
Although the latter introduces a systematic shift of the threshold distribution towards lower energies, it does not affect the channel-to-channel dispersion, as the parasitic contribution is largely uniform across channels within the same module, as shown in~\Cref{fig:ch3-400_thresholds_parinj}.

\par
These results are consistent with those from the ASIC validation campaign, in which threshold dispersion was evaluated across 9920 channels from 310 flight-selected ASICs at room temperature.
In that study, the untrimmed threshold distribution exhibited a standard deviation of \SI{25.33}{\kilo\electronvolt}, which was reduced to \SI{14.33}{\kilo\electronvolt} after fine trimming~\cite{manghisoni32channelsReadoutASIC2023}.
While slight differences in final dispersion are expected due to the reduced channel statistics and the presence of detector capacitance and system-level effects, the integrated tracker results confirm the expected performance of the fine-trimming procedure.
Despite the smaller number of channels in the tracker dataset, the measured threshold dispersion is consistent with the ASIC-level results obtained at room temperature.
This agreement is attributed to the fact that threshold dispersion primarily originates from mismatch in a pair of transistors within the shaping stage, a phenomenon that is largely insensitive to temperature variations~\cite{mennilloAnalysisTemperatureImpact2009}.

\begin{figure}[tbp]
    \centering
    \includegraphics[width=0.75\textwidth]{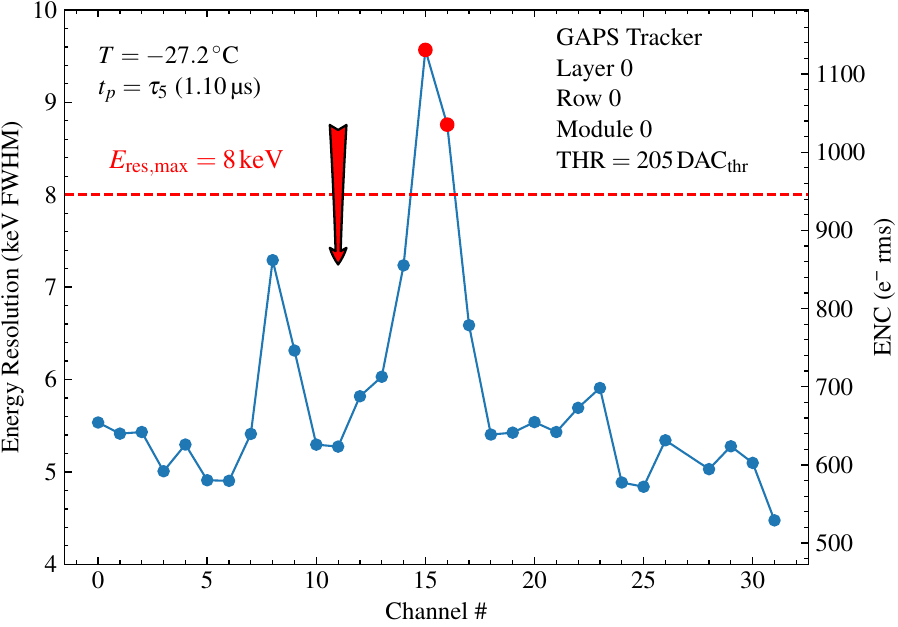}
    \caption{Energy resolution FWHM and ENC for all 32 channels of Layer~4, Row~0, Module~0, measured at \SI{-27.2}{\celsius} with a global threshold setting of $\SI{205}{DAC_{thr}}$ and a peaking time $\tau_5 = \SI{1.10}{\micro\second}$. The horizontal dashed line at \SI{8}{\kilo\electronvolt} indicates the upper limit for acceptable resolution in x-ray detection mode.}
    \label{fig:ch3-enc_channels}
\end{figure}

\par
The fine-trimming procedure was systematically applied to all modules of the GAPS tracker.
For each module, the optimal fine threshold setting was determined by scanning multiple configurations and selecting the one that minimized the threshold dispersion across its 32 channels.
The resulting 3-bit fine threshold trimming settings are now included in the configuration of the flight software.

\begin{figure}[tbp]
    \centering
    \includegraphics[width=\textwidth]{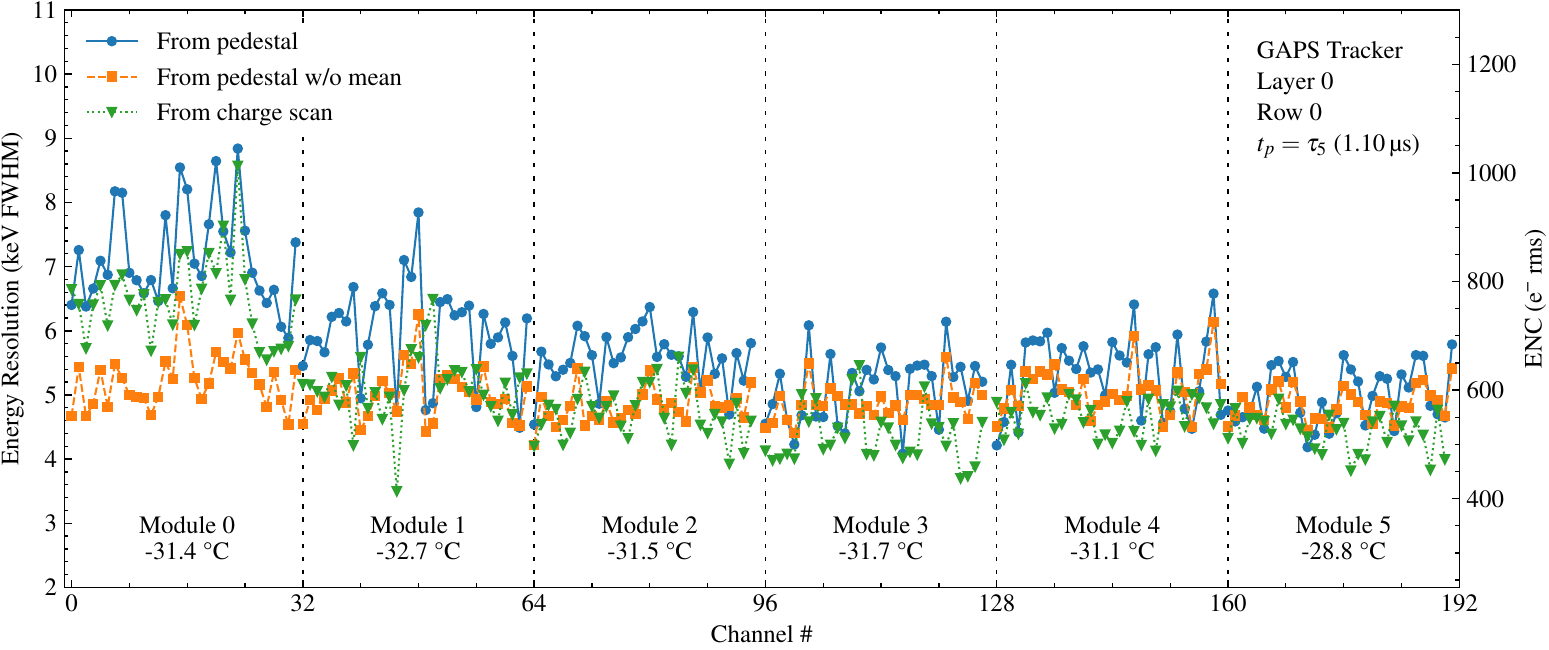}
    \caption{Comparison of energy resolution FWHM and ENC for all 192 channels of Layer~0, Row~0, Modules~0--5, obtained using three different methods: from pedestal without \hbox{common-mode} subtraction (blue circles), from pedestal with \hbox{module-level} \hbox{common-mode} subtraction (orange squares), and from charge scan (green triangles). The improvement in noise estimation after removing \hbox{common-mode} noise is evident.}
    \label{fig:ch3-L0R0M012345_enc_channels}
\end{figure}

%-------------------------------------------------------------------------------
%   Energy resolution and ENC evaluation
%-------------------------------------------------------------------------------
\section{Energy resolution and ENC evaluation}

\par
In addition to threshold dispersion and parasitic injection effects, energy resolution and electronic noise were also evaluated to identify hot channels.
Figure~\ref{fig:ch3-enc_channels} shows the channel-by-channel ENC and corresponding FWHM energy resolution for Layer~4, Row~0, Module~0, measured under the same configuration used for threshold characterization.
An energy resolution upper limit of \SI{8}{\kilo\electronvolt} was defined as the benchmark for acceptable performance in x-ray detection mode.
Channels exceeding this threshold were classified as hot and deactivated.

\par
The ENC and energy resolution values shown in~\Cref{fig:ch3-enc_channels} were extracted using the pedestal sampling method described in~\Cref{ch2-module_testing}, originally developed for \hbox{single-module} testing under laboratory conditions.
However, in this case, rather than performing common-mode subtraction within each detector group (i.e., across the eight channels connected to the same Si(Li) sensor), the correction was applied at the module level by computing the common-mode component across all 32 channels of the module.
This approach was motivated by the need to identify broader noise patterns potentially affecting the full module, which may not be visible at the single-sensor level, and by the observation that the pedestal analysis did not reveal any consistent sensor-based correlation patterns.
The corrected pedestal standard deviation was then divided by the gain of each channel---obtained from a linear interpolation of the low-energy portion of the input--output transcharacteristic---to obtain the ENC.
The resulting energy resolution was calculated using~\eqref{eq:ch2-enc_pedestal}.

\begin{figure}[tbp]
    \centering
    \captionsetup[subfigure]{justification=justified,singlelinecheck=false}
    \begin{subfigure}{0.49\textwidth}
        \centering
        \includegraphics[width=\linewidth]{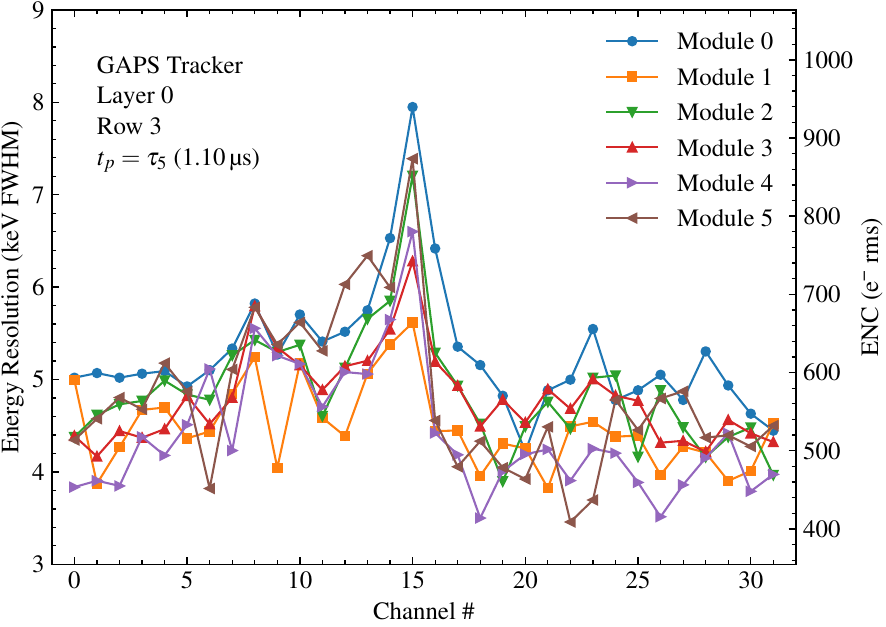}
        \caption{Channel-by-channel energy resolution FWHM and ENC for all six modules of Layer~0, Row~3. Channels 14 to 16 show systematically higher noise levels, a pattern that repeats across all modules in this row. No \hbox{common-mode} subtraction was applied.}
        \label{fig:ch3-enc_L0R3_allmodules}
    \end{subfigure}
    \hfill
    \begin{subfigure}{0.49\textwidth}
        \centering
        \includegraphics[width=\linewidth]{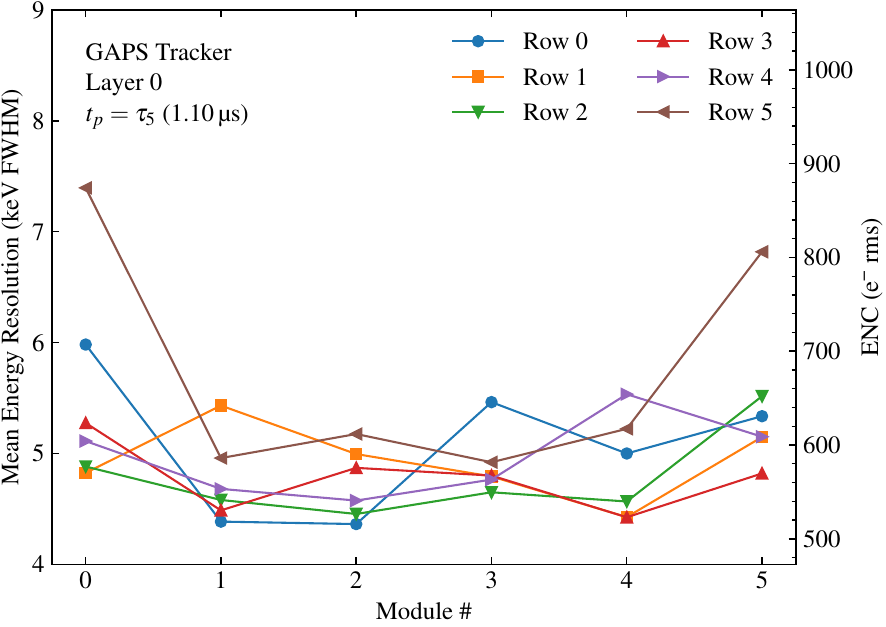}
        \caption{Mean energy resolution FWHM and ENC evaluated across Rows~0--5 of Layer~0. The increase observed at the tracker edges is consistent with the higher temperatures measured on outer modules due to weaker thermal insulation.}
        \label{fig:ch3-enc_L0_allrows}
    \end{subfigure}
    \caption{Noise performance in Layer~0 of the tracker. (\subref{fig:ch3-enc_L0R3_allmodules}) shows the energy resolution and ENC per channel for Row~3, while (\subref{fig:ch3-enc_L0_allrows}) compares the average values across all rows. Edge modules systematically show higher ENC due to thermal gradients.}
    \label{fig:ch3-enc_summary}
\end{figure}

\par
When combined with threshold analysis, these results help refine the classification of hot channels.
For example, channels with low effective thresholds but poor resolution---due to elevated leakage current or local noise pickup---may fire excessively without contributing meaningful data.
In Module~0, channels such as 15 and 16 do not fall below the \SI{20}{\kilo\electronvolt} threshold limit but exhibit elevated ENC and resolution values.
These channels were confirmed as hot and excluded from the final configuration to reduce noise occupancy and preserve bandwidth for valid events.

\par
Figure~\ref{fig:ch3-L0R0M012345_enc_channels} compares the ENC and corresponding FWHM energy resolution across all 192 channels of Layer~0, Row~0, Modules~0--5, as evaluated using three complementary methods.
The first method (blue line) involved extracting the ENC directly from the pedestal distribution without applying any common-mode noise subtraction.
In this case, the measured noise includes both the intrinsic channel fluctuations and any baseline shifts common to the full module.
A second evaluation (orange dashed line) was performed by applying module-level common-mode subtraction, where the average pedestal value across all 32 channels of each module is removed from each sample.
This approach isolates the stochastic noise component by suppressing coherent noise sources that affect the entire module, resulting in a systematically lower ENC estimate.

\par
The third method (green dotted line) involved deriving the ENC from the charge scan procedure by fitting the channel trigger efficiency curve with a Gaussian error function.
The slope of this turn-on curve provides a direct measure of the input-referred noise and is independent of pedestal-based methods.
The comparison demonstrates that common-mode subtraction significantly reduces the estimated ENC and brings it into closer agreement with the values obtained from charge scan analysis.
The agreement between these independent techniques reinforces the validity of both approaches and underscores the necessity of common-mode correction in characterizing the tracker's electronic noise performance.

\par
An interesting feature emerges from the comparison: with the exception of Module~0, the ENC values derived from the charge scan are nearly identical to those obtained from pedestal analysis after common-mode subtraction.  
This suggests that the residual excess noise observed in the pedestal-based evaluation may be associated with subsequent stages of the readout chain, such as the sample-and-hold (S\&H) circuit or the ADC.

\par
Two additional figures provide a more detailed view of noise patterns across the tracker, both at the local (row) and global (layer) levels.  
Figure~\ref{fig:ch3-enc_L0R3_allmodules} shows the ENC and energy resolution, evaluated channel-by-channel for all six modules of Layer~0, Row~3.
This row exhibits a systematic behavior in which channels 14 to 16 consistently show increased noise levels. % associated with longer PCB traces
The pattern is repeated across all modules in the row, suggesting that trace length and related parasitic coupling may significantly impact noise performance.
Specifically, no common-mode noise subtraction was applied.

\begin{figure}[t!]
    \centering
    \includegraphics[width=0.78\textwidth]{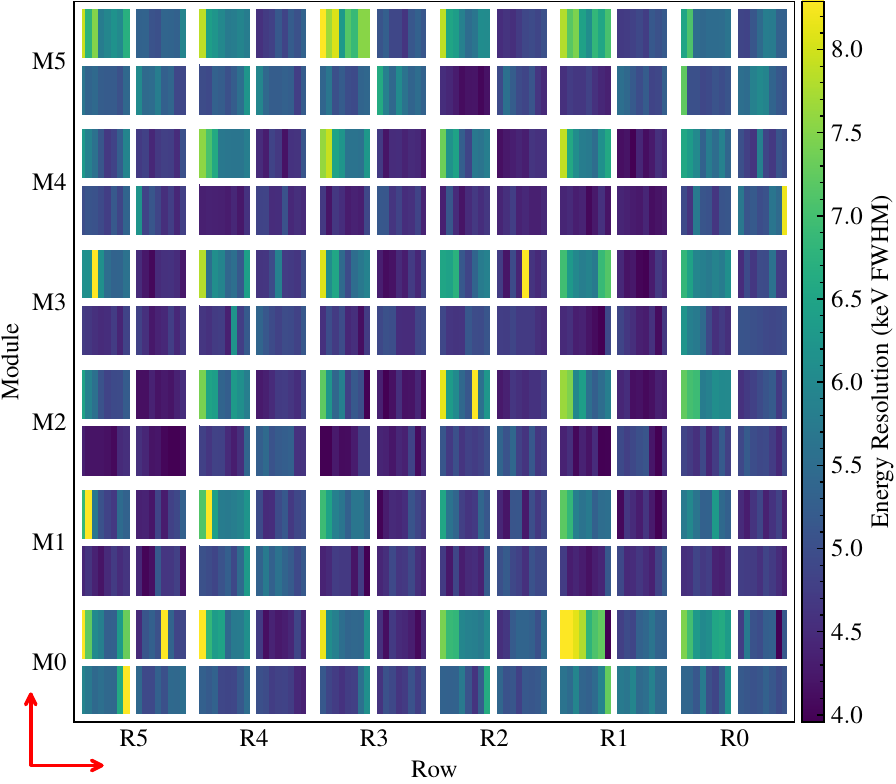}
    \caption{Energy resolution FWHM map for all readout strips in Layer~0 of the GAPS tracker. The layout reproduces the physical arrangement of the tracker, as shown in~\Cref{fig:ch3-gaps_ssl}: each of the six rows (R0--R5) contains six modules (M0--M5), and every module includes four detectors, each with eight vertically oriented strips. Energy resolution values are extracted from the charge scan procedure by fitting the trigger efficiency curve.}
    \label{fig:ch3-layer0_enc_map}
\end{figure}

\par
Figure~\ref{fig:ch3-enc_L0_allrows} presents the mean energy resolution FWHM and ENC across Rows~0--5 of Layer~0, plotted as a function of module position within each row.
A systematic increase in ENC is observed at the edges of the tracker, attributed to temperature gradients.
Edge modules are less thermally insulated from the environment and therefore tend to operate at higher temperatures.
Since electronic noise---particularly the parallel noise contributions from detector leakage current and white noise---is strongly temperature-dependent, this spatial profile explains the elevated ENC measured in peripheral modules.
In particular, Row~5 exhibits a notable increase in noise, most likely due to higher operating temperatures.
Furthermore, the modules located at the two extremities of Row~5 show the highest ENC values within the row, consistent with their correspondingly higher measured temperatures.

\par
\begin{figure}[t!]
    \centering
    \includegraphics[width=0.75\textwidth]{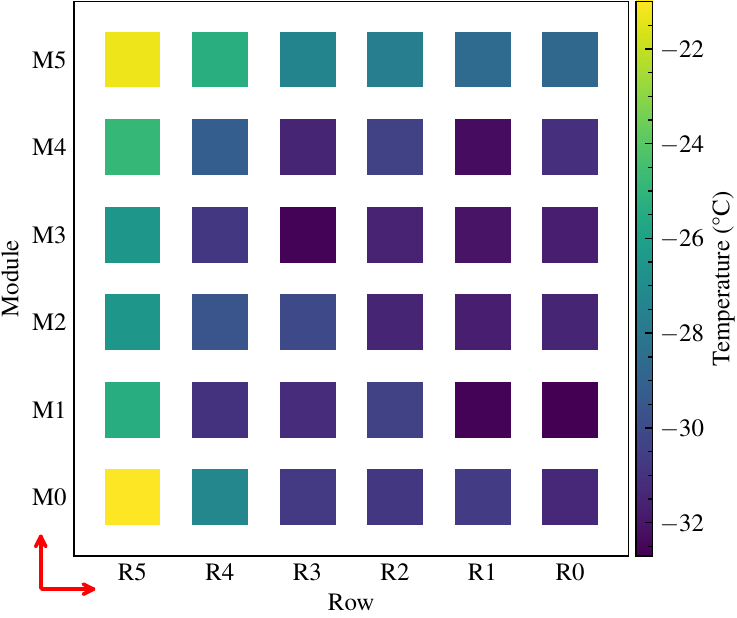}
    \caption{Measured temperature map of the modules in Layer~0 of the GAPS tracker. The values are extracted from the temperature sensors mounted on each FEB and read out by the SLIDER32 ASIC. A clear gradient is visible from right to left, with the lowest temperatures near the radiator (right side) and progressively higher values farther away. Higher temperatures are also observed at the row ends along the layer edges, where thermal insulation is less effective.}
    \label{fig:ch3-layer0_temperature_map}
\end{figure}

Figure~\ref{fig:ch3-layer0_enc_map} provides a comprehensive overview of the energy resolution FWHM across all readout strips of Layer~0, displayed in a layout that mirrors the physical geometry of the tracker.
Each module includes four detectors with vertically oriented strips, and modules are arranged in six rows (R0--R5) and six columns (M0--M5).
The energy resolution values, expressed in \si{\kilo\electronvolt} FWHM, were derived from the charge scan procedure by fitting the trigger efficiency curve.
Consistent with the results shown in~\Cref{fig:ch3-enc_L0_allrows}, higher noise levels are again evident at the outer edges of the layer, corresponding to regions subject to higher temperatures due to weaker thermal insulation, as illustrated by the measured temperature distribution in~\Cref{fig:ch3-layer0_temperature_map}.

%-------------------------------------------------------------------------------
%   Hot channel masking
%-------------------------------------------------------------------------------
\section{Hot channel masking}

\par
Following the evaluation of energy thresholds, parasitic injection effects, and noise performance, a masking strategy was implemented to exclude hot channels from the acquisition and event reconstruction process.

\par
A channel was classified as hot if it satisfied one or more of the following conditions: an effective self-trigger energy threshold, after parasitic injection compensation, below the minimum threshold limit of \SI{20}{\kilo\electronvolt}; an energy resolution FWHM exceeding \SI{8}{\kilo\electronvolt} or a non-functional or abnormal response during charge scan.
These criteria were applied uniformly across the full tracker. For each module, a 32-bit bitmask was generated to represent the status of its channels, with each bit set to 0 for masked (disabled) channels.
These bitmasks were incorporated into the tracker configuration files and are now used by the flight control software to enforce masking during self-trigger operation.

\begin{figure}[t!]
    \centering
    \includegraphics[width=0.78\textwidth]{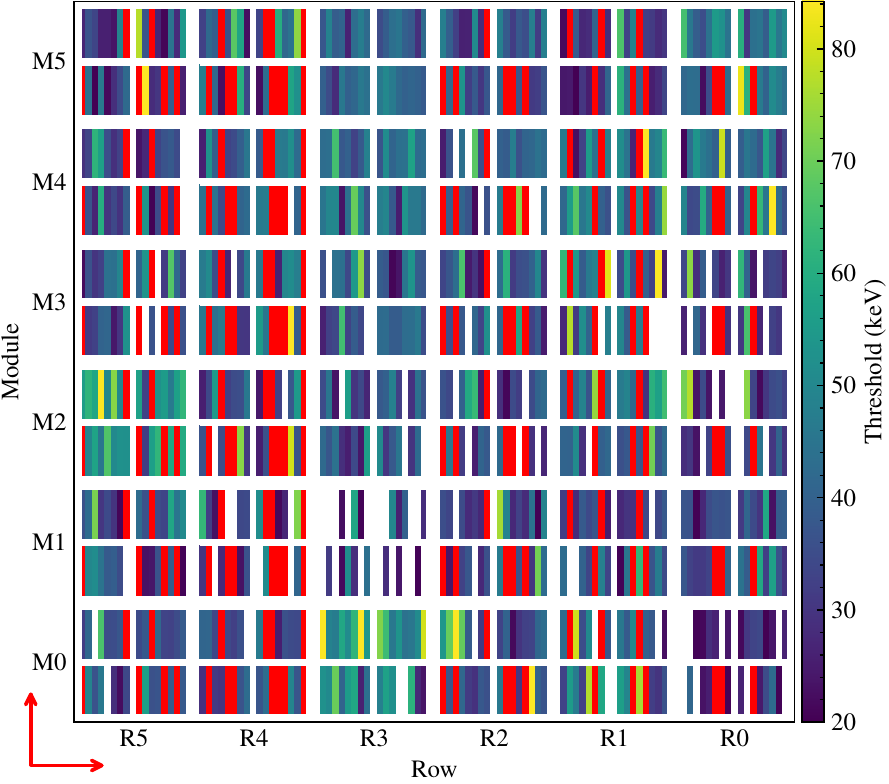}
    \caption{Self-trigger energy threshold map for all readout strips in Layer~0 of the GAPS tracker, extracted from the charge scan procedure. Hot channels are shown in red, while blank (missing) strips correspond to non-functional channels that did not respond to charge injection.}
    \label{fig:ch3-layer0_thr_map}
\end{figure}

\par
As an example, the hot channel mask for Layer~4, Row~0, Module~0 is \texttt{0x75FE7FDF}.
This corresponds to six masked channels: 5, 15, 16, 25, 27, and 31. 
Channel~27 was deactivated due to its anomalous behavior, consistently remaining above threshold and therefore unsuitable for charge injection measurements.
Channels~15 and~16 were masked due to energy resolution values exceeding the \SI{8}{\kilo\electronvolt} imposed limit (\Cref{fig:ch3-charge_scan_comp}). 
Channels~5, 25, and~31 were disabled because their corrected energy thresholds, after parasitic injection compensation, fell below the minimum threshold of \SI{20}{\kilo\electronvolt} (\Cref{fig:ch3-enc_channels}).

\par
Figure~\ref{fig:ch3-layer0_thr_map} shows the self-trigger energy threshold map for all 1152 readout strips in Layer~0 of the GAPS tracker, obtained from the charge scan procedure.  
Strips marked in red correspond to hot channels, which were deactivated either because their thresholds fell below the \SI{20}{\kilo\electronvolt} limit or their energy resolution exceeded \SI{8}{\kilo\electronvolt}.  
Blank (missing) strips indicate non-functional channels that did not respond to charge injection.  
In total, 192 strips were masked as hot or non-functional, corresponding to approximately \SI{16.7}{\percent} of Layer~0.

\par
This evaluation was performed during on-ground calibration, where modules operated at higher temperatures than those expected during flight, as shown in~\Cref{fig:ch3-layer0_temperature_map}.
The measured temperature map indicates a clear gradient across Layer~0, with the lowest temperatures (\SI{-32.7}{\celsius}) recorded near the radiator on the right-hand side and progressively higher values---reaching up to \SI{-21}{\celsius}---observed toward the left edge, due to the increasing distance from the radiator~\cite{okazakiPerformanceCircularOscillating2021}.
Given the strong temperature dependence of electronic noise---particularly the parallel noise component arising from leakage current and white noise---this assessment was intentionally conservative.

\begin{figure}[t!]
    \centering
    \includegraphics[width=0.75\textwidth]{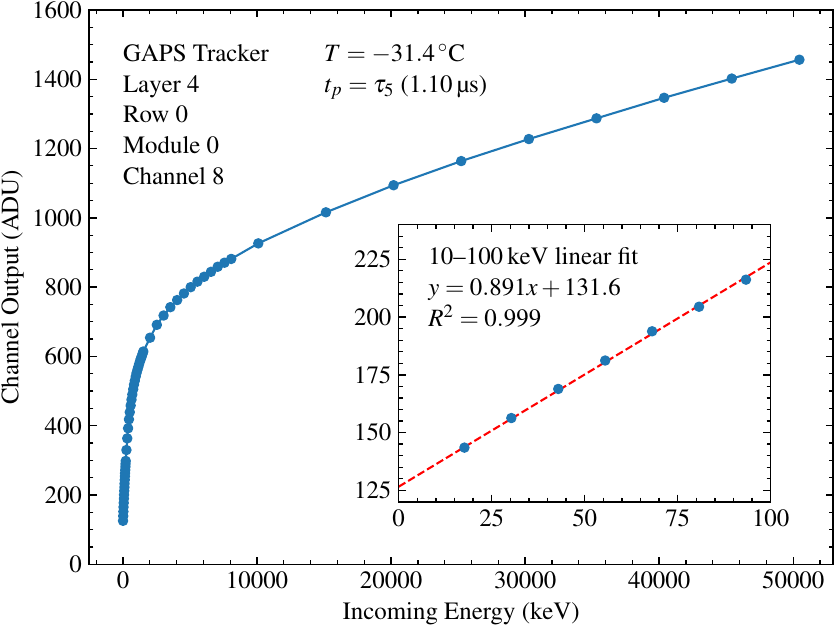}
    \caption{Measured input--output transcharacteristic for Channel~8 of Layer~4, Row~0, Module~0, acquired at \SI{-31.4}{\celsius} with peaking time $\tau_5$ (\SI{1.10}{\micro\second}). The transcharacteristic, reconstructed from 55 injection steps, highlights the bilinear nature of the response, with a distinct kink separating the low-energy and \hbox{high-energy} linear regions. A linear fit is overlaid in the \SIrange{10}{100}{\kilo\electronvolt} range, relevant for x-ray calibration.}
    \label{fig:ch3-400_fdt_linear-fit}
\end{figure}

\par
While some currently masked channels may meet acceptance criteria under nominal flight conditions, the correlation between temperature and hot channel incidence is evident.
The modules with the highest number of hot channels in~\Cref{fig:ch3-layer0_thr_map} correspond to the warmer regions identified in~\Cref{fig:ch3-layer0_temperature_map}.
Notably, Row 5 again stands out, exhibiting both the highest ENC values and the warmest peripheral modules, as also reported in~\Cref{fig:ch3-enc_L0_allrows}.

\par
Applying this masking procedure ensures that only channels with acceptable noise and threshold are included in the acquisition, reducing noise-induced events and improving overall data quality.
The procedure was validated across all tested modules and is now part of the standard for flight configuration.

%-------------------------------------------------------------------------------
%   Channel input--output transcharacteristic calibration
%-------------------------------------------------------------------------------
\section{Channel input--output transcharacteristic calibration}\label{sec:ch3-fdt_calibration}

\par
This section describes the procedure developed to calibrate the channel input--output transcharacteristic, enabling the conversion of acquired energy deposits from ADC units (ADU) to \si{\kilo\electronvolt}.
The calibration is based on the transcharacteristic introduced in~\Cref{chapter2} and is obtained through a dedicated charge injection process.
This process involves injecting known amounts of charge---corresponding to energies ranging from \SI{10}{\kilo\electronvolt} to approximately \SI{55}{\mega\electronvolt}---in 55 discrete steps using the on-chip charge injection circuit and a dedicated 16-bit calibration DAC mounted on the FEB.

\begin{figure}[tbp]
    \centering
    \includegraphics[width=0.75\textwidth]{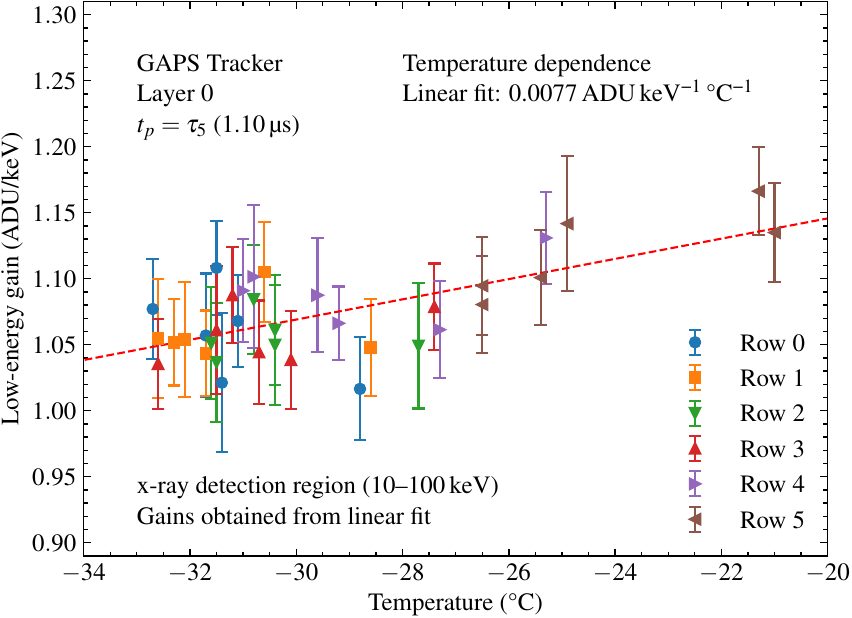}
    \caption{Mean module low-energy gain as a function of module temperature for all modules in Layer~0. The gain values are extracted by interpolation in the \SIrange{10}{100}{\kilo\electronvolt} range, corresponding to the low-energy x-ray detection region. Each point represents the average gain of a module, with vertical error bars indicating the spread among its channels. A linear fit (red dashed line) reveals a temperature dependence of \SI{0.0077}{ADU\,\kilo\electronvolt^{-1}\,\celsius^{-1}}, consistent with previous ASIC measurements.}
    \label{fig:ch3-gain_vs_temperature}
\end{figure}

\par
Although the full DAC range nominally allows 65536 discrete steps, a representative subset of 55 values was selected to balance coverage of the transcharacteristic with reasonable acquisition times.
Each injection point is sampled 1000 times, and the mean ADC response is computed to minimize the impact of spurious variations during injection and sampling.
For each module, the complete procedure requires about \SI{60}{\second} for a single peaking time setting.

\par
For the on-ground calibration of the GAPS Si(Li) tracker, two energy regions are of particular interest for energy reconstruction: the \SIrange{10}{100}{\kilo\electronvolt} range, relevant for x-ray acquisitions using radioactive sources such as Americium-241 (used in laboratory tests, see~\Cref{sec:ch2-americium}) and Cadmium-109 (used during full tracker calibration), and the \SI{500}{\kilo\electronvolt} to \SI{1}{\mega\electronvolt} range, corresponding to energy deposits from MIPs, such as cosmic muons at ground level.

\par
The injection steps are spaced non-uniformly, with a higher density of points in the ``low-energy'' region (up to about \SI{2}{\mega\electronvolt}) to resolve the curvature of the transcharacteristic near the ``kink'' where the linear regime transitions from x-ray to high-energy particle detection.
An example transcharacteristic is shown in~\Cref{fig:ch3-400_fdt_linear-fit} for Channel~8 of Module~0, Row~0, Layer~4, which is used here as a reference for illustrative purposes.
The plot also includes a linear fit in the \SIrange{10}{100}{\kilo\electronvolt} region, demonstrating the high linearity of the response in the low-energy x-ray detection range.

\begin{figure}[tbp]
    \centering
    \includegraphics[width=0.75\textwidth]{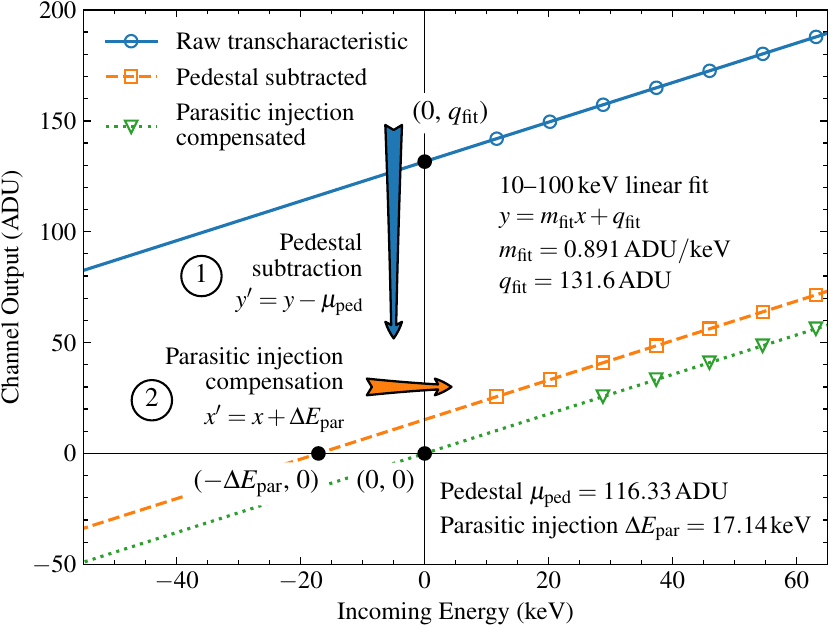}
    \caption{Graphical representation of the two-step calibration applied to the channel transcharacteristic, based on a linear fit in the \SIrange{10}{100}{\kilo\electronvolt} region. Step~(1) subtracts the pedestal value $\mu_{\text{ped}}$ from all ADC values to align the baseline. Step~(2) compensates for parasitic injection by shifting the energy axis by $\Delta E_{\text{par}}$, correcting the systematic overestimation caused by parasitic capacitance in the charge injection circuit. The final corrected curve exhibits a linear response intersecting the origin in the low-energy region. The original raw data (open circles), pedestal-subtracted curve (open squares), and fully corrected transcharacteristic (open triangles) are shown.}
    \label{fig:ch3-fdt_calibration}
\end{figure}

\par
A clear temperature dependence of the low-energy gain is observed when averaging across all channels of Layer~0, as shown in~\Cref{fig:ch3-gain_vs_temperature}.
Each point represents the mean gain of a module, extracted from a linear fit in the \SIrange{10}{100}{\kilo\electronvolt} range, with vertical error bars spanning the gain range across its channels.
The data reveal an increase in gain with temperature, with a fitted slope of \SI{0.0077}{ADU\,\kilo\electronvolt^{-1}\,\celsius^{-1}}.

\par
The temperature coefficient $\mathrm{TC}_{G_{\text{le}}}$, calculated according to

\begin{equation}
    \mathrm{TC}_{G_{\text{le}}} = 
    \left[
    \frac{G_{\text{le},\text{max}} - G_{\text{le},\text{min}}}{T_{\text{max}} - T_{\text{min}}}
    \frac{10^6}{G_{\text{le}}(-40\,^{\circ}\mathrm{C})}
    \right]
    \quad \left(\frac{\mathrm{ppm}}{^{\circ}\mathrm{C}}\right)
    \label{eq:ch3-tc_gain}
\end{equation}

\noindent
with reference to \SI{-40}{\celsius}, was found to be \SI{7723}{ppm/\celsius}, corresponding to a relative variation of \SI{0.77}{\percent} per degree Celsius. 
Over the observed temperature range of \SIrange{-32.7}{-21.0}{\celsius}, this results in a total gain change of about \SI{8.6}{\percent}. 
This behavior is consistent with previous ASIC measurements and confirms the expected temperature dependence of the analog front-end~\cite{ghislottiCharacterisationReadoutElectronics2022}.

\par
In the higher-energy range, a cubic interpolation is employed to derive a continuous function for converting ADU to \si{\kilo\electronvolt}.
Before either interpolation can be applied, however, the raw transcharacteristic must undergo a two-step correction process, as illustrated in~\Cref{fig:ch3-fdt_calibration}.
The first correction step, labeled (1) in~\Cref{fig:ch3-fdt_calibration}, adjusts the baseline of the transcharacteristic using the pedestal value $\mu_{\text{ped}}$ measured through a dedicated acquisition and Gaussian fitting procedure.
This accounts for the baseline subtraction applied to all energy events and involves shifting the transcharacteristic along the \hbox{$y$-axis} by subtracting $\mu_{\text{ped}}$ from all ADU values.
If no systematic errors are present, this adjustment would align the transcharacteristic such that the intercept of the linear fit in the x-ray range matches the pedestal.

\begin{figure}[tbp]
    \centering
    \includegraphics[width=0.75\textwidth]{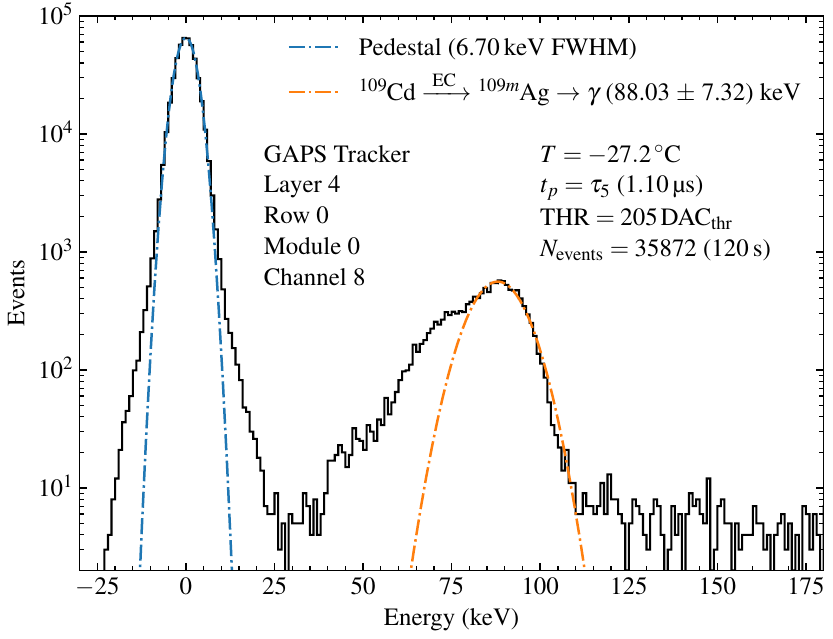}
    \caption{Energy spectrum acquired in self-trigger mode using a \ce{^{109}Cd} source, with the global threshold set to \SI{205}{DAC_{thr}} on Channel~8 of Module~0, Row~0, Layer~4. The spectrum shows the pedestal peak and a photopeak associated with the internal conversion decay of \ce{^{109}Cd} to \ce{^{109m}Ag}, with the main $\gamma$ emission observed at $(88.03 \pm 7.32)\,\si{\kilo\electronvolt}$. The acquisition was performed with the module at \SI{-27.2}{\celsius} over a 120-second interval with a peaking time of $\tau_5 = \SI{1.10}{\micro\second}$. The histogram bin width is \SI{1}{\kilo\electronvolt}.}
    \label{fig:ch3-cadmium109_ch8_spectrum}
\end{figure}

\par
However, in practice, the intercept $q_{\text{fit}}$ of the fitted line is consistently greater than $\mu_{\text{ped}}$, due to the parasitic injection effects discussed in~\Cref{sec:ch3-par_inj}.
This discrepancy arises from parasitic capacitance in the on-chip charge injection circuit, leading to a systematic overestimation of the injected charge.
To correct for this, a second compensation step (labeled (2) in~\Cref{fig:ch3-fdt_calibration}) is applied: the $x$-axis is adjusted by shifting each energy point by the parasitic injection offset $\Delta E_{\text{par}}$.

\par
The result of this two-step process is a calibrated transcharacteristic with a linear response passing through the origin in the low-energy region, that is, with intercept at $(0, 0)$.
This correction is essential for accurate cubic interpolation in the MIP region, although it has negligible impact on x-ray gain estimation since the slope remains unaffected.

%-------------------------------------------------------------------------------
%   X-ray and cosmic muon detection
%-------------------------------------------------------------------------------
\section{X-ray and cosmic muon detection}

\begin{figure}[tbp]
    \centering
    \includegraphics[width=0.75\textwidth]{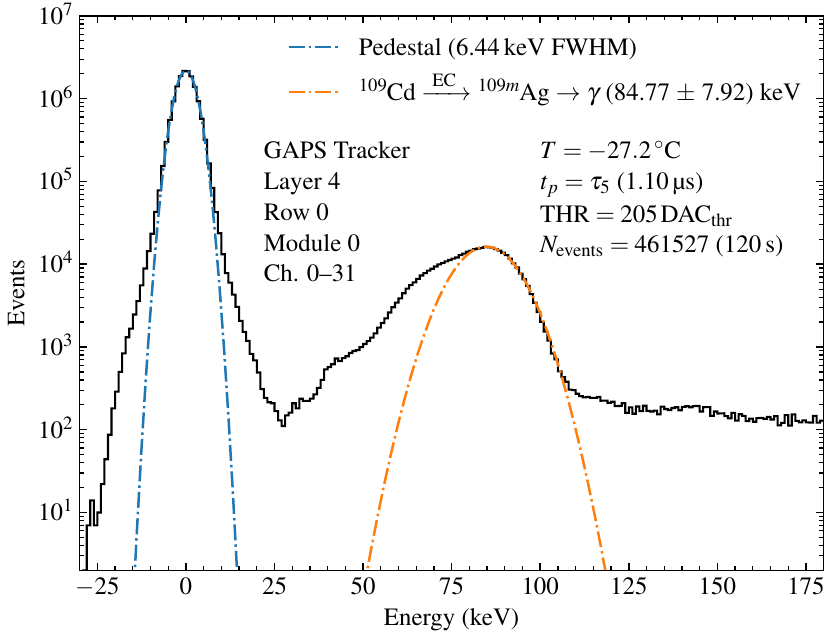}
    \caption{Energy spectrum acquired in self-trigger mode using a \ce{^{109}Cd} source, with the global threshold set to \SI{205}{DAC_{thr}} on Channels~0--31 of Module~0, Row~0, Layer~4. The spectrum shows the pedestal peak and a photopeak associated with the internal conversion decay of \ce{^{109}Cd} to \ce{^{109m}Ag}, with the main $\gamma$ emission observed at $(84.77 \pm 7.92)\,\si{\kilo\electronvolt}$. The acquisition was performed with the module at \SI{-27.2}{\celsius} over a 120-second interval with a peaking time of $\tau_5 = \SI{1.10}{\micro\second}$. The histogram bin width is \SI{1}{\kilo\electronvolt}.}
    \label{fig:ch3-cadmium109_allch_spectrum}
\end{figure}

\par
This section presents representative acquisitions performed with the GAPS Si(Li) tracker operating in self-trigger mode, targeting two distinct physical processes: low-energy x-ray interactions from a calibrated radioactive source and high-energy deposits from atmospheric muons~\cite{heyburnCommissioningCMSPixel2010}.
These tests validate the full readout chain and demonstrate the tracker's ability to resolve and reconstruct both signal types under different operating conditions.

\par
An initial acquisition was performed using a \ce{^{109}Cd} radioactive source~\cite{hutchinsonStandardizationCadmium109Sources1971} to validate the tracker's x-ray response.
The first dataset was collected from Channel~8 of Module~0, Row~0, Layer~4, over a 120-second interval with a peaking time setting of $\tau_5 = \SI{1.10}{\micro\second}$ and a global threshold of \SI{205}{DAC_{thr}}.
As shown in~\Cref{fig:ch3-cadmium109_ch8_spectrum}, the resulting energy spectrum exhibits a well-resolved pedestal peak and a photopeak at $(88.03 \pm 7.32)\,\si{\kilo\electronvolt}$, corresponding to the internal conversion decay of \ce{^{109}Cd} to \ce{^{109m}Ag}.

\par
To improve statistics and assess the response of the full module, a second acquisition was performed using the same configuration, this time recording from all 32 channels.
The combined spectrum, shown in~\Cref{fig:ch3-cadmium109_allch_spectrum}, also features the pedestal and a photopeak at $(84.77 \pm 7.92)\,\si{\kilo\electronvolt}$.
In both cases, the data were baseline corrected by subtracting the individual pedestal values for each channel.

\par
Although the measured peak is slightly lower than the nominal value, this deviation can be attributed to two factors: the inclusion of channels with slightly different gain calibration and the elevated module temperature of \SI{-27.2}{\celsius} during the acquisition, which was higher than the nominal operating temperature of \SI{-40}{\celsius}.
Taken together, these results confirm the tracker's capability to detect and resolve x-ray energies relevant for detecting exotic atom de-excitation signatures.

\begin{figure}[tbp]
    \centering
    \includegraphics[width=0.8\textwidth]{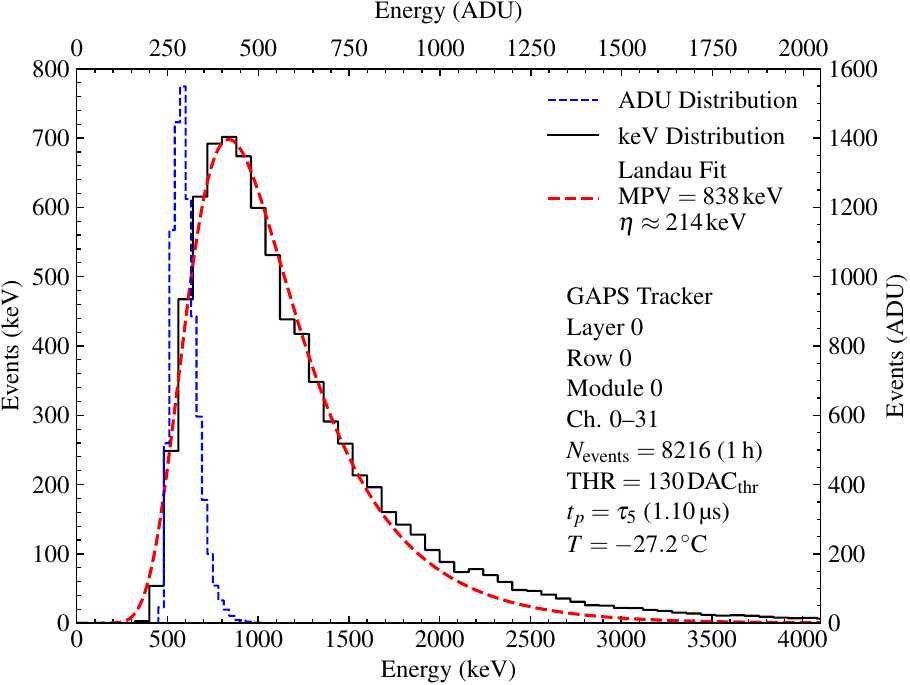}
    \caption{Muon energy deposition spectrum acquired in self-trigger mode on Module~0, Row~0, Layer~0. The histogram shows both the ADC distribution in ADU, spanning the full dynamic range of the 11-bit ADC (\SI{0}{ADU} to \SI{2047}{ADU}), and the corresponding energy distribution in \si{\kilo\electronvolt} obtained via calibrated transcharacteristics. A Landau fit to the energy distribution yields an MPV of $\SI{838}{\kilo\electronvolt}$ with a scale parameter $\eta \approx \SI{214}{\kilo\electronvolt}$. The pedestal peak is excluded from the plot. The histogram bin width is \SI{30}{ADU} for the ADC spectrum and \SI{80}{\kilo\electronvolt} for the calibrated energy spectrum.}
    \label{fig:ch3-muons_energy_deposition}
\end{figure}

\par
The detection of MIPs, such as cosmic muons, provides an independent means of verifying the tracker's response in the high-energy regime~\cite{groomPassageParticlesMatter2000}.
To this end, a dedicated self-trigger acquisition was conducted using all 32 channels of Module~0, Row~0, Layer~0, with a global threshold of \SI{130}{DAC_{thr}}.
The histogram in~\Cref{fig:ch3-muons_energy_deposition} displays both the raw ADC spectrum in ADU, covering the entire dynamic range of the 11-bit ADC (\SI{0}{ADU} to \SI{2047}{ADU}), and the corresponding energy distribution in \si{\kilo\electronvolt}, obtained by applying the calibrated channel transcharacteristics.

\par
A Landau fit to the energy distribution, based on~\eqref{eq:ch2-moyal}, yields an MPV of \SI{838}{\kilo\electronvolt} and a scale parameter $\eta \approx \SI{214}{\kilo\electronvolt}$, consistent with the expected energy loss of MIPs in \hbox{\SI{2.29}{\milli\meter}-thick} silicon sensors.
The larger scale parameter, compared to that obtained during laboratory characterization, suggests a broader distribution likely due to increased charge dispersion at the higher operating temperature.  
This may impact the detector's depletion region, thereby altering the effective active thickness available for charge collection.
The pedestal peak is excluded from the plot.  
Histogram bin widths are \SI{30}{ADU} in the ADC domain and \SI{80}{\kilo\electronvolt} in the calibrated energy domain.

%-------------------------------------------------------------------------------
%   Summary and outlook
%-------------------------------------------------------------------------------
\section{Summary and outlook}

\par
The calibration campaign performed on the GAPS Si(Li) tracker during full system integration confirmed the successful operation of the readout electronics within the assembled instrument. 
This activity represented the final step of the characterization program that began with ASIC-level validation~\cite{manghisoni32channelsReadoutASIC2023, ghislottiCharacterisationReadoutElectronics2022} and continued with the module-level testing presented in~\Cref{chapter2}. 

\par
A comprehensive set of procedures was established to determine the self-trigger energy thresholds, account for parasitic injection effects, and evaluate electronic noise, enabling the identification and masking of hot channels.
These procedures have since been implemented in the flight software of the experiment and form part of the standard calibration workflow.
The per-channel threshold fine-trimming capability of the ASIC proved effective in reducing threshold dispersion across modules, demonstrating its functionality in a large-scale tracker environment.

\par
An essential outcome of this campaign was the assessment of temperature effects on energy resolution and gain, found to be consistent with previous ASIC-level measurements. 
The calibration of the input--output transcharacteristic, including parasitic injection compensation, enabled accurate conversion from ADC units to deposited energy in \si{\kilo\electronvolt}, a procedure now adopted as standard for flight operations.

\par
Finally, the successful detection of both low-energy x-rays from a \ce{^{109}Cd} source and high-energy deposits from atmospheric muons validated the full readout chain and demonstrated the tracker's capability to reconstruct signals across its operating energy range.

%% file: maintext/chapters/chapter4.tex
\chapter{\texorpdfstring
  {Design of the \antareslogo readout ASIC}
  {Design of the ANTARES4 readout ASIC}
}\label{chapter4}

\par
This chapter presents the design of the \antareslogo (ANTiparticle Asic REadout for Si(Li) detectors with four CSA feedback configurations) ASIC, developed as a research and development (R\&D) prototype for upgrading the SLIDER32 ASIC used in the GAPS Si(Li) detector readout system, in preparation for the second flight of the experiment.

\par
The chapter is organized as follows.
First, the reasons behind the development of a new ASIC are presented, highlighting the limitations of the previous design and the new requirements introduced by the second GAPS flight.
Next, the architecture of the analog readout channel is described in detail. This includes the CSA with its feedback network, the CR-RC shaper, the charge injection circuit for calibration, and the leakage current generator used to simulate detector leakage.
Special attention is given to the CSA feedback capacitor, implemented using a dynamic threshold MOSFET (DTMOS), which enables dynamic signal compression. The design choices are discussed with respect to the strict power and performance constraints of the experiment.
Then, the chapter describes the other circuit blocks integrated on the chip, such as the analog output drivers, the shift register used for slow control, and the binary-to-thermometric decoder for peaking time selection of the shaper.
Finally, the complete eight-channel ASIC is introduced, along with layout strategies and packaging considerations.

%-------------------------------------------------------------------------------
%   Motivations
%-------------------------------------------------------------------------------
\section{Motivations}

\par
The Si(Li) tracker is one of the core components of the GAPS instrument, comprising up to 11520 readout channels in its fully populated configuration with ten layers of detector modules.
This high channel count imposes stringent constraints on power consumption, mechanical integration, and data throughput. Consequently, any modification to the \hbox{front-end} electronics significantly impacts the overall system architecture and its performance.

\par
For the first flight, 600 SLIDER32 ASICs were fabricated, and this highlights the scale of production required to equip the tracker.
After validation and flight-qualification procedures, 310 devices were selected and deployed in a configuration with seven instrumented layers~\cite{manghisoni32channelsReadoutASIC2023}.
Given the volume and centrality of these components, any redesign of the front-end ASIC significantly affects the performance of the overall experiment.

\par
To address both technological limitations and evolving requirements for the second GAPS flight, a new ASIC has been developed in a \SI{65}{\nano\meter} CMOS process.
The \antareslogo prototype represents an upgrade over the previously employed SLIDER32 ASIC~\cite{manghisoniLowNoiseAnalogChannel2021}, which was implemented in a \SI{180}{\nano\meter} CMOS technology.
While SLIDER32 totally fulfilled the noise and dynamic range specifications of the first mission, several factors motivate the transition to a more advanced node.

\par
First, the long-term availability of the \SI{180}{\nano\meter} process is uncertain, and its integration capabilities are limited compared to more modern alternatives. 
In contrast, the \SI{65}{\nano\meter} process provides higher integration density, allowing for reduced silicon area per channel—particularly in digital blocks—and enabling more compact layouts~\cite{manghisoniIntroducing65Nm2010}.
A smaller per-channel footprint may also enable higher strip segmentation in future detector modules, thereby improving the spatial resolution of the tracker.

\par
The \SI{65}{\nano\meter} node was selected as the optimal compromise between performance and fabrication cost.
Although more advanced technology nodes, such as \SI{28}{\nano\meter}~\cite{gaioni28NmCMOS2023}, were considered, the combination of higher fabrication costs and reduced voltage headroom---due to the lower nominal supply voltage of \SI{0.9}{\volt}---rendered them less suitable for this application.
In addition, the increased complexity of analog design and layout in such scaled nodes—primarily due to stricter design rule constraints—further limited their suitability for this application.
The choice of a less aggressive scaling node like \SI{65}{\nano\meter} thus offers a better balance between analog design flexibility and long-term technological availability.

\par
In the event of failed payload recovery, the ability to reproduce readout ASICs is essential.
The adoption of a more recent and widely supported technology mitigates risks associated with the obsolescence of the SLIDER32 \SI{180}{\nano\meter} implementation and enhances the long-term sustainability of the experiment, for which at least three flights are planned before 2030. 

\par
While the \SI{28}{\nano\meter} node is increasingly emerging as the next standard for front-end electronics in high-energy physics experiments~\cite{traversi65Nm282024}, the \SI{65}{\nano\meter} node remains the current reference technology in this field.
At CERN, for example, the RD53 collaboration adopted this technology for the ATLAS and CMS pixel detectors at the High-Luminosity LHC (HL-LHC)~\cite{gaioniDesignAnalogFrontends2017, gaioniTestResultsProspects2019}.
The ALICE experiment has similarly employed monolithic active pixel sensors (MAPS) fabricated in a \SI{65}{\nano\meter} process for its ITS3 upgrade~\cite{gautamCharacterisationAnalogueMAPS2024}.
In addition, this technology has demonstrated great radiation tolerance in both laboratory~\cite{manghisoniAssessmentLowPower652014a,rattiChargePreamplifier652016} and collider environments~\cite{mekkaouiResults65Nm2013}. 
Although such tolerance is not required for GAPS, which operates in the stratosphere, it further supports the use of a well-qualified and robust technology platform.

\par
Second, the transition from the \SI{180}{\nano\meter} to the \SI{65}{\nano\meter} process enables improved power efficiency.
In a balloon-borne experiment, available power is limited, and thermal dissipation must be minimized to reduce the load on the cooling system~\cite{fukeOperationConceptGAPS2024}.
Reducing the overall power demand simplifies the design of the onboard power system, which relies on batteries with limited capacity due to weight constraints, and on solar panels, whose output is restricted by the available surface area.
A key factor in this optimization is the reduction of the core supply voltage imposed by the technology node, from \SI{1.8}{\volt} in the \SI{180}{\nano\meter} process to \SI{1.2}{\volt} in the \SI{65}{\nano\meter} node.
On the other hand, at constant power supply, lowering the supply voltage, in turn, leaves room for an increase in the current of the CSA input device, and therefore creates the opportunity to improve the noise performance of the analog front-end.
In a properly designed channel, energy resolution is primarily limited by the noise contribution of the input MOS device in the CSA.
As will be discussed later in this chapter, increasing the drain current of the input transistor improves its transconductance, thereby reducing the series noise contribution.

\par
The \antareslogo ASIC has been developed as a research and development prototype to evaluate the analog front-end architecture up to the shaping stage. It is a purely analog implementation designed to characterize different CSA feedback configurations, and to assess the performance of the integrated leakage current compensation circuit.
The upgraded flight-ready version is intended to reuse the same 128-pin TQFP package employed for SLIDER32, allowing compatibility with the existing FEBs.
Only minor modifications to discrete components supplying bias and power are expected, thereby minimizing system requalification costs for the second flight.

%-------------------------------------------------------------------------------
%   Technology overview
%-------------------------------------------------------------------------------
\subsubsection*{\SI{65}{\nano\meter} CMOS technology overview}

\par
The design presented in this work has been carried out in a \SI{65}{\nano\meter} CMOS process developed for mixed-signal and RF applications. 
The technology features a high-resistivity epitaxial substrate, shallow trench isolation (STI), and two gate oxide options for core and I/O devices, with both $n^+$ and $p^+$ polysilicon gate materials. 
It supports deep-$n$-well devices, as well as low- (LVT), standard- (SVT), and high-threshold (HVT) voltage options. 
The back-end of line (BEOL) offers 4 to 9 metal layers.

\begin{table}[tbp]
    \centering
    \renewcommand{\arraystretch}{1.1}
    \setlength{\tabcolsep}{10pt}
    \begin{tabular}{@{}ll@{}}
        \toprule
        Parameter & Specification \\
        \midrule
        Nominal operating temperature & \SI{-40}{\celsius} \\
        Power dissipation & $\leq\SI{10}{\milli\watt\per{channel}}$ \\
        Signal polarity & electrons \\
        Detector capacitance & \SI{38}{\pico\farad} \\
        Input dynamic energy range & 10--\SI{100}{\mega\electronvolt} \\
        Energy resolution ($C_d = \SI{40}{\pico\farad}$) & 4--\SI{5}{\kilo\electronvolt}~FWHM \\
        ENC ($C_d = \SI{40}{\pico\farad}$) & 480--\SI{600}{e^{\text{-}}} rms \\
        Low-energy gain ($G_{\text{le}}$) & $\ge\SI{250}{\micro\volt\per\kilo\electronvolt}$ \\
        High-energy gain ($G_{\text{le}}$) & $\ge\SI{2}{\micro\volt\per\kilo\electronvolt}$ \\
        CSA output dynamic range & $\le\SI{500}{\milli\volt}$ \\
        Shaper output dynamic range & $\le\SI{600}{\milli\volt}$ \\
        Compression factor $k = G_\mathrm{le}/G_\mathrm{he}$ & $\geq80$ \\
        Kink energy & $\geq\SI{1.5}{\mega\electronvolt}$ \\
        Detector leakage current ($T = \SI{-40}{\celsius}$) & 2--\SI{10}{\nano\ampere} \\
        Detector leakage current ($T = \SI{\sim-10}{\celsius}$) & $\mathord{\lesssim}\,200\,\si{\nano\ampere}$ \\
        Event rate & \SI{100}{\hertz} \\
        \bottomrule
    \end{tabular}
    \caption{Front-end electronics requirements for the GAPS Si(Li) tracker, covering operating conditions, power budget, signal and detector parameters, and performance targets in terms of energy resolution, noise, and dynamic range.} % The detector leakage current is constrained at both the nominal \SI{-40}{\celsius} and the on-ground testing temperature of about \SI{-10}{\celsius}.
    \label{tab:ch4-frontend_requirements}
\end{table}

\par
Compared to other flavors commonly available for CMOS technologies, such as the general purpose (GP) and high speed (HS) variants, the low power (LP) option employed here provides an increased supply voltage and a larger equivalent oxide thickness for the core devices, leading to a reduced gate leakage current.

%-------------------------------------------------------------------------------
%   Analog front-end electronics requirements
%-------------------------------------------------------------------------------
\section{Analog front-end electronics requirements}

\par
The design of the analog front-end electronics for the GAPS Si(Li) tracker is driven by the scientific goals of the experiment and by the specific operating characteristics of the silicon strip detectors, detailed in~\Cref{sec:ch2-sili_detector}.
The main requirements that the analog readout chain must satisfy are summarized in~\Cref{tab:ch4-frontend_requirements}~\cite{ghislottiDesignANTARES4Readout2025a}.

\par
The tracker modules operate at a nominal temperature of \SI{-40}{\celsius}, chosen to minimize the leakage current in the Si(Li) sensors, thereby guaranteeing the lowest parallel noise contribution achievable by the detectors~\cite{perezFabricationLowcostLargearea2018}.
At this temperature, and under a high-voltage bias of \SI{-250}{\volt}, the detectors are fully depleted, ensuring efficient charge collection and minimal leakage current. This operating condition enables the best achievable energy resolution from the sensors.
Within this configuration, the leakage current is expected to remain in the range of \SIrange{2}{10}{\nano\ampere} per strip, which defines the baseline for acceptable noise performance in the front-end design.
Although the nominal detector capacitance is \SI{38}{\pico\farad}, as reported in \Cref{tab:ch4-frontend_requirements}, a value of \SI{40}{\pico\farad} is used throughout this chapter as a conservative reference for simulations.

\par
An additional requirement emerged from the extensive on-ground calibration campaign of the experiment: the tracker readout electronics must be capable of functioning at temperatures significantly above the nominal value, including scenarios near \SI{-15}{\celsius} or \SI{-10}{\celsius}, as encountered during integration and testing phases.  
Although the tracker is never operated at such temperatures during scientific data acquisition in flight, these conditions are typical during on-ground calibration and functional validation.  
The primary limitation in these scenarios is the increased leakage current of the Si(Li) detectors, which can exceed \SI{150}{\nano\ampere} per strip.  
The current ASIC architecture allows compensation up to approximately \SI{50}{\nano\ampere}, beyond which many channels become inoperable during testing.  
To address this limitation, the next-generation ASIC must support leakage current compensation up to at least \SI{200}{\nano\ampere} per strip.  
It should be emphasized that the energy resolution requirements apply only under nominal flight conditions at \SI{-40}{\celsius}, while at higher temperatures the goal is limited to ensuring detector operability for calibration and system validation.  
The ability to operate at higher temperatures with increased detector leakage currents enables a broader range of calibration activities, preventing channels from becoming saturated and inoperable under these conditions.

\par
Due to the exotic atom formation and charged particle tracking techniques employed in GAPS, the readout channel must accommodate a wide input dynamic energy range spanning over four decades, from low-energy x-rays in the \SIrange{10}{100}{\kilo\electronvolt} range to charged particles with energy deposits up to \SI{100}{\mega\electronvolt}.  
Historically, an energy resolution better than \SI{4}{\kilo\electronvolt} FWHM at a detector capacitance of \SI{40}{\pico\farad}---corresponding to a maximum ENC of \SI{480}{e^{\text{-}}} rms---was targeted to enable identification of characteristic x-rays from antiprotonic and antideuteronic exotic atoms.
This resolution was originally regarded as a critical discriminator between antiprotons and antideuterons.  
However, the experimental strategy has evolved to leverage additional observables such as pion multiplicity and ionization energy loss ($\mathrm{d}E/\mathrm{d}x$), which contribute significantly to particle identification. In this broader context, the x-ray resolution remains valuable but is no longer the sole or limiting factor.  
Consequently, while the design goal remains at \SI{4}{\kilo\electronvolt} FWHM, adequate discriminating power is achieved with a relaxed requirement in the \SIrange{4}{5}{\kilo\electronvolt} range, corresponding to an ENC of 480--\SI{600}{e^{\text{-}}} rms. 
The polarity of the signal is determined by the Si(Li) sensor structure, which collects electrons at the readout node, requiring the analog front-end to process negative charge signals.

\par
Given the balloon-borne nature of the experiment and the limitations in cooling~\cite{fukeDevelopmentCoolingSystem2016} and power delivery, the front-end must operate with a power dissipation below \SI{10}{\milli\watt\per{channel}}.
The SLIDER32 ASIC, developed in \SI{180}{\nano\meter} CMOS technology, achieves a total consumption of \SI{264}{\milli\watt}, or approximately $\SI{8.3}{\milli\watt}$ per channel, which is already close to this upper limit~\cite{manghisoni32channelsReadoutASIC2023}.
The development of the new ASIC aims to further reduce this figure while improving the energy resolution performance.

\par
Finally, the readout must be capable of sustaining a nominal event rate of \SI{100}{\hertz} per channel without introducing pile-up. This requirement imposes constraints on the design of the analog front-end, particularly on the return-to-baseline time of the CSA.

\par
These specifications were successfully met in the previous generation ASIC, but transitioning to \SI{65}{\nano\meter} CMOS technology introduces additional design challenges.
The reduced supply voltage (\SI{1.2}{\volt} compared to \SI{1.8}{\volt}) limits the available signal swing and impacts the biasing of analog blocks.
The \antareslogo ASIC was developed specifically to address these constraints while fulfilling all requirements for integration into the GAPS tracker, as detailed in the following sections.

%-------------------------------------------------------------------------------
%   Readout architectures for silicon strip detectors
%-------------------------------------------------------------------------------
\section{Readout architectures for silicon strip detectors}

\par
Multiple front-end architectures have been developed for the readout of silicon strip detectors, each suited to different detector characteristics and system-level constraints.
Key considerations in selecting a readout architecture include detector capacitance, dynamic range of the input charge signals, noise performance, timing resolution, and power consumption~\cite{kaplonFrontendElectronicsSilicon2023}.

\par
CSAs followed by shaping amplifiers represent the most widely adopted architecture in high-energy physics and spectrometry applications involving silicon detectors.
In this configuration, the input charge collected on the detector electrode is integrated on a feedback capacitor, generating a voltage step proportional to the deposited energy.
The shaped output provides improved SNR pulse shaping for pile-up rejection, and compatibility with analog multiplexing or digitization~\cite{monteScientificPerformancesXAA122007, michailidisChargesensitiveAmplifierDesign2024}.
This architecture supports calibration via dedicated charge injection circuits and enables the implementation of leakage current compensation schemes, allowing the front-end electronics to operate under varying detector leakage current conditions induced by temperature changes or variations in detector bias~\cite{seidelSiliconStripPixel2019}.
These features make CSA-based readout suitable for detectors with moderate to large capacitance and energy deposits spanning a wide interval, such as those encountered in high-energy astrophysics or particle physics experiments.

\par
Current-mode architectures, including transimpedance amplifiers (TIAs) or resistive-feedback preamplifiers, operate by converting the detector current directly into a voltage signal across a feedback element.
While they offer high-speed response and are relatively simple to implement, their performance degrades in the presence of large input capacitance and in applications involving small charge deposits, such as low-energy x-ray detection.
These limitations reduce their suitability for systems with large-area detectors or when high charge sensitivity is required.

\par
Hybrid charge-current readout architectures have also been explored, combining \hbox{CSA-like} integration with current-mode operation to enable pulse-shape discrimination or improved timing resolution~\cite{dormardPulseShapeDiscrimination2021}.
These designs can provide complementary information on signal dynamics but typically involve higher circuit complexity and power consumption.

\par
Time-over-threshold (ToT) architectures employ a fast comparator to detect when the input signal crosses a fixed threshold, and then encode the signal amplitude into the duration for which the signal remains above threshold~\cite{mazzaToASt64channelASIC2025}.
This approach is commonly used in high-rate tracking systems where per-channel digitization and compact data representation are needed.
However, the ToT method introduces nonlinearity in the amplitude-to-time conversion and is sensitive to input pulse shape variations, which require calibration and compensation strategies~\cite{vonschmidFirstApplicationPulseshape2011}.

\par
For the GAPS experiment, the selected architecture is based on a CSA followed by a CR-RC shaper. This choice is driven by the need to resolve energy deposits ranging from \SI{10}{\kilo\electronvolt} x-rays to energy deposits up to \SI{100}{\mega\electronvolt} from charged particles. 
The CSA provides the necessary charge integration and low-noise performance, while the shaping amplifier improves timing characteristics and SNR, enabling compatibility with \hbox{threshold-based} triggering and analog-to-digital conversion.
This architecture enables the inclusion of leakage current compensation structures and the implementation of the dynamic signal compression technique, necessary to support the whole input dynamic range while guaranteeing the necessary resolution for resolving x-rays below \SI{100}{\kilo\electronvolt}.

\begin{figure}[tbp]
    \centering
    \includegraphics[width=.95\textwidth]{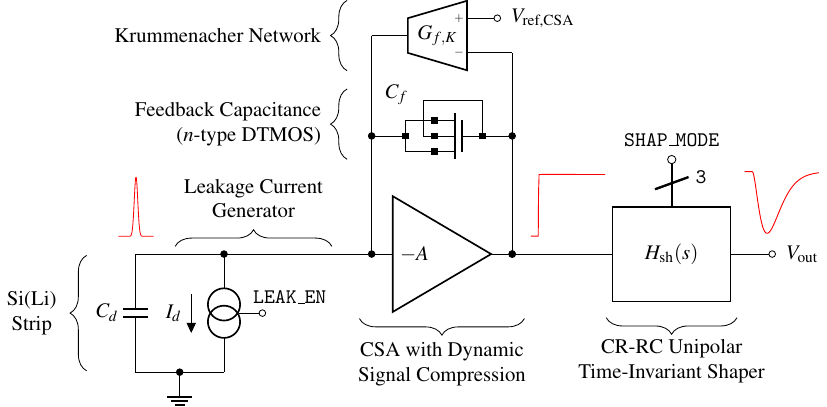}
    \caption{Simplified block diagram of the proposed readout channel. The CSA employs dynamic signal compression using a DTMOS-based NMOS feedback capacitor. A Krummenacher network provides continuous charge restoration and compensates for detector leakage current, which is emulated during testing by an integrated current generator. A charge injection circuit (not shown) is included for calibration. The CSA output is shaped by a unipolar CR-RC filter with semi-Gaussian response.}
    \label{fig:ch4-antares4_channel_diagram_simple}
\end{figure}

\par
To illustrate the implementation of the selected architecture, a simplified block diagram of the analog readout channel developed for the GAPS experiment is shown in~\Cref{fig:ch4-antares4_channel_diagram_simple}.
The diagram highlights the core components of the signal chain, including the CSA with dynamic signal compression, the Krummenacher feedback network for charge restoration and detector leakage current compensation, and the CR-RC shaping stage.
Additional elements, such as a leakage current emulator and a charge injection circuit for calibration (not shown in~\Cref{fig:ch4-antares4_channel_diagram_simple}) are also integrated into the design.
During all phases of design, the Si(Li) strip has been modeled as a \SI{40}{\pico\farad} capacitor in parallel with a current source to emulate the detector leakage current, that in the final design has been implemented as a dedicated circuit for leakage current emulation up to about \SI{500}{\nano\ampere}.

%-------------------------------------------------------------------------------
%   The charge-sensitive amplifier (CSA)
%-------------------------------------------------------------------------------
\section{The charge-sensitive amplifier (CSA)}

\begin{figure}[tbp]
    \centering
    \includegraphics[width=\textwidth]{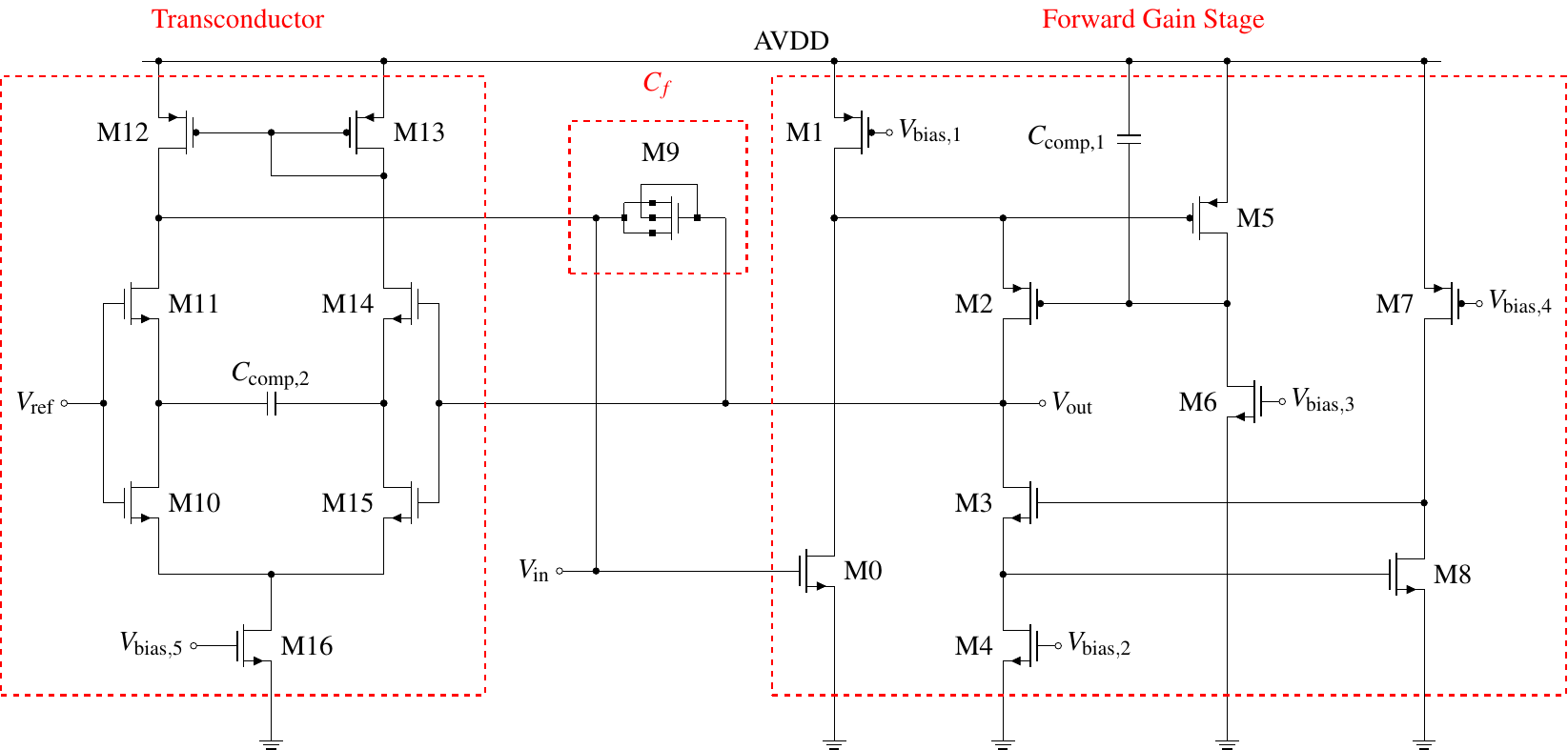}
    \caption{Transistor-level schematic of the CSA used during the initial analysis and design phase. The circuit includes the forward gain stage (M0--M8) and a temporary transconductance amplifier (M10--M16) to study the behavior of the CSA input device. The feedback capacitor (M9) is realized using an $n$-type DTMOS device. The bias transistors implementing the current mirrors are not reported.}
    \label{fig:ch4-antares4_csa_schematic}
\end{figure}

\par
This section presents the design and layout of the CSA.
It begins with an overview of the forward gain stage topology and the associated design choices.
Next, the noise optimization strategy is discussed, with particular attention to the sizing and biasing of the CSA forward gain stage input transistor.
Finally, the feedback network is detailed, including both the MOS- and DTMOS-based capacitors used for dynamic signal compression, and the Krummenacher circuit for continuous charge restoration and compensation of the detector leakage current.

%   Operating principle
\subsection{Operating principle}

\par
The CSA is the first block of the analog front-end and plays a key role in the readout of semiconductor radiation detectors.  
It collects the current pulse generated by charge carriers in the detector and integrates it on a feedback capacitance \( C_f \), producing an output voltage proportional to the total collected charge \( Q_{\text{tot}} \), and thus to the energy deposited by the incoming particle~\cite{turchettaAnalogElectronicsRadiation2017a}.  
It is based on a low-noise inverting voltage amplifier, usually implemented as a cascode or folded cascode stage, configured as an integrator by means of \( C_f \).
Assuming an ideal inverting voltage amplifier with gain \( A \gg 1 \), the output voltage is given by:

\begin{equation}
    V_{\text{out}} = \frac{1}{C_f}Q_{\text{tot}}.
\end{equation}
\noindent

\par
In this configuration, the output pulse amplitude becomes independent of the input capacitance \( C_{\text{in}} \), provided that \( (C_{\text{in}} + C_f)/C_f \ll A \), a condition typically satisfied in practical implementations~\cite{knollRadiationDetectionMeasurement2010a}.  
Here, \( C_{\text{in}} \) represents the total capacitance shunting the input node of the charge amplifier, primarily composed of the detector capacitance \( C_d \) and the input transistor's gate capacitance \( C_{\text{G,0}} \), which depends on its geometry, namely the width \( W_0 \) and the length \( L_0 \), and the device operating region.

\par
A more accurate estimate of \( C_{\text{G,0}} \) should account for both intrinsic and extrinsic contributions, including the gate-to-source and gate-to-drain overlap capacitances~\cite{tsividisOperationModelingMOS2011}.  
An improved model~\cite{degeronimoMOSFETOptimizationDeep2005} expresses the input device total gate capacitance as:

\begin{equation}
    C_\text{G,0} = 2C_{\text{ov}}W_0 + \frac{2}{3} C_{\text{ox}} W_0 L_0,
    \label{ch4:eq-gate_capacitance}
\end{equation}

\noindent
where \( C_{\text{ov}} \) is the overlap capacitance density (assumed equal for gate-to-source and \hbox{gate-to-drain} capacitances), and \( C_{\text{ox}} \) is the gate oxide capacitance per unit area.

\begin{table}[tbp]
    \centering
    \renewcommand{\arraystretch}{1.1}
    \setlength{\tabcolsep}{8pt}
    \begin{tabular}{cccccc}
        \toprule
        \multirow{2}{*}{Device} & \multirow{2}{*}{Type} & $L$ & $W_\text{tot}$ & $W_\text{finger}$ & Fingers \\
                                &                        & (\si{\micro\meter}) & (\si{\micro\meter}) & (\si{\micro\meter}) & \# \\
        \midrule
        \textbf{M0} & \textbf{NMOS, SVT} & \textbf{0.2} & \textbf{2100} & \textbf{28} & \textbf{75} \\
        M1 & PMOS, LVT & 0.7 & 250  & 25 & 10 \\
        M2 & PMOS, LVT & 0.7 & 100  & 10 & 10 \\
        M3 & NMOS, LVT & 0.5 & 100  & 10 & 10 \\
        M4 & NMOS, LVT & 2.0 & 70   & 14 & 5  \\
        M5 & PMOS, LVT & 0.1 & 20   & 10 & 2  \\
        M6 & NMOS, LVT & 1.0 & 10   & 10 & 1  \\
        M7 & PMOS, LVT & 1.0 & 10   & 10 & 1  \\
        M8 & NMOS, LVT & 0.3 & 10   & 10 & 1  \\
        \bottomrule
    \end{tabular}
    \caption{Geometrical parameters of the MOS transistors used in the gain stage of the CSA in the \SI{65}{\nano\meter} implementation developed in this work. The table reports the channel length \( L \), total width \( W_\text{tot} \), width per finger \( W_\text{finger} \), and number of fingers for each device. Bias transistors implementing the current mirrors are not included.}
    \label{tab:ch4-csa_gain_stage_size}
\end{table}

\par
A resistor \( R_f \) can be connected in parallel with \( C_f \) to implement a continuous reset mechanism, forming a discharge path for the integrated charge.  
The resulting exponential decay of the output signal is characterized by the time constant \( \tau = R_f C_f \), which determines the tail of the CSA output pulse, guaranteeing baseline restoration for successive events.  
However, eliminating the resistor from the feedback loop offers significant advantages, including reduced noise due to the absence of resistor-related Johnson noise, and improved performance at high event rates by mitigating pile-up associated with resistive feedback configurations~\cite{spielerSemiconductorDetectorSystems2005a}.
The inclusion of a transconductance amplifier \( G_f \) or more sophisticated circuit topologies in place of the reset resistor~\cite{krummenacherPixelDetectorsLocal1991,gallianiZeroDeadtimeFrontend2023a} also allows controlling the baseline voltage at the output of the CSA and compensating for the detector leakage current, as is the case of silicon-strip detectors.

%   Forward gain stage
\subsection{Forward gain stage}\label{sec_ch4-csa_forward_gain_stage}

\par
The schematic diagram of the CSA is shown in~\Cref{fig:ch4-antares4_csa_schematic}, where the forward gain stage is implemented by transistors \hbox{M0--M8}.
In this design, the feedback element is implemented using an $n$-type MOS transistor in dynamic threshold (DTMOS) configuration, which enables dynamic signal compression, as discussed later in~\Cref{sec:ch4-csa-feedback-capacitor}.
Charge restoration is handled by a transconductance amplifier \hbox{(M10--M16)}, which sets the output DC level via the reference voltage \( V_{\text{ref}} \).
It is implemented with NMOS input devices and PMOS load, achieving a transconductance $G_f\simeq\SI{90}{\nano\siemens}$.
This auxiliary stage does not compensate for detector leakage current, which is not yet considered in this analysis, and was used only during the initial phase of the CSA design.
The forward gain stage is based on a folded cascode architecture with local feedback and active load, serving the purpose of increasing the impedance seen at the amplifier output node.

\begin{figure}[tbp]
    \centering
    \includegraphics[width=0.8\textwidth]{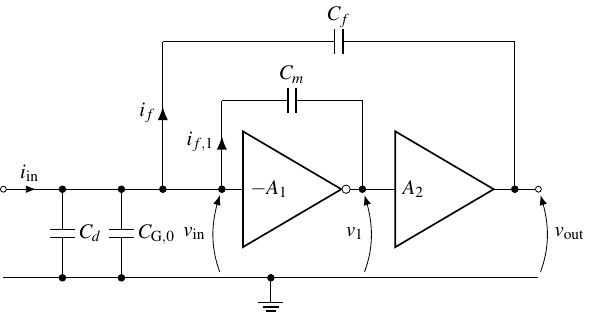}
    \caption{Simplified equivalent diagram of a CSA employing a cascode configuration, adapted from~\cite{kaplonFrontendElectronicsSilicon2023}. The model includes detector capacitance \( C_d \), input device gate capacitance \( C_{\text{G},0} \), feedback capacitance \( C_f \), and Miller capacitance \( C_m \). The forward path is modeled with two amplifier blocks: the first stage \( H_1(s) \) has gain \( -A_1 \), dominant pole time constant \( \tau_p \), and zero time constant \( \tau_z \); the second stage \( H_2(s) \) serves to cancel the zero introduced by \( H_1(s) \) and has unitary gain \( A_2 \).}
    \label{fig:ch4-csa_miller_kaplon}
\end{figure}

\par
A list of the geometrical parameters for all transistors in the gain stage is provided in~\Cref{tab:ch4-csa_gain_stage_size}. 
The table specifies the transistor type (NMOS or PMOS) along with its threshold voltage classification, distinguishing between SVT (standard threshold voltage) and LVT (low threshold voltage) devices. 
For each transistor, the channel length, total width, per-finger width, and number of fingers are reported.

\par
The amplifier input device (M0) is an SVT NMOS with \hbox{$W_0/L_0=2100/0.2$\,\si{\micro\meter/\micro\meter}} biased with a drain current $I_{\text{D},0}=\SI{1.8}{\milli\ampere}$.
The selection of the input device and its bias current was guided by noise optimization criteria for deep sub-micron technologies~\cite{rattiDesignOptimizationCharge2009}, while also considering power dissipation constraints, which impose a limit on the allowable drain current, as discussed later in~\Cref{ch4:sec-csa_input_device}.
The CSA input device design was validated through both schematic-level and post-layout simulations, confirming that the forward gain stage achieves the expected performance in terms of noise, gain, and bandwidth, and highlighting that a proper choice of device geometry and biasing is critical to optimize noise, linearity, gain, and bandwidth.

\begin{table}[tbp]
    \centering
    \renewcommand{\arraystretch}{1.2}
    \setlength{\tabcolsep}{10pt}
    \begin{tabular}{@{}ccccccc@{}}
        \toprule
        \multirow{2}{*}{Corner} &
        \multicolumn{2}{c}{DC gain (\si{\decibel})} &
        \multicolumn{2}{c}{$f_{\SI{-3}{\decibel}}$ (\si{\kilo\hertz})} &
        \multicolumn{2}{c}{GBP (\si{\giga\hertz})} \\
        \cmidrule(lr){2-3} \cmidrule(lr){4-5} \cmidrule(lr){6-7}
        & \SI{65}{\nano\meter} & \SI{180}{\nano\meter}
        & \SI{65}{\nano\meter} & \SI{180}{\nano\meter}
        & \SI{65}{\nano\meter} & \SI{180}{\nano\meter} \\
        \midrule
        TT & \SI{88.7}{} & \SI{105.6}{} & \SI{710.0}{} & \SI{100.0}{} & \SI{19.3}{} & \SI{19.1}{} \\
        SS & \SI{88.5}{} & \SI{108.4}{} & \SI{742.7}{} & \SI{70.8}{}  & \SI{19.8}{} & \SI{18.6}{} \\
        FF & \SI{87.6}{} & \SI{101.3}{} & \SI{757.5}{} & \SI{158.5}{} & \SI{18.1}{} & \SI{18.4}{} \\
        \midrule
        TT (\SI{27}{\celsius}) & \SI{83.7}{} & -- & \SI{710.7}{} & -- & \SI{15.3}{} & -- \\
        TT ($V_\text{DD} + \SI{10}{\percent}$) & \SI{89.1}{} & -- & \SI{771.8}{} & -- & \SI{21.8}{} & -- \\
        TT ($V_\text{DD} - \SI{10}{\percent}$) & \SI{84.7}{} & -- & \SI{563.9}{} & -- & \SI{15.2}{} & -- \\
        \bottomrule
    \end{tabular}
    \caption{Small-signal performance metrics of the forward gain stage across process corners (TT, SS, FF) for both the \SI{65}{\nano\meter} implementation developed in this work and the \SI{180}{\nano\meter} version. The table reports the open-loop DC gain, the \SI{-3}{\decibel} bandwidth, and the GBP from schematic-level simulations at \SI{-40}{\celsius} with $C_d=\SI{40}{\pico\farad}$ for the nominal $V_{\text{DD}}=\SI{1.2}{\volt}$, unless otherwise specified.}
    \label{tab:ch4-gain_gbp_csa}
\end{table}

\par
The folded cascode architecture was chosen over a cascade of common-source stages because it enables high gain through impedance boosting within a single stage, avoiding the need for multiple amplification steps.
The choice between these two options requires careful consideration of several factors, including power consumption, achievable \hbox{open-loop} gain and gain-bandwidth product, and potential speed limitations due to the Miller effect.

\par
The cascode topology offers three key advantages.
First, it suppresses the Miller effect, which could otherwise degrade the time response of the preamplifier.
Second, it behaves as a single-stage amplifier with one dominant, low-frequency pole, simplifying frequency compensation and ensuring excellent stability margins without additional power consumption.
Third, it provides favorable noise versus power performance, as the input device remains the dominant contributor to the series noise, typically accounting for about \SI{60}{\percent}--\SI{70}{\percent} of the total. 

%   Miller effect in the CSA
\subsubsection{Miller effect in the CSA}

The time response of the CSA is fundamentally limited by the total effective input capacitance seen at the preamplifier node~\cite{kaplonFrontendElectronicsSilicon2023}.
One of the dominant contributors to this capacitance is the Miller effect, which arises from the amplification of the parasitic capacitance \( C_m \) bridging the input and output of the forward amplifier stage \( H_1(s) \).
This effect is visually modeled in the diagram shown in~\Cref{fig:ch4-csa_miller_kaplon}.
It is important to note that this analysis assumes a fixed feedback capacitor \( C_f \) and does not include any charge restoration feedback.

The forward path of the CSA is modeled using two amplifier stages to reflect the cascode topology adopted in this work.
The first stage, \( H_1(s) \), includes a dominant pole with time constant \( \tau_{p,1} \) and a high-frequency zero with time constant \( \tau_z \), and is represented by the transfer function:

\begin{equation}
    H_1(s) = A_1\,\frac{1 + s\tau_z}{1 + s\tau_{p,1}}.
\end{equation}

\noindent
The non-dominant, high-frequency pole with time constant \( \tau_{p,2} \) is omitted from the analysis.
The second stage, $H_2(s)$, is assumed to provide unity gain ($A_2 = 1$) and is introduced exclusively to cancel the high-frequency zero. As a result, it is modeled using the following simplified expression:

\begin{equation}
    H_2(s) = \frac{1}{1 + s\tau_z}.
\end{equation}

\begin{table}[t!]
    \centering
    \renewcommand{\arraystretch}{1.2}
    \setlength{\tabcolsep}{10pt}
    \begin{tabular}{@{}l cc@{}}
        \toprule
        Parameter & \SI{65}{\nano\meter} (this work) & \SI{180}{\nano\meter} (SLIDER32 ASIC) \\
        \midrule
        $L_0$ & \SI{0.2}{\micro\meter} & \SI{0.5}{\micro\meter} \\
        $W_{\text{tot},0}$ & \SI{2100}{\micro\meter} & \SI{2600}{\micro\meter} \\
        $W_{\text{finger},0}$ & \SI{28}{\micro\meter} & \SI{20}{\micro\meter} \\
        \# Fingers & 75 & 130 \\
        \midrule
        $I_{\text{D},0}$ & 
            \SI{1.8}{\milli\ampere} & \SI{1.5}{\milli\ampere} \\
        $g_{m,0} = \left. \frac{\partial I_{\text{D},0}}{\partial V_{\text{GS},0}} \right|_{V_{\text{DS},0} = \text{const}}$ &
            \SI{38.21}{\milli\ampere\per\volt} & \SI{44.71}{\milli\ampere\per\volt} \\
        $g_{ds,0} = \left. \frac{\partial I_{\text{D},0}}{\partial V_{\text{DS},0}} \right|_{V_{\text{GS},0} = \text{const}}$ &
            \SI{648.68}{\micro\ampere\per\volt} & \SI{302.75}{\micro\ampere\per\volt} \\
        $A_v^{\text{int}} = \tfrac{g_{m,0}}{g_{ds,0}}$ &
            \num{58.90} & \num{147.67} \\
        $C_{\text{G},0}$~(\Cref{ch4:eq-gate_capacitance}) &
            \SI{3.22}{\pico\farad} & \SI{5.00}{\pico\farad}~\cite{manghisoniLowNoiseAnalogChannel2021} \\
        \bottomrule
    \end{tabular}
    \caption{Geometry and small-signal parameters of the input transistor M0 in the CSA forward gain stage for both the \SI{65}{\nano\meter} and \SI{180}{\nano\meter} implementations at \SI{-40}{\celsius}. Transistor dimensions include channel length \( L_0 \), total width \( W_{\text{tot},0} \), per-finger width \( W_{\text{finger},0} \), and the number of fingers. The table also reports bias current \( I_{\text{D},0} \), transconductance \( g_{m,0} \), output conductance \( g_{ds,0} \), intrinsic gain \( A_v^{\text{int}} \), and gate capacitance \( C_{\text{G},0} \). All performance parameters have been extracted from schematic-level simulations.}
    \label{tab:ch4-gm_gds_gain_single}
\end{table}

\noindent
The overall open-loop gain is thus \( K = A_1 \), and the combined transfer function preserves the low-frequency behavior while mitigating the effect of the high-frequency zero.
The output response of the CSA in the time domain to a Dirac delta charge pulse at the preamplifier input \( i_{\text{in}}(t) \) is given by:

\begin{equation}
    v_{\text{out}}(t) = A_{\text{CSA}} \left( 1 - e^{-t/\tau_{\text{CSA}}} \right),
\end{equation}

\noindent
where the amplitude of the output voltage step and the time constant are:

\begin{align}
    A_{\text{CSA}} &= \frac{1}{C_f + C'_{\text{in}} / A_1}, \\
    \tau_{\text{CSA}} &= \frac{C''_{\text{in}}}{\text{GBP}\,C_f}.
\end{align}

\noindent
The effective input capacitance \( C'_{\text{in}} \) depends on the amplifier topology.
For a single-stage forward gain block, it is given by:

\begin{equation}
    C'_{\text{in}} = C_d + C_{\text{G},0} + C_f + (1 + A_1) C_m.
\end{equation}

\noindent
Assuming sufficient open-loop gain such that \( C'_{\text{in}} / A_1 \ll C_f \), the CSA output in response to a Dirac delta input current is a voltage step of amplitude approximately equal to:

\begin{equation}
    A_{\text{CSA}} \approx \frac{1}{C_f}.
\end{equation}

\noindent
The rising edge of the output is characterized by a time constant determined by the total input capacitance \( C'_{\text{in}} \), which includes the Miller capacitance \( C_m \) multiplied by \( (1 + A_1) \).
This time constant is inversely proportional to the gain-bandwidth product (GBP) of the amplifier and the feedback capacitance \( C_f \), leading to a slower response for larger input capacitance.
In contrast, the cascode configuration introduces a high-frequency zero and modifies the Miller multiplication factor to:

\begin{equation}
    1 + A_1 \frac{\tau_z}{\tau_{p,1}},
\end{equation}

\noindent
resulting in a reduced effective input capacitance:

\begin{equation}
    C''_{\text{in}} = C_d + C_{\text{G},0} + C_f + \left(1 + A_1 \frac{\tau_z}{\tau_{p,1}} \right) C_m.
\end{equation}

\noindent
Since \( \tau_z \ll \tau_{p,1} \), the term \( \tau_z / \tau_{p,1} \ll 1 \), and the Miller contribution becomes significantly smaller than in the case of a cascade of common-source amplifiers.
This enables a faster CSA response without the need to increase \( C_f \), which would degrade noise performance.

\begin{figure}[t!]
    \centering
    \includegraphics[width=0.75\textwidth]{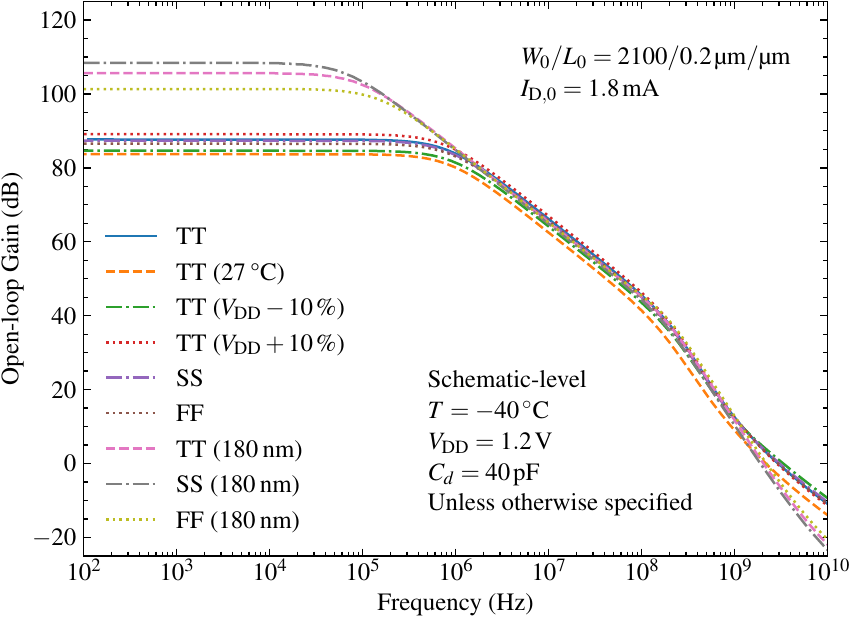}
    \caption{Open-loop gain versus frequency for the CSA forward amplifier stage, obtained from schematic-level simulations across process corners (TT, SS, FF), supply voltage variations (\( V_\text{DD} \pm 10\% \)), and temperatures (\SI{-40}{\celsius} and \SI{27}{\celsius}). The comparison includes both the upgraded \SI{65}{\nano\meter} implementation developed in this work and the \SI{180}{\nano\meter} version, simulated at \SI{-40}{\celsius} with a detector capacitance of \( C_d = \SI{40}{\pico\farad} \).}
    \label{fig:ch4-csa_gain_stage_comparison}
\end{figure}

\par
For the cascode topology, the input capacitance converges to:

\begin{equation}
    C''_{\text{in}} = C_d + C_{\text{G},0} + C_f + \left(1 + \frac{g_{m,0}}{g_{m,2}} \right) C_m,
    \label{eq:cin_transconductance}
\end{equation}

\noindent
with \( g_{m,0} \) and \( g_{m,2} \) denoting the transconductance of devices M0 and M2, respectively, in the schematic of~\Cref{fig:ch4-antares4_csa_schematic}. This expression shows that the Miller effect is ultimately governed by the transconductance ratio of the cascode pair.
Furthermore, the input impedance of the CSA in a purely capacitive feedback configuration is:

\begin{equation}
    Z_{\text{in}}(s) \approx \frac{1}{\text{GBP}\,C_f}.
\end{equation}

\noindent
This confirms that the cascode topology mitigates the Miller effect and supports a faster time response while preserving low input impedance and efficient charge collection.
In the specific case of this work, the dominant contribution to the total input capacitance is given by the detector capacitance \( C_d = \SI{40}{\pico\farad} \), which is by far the main contributor, followed by the input device gate capacitance \( C_{\text{G},0} = \SI{3.22}{\pico\farad} \) obtained from simulations and calculated using~\eqref{ch4:eq-gate_capacitance} assuming  \( C_{\text{ox}}=\SI{14.8}{\femto\farad/\micro\meter\squared} \).

\par
It should be emphasized that the degradation of CSA timing performance due to the Miller effect is not caused by the conventional \( RC \) bandwidth limitation at the input.
Although the detector can be modeled as a high-impedance current source in parallel with a parasitic capacitance, the current signal is not filtered by the input capacitance itself.
Instead, the slowing of the rising edge is due to the limited GBP of the amplifier and the increase in the effective input capacitance caused by the Miller effect, as defined by the time constant \( \tau_{\text{CSA}} \).
While increasing the feedback capacitance \( C_f \) can partially compensate for insufficient GBP and improve timing, this strategy increases the power required in the shaping stage and may compromise noise performance, since a larger \( C_f \) increases the relative contribution of the shaper noise to the total noise.
Therefore, maximizing the GBP and minimizing the input capacitance are the primary design strategies for controlling the CSA timing response.

\begin{figure}[tbp]
    \centering
    \includegraphics[width=0.75\textwidth]{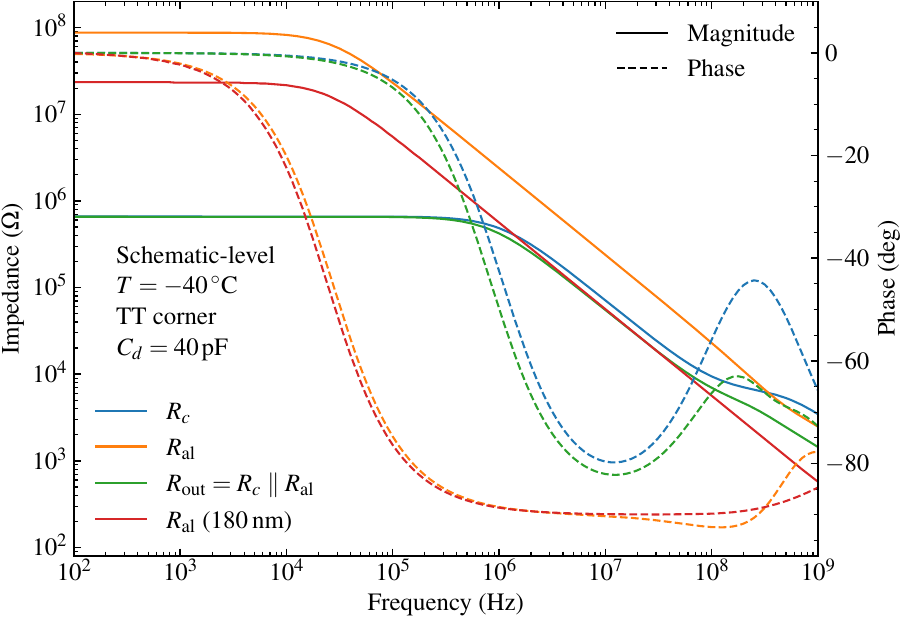}
    \caption{Simulated output impedance of the CSA forward gain stage, decomposed into its dominant contributions: the cascode output resistance \( R_c \), the active load \( R_{\text{al}} \), and their parallel combination \( R_{\text{out}} = R_c \parallel R_{\text{al}} \). Simulations were performed in the typical process corner at \SI{-40}{\celsius}, assuming a detector capacitance of \( C_d = \SI{40}{\pico\farad} \).
    The plot also includes the value of \( R_{\text{al}} \) from the \SI{180}{\nano\meter} implementation under the same operating conditions.}
    \label{fig:ch4-csa_Z0_contribs}
\end{figure}

%   Open-loop gain and related metrics
\subsubsection{Open-loop gain and related metrics}

\begin{figure}[tbp]
    \centering
    \includegraphics[width=0.75\textwidth]{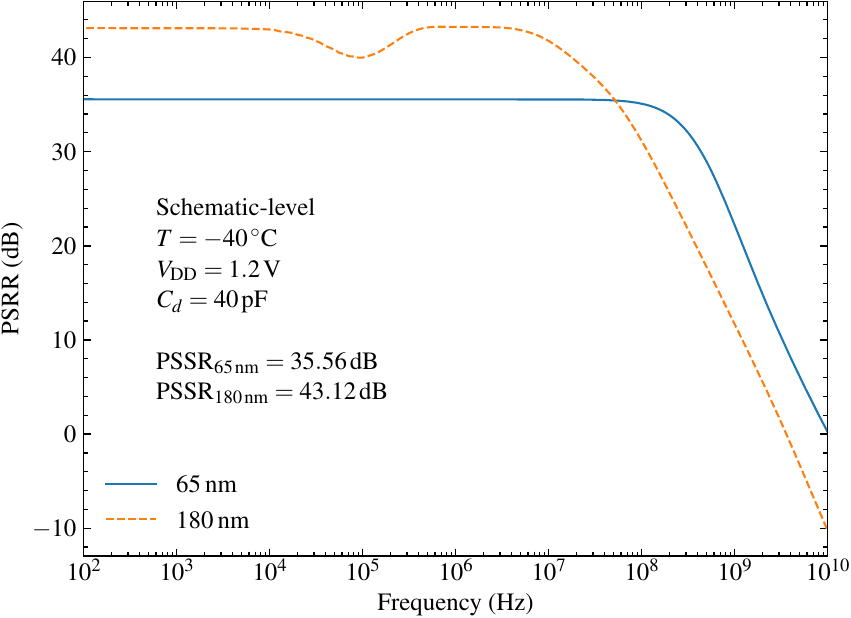}
    \caption{Power supply rejection ratio (PSRR) versus frequency for the CSA forward gain stage, obtained from schematic-level simulations in the typical process corner at \SI{-40}{\celsius}, with a nominal supply voltage of \( V_{\text{DD}} = \SI{1.2}{\volt} \) and a detector capacitance of \( C_d = \SI{40}{\pico\farad} \). The plot compares the \SI{65}{\nano\meter} implementation developed in this work with the previous \SI{180}{\nano\meter} design. The extracted low-frequency PSRR is \SI{35.56}{\decibel} for the \SI{65}{\nano\meter} version and \SI{43.12}{\decibel} for the \SI{180}{\nano\meter} implementation.}
    \label{fig:ch4-CSA_pssr}
\end{figure}

\par
The performance metrics reported in~\Cref{tab:ch4-gain_gbp_csa} across the three main process corners---typical-typical (TT), slow-slow (SS), and fast-fast (FF)~\cite{mullerDeviceElectronicsIntegrated2002a}---show that the CSA forward gain stage, designed with the same architecture as the current \SI{180}{\nano\meter} implementation, provides better GBP in almost all corners, extending the amplifier bandwidth.
This is also visible in the frequency response plotted in~\Cref{fig:ch4-csa_gain_stage_comparison}, where the \SI{65}{\nano\meter} version consistently shows a higher \SI{-3}{\decibel} bandwidth across all operating conditions compared to the \SI{180}{\nano\meter} version.

\par
The DC gain is systematically lower due to design constraints, mainly related to area occupation, power consumption, and technology parameters at the more scaled node.
As shown in~\Cref{tab:ch4-gm_gds_gain_single}, which reports the relevant parameters for the CSA input device in both implementations, although the drain current \( I_{\text{D},0} \) has been increased from \SI{1.5}{\milli\ampere} to \SI{1.8}{\milli\ampere}, the simulated transconductance is lower than in the previous case.
This effect is reflected in a reduced intrinsic gain \( A_v^{\text{int}} \), consistent with the lower DC gain observed in~\Cref{fig:ch4-csa_gain_stage_comparison}.
The latter can be assumed as:

\begin{equation}
    \left| A_{\text{CSA}} \right| = \frac{V_{\text{out}}}{V_{\text{in}}} \approx g_{m,0} R_{\text{out}}
\end{equation}

\noindent
where \( g_{m,0} \) is the transconductance of the input transistor M0, and \( R_{\text{out}} \) is the output resistance, which, to a first-order approximation, can be taken as the parallel combination of the cascode impedance \( R_c \) and the active load impedance \( R_{\text{al}} \).
The cascode impedance \( R_c \) is evaluated at the drain of M2 and is given by:

\begin{equation}
    R_c = \frac{g_{m2} g_{m5}}{g_{ds2} (g_{ds0} + g_{ds1})(g_{ds5} + g_{ds6})},
\end{equation}

\noindent
with the active load impedance \( R_{\text{al}} \) being the impedance seen at the drain on M3 and is defined as:

\begin{equation}
    R_{\text{al}} = \frac{1}{g_{ds,4}} \frac{g_{m,3}}{g_{ds,3}} \frac{g_{m,8}}{g_{ds,7} + g_{ds,8}}.
\end{equation}

\par
From schematic-level simulation in the typical process corner at \SI{-40}{\celsius}, \hbox{\( R_c = \SI{599.42}{\kilo\ohm} \)} and \( R_{\text{al}} = \SI{89.42}{\mega\ohm} \), from which \( R_{\text{out}} = \SI{595.42}{\kilo\ohm} \) is derived.
The \hbox{small-signal} open-loop DC gain is therefore \SI{87.14}{\decibel}, in agreement with the DC simulation result of \SI{87.67}{\decibel}.
The impedance contributions of the two series stages are illustrated in~\Cref{fig:ch4-csa_Z0_contribs}, showing the frequency response of the cascode output resistance \( R_c \), the active load \( R_{\text{al}} \), and their parallel combination \( R_{\text{out}} = R_c \parallel R_{\text{al}} \).
As expected, the high impedance of \( R_{\text{al}} \) dominates at low frequency, while \( R_c \) defines the output impedance at high frequency due to the pole introduced by the active load.
A comparison with the \SI{180}{\nano\meter} implementation confirms the substantial improvement in bandwidth without significant penalty in output resistance.

\par
Beyond output impedance and open-loop gain, other key metrics have been evaluated to characterize the analog performance of the forward gain stage.
In terms of power supply rejection, the simulated low-frequency PSRR is \SI{35.56}{\decibel} for the \SI{65}{\nano\meter} implementation, compared to \SI{43.12}{\decibel} in the previous \SI{180}{\nano\meter} design, as reported in~\Cref{fig:ch4-CSA_pssr}.
Additionally, a closed-loop stability analysis with the amplifier configured in unity-gain buffer mode yields a phase margin of \SI{31.65}{\degree} at a unity-gain frequency of \SI{321}{\mega\hertz}.
This \hbox{unity-gain} configuration is adopted solely as a worst-case scenario for stability assessment and is not representative of the amplifier operating conditions in the actual CSA architecture.

\begin{figure}[tbp]
    \centering
    \includegraphics[width=\textwidth]{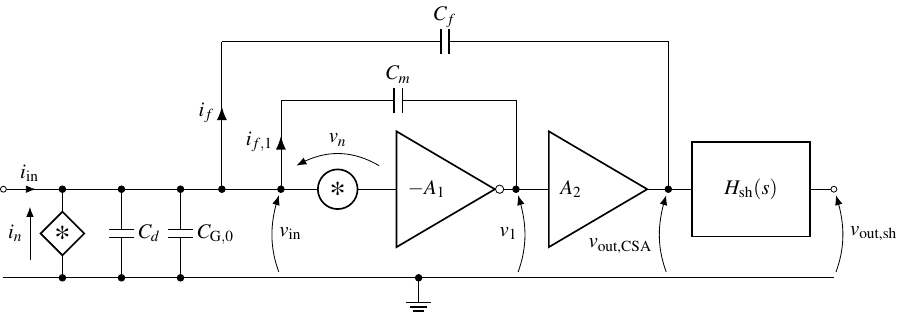}
    \caption{Simplified noise model of a CSA followed by a \( \text{CR-(RC)}^n \) shaper, including series noise \( v_n \) and parallel noise \( i_n \). The model incorporates the Miller effect through capacitance \( C_m \), and assumes uncorrelated white noise sources. Adapted from~\cite{kaplonFrontendElectronicsSilicon2023}.}
    \label{fig:ch4-noise_model_kaplon}
\end{figure}

%   CSA noise modelling
\subsubsection{CSA noise modelling}

\par
The noise performance of the CSA determines the energy resolution of the entire analog front-end and, in deep-submicron CMOS implementations, is often dominated by the \hbox{low-frequency} \(1/f\) noise of the input transistor. 
The noise contributions can be categorized into series and parallel noise sources. 
A simplified model of the CSA used for analytical evaluation is shown in~\Cref{fig:ch4-noise_model_kaplon}, where the front-end readout chain is represented by an ideal CSA followed by a CR-RC\textsuperscript{n} shaper with transfer function $H_{\text{sh}}(s)$.

\par
According to the original Van der Ziel noise model~\cite{vanderzielNoiseSolidState1978}, the series noise arises mainly from the channel thermal noise of the input MOS transistor, which can be represented by an equivalent input voltage source \( v_n \) with spectral density~\cite{turchettaAnalogElectronicsRadiation2017a}:

\begin{equation}
    S^2_\text{W} = \frac{4k_B T\gamma n}{g_{m,0}},
\end{equation}

\noindent
where \( k_B \) is Boltzmann's constant, \( T \) is absolute temperature, \( \gamma \) and \( n \) are bias-dependent parameters~\cite{enzAnalyticalMOSTransistor1995a}, and \( g_{m,0} \) is the transconductance of the input transistor.
The corresponding ENC contribution from the thermal series noise is:

\begin{equation}
    \text{ENC}_{\mathrm{S,W}} = F_V\, C_{\text{in}}\, \frac{1}{\sqrt{t_p}}\,v_n,
\end{equation}

\noindent
where \( t_p \) is the peaking time of the shaper, \( C_{\text{in}} = C_d + C_{\text{G},0} + C_f + C_m \), and \( F_V \) is a shaping-dependent form factor~\cite{kaplonFrontendElectronicsSilicon2023}.

\par
In addition to the white thermal contribution, the input device also generates a \hbox{low-frequency} \(1/f\) noise component. 
In the general ENC formulation of~\cite{rattiDesignOptimizationCharge2009}, this appears as a separate series term with frequency exponent \(\alpha_{fs}\). 
For simplicity, in this work a purely flicker spectrum is assumed by setting \(\alpha_{fs}=1\), leading to a contribution that scales as:

\begin{equation}
    \text{ENC}_{\mathrm{S},1/f} \propto C_{\text{in}} \, \sqrt{A_f},
\end{equation}

\noindent
where \(A_f\) is the flicker noise coefficient of the device.
Under the assumption \(\alpha_{fs}=1\), this term does not explicitly depend on the shaping time.
Its relative weight may nevertheless become more apparent at longer peaking times, but only because the series white noise contribution decreases with \(t_p\), and provided that parallel noise sources remain negligible.

\par
Parallel noise sources include contributions from the detector leakage current \( I_{\text{leak}} \), the feedback network (e.g., a Krummenacher circuit or a resistive element), and any bias circuitry.
These are modeled as current noise sources at the input node of the CSA.
The first major contribution originates from the detector leakage current, which induces shot noise due to the statistical fluctuations of carrier injection across the reverse-biased junction.
This noise source is characterized by a white spectral density given by:

\begin{equation}
    S_{\text{leak}}^2 = 2qI_{\text{leak}},
\end{equation}

\noindent
where \( q \) is the elementary charge and \( I_{\text{leak}} \) is the detector leakage current flowing through the detector.
This contribution increases with temperature and becomes particularly relevant during on-ground testing, when the leakage current can exceed nominal values.
A second source of parallel noise arises from the thermal (Johnson-Nyquist) noise of the feedback resistor \( R_f \), if present.
This component is modeled as a white current noise source with spectral density:

\begin{equation}
    S_{\text{fdb}}^2 = \frac{4k_B T}{R_f}.
\end{equation}

\begin{figure}[tbp]
    \centering
    \includegraphics[width=0.75\textwidth]{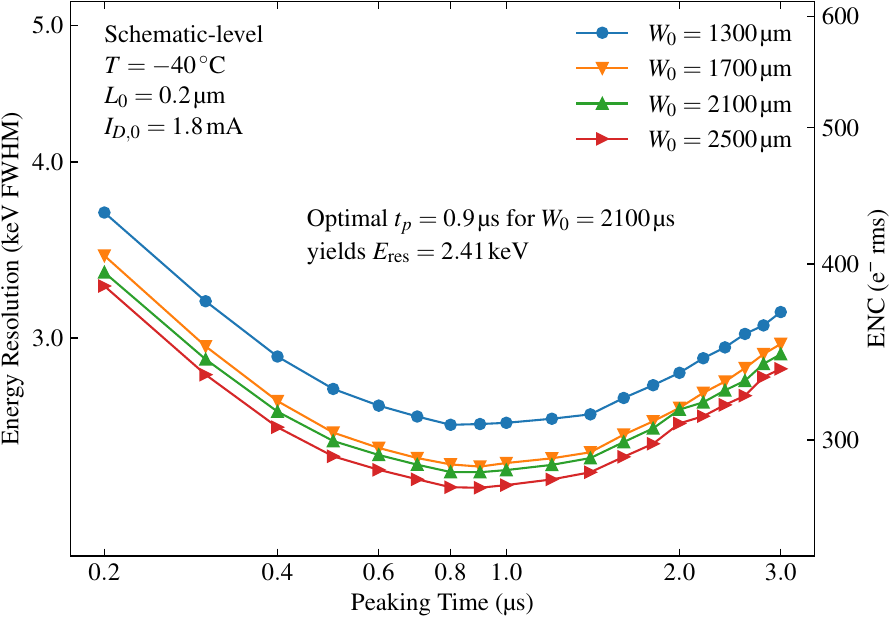}
    \caption{Energy resolution and ENC as a function of the peaking time for four values of the input transistor width \( W_0 \), obtained from schematic-level simulations at \SI{-40}{\celsius} with \( L_0 = \SI{0.2}{\micro\meter} \) and \( I_{D,0} = \SI{1.8}{\milli\ampere} \). The optimal peaking time of \SI{0.9}{\micro\second} yields an energy resolution of \SI{2.41}{\kilo\electronvolt} for \( W_0 = \SI{2100}{\micro\meter} \) and \SI{2.35}{\kilo\electronvolt} for \( W_0 = \SI{2500}{\micro\meter} \).}
    \label{fig:ch4_CSA_ENC_vs_tp_W}
\end{figure}

\par
In practical implementations, particularly in integrated circuits where large resistors are not feasible, this element is often replaced by a transconductance stage (e.g., Krummenacher circuit), which achieves the same charge restoration function while reducing parallel noise.
The corresponding parallel noise contribution to the ENC is:

\begin{equation}
    \text{ENC}_{\mathrm{P,W}} = F_I\, \sqrt{t_p}\,i_n,
\end{equation}

\noindent
with \( F_I \) is another shaping-dependent factor.
In advanced CMOS processes, an additional parallel \(1/f\) contribution may arise from low-frequency fluctuations in the gate current of the input transistor or in bias circuitry.
This term, usually modeled with a spectral density scaling as \(1/f^{\alpha_{fp}}\) with \(\alpha_{fp}\approx 1\), is generally negligible in older technologies but can become relevant at long shaping times in deep-submicron nodes.

\par
The total noise is obtained by combining the contributions from series white noise, series \(1/f\) noise, and parallel noise sources. 
Following the general formulation of~\cite{rattiDesignOptimizationCharge2009}, the ENC can be expressed as:

\begin{equation}
    \text{ENC}^2 = \text{ENC}_{\mathrm{S,W}}^2 + \text{ENC}_{\mathrm{S},1/f}^2 + \text{ENC}_{\mathrm{P,W}}^2 + \text{ENC}_{\mathrm{P},1/f}^2 ,
    \label{eq:ch4-enc_series_parallel}
\end{equation}

\noindent
where \(\text{ENC}_{\mathrm{S,W}}\) is the thermal series (white) contribution, \(\text{ENC}_{\mathrm{S},1/f}\) accounts for the series flicker noise component (here with \(\alpha_{fs}=1\)), \(\text{ENC}_{\mathrm{P,W}}\) represents the parallel white contributions (detector leakage and feedback network), and \(\text{ENC}_{\mathrm{P},1/f}\) denotes the parallel flicker term arising from gate current fluctuations.

\par
An important conclusion of this analysis is that although the Miller capacitance \( C_m \) affects both signal gain and noise bandwidth, its impact cancels out in the final expression of the ENC, assuming the amplifier does not contribute to shaping. 
In practical scenarios, where the CSA has non-negligible bandwidth limitations, this ideal assumption introduces deviations that must be accounted for when matching simulations to silicon measurements.

\par
To meet the target energy resolution of \SI{4}{\kilo\electronvolt} FWHM at \( C_d = \SI{40}{\pico\farad} \), the input transistor was dimensioned and biased to achieve an optimal trade-off among the different noise components. 
The large device width and operating current were selected to reduce the series white noise while keeping the impact of series \(1/f\) noise under control, and the bias conditions were chosen to limit the contribution of parallel sources, all within the \SI{10}{\milli\watt\per{channel}} power budget, as reported in~\Cref{tab:ch4-gm_gds_gain_single} and detailed in the next section.

%   Input transistor sizing and noise optimization
\subsubsection{Input transistor sizing and noise optimization}\label{ch4:sec-csa_input_device}

\begin{figure}[tbp]
    \centering
    \includegraphics[width=0.75\textwidth]{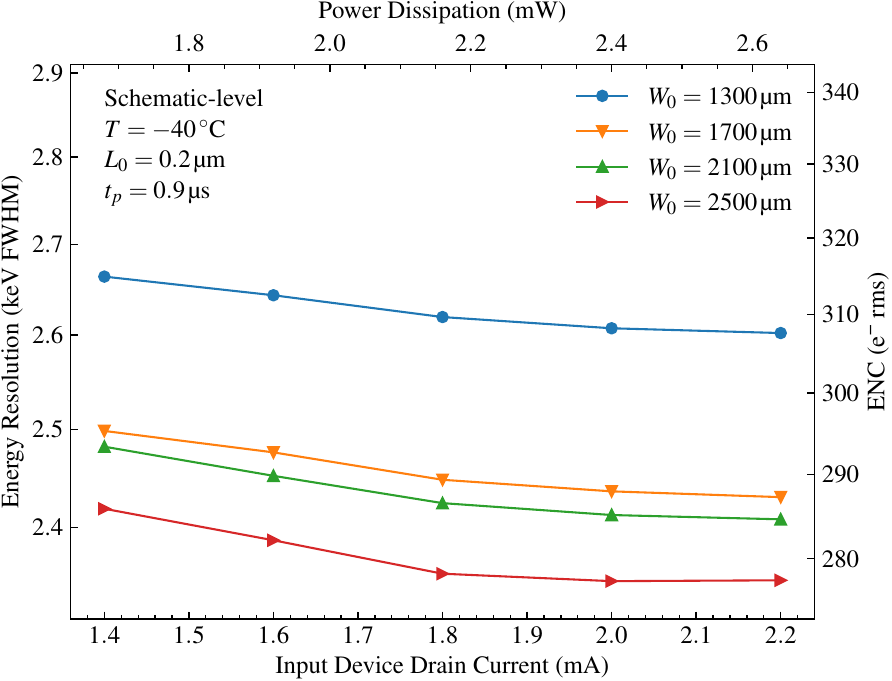}
    \caption{Energy resolution and ENC as a function of the input device drain current for four values of \( W_0 \), at \( t_p = \SI{0.9}{\micro\second} \). Power dissipation is also reported. Simulations were performed at schematic level with \( L_0 = \SI{0.2}{\micro\meter} \) and \( T = \SI{-40}{\celsius} \).}
    \label{fig:ch4_CSA_ENC_vs_ID_W}
\end{figure}

In a properly designed readout channel, the resolution is mainly determined by the noise performance of the CSA input device.
Therefore, its dimensions ($W_0/L_0$) and the bias condition ($I_{D,0}$) are determined by evaluating the performance of the channel in terms of energy resolution FWHM and ENC.

\par
Since simulations were performed at the schematic level, in this analysis an additional noise source contribution has been introduced in series with the gate of the input device to take the thermal noise contribution of the gate distributed resistance $R_{\text{GG}}$ of M0 into account, which is non-negligible in wide gate transistors as the one considered.
It is evaluated as the sum of the resistance of the polysilicon gate and the gate interconnections, assuming an interdigitated multifinger layout with each gate finger contacted on both sides~\cite{tsakasNoiseAssociatedInterdigitated2000}:

\begin{equation}
    R_{\text{GG}} = \frac{1}{12n}R_i + \frac{(n+1)(2n+1)}{12n}R_{\text{int}},
\end{equation}

\noindent
where $R_i = \SI{74.5}{\ohm}$ is the resistance of one gate finger, $n$ is the number of fingers, and $R_{\text{int}}=\SI{0.32}{\ohm}$ is the resistance of the interconnections between adjacent fingers.
These values come from layout simulations and process parameters~\cite{manghisoniLowNoiseAnalogChannel2021}.
A noise analysis was performed at \SI{-40}{\celsius} with $C_{d}=\SI{40}{\pico\farad}$, using an ideal CR-RC$^2$ semi-Gaussian unipolar shaper and varying $t_p$ from \SIrange{0.2}{3}{\micro\second}.
According to a \hbox{well-established} practice, the channel length of M0 was set to \SI{0.2}{\micro\meter}, i.e., more than three times the minimum feature length, to avoid the increase of $1/f$ noise exhibited by short-channel devices~\cite{manfrediResolutionLimitsAchievable2003}.

\par
In~\Cref{fig:ch4_CSA_ENC_vs_tp_W}, the energy resolution and the ENC are plotted for four values of $W_0$ as a function of $t_p$ with $I_{D,0}=\SI{1.8}{\milli\ampere}$.
As expected from~\eqref{eq:ch4-enc_series_parallel}, the ENC initially decreases due to the reduction of series white noise, reaching a minimum between peaking times of \SI{0.8}{\micro\second} and \SI{1.0}{\micro\second}, where the noise is dominated by the $1/f$ term.
Results in~\Cref{fig:ch4_CSA_ENC_vs_tp_W} show that the minimum ENC decreases with the increase of $W_0$ thanks to a reduction of $1/f$ noise, whose power spectral density is proportional to the inverse of the gate area of M0.
The increment for longer peaking times is associated to the increase of parallel noise.

\begin{figure}[tbp]
    \centering
    \includegraphics[width=0.8\textwidth]{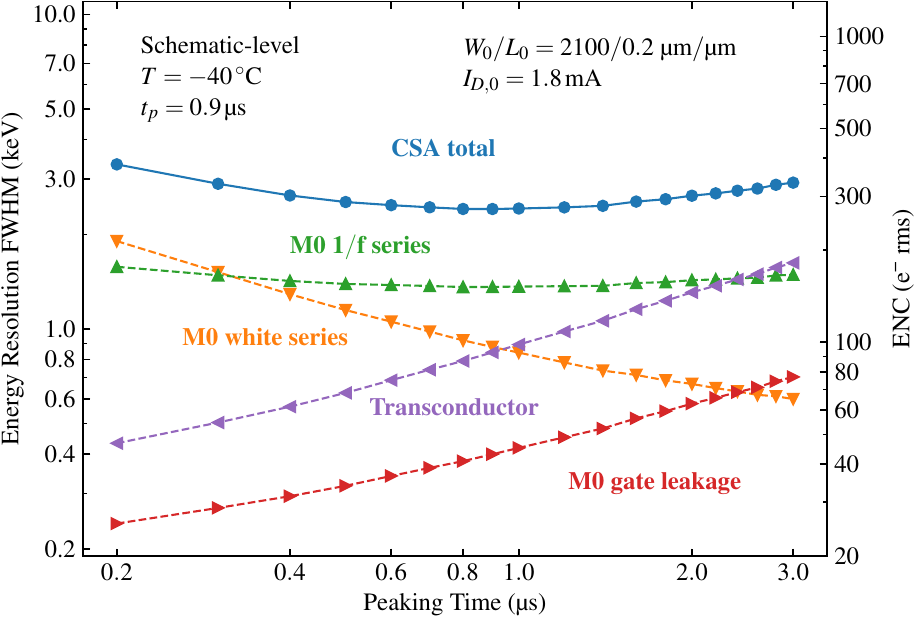}
    \caption{Energy resolution and ENC as a function of the peaking time, highlighting the individual noise contributions from the input transistor M0 and the transconductor. Simulations were performed at schematic level with \( W_0/L_0=2100/0.2~\si{\micro\meter/\micro\meter} \), \hbox{\( I_{D,0} = \SI{1.8}{\milli\ampere} \)}, and \( T = \SI{-40}{\celsius} \). The contribution from gate leakage, $1/f$ and white series noise, and the transconductor parallel noise are shown separately.}
    \label{fig:ch4-CSA_ENC_contribs}
\end{figure}

\par
The input device drain current was varied from \SIrange{1.4}{2.2}{\milli\ampere} and no significant noise reduction was achieved above \SI{1.8}{\milli\ampere}, as reported in~\Cref{fig:ch4_CSA_ENC_vs_ID_W}.
Consequently, it was decided to better evaluate the noise performance of the amplifier with M0 having \hbox{$W_0/L_0=2100/0.2$~\si{\micro\meter/\micro\meter}} and \hbox{$I_{D,0}=\SI{1.8}{\milli\ampere}$}.
The ENC of the CSA is plotted in~\Cref{fig:ch4-CSA_ENC_contribs} as a function of $t_p$. In the same plot, contributions coming from the input device and the restoration network are highlighted.
A minimum of \SI{285}{e^{\text{-}}}~rms, corresponding to an energy resolution of \SI{2.41}{\kilo\electronvolt}~FWHM, is achieved at \hbox{$t_p=\SI{0.9}{\micro\second}$}, with M0 contributing for \SI{69}{\percent} of the overall noise.

%   Gate leakage current as a noise source
\subsubsection{Gate leakage current as a noise source}

\par
As shown in~\Cref{fig:ch4-CSA_ENC_contribs}, a \hbox{non-negligible} parallel noise contribution arises from the input device gate leakage current, accounting for \SI{16}{\percent} of the total CSA noise at $t_p=\SI{0.9}{\micro\second}$, increasing to \SI{24}{\percent} at $t_p=\SI{3.0}{\micro\second}$.
Gate leakage current cannot be neglected in the process of designing analog circuits in nanoscale CMOS technologies, as it results from discrete charges randomly crossing a potential barrier, leading to increased static power consumption in digital circuits and possible degradation of analog noise performance because of the associated shot and noise contributions~\cite{manghisoniAssessmentLowPower652014a}. 

\par
In the \SI{65}{\nano\meter} low-power option, the increased equivalent oxide thickness of the core devices results in a lower gate tunneling probability and hence reduced leakage currents. 
Data reported in~\cite{manghisoniAssessmentLowPower652014a} show that LP MOSFETs exhibit gate current densities comparable to those measured in some \SI{90}{\nano\meter} and \SI{130}{\nano\meter} processes, and remain well below the commonly adopted critical limit of \SI{1}{\ampere\per\centi\meter\squared}.

\begin{figure}[tbp]
    \centering
    \includegraphics[width=0.95\textwidth]{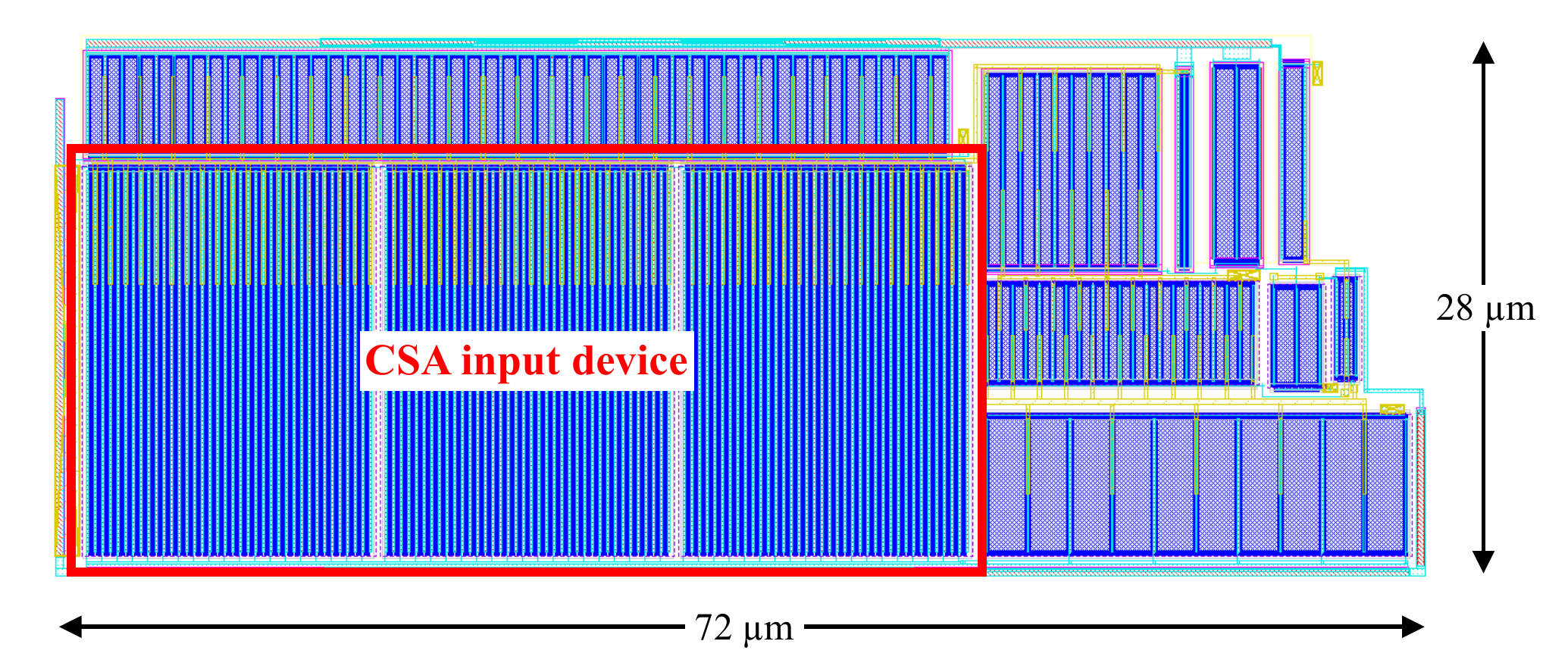}
    \caption{Layout of the CSA forward gain stage, occupying an area of \(\SI{72}{\micro\meter} \times \SI{28}{\micro\meter}\). 
    The input transistor is implemented as a parallel combination of three identical devices to meet the required total gate width of \(\SI{2100}{\micro\meter}\), due to the technological constraint limiting the maximum gate width of a single device to \(\SI{900}{\micro\meter}\).}
    \label{fig:ch4-csa_gain_stage_layout}
\end{figure}

\par
The relevance of gate leakage current as a noise source is further confirmed by earlier investigations in a \SI{90}{\nano\meter} CMOS process~\cite{manghisoniGateCurrentNoise2008a}, where static and noise measurements on NMOS devices demonstrated that direct tunneling currents may generate a significant noise contribution that must be taken into account during circuit design and optimization. 
That study also confirmed the presence of both white and $1/f$ components in the power spectral density of the gate current noise, and modeled this effect as a parallel current noise source at the device gate.

%   Layout of the forward gain stage
\subsubsection{Layout of the forward gain stage}

\par
The layout of the CSA forward gain stage occupies an area of $\SI{72}{\micro\meter}\times\SI{28}{\micro\meter}$, equivalent to \SI{2016}{\micro\meter\squared}.
The major area occupation comes from the input device M0, which has been implemented as the parallel combination of three identical devices with \hbox{\( W_0'/L_0' = 700/0.2~\si{\micro\meter/\micro\meter} \)} to achieve the required width \( W_0 = \SI{2100}{\micro\meter} \).
This configuration was needed due to the constraint imposed by the technology, that limited the maximum gate width to \SI{900}{\micro\meter} for a single device.

%   Feedback capacitor
\subsection{Feedback capacitor}\label{sec:ch4-csa-feedback-capacitor}

\par
The feedback capacitor \( C_f \) integrates the input charge and defines the charge-to-voltage gain of the CSA as \( V_{\text{out}} = Q_{\text{in}} / C_f \).
Its value is determined based on the required gain and the expected input charge range.
In the design of the \antareslogo CSA, \( C_f \) is responsible for the nonlinear response introduced by the dynamic signal compression mechanism.

%   Dynamic signal compression
\subsubsection{Dynamic signal compression}

\begin{figure}[tbp]
    \centering
    \includegraphics[width=0.75\textwidth]{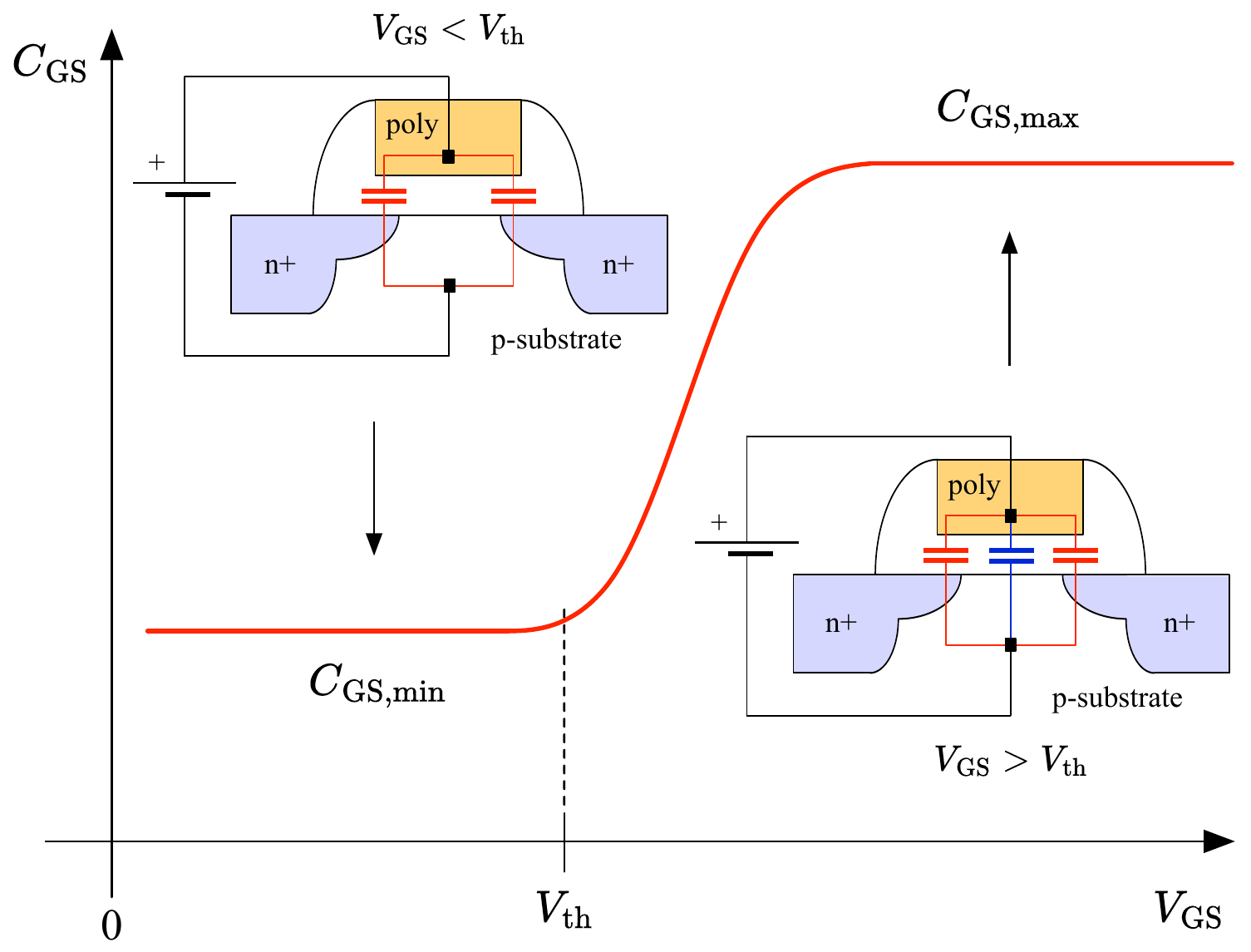}
    \caption{Qualitative behavior of the gate capacitance \( C_{\text{GS}} \) as a function of the \hbox{gate-source} voltage \( V_{\text{GS}} \) for an inversion-mode NMOS capacitor with fixed gate width. The transition from accumulation to inversion near \( V_{\text{th}} \) results in a sharp capacitance increase, which is exploited for signal compression. Figure adapted from~\cite{manghisoniDynamicCompressionSignal2015}.}
    \label{fig:ch4-dynamic_signal_compression}
\end{figure}

\par
In conventional CSAs, the charge-to-voltage gain is set by the fixed feedback capacitance \( C_f \), resulting in a linear input--output relationship.
To extend the dynamic range without compromising resolution, a voltage-dependent feedback capacitance is introduced.
This is achieved using an inversion-mode MOS capacitor with shorted source and drain terminals and the gate connected to the CSA output~\cite{girolettiADA5Dv0AnalogFrontEnd2024}.
The capacitance \( C_{\text{GS}} \) varies with the \hbox{gate-to-source} voltage \( V_{\text{GS}} \).
As evident from~\Cref{fig:ch4-dynamic_signal_compression}, two regimes can be identified.
For \( V_{\text{GS}} \ll V_{\text{th}} \), the MOS capacitor operates in depletion or weak inversion, and the total capacitance is dominated by gate-to-source $C_{\text{GS,ov}}$ and gate-to-drain $C_{\text{GD,ov}}$ overlap components:

\begin{equation}
    C_{f,\text{min}} \approx C_{\text{GS,ov}} + C_{\text{GD,ov}} = 2W_f \Delta L C_{\text{ox}},
\end{equation}

\noindent
where \( W_f \) and \( \Delta L \) are the gate width and overlap length, and \( C_{\text{ox}} \) is the oxide capacitance per unit area.

\par
For \( V_{\text{GS}} \gg V_{\text{th}} \), strong inversion occurs, and the capacitance reaches a maximum dominated by the gate-to-channel $C_{\text{GC}}$ component:

\begin{equation}
    C_{f,\text{max}} \approx C_{\text{GC}} = W_f L_f C_{\text{ox}},
\end{equation}

\noindent
where \( L_f \) is the channel length. The transition occurs near \( V_{\text{th}} \), introducing a nonlinear gain:

\begin{figure}[tbp]
    \centering
    \includegraphics[width=0.9\textwidth]{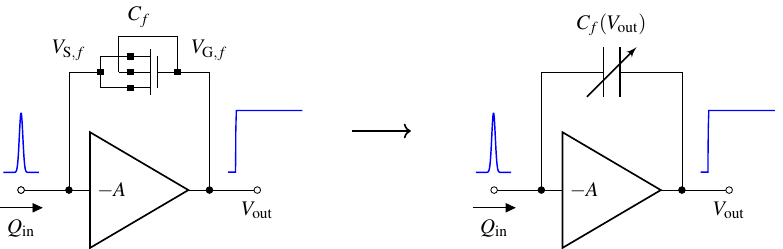}
    \caption{Circuit-level representation of dynamic signal compression in the CSA feedback path. Left: implementation using an inversion-mode MOS capacitor with the gate driven by the CSA output. Right: equivalent representation that emphasizes the voltage-dependent nature of the feedback capacitance with respect to the output voltage, shown as \( C_f(V_{\text{out}}) \). The configuration shown corresponds to an \( n \)-type DTMOS device.}
    \label{fig:ch4-compression_csa}
\end{figure}

\begin{equation}
    V_{\text{out}} = \frac{Q_{\text{in}}}{C_f(V_{\text{out}})},
\end{equation}

\noindent
where \( C_f(V_{\text{out}}) \) increases with the integrated signal, resulting in gain compression.

\par
This behavior has been verified using both standard MOS and DTMOS devices. In all configurations, the source and drain terminals are shorted to form one terminal of the feedback capacitor. 
For \( n \)-type devices, the gate is connected to the CSA output, while for \( p \)-type devices, the gate is connected to the input node. 
The body terminal is either grounded for NMOS devices or tied to \(V_{\mathrm{DD}}\) for PMOS devices in the case of standard MOS operation. 
In the DTMOS configuration, instead, the body is shorted to the gate, so that the threshold voltage follows the gate bias dynamically. 
This provides an additional modulation of the capacitance--voltage characteristic, enhancing the effectiveness of the compression mechanism compared to standard MOS devices. 

\par
The resulting transcharacteristic exhibits a bilinear profile with a transition near \( V_{\text{th}} \). 
This configuration enables the front-end to handle the full detector input range, from a few \si{\kilo\electronvolt} up to \SI{100}{\mega\electronvolt}, while maintaining the required resolution in the low-energy region.
Moreover, the compression profile is tunable via geometry (\( W_f \), \( L_f \)), device polarity, threshold voltage flavor, and body connection, with the DTMOS option introducing an extra degree of control not available in conventional MOS implementations.

\begin{figure}[tbp]
    \centering
    \includegraphics[width=0.85\textwidth]{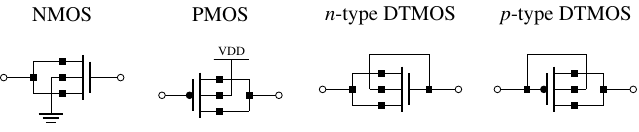}
    \caption{Standard and dynamic threshold MOSFET capacitors implemented as feedback elements. From left to right: NMOS, PMOS, \( n \)-type DTMOS, and \( p \)-type DTMOS configurations. All devices are operated with shorted source and drain terminals to form \hbox{voltage-dependent} capacitors.}
    \label{fig:ch4-DTMOS_configurations}
\end{figure}

%   Design implications of a non-linear channel transcharacteristic
\subsubsection*{Design implications of a nonlinear channel transcharacteristic}

\par
While the nonlinearity introduced by the dynamic signal compression mechanism is the key feature enabling both high energy resolution in the spectroscopy range and the extension of the input dynamic range up to \SI{100}{\mega\electronvolt} in the particle-tracking regime, this mechanism also introduces additional challenges compared to a conventional linear CSA.
Building on the experience gained during the experimental characterization of the SLIDER32 \hbox{front-end}, a fine-tuning of the CSA input--output transcharacteristic was identified as a possible approach to mitigate some limitations of the current implementation, in particular those related to the position of the kink energy and to the width of the transition region between the low- and high-energy regimes.

\par
As reported in~\Cref{tab:ch4-frontend_requirements}, when discussing the front-end electronics requirements, the kink energy position should be greater than \SI{1.5}{\mega\electronvolt} in order to ensure that, during calibration with on-ground cosmic muons, the MIP energy does not fall within the ``bend'' of the energy-to-voltage transcharacteristic, which would complicate the energy reconstruction process.
As described in \Cref{sec:ch3-fdt_calibration}, a dedicated calibration procedure had to be implemented to reconstruct the deposited energy in \si{\kilo\electronvolt} from the ADC output expressed in ADU.
This procedure proved particularly delicate in the proximity of the kink energy, where a cubic interpolation model was required to fit the non-linear behaviour of the front-end.
For the new design, this complication was deliberately avoided by shifting the kink location to higher energies ($\SI{\geq1.5}{\mega\electronvolt}$), thereby guaranteeing a linear interpolation in the MIP energy region.

\par
These challenges directly translated into design constraints for the feedback MOS device, whose geometric parameters, bias conditions, polarity, and bulk connection were systematically investigated to identify the most suitable configuration for the new ASIC design.
The adoption of the DTMOS concept for the CSA feedback capacitor provided an additional design degree of freedom, which was exploited to improve the dynamic signal compression mechanism also in the context of the more scaled \SI{65}{\nano\meter} technology node, where the available dynamic range is reduced due to the lower \SI{1.2}{\volt} supply voltage, compared to the \SI{1.8}{\volt} supply of the \SI{180}{\nano\meter} technology.

%   Dynamic threshold MOSFET (DTMOS)
\subsubsection{Dynamic threshold MOSFET (DTMOS)}

\begin{figure}[tbp]
    \centering
    \includegraphics[width=0.75\textwidth]{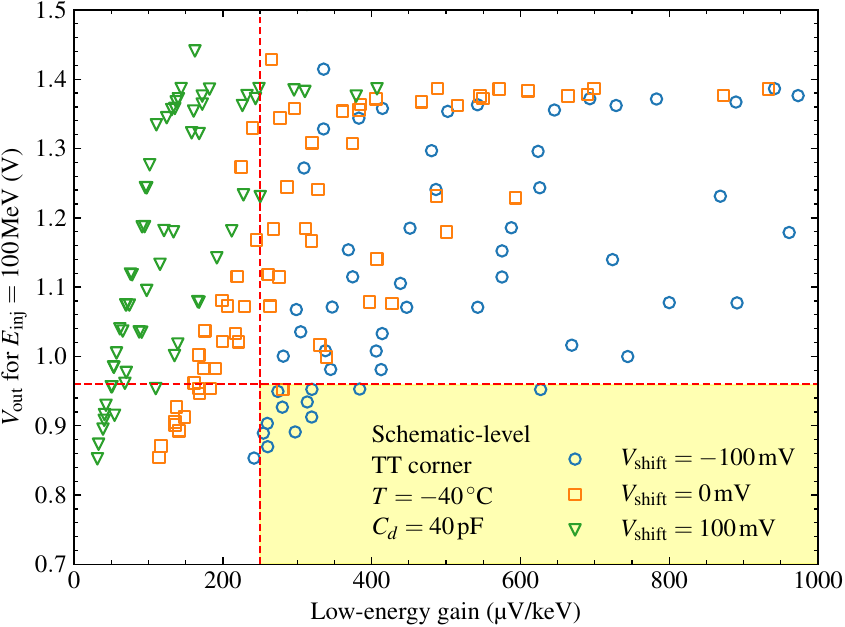}
    \caption{Simulated low-energy gain versus CSA output voltage at \SI{100}{\mega\electronvolt} for all feedback device configurations under test. Simulations were performed at schematic level in the typical process corner at \SI{-40}{\celsius}, with \( C_d = \SI{40}{\pico\farad} \). Each point represents a different combination of polarity, configuration, and geometry, for three values of \( V_{\text{shift}} \). The yellow acceptance area is defined by \( V_{\text{out}} \le \SI{980}{\milli\volt} \) (corresponding to \( \Delta V_{\text{out}} \le \SI{500}{\milli\volt} \)) and \( G_{\text{le}} \ge \SI{250}{\micro\volt\per\kilo\electronvolt} \).}
    \label{fig:ch4-scatter_W20_100_L5_20_allconf_vgs}
\end{figure}

\par
The DTMOS was first introduced in 1994 by Assaderaghi~\cite{assaderaghiDynamicThresholdVoltage1994a} as a device concept aimed at extending the lower bound of supply voltage in CMOS digital circuits.
With technology scaling and the associated drive for reduced power consumption, lowering the supply voltage became essential.
DTMOS addresses this limitation by dynamically modulating the threshold voltage through gate-body shorting: at zero gate bias, a high threshold suppresses off-state leakage, while under forward body bias (i.e., as the gate voltage rises), the threshold decreases, resulting in significantly higher drive current—even at supply voltages as low as 0.2--\SI{0.3}{\volt} in early silicon-on-insulator (SOI) implementations~\cite{jimenez-pModellingDynamicThreshold2005}.

\par
In this work, DTMOS devices are employed as voltage-dependent capacitors in the feedback network of the CSA to implement dynamic signal compression.
The mechanism was tested using both standard MOS and DTMOS devices, with all variants configured as two-terminal capacitors by shorting source and drain.
The gate connection depends on the device polarity: it is connected to the CSA output for \( n \)-type devices and to the input node for \( p \)-type devices.

\par
In the DTMOS configuration, the gate and body terminals are shorted, so that the threshold voltage \( V_{\text{th}} \) varies with the gate-to-source voltage \( V_{\text{GS}} \) due to the body effect. The threshold voltage is expressed as:

\begin{equation}
    V_{\text{th}} = V_{\text{th,0}} + \gamma \left( \sqrt{2\phi_F + V_{\text{SB}}} - \sqrt{2\phi_F} \right),
\end{equation}

\noindent
where \( V_{\text{th,0}} \) is the zero-bias threshold voltage, \( \gamma \) is the body-effect coefficient, \( \phi_F \) is the Fermi potential, and \( V_{\text{SB}} = -V_{\text{GS}} \) in this configuration~\cite{assaderaghiDynamicThresholdVoltage1994}.

\begin{table}[t!]
    \centering
    \renewcommand{\arraystretch}{1.1}
    \setlength{\tabcolsep}{4.9pt}
    \begin{tabular}{lcccccccc}
        \toprule
        \multirow{2}{*}{Configuration} & $L_f$ & $W_f$ & $A_{\text{gate}}$ & $G_{\text{le}}$ & $G_{\text{he}}$ & $\Delta V_{\text{out}}$ & $k$ & Kink \\
                     & (\si{\micro\meter}) & (\si{\micro\meter}) & (\si{\micro\meter\squared}) & (\si{\micro\volt/\kilo\electronvolt}) & (\si{\micro\volt/\kilo\electronvolt}) & (\si{\milli\volt}) & & (\si{\kilo\electronvolt}) \\
        \midrule
        $n$-type DTMOS  & 20 & 80  & 1600 & 297 & 2.14 & 503 & 139 & 980 \\
        $p$-type DTMOS  & 20 & 100 & 2000 & 262 & 2.01 & 482 & 130 & 1095 \\
        NMOS            & 20 & 100 & 2000 & 255 & 2.09 & 502 & 123 & 1155 \\
        \midrule
        NMOS (\SI{180}{\nano\meter})    & 20 & 115 & 2300 & 312 & 2.53 & 688 & 123 & 887 \\
        \bottomrule
    \end{tabular}
    \caption{Candidate configurations satisfying a low-energy gain \( G_{\text{le}} \) of at least \SI{250}{\micro\volt/\kilo\electronvolt} and a CSA output voltage \( V_{\text{out}} \) below \SI{980}{\milli\volt} for an incoming energy \( E_{\text{inj}} = \SI{100}{\mega\electronvolt} \), corresponding to a maximum output swing \( \Delta V_{\text{out}} \) not exceeding \SI{500}{\milli\volt}. The gate voltage shift is set to \( V_{\text{shift}} = \SI{-100}{\milli\volt} \). The compression factor is defined as \( k = G_{\text{le}} / G_{\text{he}} \), and \( \Delta V_{\text{out}} \) denotes the CSA output voltage swing.}
    \label{tab:ch4-feedback_device_selection}
\end{table}

\begin{table}[t!]
    \centering
    \renewcommand{\arraystretch}{1.1}
    \setlength{\tabcolsep}{8pt}
    \begin{tabular}{lccccc}
        \toprule
        \multirow{2}{*}{Configuration} & $L_f$ & $W_{f,\text{tot}}$ & $W_{f,\text{finger}}$ & Fingers & $A_{\text{gate}}$ \\
                     & (\si{\micro\meter}) & (\si{\micro\meter}) & (\si{\micro\meter}) & \# & (\si{\micro\meter\squared}) \\
        \midrule
        $n$-type MOS    & 14.0 & 114.3 & 22.86 & 5 & 1600.2 \\
        $n$-type DTMOS  & 14.0 & 114.3 & 22.86 & 5 & 1600.2 \\
        $p$-type MOS    & 20.0 & 80.0  & 20.00 & 4 & 1600.0 \\
        $p$-type DTMOS  & 20.0 & 80.0  & 20.00 & 4 & 1600.0 \\
        \bottomrule
    \end{tabular}
    \caption{Geometrical parameters of the four feedback MOS configurations integrated in the final design. Each configuration was dimensioned to match the gate area of the selected $n$-type DTMOS (\(80/20\,\si{\micro\meter/\micro\meter}\)) configuration, enabling a fair comparison of their performance.}
    \label{tab:ch4-antares4_feedback_capacitors}
\end{table}

\par
The compressive behavior arising from a voltage-dependent feedback capacitance is a general feature of MOS capacitors operated with shorted source and drain terminals~\cite{manghisoniNovelActiveSignal2015}.
As the CSA output voltage increases, the gate-to-source voltage of the feedback device rises, enhancing channel inversion and increasing the gate capacitance.
In the DTMOS configuration, this effect is further amplified by the dynamic reduction of \( V_{\text{th}} \) under forward gate-body bias, which accelerates the onset of strong inversion~\cite{assaderaghiDTMOSItsDerivatives2000}.
As a result, DTMOS devices exhibit a steeper capacitance increase and stronger signal compression compared to standard MOS devices with fixed body potential.
This enhancement is evaluated in the next section through simulation-based comparisons of standard and dynamic threshold MOS configurations, aimed at selecting the optimal device for implementing the \hbox{voltage-dependent} feedback capacitance of the CSA.

\par
To the best of the author's knowledge, this work presents the first implementation of DTMOS devices---of both polarities---as voltage-dependent feedback capacitors in the feedback network of a CSA for a particle detector front-end ASIC.
Prior works have investigated DTMOS behavior in relation to threshold voltage modulation~\cite{deenDynamicThresholdVoltage2000}, \hbox{low-voltage} analog design techniques~\cite{maymandi-nejadDTMOSTechniqueLowvoltage2006}, and low-frequency noise characteristics~\cite{hsuLowfrequencyNoiseProperties1999}.
However, no studies have explored the use of DTMOS devices as voltage-dependent feedback capacitors in charge-sensitive amplifiers, nor their application to dynamic signal compression in silicon-strip front-end readout electronics.

%   Feedback device sizing and geometry
\subsubsection{Feedback device sizing and geometry} \label{sec:ch4-feedback_device_dimensioning}

\begin{figure}[t!]
    \centering
    \includegraphics[width=0.75\textwidth]{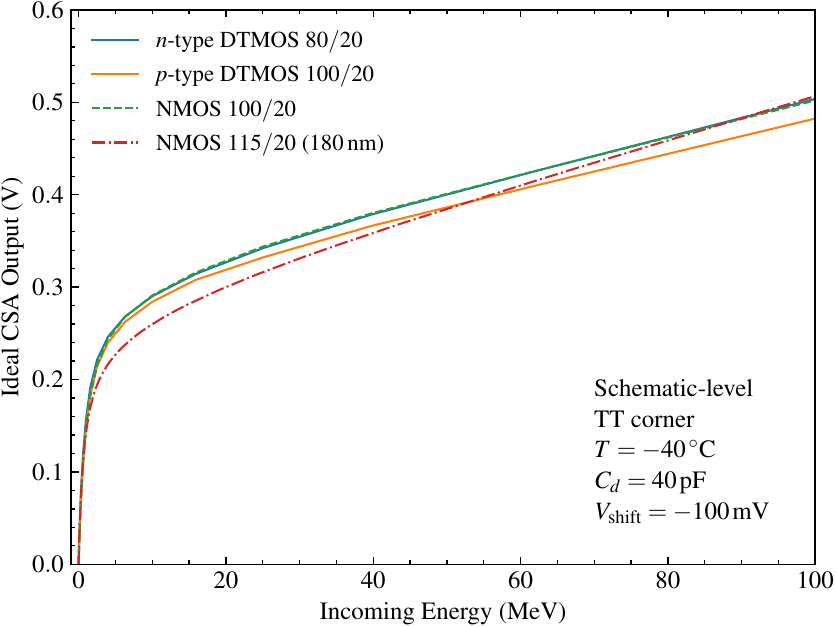}
    \caption{Schematic-level simulation of the CSA output transcharacteristic as a function of incoming energy for different feedback device configurations. Results are shown for an $n$-type DTMOS ($80/20$), a $p$-type DTMOS ($100/20$), and a standard NMOS ($100/20$). For comparison, the NMOS ($115/20$) from the previous \SI{180}{\nano\meter} implementation is also reported. Simulations were performed in the typical process corner at \SI{-40}{\celsius}, with \( C_d = \SI{40}{\pico\farad} \) and \( V_{\text{shift}} = \SI{-100}{\milli\volt} \).}
    \label{fig:ch4-CSA_fdt}
\end{figure}

\par
The design of the feedback element required a careful trade-off among multiple constraints. 
The objective was to define a configuration of the feedback MOS device that met the requirements on low- and high-energy gain, output dynamic range, and area occupation.

\par
The selection criteria were: a gain \( G_{\text{le}} \ge \SI{250}{\micro\volt/\kilo\electronvolt} \) in the low-energy x-ray region, a sensitivity \( G_{\text{he}} \ge \SI{2}{\micro\volt/\kilo\electronvolt} \) in the \SIrange{25}{100}{\mega\electronvolt} range, a maximum output swing \( \Delta V_{\text{out}} \le \SI{0.5}{\volt} \), and minimal gate area.
A preliminary sweep was performed using an ideal CSA testbench composed of an ideal voltage-controlled gain stage (\( A_{\text{ideal}} = \SI{-e6}{\volt/\volt} \)), an ideal transconductor (\( G_{f,\text{ideal}} = \SI{100}{\nano\siemens} \)), and an ideal current pulse generator for charge injection.

\begin{figure}[t!]
    \centering
    \includegraphics[width=0.75\textwidth]{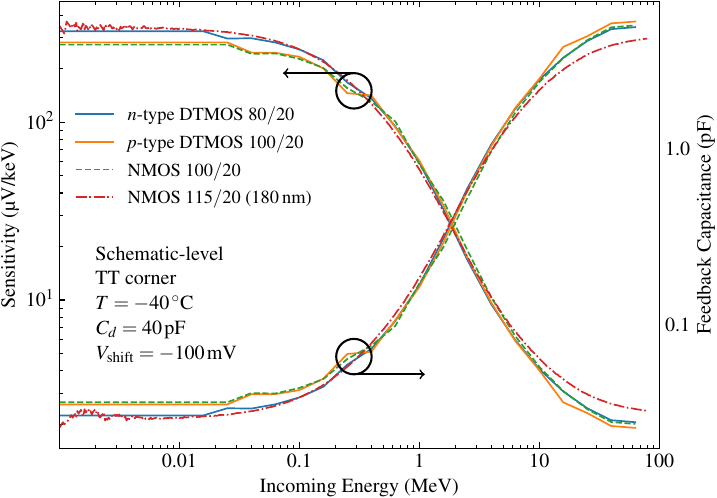}
    \caption{Schematic-level simulation of CSA sensitivity (left $y$-axis) and feedback capacitance (right $y$-axis) as a function of incoming energy for the same feedback device configurations: $n$-type DTMOS ($80/20$), $p$-type DTMOS ($100/20$), standard NMOS ($100/20$), and the reference NMOS ($115/20$) from the \SI{180}{\nano\meter} implementation. The results highlight the effect of device geometry and DTMOS configuration on the dynamic compression behavior. Simulations were performed in the typical process corner at \SI{-40}{\celsius}, with \( C_d = \SI{40}{\pico\farad} \) and \( V_{\text{shift}} = \SI{-100}{\milli\volt} \).}
    \label{fig:ch4-CSA_gain_cap}
\end{figure}

\par
This initial analysis aimed to evaluate the behavior of the feedback device as a function of its DC operating point under quiescent conditions, i.e., without injected charge.
The source terminal of the feedback device was biased at \( V_{\text{S},f} \), corresponding to the quiescent gate voltage of the gain stage input device described in~\Cref{sec_ch4-csa_forward_gain_stage}, while the gate terminal was set to \( V_{\text{G},f} \), which corresponds to the CSA output baseline voltage imposed by the feedback transconductor via \( V_{\text{ref,CSA}} \). 
\( V_{\text{S},f} \) was set to \SI{480}{\milli\volt} to match the forward stage input bias point, and \( V_{\text{G},f} \) was varied by \( V_{\text{shift}} = \SI{\pm100}{\milli\volt} \) around \( V_{\text{S},f} \) to explore the typical range of output operating voltages expected in steady-state conditions.
In the case of PMOS devices, the source and gate terminals are reversed due to polarity, but the voltage nomenclature \( V_{\text{S},f} \) and \( V_{\text{G},f} \) is preserved for consistency and to avoid confusion across device types.

\par
The analysis included LVT NMOS and PMOS devices in both standard and DTMOS configurations.
Widths \( W_f \) ranged from \SI{20}{\micro\meter} to \SI{100}{\micro\meter} and lengths \( L_f \) from \SI{5}{\micro\meter} to \SI{20}{\micro\meter}.
Simulations were performed in the typical process corner at \SI{-40}{\celsius} with an input \SI{40}{\pico\farad} capacitor to account for the Si(Li) strip detector capacitance.

\begin{figure}[tbp]
    \centering
    \captionsetup[subfigure]{justification=justified,singlelinecheck=false}

    \begin{subfigure}{0.99\textwidth}
        \centering
        \includegraphics[width=0.81\linewidth]{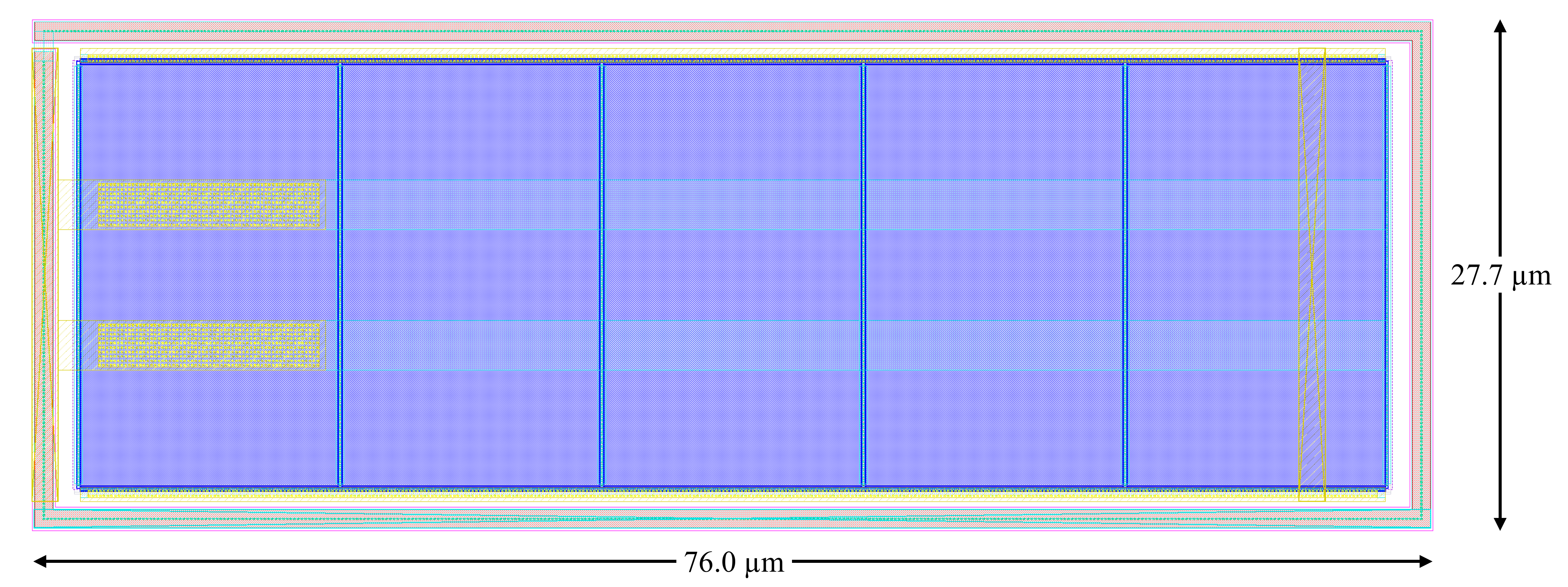}
        \caption{$n$-type MOS capacitor with $L = \SI{14}{\micro\meter}$, 5 fingers of width \SI{22.86}{\micro\meter}, total width \SI{114.3}{\micro\meter}, gate area \SI{1600.2}{\micro\meter\squared}.}
        \label{fig:ch4-NMOS_layout}
    \end{subfigure}

    \vspace{0.5em}

    \begin{subfigure}{0.99\textwidth}
        \centering
        \includegraphics[width=0.81\linewidth]{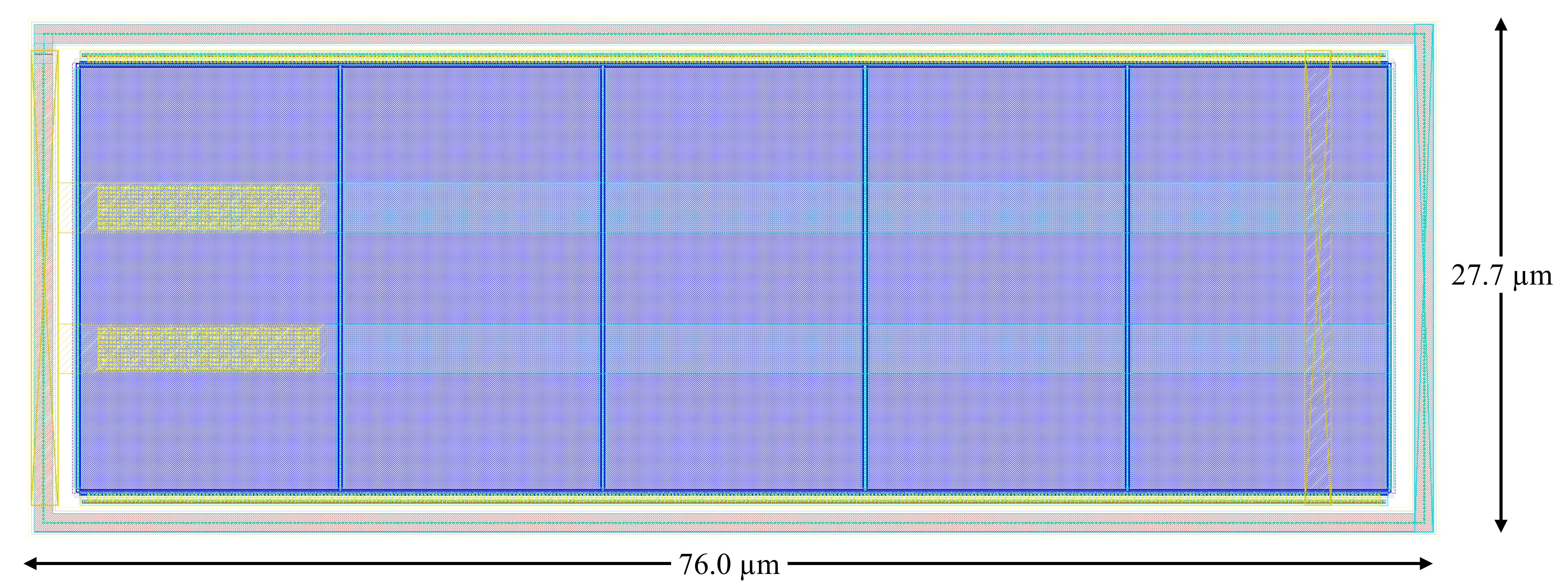}
        \caption{$n$-type DTMOS capacitor with $L = \SI{14}{\micro\meter}$, 5 fingers of width \SI{22.86}{\micro\meter}, total width \SI{114.3}{\micro\meter}, gate area \SI{1600.2}{\micro\meter\squared}.}
        \label{fig:ch4-PMOS_layout}
    \end{subfigure}

    \vspace{0.5em}

    \begin{subfigure}{0.99\textwidth}
        \centering
        \includegraphics[width=0.81\linewidth]{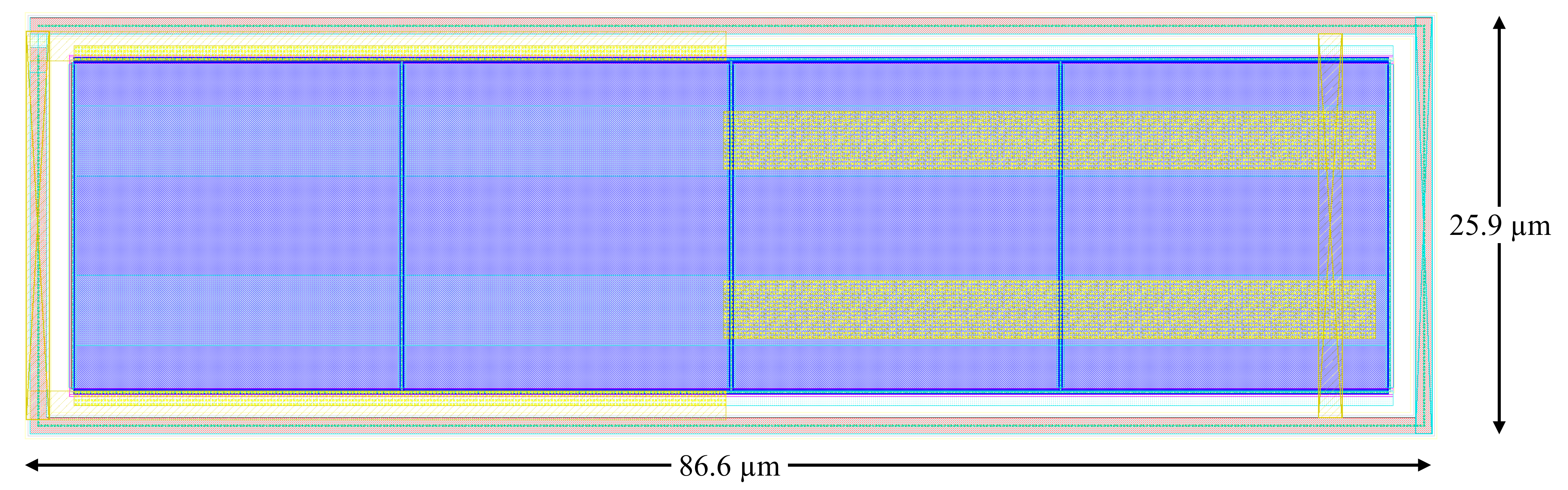}
        \caption{$p$-type MOS capacitor with $L = \SI{20}{\micro\meter}$, 4 fingers of width \SI{20.00}{\micro\meter}, total width \SI{80.0}{\micro\meter}, gate area \SI{1600}{\micro\meter\squared}.}
        \label{fig:ch4-NDTMOS_layout}
    \end{subfigure}

    \vspace{0.5em}

    \begin{subfigure}{0.99\textwidth}
        \centering
        \includegraphics[width=0.81\linewidth]{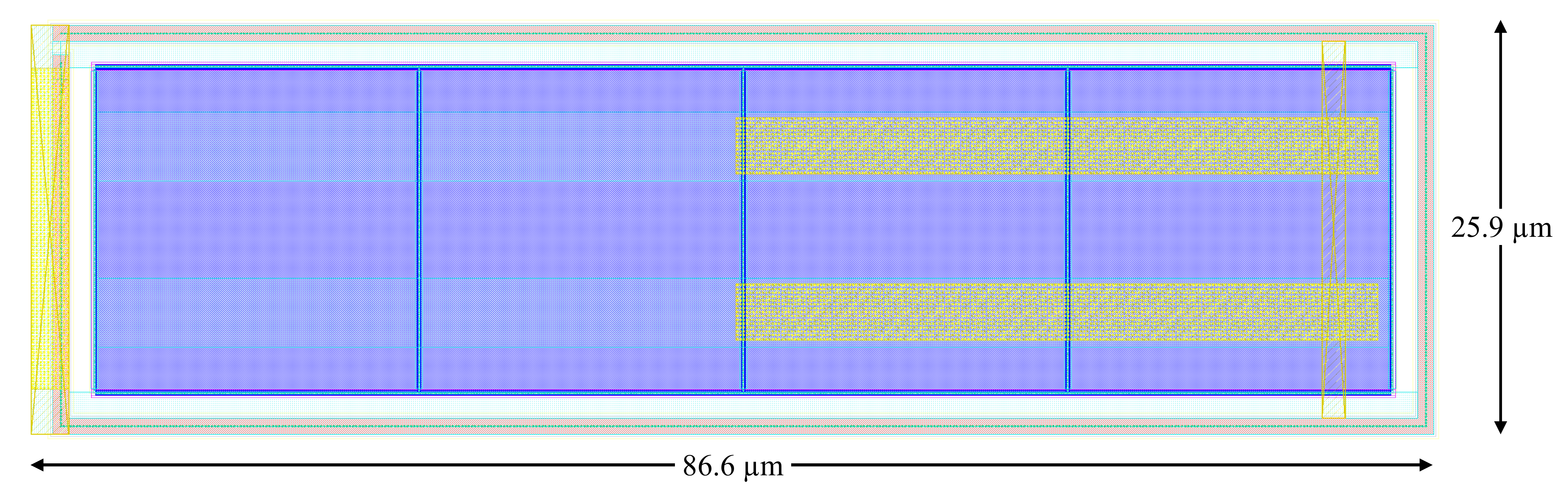}
        \caption{$p$-type DTMOS capacitor with $L = \SI{20}{\micro\meter}$, 4 fingers of width \SI{20.00}{\micro\meter}, total width \SI{80.0}{\micro\meter}, gate area \SI{1600}{\micro\meter\squared}.}
        \label{fig:ch4-PDTMOS_layout}
    \end{subfigure}

    \caption{Layout views of the four feedback MOS capacitor configurations implemented in the CSA design. All layouts are dimensioned to ensure a common gate area of \SI{1600}{\micro\meter\squared} to enable a consistent comparison across device types and configurations.}
    \label{fig:ch4-feedback_layouts}
\end{figure}

\par
Initial filtering based on \( G_{\text{le}} \ge \SI{250}{\micro\volt/\kilo\electronvolt} \) and \( V_{\text{out}} \) for \( E_{\text{inj}} = \SI{100}{\mega\electronvolt} \le \SI{980}{\milli\volt} \)---corresponding to a maximum output voltage swing \( \Delta V_{\text{out}} \le \SI{500}{\milli\volt} \)---identified twelve candidate configurations.
The results of the parameter sweep are summarized in~\Cref{fig:ch4-scatter_W20_100_L5_20_allconf_vgs}, which shows the trade-off between low-energy gain and output voltage at \SI{100}{\mega\electronvolt} for all simulated configurations.
Among the configurations satisfying the selection criteria, eleven belong to the set with \( V_{\text{shift}} = \SI{-100}{\milli\volt} \), of which seven are implemented using DTMOS devices and correspond to the smallest gate area with respect to the standard configuration falling in the same area.
The spread across configurations and bias points clearly indicates that only devices with reduced \( V_{\text{G},f} \) are able to simultaneously meet the constraints on gain and output swing.
This observation motivated the choice of a reduced CSA output baseline voltage in the final design.

\par
The selection of \( V_{\text{ref,CSA}} \) in the transconductor feedback path---discussed later in the chapter---was guided by the need to reduce \( V_{\text{G},f} \) and thereby shift the operating point of the feedback device into the optimal operating region, whilst guaranteeing a sufficient voltage headroom to account for the expected output voltage swing.

\par
A further selection was performed to identify configurations that additionally satisfy the high-energy gain requirement \( G_{\text{he}} \ge \SI{2}{\micro\volt/\kilo\electronvolt} \) in the \SIrange{25}{100}{\mega\electronvolt} energy range.
The subset of configurations meeting both the low-energy and high-energy gain constraints, along with the output swing and area requirements, is reported in~\Cref{tab:ch4-feedback_device_selection} compared to the NMOS configuration used in the \SI{180}{\nano\meter} implementation.
The best performance in terms of both low- and high-energy gain is achieved with the \( n \)-type DTMOS device with \( W_f/L_f = 80/20~\si{\micro\meter/\micro\meter} \), which exhibits a low-energy gain of \SI{297}{\micro\volt/\kilo\electronvolt} and a high-energy gain of \SI{2.14}{\micro\volt/\kilo\electronvolt}, while maintaining an output swing of \SI{503}{\milli\volt} and a compression factor \( k = 139 \). 

\par
Based on the selected configuration, the four final feedback device implementations integrated in the chip were designed to match the gate area of the reference \( n \)-type DTMOS configuration, enabling a direct comparison of their performance.
While maintaining identical gate area, different combinations of gate length and width were adopted to satisfy layout constraints and to ensure a modular and symmetric implementation of the CSA in the final floorplan.
The geometrical parameters of these four configurations are summarized in~\Cref{tab:ch4-antares4_feedback_capacitors}.

\par
\Cref{fig:ch4-CSA_fdt} shows the CSA output voltage as a function of the incoming energy over the whole \SI{10}{\kilo\electronvolt}--\SI{100}{\mega\electronvolt} input dynamic range, obtained from schematic-level simulations in the typical process corner at \SI{-40}{\celsius} with \( C_d = \SI{40}{\pico\farad} \) and \( V_{\text{shift}} = \SI{-100}{\milli\volt} \). 
The selected candidates---$n$-type DTMOS with \( W_f/L_f = 80/20~\si{\micro\meter/\micro\meter} \), $p$-type DTMOS with \( 100/20 \), and standard NMOS with \( 100/20 \)---are compared to the NMOS (\( 115/\SI{20}{\micro\meter/\micro\meter} \)) used in the \SI{180}{\nano\meter} implementation.
The transcharacteristics are baseline-subtracted to allow for a fair comparison with the NMOS of the \SI{180}{\nano\meter} implementation.
The $n$-type DTMOS exhibits the steepest slope at low energy, corresponding to the highest low-energy gain, demonstrating the best compression factor $k$ among the selected configurations.

\par
\Cref{fig:ch4-CSA_gain_cap} presents the corresponding sensitivity and feedback capacitance versus incoming energy. 
The left \( y \)-axis shows the gain, while the right \( y \)-axis reports the equivalent capacitance. 
As expected with the dynamic signal compression technique, all configurations show an increasing feedback capacitance with energy, leading to a reduction in gain, with a change in gain corresponding to the threshold voltage of the feedback device. 
The $n$-type DTMOS provides the largest dynamic variation in capacitance and the highest compression factor, followed by the $p$-type DTMOS.
The standard NMOS and the \SI{180}{\nano\meter} NMOS exhibit the smallest compression factor.

\par
The effectiveness of the dynamic threshold technique is demonstrated by its ability to achieve comparable low- and high-energy gain performance, while providing a higher compression factor relative to the NMOS configuration adopted in the \SI{180}{\nano\meter} implementation, which featured a gate area of \SI{2300}{\micro\meter\squared}.
This approach allows for a gate area reduction of approximately \SI{30}{\percent}, along with a decrease of about \SI{27}{\percent} in the output voltage dynamic range.

\begin{figure}[t!]
    \centering
    \includegraphics[width=0.75\textwidth]{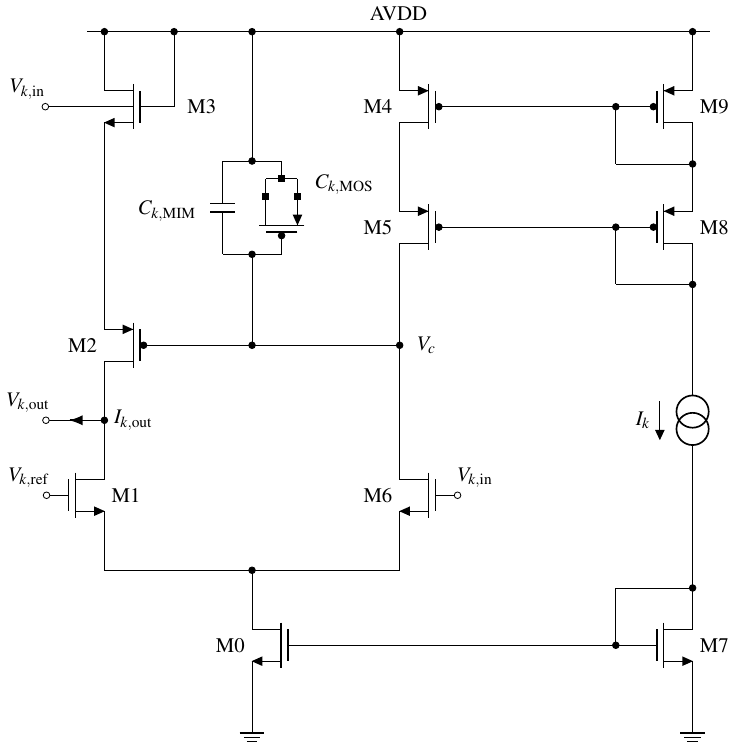}
    \caption{Schematic of the Krummenacher feedback circuit used in the CSA. The circuit employs a cascoded current mirror (M7--M9) to replicate the bias current \( I_k \), which is evenly split into two branches: (M1--M3) and (M4--M6). The bulk-driven device M3 has its bulk terminal tied to the CSA output \( V_{k,\text{in}} \). The baseline reference voltage is applied to the gate of M1 via \( V_{k,\text{ref}} \). The capacitor \( C_k \) is implemented as a combination of MIM (\(C_{k,\text{MIM}}\)) and MOS (\(C_{k,\text{MOS}}\)) capacitances. The current bias network is not reported.}
    \label{fig:ch4-krummenacher_schematic}
\end{figure}

%   Layout of the feedback capacitor
\subsubsection{Layout of the feedback capacitor}

\par
The four feedback MOS capacitor configurations integrated in the final design were designed to have an identical gate area of approximately \SI{1600}{\micro\meter\squared}, enabling a direct comparison of their performance.
The layout parameters are summarized in~\Cref{tab:ch4-antares4_feedback_capacitors}.

\par
For the $n$-type MOS and DTMOS configurations, the layout consists of 5 fingers of width \SI{22.86}{\micro\meter} and length \SI{14.0}{\micro\meter}, resulting in a total width of \SI{114.3}{\micro\meter}.
For the $p$-type MOS and DTMOS configurations, 4 fingers of width \SI{20.00}{\micro\meter} and length \SI{20.0}{\micro\meter} are used, yielding a total width of \SI{80.0}{\micro\meter}.
\Cref{fig:ch4-feedback_layouts} shows the implemented layouts of the four configurations: $n$-type MOS, $n$-type DTMOS, $p$-type MOS, and $p$-type DTMOS.

%   Krummenacher feedback network
\subsection{Krummenacher feedback network}

%   Operating principle
\subsubsection{Operating principle}

\begin{table}[t!]
    \centering
    \renewcommand{\arraystretch}{1.1}
    \setlength{\tabcolsep}{8pt}
    \begin{tabular}{cccccc}
        \toprule
        \multirow{2}{*}{Device} & \multirow{2}{*}{Type} & $L$ & $W_\text{tot}$ & Fingers \\
                                &                        & (\si{\micro\meter}) & (\si{\micro\meter}) & \# \\
        \midrule
        M0 & NMOS, LVT & 4.0  & 0.22 & 1 \\
        M1 & NMOS, SVT & 1.0  & 3.00 & 1 \\
        M2 & PMOS, LVT & 10.0 & 0.50 & 1 \\
        M3 & NMOS, SVT & 5.0  & 0.50 & 1 \\
        M4 & PMOS, SVT & 10.0 & 0.50 & 1 \\
        M5 & PMOS, SVT & 10.0 & 0.50 & 1 \\
        M6 & NMOS, SVT & 1.0  & 3.00 & 1 \\
        M7 & NMOS, LVT & 4.0  & 0.22 & 1 \\
        M8 & PMOS, SVT & 10.0 & 1.00 & 2 \\
        M9 & PMOS, SVT & 10.0 & 1.00 & 2 \\
        \bottomrule
    \end{tabular}
    \caption{Geometrical parameters of the transistors used in the Krummenacher feedback network, as shown in~\Cref{fig:ch4-krummenacher_schematic}. The table reports the channel length \( L \), total width \( W_\text{tot} \), and number of fingers for each device. The current bias devices are not included.}
    \label{tab:krummenacher_device_sizes}
\end{table}

\par
The CSA employs a Krummenacher feedback circuit, shown in Figure~\ref{fig:ch4-krummenacher_schematic}, to provide a continuous reset path for the integrated charge.
This topology was selected over other baseline restoration techniques due to its ability to compensate for the leakage current of the Si(Li) detectors, which typically ranges from \SI{2.5}{\nano\ampere} to \SI{10}{\nano\ampere} at \SI{-40}{\celsius}, and can increase to tens of microamperes at room temperature.
The geometry of the transistors used in the Krummenacher circuit designed in this work is summarized in~\Cref{tab:krummenacher_device_sizes}.

\par
The bias current \( I_k \) is mirrored through a cascoded current mirror (M7--M9) within the M4--M6 branch, providing a constant current of \( I_k/2 \) to each of the two branches: M1--M3 and M4--M6.
Under stationary conditions, both branches conduct \( I_k/2 \).
The baseline reference voltage \( V_{k,\text{ref}} \) is applied to the gate of M1 and, under steady-state conditions, is replicated at the gate of M6 thanks to the negative feedback, thus defining the DC voltage at the output of the CSA.
The value of \( V_{k,\text{ref}} \) has been chosen as \SI{360}{\milli\volt}, in accordance with the discussion on the optimal gate-source bias \( V_{\text{GS},f} \) for the feedback capacitance device presented in~\Cref{sec:ch4-feedback_device_dimensioning}.
In this configuration, the input of the CSA is connected to \( V_{k,\text{out}} \), which corresponds to the output of the Krummenacher network, while the output of the CSA is connected to \( V_{k,\text{in}} \), serving as the input node of the Krummenacher.

\par
A distinctive feature of this implementation is the introduction of a bulk-driven NMOS transistor M3, representing a novel enhancement to the classical Krummenacher topology. 
The body terminal of M3 is connected to the CSA output node \( V_{k,\text{in}} \), allowing the feedback current in the M1--M3 branch to be modulated via the body effect. 
This mechanism becomes particularly relevant when the differential structure becomes unbalanced due to an increase in detector leakage current, which adds additional current to the M1--M3 branch on top of the nominal \( I_k/2 \). 
The resulting asymmetry can shift the operating point of the circuit at the \( V_c \) node. 
The purpose of this modification is to minimize the component of the output current that depends on the control node \( V_c \). 
This is achieved through a twofold action: a sufficiently large capacitor \( C_k \) is employed to limit the variation of \( V_c \), while the bulk-driven control path reduces, or ideally cancels, the residual effect of such variation. 

\begin{figure}[tbp]
    \centering
    \includegraphics[width=0.99\textwidth]{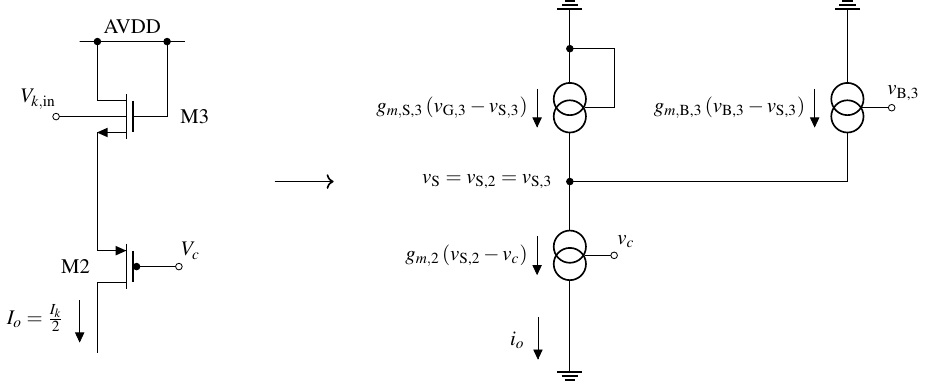}
    \caption{Detail of the Krummenacher feedback circuit showing devices M2 and M3 and their equivalent small-signal model. The analysis focuses on the bulk-driven operation of M3, whose body terminal is connected to the CSA output \( V_{k,\text{in}} \). Both the source and bulk contributions are modeled to evaluate the impact of the body voltage \( v_{\text{B},3} \) on the output current \( i_o \). The small-signal currents through M3 and M2 are expressed as \hbox{\( g_{m,\text{S},3}(v_{\text{G},3} - v_{\text{S},3}) \)}, \( g_{m,\text{B},3}(v_{\text{B},3} - v_{\text{S},3}) \), and \( g_{m,2}(v_{\text{S},2} - v_c) \), respectively, where \hbox{\( v_{\text{S}} = v_{\text{S},2} = v_{\text{S},3} \)}.}
    \label{fig:ch4-krummenacher_bulk_analysis}
\end{figure}

\par
This architectural upgrade guarantees a leakage current compensation capability up to \SI{200}{\nano\ampere}, which is a factor of four higher than the \SI{50}{\nano\ampere} limit of the previous \SI{180}{\nano\meter} implementation.
The extended range was motivated by the results of the on-ground calibration campaign of the GAPS tracker readout electronics for the first flight, where the necessity to compensate leakage currents up to \SI{200}{\nano\ampere} during operation at temperatures as high as \SI{-10}{\celsius} became evident.

\par
The analytical evaluation of this behavior is based on the small-signal equivalent model of transistors M2 and M3, shown in Figure~\ref{fig:ch4-krummenacher_bulk_analysis}, where both the gate and body contributions of M3 are explicitly included to assess the impact of body modulation on the output current.
The small-signal contributions to the output current from M3 can be written separately as:

\begin{equation}
    i_{o,\text{S}} = g_{m,\text{S},3}\,(v_{\text{G},3} - v_{\text{S}}),
\end{equation}

\begin{equation}
    i_{o,\text{B}} = g_{m,\text{B},3}\,(v_{\text{B},3} - v_{\text{S}}),
\end{equation}

\noindent
where \( i_{o,\text{S}} \) and \( i_{o,\text{B}} \) denote the source- and body-transconductance contributions of M3 to the output current \( i_o \), respectively.
Given that the two transistors share a common source node (\( v_{\text{S}} = v_{\text{S,2}} = v_{\text{S,3}} \)), the resulting expression for the output current is:

\begin{equation}
    i_o = i_{o,\text{S}} + i_{o,\text{B}} = \frac{g_{m,2}g_{m,\text{B},3}}{g_{m,2}-g_{m,\text{S},3}-g_{m,\text{B},3}}\left[ v_{\text{B,3}} - v_c\left(\frac{g_{m,\text{S},3}}{g_{m,\text{B},3}} + 1\right) \right],
    \label{eq:ch4-krum_small_signal}
\end{equation}

\noindent
expressing the total current deviation in the M1--M3 branch as a function of both the voltage at the bulk terminal of M3 and the voltage at the \( v_c \) node.

\begin{figure}[t!]
    \includegraphics[width=0.85\columnwidth]{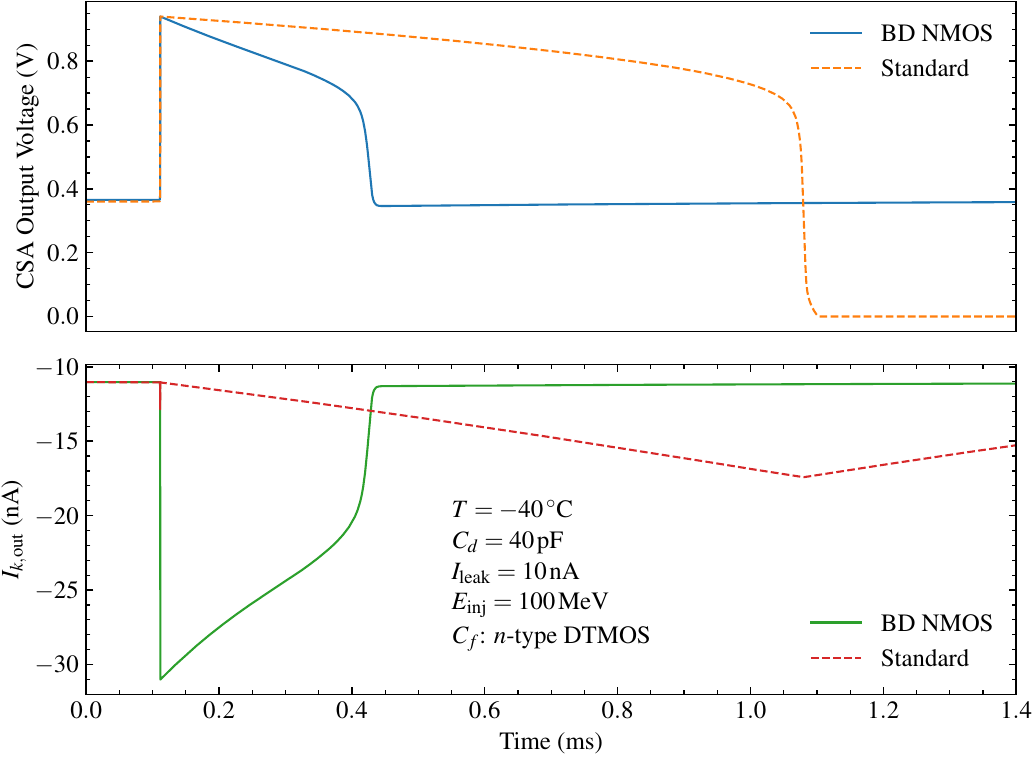}
    \caption{Comparison of the CSA transient response for the standard and \hbox{bulk-driven} Krummenacher networks. The CSA output voltage (top) and discharge current \( I_{k,\text{out}} \) (bottom) are shown for \SI{-40}{\celsius}, \hbox{\( C_d = \SI{40}{\pico\farad} \)}, \hbox{\( I_{\text{leak}} = \SI{10}{\nano\ampere} \)}, and an injected energy of \SI{100}{\mega\electronvolt}. The feedback capacitance \( C_f \) is implemented using an \hbox{\emph{n}-type} DTMOS device.}
    \label{fig:krummenacher_response}
\end{figure}

\par
In the traditional implementation of this circuit topology, the discharge current is determined soley by the drift if the node \(V_c\) that evolve as

\begin{equation}
    V_c(t) = \frac{I_k}{2C_k}\,t ,
\end{equation}

\noindent
which leads to

\begin{equation}
    I_{k,\text{out}}(t) = g_{m,2} V_c(t) = g_{m,2}\,\frac{I_k}{2C_k}\,t ,
    \label{eq:ch4-krum_standard}
\end{equation}

\begin{figure}[t!]
    \centering
    \includegraphics[width=0.75\textwidth]{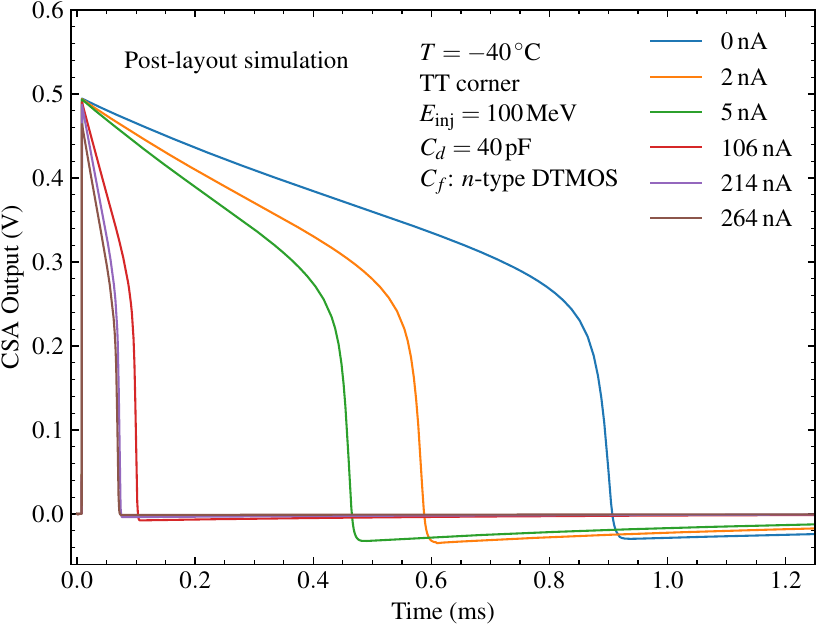}
    \caption{Post-layout transient simulation of the CSA output for varying leakage current levels up to \SI{264}{\nano\ampere}, with a fixed detector capacitance of \SI{40}{\pico\farad}. 
    A baseline shift of approximately \SI{30}{\milli\volt} is observed between 0 and \SI{264}{\nano\ampere}. 
    The feedback capacitance is implemented using an $n$-type DTMOS device. 
    Simulations were performed at \SI{-40}{\celsius} in the typical process corner for an incoming energy of \SI{100}{\mega\electronvolt}.}
    \label{fig:ch4-krummenacher_fdt_leakage_100MeV}
\end{figure}

\noindent
showing that the discharge current increases approximately linearly with time and directly reflects the drift of \( V_c \).
With the introduction of the \hbox{bulk-driven} device, the discharge current includes an additional \hbox{body-controlled} term provided by transistor M3.
The \hbox{small-signal} analysis presented in Fig.~\ref{fig:ch4-krummenacher_bulk_analysis} follows~\eqref{eq:ch4-krum_small_signal}, yielding:

\begin{equation}
    I_{k,\text{out}}(V_{k,\text{in}}(t),t) \approx g_{m2}\!\left( V_c(t) - \frac{g_{m,\text{B},3}}{g_{m,\text{S},3}}\,V_{k,\text{in}}(t) \right),
\end{equation}

\noindent
where \( V_{k,\text{in}}(t) \) is the CSA output voltage that drives the body of M3. 
The term \( \tfrac{g_{m,\text{B},3}}{g_{m,\text{S},3}} V_{k,\text{in}}(t) \) counteracts the shift of \( V_c(t) \), reducing the sensitivity of the discharge current to variations at this node.

\par
The difference between the two architectures is illustrated in Fig.~\ref{fig:krummenacher_response}, which compares the CSA output voltage and the corresponding \( I_{k,\text{out}} \) transient for a temperature of \SI{-40}{\celsius}, a detector capacitance of \SI{40}{\pico\farad}, a leakage current of \SI{10}{\nano\ampere}, an incoming energy of \SI{100}{\mega\electronvolt}, and a feedback capacitor implemented as an \hbox{\( n \)-type} DTMOS device.
In the conventional implementation, \( I_{k,\text{out}} \) follows the dependence predicted in~\eqref{eq:ch4-krum_standard}, resulting in an almost linear decay of the CSA output.
When the \hbox{bulk-driven} device M3 is introduced, \( I_{k,\text{out}} \) follows the CSA output voltage \( V_{k,\text{in}} \) and exhibits a larger transient magnitude, leading to a faster restoration of the CSA baseline.

%   Implementation of the compensation capacitance
\subsubsection{Implementation of the compensation capacitance}

\begin{figure}[t!]
    \centering
    \includegraphics[width=0.75\textwidth]{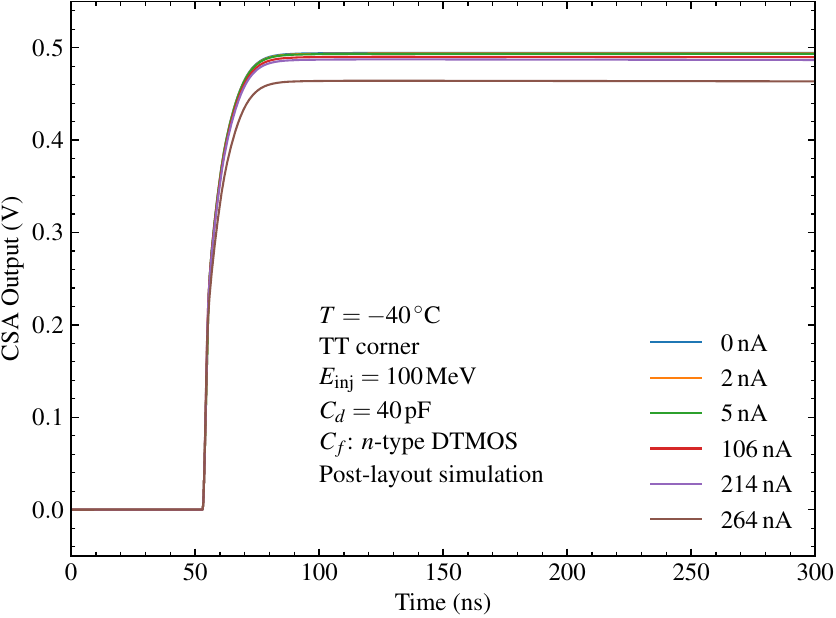}
    \caption{Zoomed-in post-layout transient simulation of the CSA output rising edge for varying leakage current levels up to \SI{264}{\nano\ampere}, with a fixed detector capacitance of \SI{40}{\pico\farad}. 
    The rise time at \SI{264}{\nano\ampere} leakage current is approximately \SI{30}{\nano\second}. 
    The feedback capacitance is implemented using an $n$-type DTMOS device. 
    Simulations were performed at \SI{-40}{\celsius} in the typical process corner for an incoming energy of \SI{100}{\mega\electronvolt}.}
    \label{fig:ch4-krummenacher_fdt_leakage_rise}
\end{figure}

\par
The compensation capacitance \( C_k \) in the Krummenacher feedback loop is dimensioned relative to the signal feedback capacitance \( C_f \) to minimize variations of the control voltage \( V_c \) under changing leakage current conditions.
This section describes the physical implementation of \( C_k \) and the design strategy adopted to maximize capacitance density while minimizing the overall area.

\par
A first version of the CSA implemented \( C_k \) as an array of four metal--insulator--metal (MIM) capacitors, achieving a total nominal capacitance of \SI{30}{\pico\farad}.
The total layout area of the array was \SI{231.25}{\micro\meter} by \SI{80.56}{\micro\meter}, corresponding to an overall footprint of

\[
A_\text{MIM} = \SI{231.25}{\micro\meter} \times \SI{80.56}{\micro\meter} = \SI{18630.5}{\micro\meter\squared}.
\]

\par
To improve capacitance density and reduce area while preserving the required leakage current compensation capability, a second version was designed using a hybrid architecture that combines stacked MIM and MOS capacitors.
The new array, composed of four identical stacked MIM+MOS units, occupies a reduced area of \SI{144.72}{\micro\meter} by \SI{97.88}{\micro\meter}, yielding

\[
A_\text{hybrid} = \SI{144.72}{\micro\meter} \times \SI{97.88}{\micro\meter} = \SI{14174.9}{\micro\meter\squared}.
\]

\par
This redesign achieves a \SI{24}{\percent} reduction in area while also introducing a dynamic capacitance behavior.
To limit gate leakage, part of \( C_k \) was implemented using \hbox{thick-oxide} PMOS devices, whose larger physical oxide thickness ensures negligible tunneling current compared to thin-oxide devices.
As a result, the capacitance becomes voltage dependent, increasing with leakage current: from \SI{48.71}{\pico\farad} at \( I_{\text{leak}} = \SI{0}{\nano\ampere} \) to \SI{94.36}{\pico\farad} at \hbox{\( I_{\text{leak}} = \SI{292}{\nano\ampere} \)}.

\par
At zero leakage, this corresponds to a \SI{37.8}{\percent} increase in capacitance compared to the MIM-only design.
In terms of density, the MIM-only implementation achieves \SI{1.61}{\femto\farad\per\square\micro\meter}, a value consistent with the typical \(\SI{\sim2}{\femto\farad\per\square\micro\meter}\) MIM density for this technology node.
The hybrid implementation increases the effective density to \SI{3.44}{\femto\farad\per\square\micro\meter} at zero leakage current, and up to \SI{6.66}{\femto\farad\per\square\micro\meter} at the maximum leakage of \SI{292}{\nano\ampere}.
Additionally, the optimization reduces the length of the capacitor array in the signal channel to \SI{86.63}{\micro\meter}, contributing to the reduced footprint while maintaining the required compensation performance under varying leakage current.

\begin{figure}[t!]
    \centering
    \includegraphics[width=0.75\textwidth]{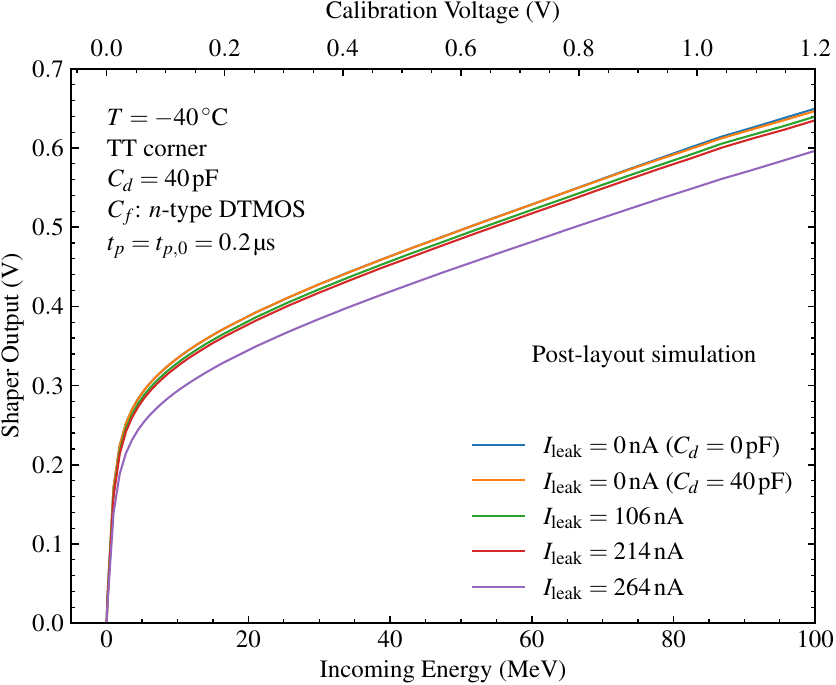}
    \caption{Simulated input-output transfer characteristics of the CSA evaluated at the output of the CR-RC shaper for varying leakage current up to \SI{264}{\nano\ampere}, with a fixed detector capacitance of \SI{40}{\pico\farad} in all cases except the zero-leakage configuration, where \( C_d = \SI{0}{\pico\farad} \). The feedback capacitance in the CSA is implemented using an $n$-type DTMOS device. Post-layout simulations are performed at \SI{-40}{\celsius} in the typical process corner with a peaking time of \SI{0.2}{\micro\second}.}
    \label{fig:ch4-krummenacher_fdt_leakage}
\end{figure}

%   Krummenacher performance evaluation
\subsubsection{Krummenacher performance evaluation} \label{sec:ch4-krummenacher_feedback}

\par
The compensation capabilities of the Krummenacher feedback circuit have been evaluated by varying the Si(Li) detector leakage current from \SI{0}{\nano\ampere} to \SI{264}{\nano\ampere} in post-layout simulations. 
An initial evaluation at the CSA output was performed through transient simulations for an input energy \( E_{\text{inj}} = \SI{100}{\mega\electronvolt} \), corresponding to the maximum voltage swing.
These simulations were carried out on the CSA employing the $n$-type DTMOS as the feedback element. 

\par
\Cref{fig:ch4-krummenacher_fdt_leakage_100MeV} shows the CSA output for different leakage currents, illustrating the discharge behavior of the feedback capacitor.
The variation in feedback capacitance is evident in the shape of the voltage discharge profile: an initial slower discharge occurs when \( V_{\text{GS},f} > V_{\text{th}} \), where the capacitance is larger, followed by a faster discharge rate as the feedback capacitance decreases below the threshold voltage.
The influence of leakage current on the discharge profile is further highlighted by the reduced discharge time, due to the increased current contribution in the input branch of the Krummenacher circuit, leading to a faster discharge of the feedback capacitor.

\par
\Cref{fig:ch4-krummenacher_fdt_leakage_rise} shows the rising edge of the CSA output signal, demonstrating consistent transient behavior, with a maximum rise time of \SI{30}{\nano\second} observed at \SI{264}{\nano\ampere} of leakage current.
Overall, the performance evaluation confirms the CSA feedback network's ability to compensate leakage currents above the requirement of \SI{200}{\nano\ampere}, thereby extending the operating capability of the GAPS tracker during on-ground calibration procedures beyond that of the current readout ASIC.

\par
The CSA input-output transcharacteristic was assessed by evaluating the output of the CR-RC shaper, described later in the chapter, by measuring the amplitude of the output waveform at the peaking time.
The result is shown in~\Cref{fig:ch4-krummenacher_fdt_leakage}, in the typical process corner at \SI{-40}{\celsius}, with a peaking time of \SI{0.2}{\micro\second}.
The transcharacteristic was obtained by injecting 50 points spanning the entire input dynamic range, from \SI{10}{\kilo\electronvolt} to \SI{100}{\mega\electronvolt}, with a fixed detector capacitance of \SI{40}{\pico\farad} in all cases except the zero-leakage configuration, where \( C_d = \SI{0}{\pico\farad} \).
The latter represents the best-case scenario, yielding the highest low-energy gain, as reported in~\Cref{tab:ch4-transferchar_vs_leak}.

\begin{table}[t!]
    \centering
    \renewcommand{\arraystretch}{1.1}
    \setlength{\tabcolsep}{8pt}
    \begin{tabular}{@{}ccccccccc@{}}
        \toprule
        $I_{\text{leak}}$ & $C_d$ & $G_{\text{le}}$ & INL$_\text{le}$ & $G_{\text{he}}$ & INL$_\text{he}$ & $k$ & Kink & $\Delta V_{\text{out}}$ \\
        (\si{\nano\ampere}) & (\si{\pico\farad}) & (\si{\micro\volt/\kilo\electronvolt}) & (\%) & (\si{\micro\volt/\kilo\electronvolt}) & (\%) & & (\si{\kilo\electronvolt}) & (\si{\milli\volt}) \\
        \midrule
        0   & 0  & 306 & 0.31 & 3.21 & 1.78 & 95 & 1104 & 650 \\
        0   & 40 & 301 & 0.44 & 3.17 & 2.02 & 95 & 1130 & 647 \\
        106 & 40 & 268 & 0.49 & 3.16 & 1.93 & 85 & 1246 & 640 \\
        214 & 40 & 252 & 0.57 & 3.15 & 1.90 & 80 & 1312 & 635 \\
        264 & 40 & 209 & 0.39 & 3.05 & 2.07 & 68 & 1440 & 596 \\
        \bottomrule
    \end{tabular}
    \caption{Performance parameters extracted from the input-output transcharacteristic of the CSA, obtained from post-layout simulations and evaluated at the output of the CR-RC shaper with a peaking time of \SI{0.2}{\micro\second} in the typical process corner at \SI{-40}{\celsius}. The table reports the low- (\( G_{\text{le}} \)) and high-energy gain (\( G_{\text{he}} \)), index of nonlinearity (INL), compression factor \( k \), kink energy, and output swing (\( \Delta V_{\text{out}} \)) for increasing leakage current.}
    \label{tab:ch4-transferchar_vs_leak}
\end{table}

\par
The low-energy gain \( G_{\text{le}} \) was evaluated by fitting a linear model to the low-energy portion of the transcharacteristic, corresponding to the \SIrange{10}{100}{\kilo\electronvolt} range.
Similarly, the high-energy gain \( G_{\text{he}} \) was extracted by a linear fit over the \SIrange{25}{100}{\mega\electronvolt} range.
In both cases, the linearity of the transcharacteristic was quantified using the index of nonlinearity (INL), defined as:

\begin{equation}
    \text{INL} = \frac{\Delta V_{\text{max}}}{\Delta E_{\text{inj}} G},
    \label{eq:ch4-index_non_linearity}
\end{equation}

\begin{figure}[tbp]
    \centering
    \includegraphics[width=\textwidth]{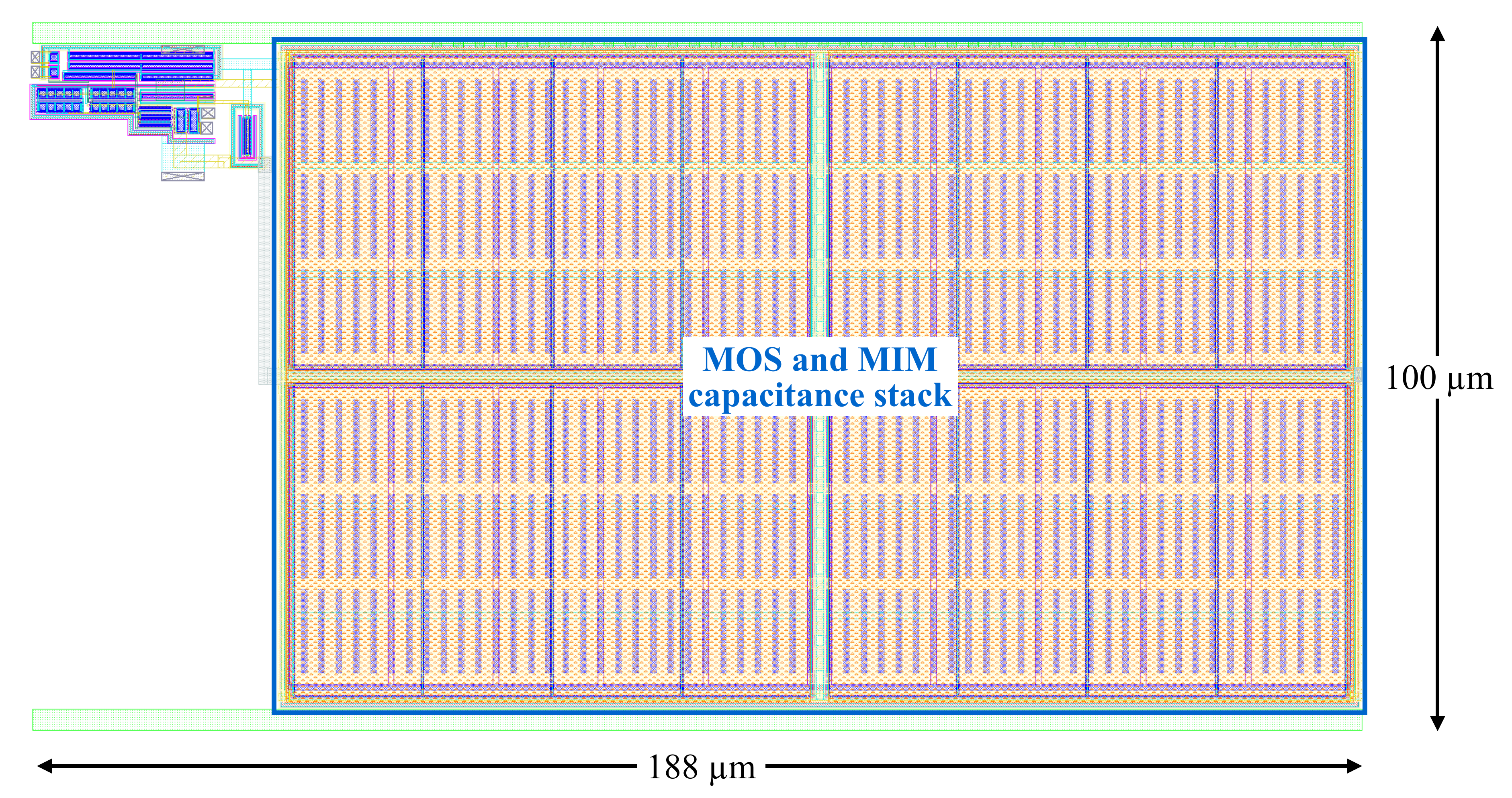}
    \caption{Layout view of the Krummenacher feedback circuit, showing the integrated stack of MOS and MIM capacitances used to implement the compensation capacitor \( C_k \). The array combines four hybrid units for a total footprint of approximately \hbox{$\SI{188}{\micro\meter} \times \SI{100}{\micro\meter}$}.}
    \label{fig:ch4-krummenacher_layout}
\end{figure}

\noindent
where \( \Delta V_{\text{max}} \) is the maximum deviation from the linear fit, \( G \) is the gain extracted from the linear interpolation (either \( G_{\text{le}} \) or \( G_{\text{he}} \)), and \( \Delta E_{\text{inj}} \) is the input energy span.
The INL expresses the deviation from ideal linear behavior as a percentage of the expected output swing.

\par
The compression factor \( k \) was computed as the ratio between the low-energy gain and the high-energy gain, representing a compact metric for the level of dynamic signal compression achieved in each configuration.
It can be observed that \( k \) decreases monotonically with increasing leakage current, as the low-energy gain is more strongly affected by leakage-induced effects than the high-energy gain.
Over the explored leakage current range, the low-energy gain decreases by approximately \SI{31.7}{\percent}, while the high-energy gain exhibits a much smaller reduction of about \SI{4.9}{\percent}.
This behavior can be attributed to the reduction of the effective linear operating region below the kink energy in the low-energy part of the transcharacteristic, where the increased feedback current alters the charge integration dynamics and progressively limits the achievable gain, whereas the high-energy region remains largely dominated by the minimum feedback capacitance.

\par
The kink energy is defined as the energy corresponding to the intersection point between the extrapolated low- and high-energy linear regions of the transcharacteristic.
It can be observed that the kink progressively shifts toward higher energies as the leakage current increases, reflecting the upward displacement of the operating point induced by the additional current injected by the Krummenacher feedback network.
Leakage current also affects the maximum output voltage swing \( \Delta V_{\text{out}} \), which decreases from \SI{650}{\milli\volt} at zero leakage to \SI{596}{\milli\volt} at \SI{264}{\nano\ampere}, corresponding to a total reduction of \SI{8.3}{\percent}.
This reduction is clearly visible in the transcharacteristic evaluated at \SI{264}{\nano\ampere} in~\Cref{fig:ch4-krummenacher_fdt_leakage}, where the output curve appears shifted upward by approximately \SI{30}{\milli\volt} with respect to the zero-leakage configuration for the same injected charge, effectively reducing the available headroom while preserving the overall shape of the characteristic.

%   Layout of the Krummenacher circuit
\subsubsection{Layout of the Krummenacher circuit}

\par
The layout of the Krummenacher feedback circuit is shown in~\Cref{fig:ch4-krummenacher_layout}.
The design is dominated by the \( \SI{144.72}{\micro\meter} \times \SI{97.88}{\micro\meter} \) capacitance stack, which occupies almost the entire area dedicated to this block.
The layout was developed according to the design constraint of maintaining a readout channel height of \SI{100}{\micro\meter}, enabling a modular approach for assembling channel submodules.

\begin{figure}[t!]
    \centering
    \includegraphics[width=0.75\textwidth]{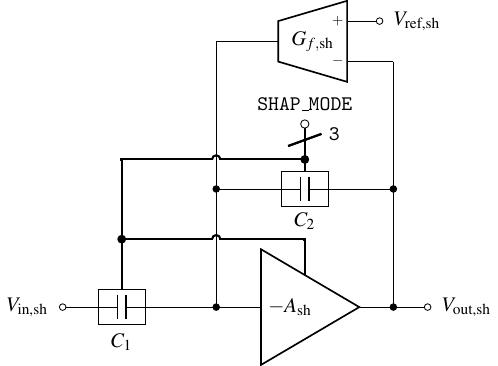}
    \caption{Simplified schematic diagram of the CR-RC shaping stage. The circuit consists of an input capacitance \( C_1 \), a forward gain stage, a feedback capacitance \( C_2 \), and a transconductance element with transconductance \( G_{f,\text{sh}} \). The transconductor sets the DC baseline voltage at the output to \( V_{\text{ref,sh}} \). The capacitors \( C_1 \), \( C_2 \), and the \hbox{pole-setting} capacitance \( C_p \) in the gain stage (not shown) are implemented as arrays of eight unit capacitors, selectively enabled by a binary-to-thermometric decoder driven by the \hbox{3-bit} digital control signal \texttt{SHAP\_MODE}. This configuration allows the shaping time to be programmed over eight selectable values ranging from \SI{0.2}{\micro\second} to \SI{1.6}{\micro\second}.}
    \label{fig:ch4-shaper_diagram}
\end{figure}

\par
At the top and bottom of the layout, two \SI{4}{\micro\meter} routing tracks---implemented in metal 7, one of the topmost among the nine available metal layers in the technology---are dedicated to the analog positive supply (AVDD) and ground, respectively.
This design choice serves a twofold purpose: it allows horizontal sharing of the analog power supply between blocks within the same readout channel and vertical sharing across adjacent channels, by mirroring the channel layout along the $x$-axis to match the power net.
This effectively doubles the track height and helps to reduce the distributed resistance of the power tracks.

\par
This modular and compact layout approach ensures that the power supply tracks maintain reasonable width while minimizing the overall distributed resistance across all channels. 
The horizontal metal 7 tracks are extended vertically to reach all required analog power nets, with PMOS devices placed near the top of the layout, close to AVDD, and NMOS devices placed near the bottom, adjacent to the ground net.
The use of a high-level metal layer for power distribution enables effective analog supply routing across the entire chip, while allowing the lower metal layers to be used for local routing within each readout channel cell.

\begin{figure}[tbp]
    \centering
    \includegraphics[width=0.75\textwidth]{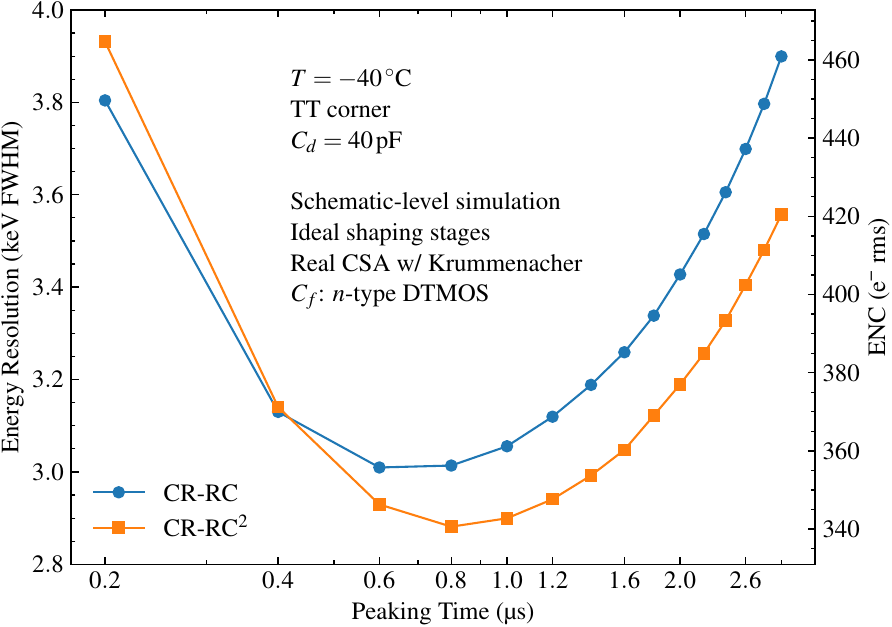}
    \caption{Comparison of energy resolution and ENC as a function of peaking time for first-order (CR-RC) and second-order (CR-RC\textsuperscript{2}) unipolar shaping filters. Simulations were performed at \SI{-40}{\celsius} using ideal shaping stages and the actual CSA with Krummenacher feedback, with a detector capacitance of \SI{40}{\pico\farad}. The CR-RC filter achieves a minimum energy resolution of \SI{3.01}{\kilo\electronvolt} at \SI{0.6}{\micro\second}, while the CR-RC\textsuperscript{2} reaches \SI{2.88}{\kilo\electronvolt} at \SI{0.8}{\micro\second}, corresponding to a relative improvement of \SI{4.26}{\percent}.}
    \label{fig:ch4-shaper_CR-RC_CR-RC2}
\end{figure}

%-------------------------------------------------------------------------------
%   The CR-RC shaper
%-------------------------------------------------------------------------------
\section{The CR-RC shaper}

%   Introduction to signal shaping
\subsection{Introduction to signal shaping}

\par
The shaping stage follows the CSA and processes its output signal.
It improves the signal-to-noise ratio, increases the signal amplitude to match subsequent stages, and shortens the signal duration to reduce the probability of pile-up.
The shaping filter type, order, and time constants are selected according to the energy resolution and rate requirements of the system.
Unipolar semi-Gaussian shaping filters of the \( \text{CR-(RC)}^n \) type are commonly implemented in integrated circuits due to their compatibility with standard analog design techniques~\cite{grybosFrontendElectronicsMultichannel2010}.
Other shaping architectures, such as bipolar semi-Gaussian filters and nearly true Gaussian filters based on synthesis methods, also exist but are less frequently adopted in integrated designs due to increased complexity or area constraints~\cite{hamritaChargeCurrentsensitivePreamplifiers2004}.

\par
A unipolar, semi-Gaussian, time-invariant filter of the \( \text{CR-(RC)}^n \) type consists of one CR differentiator followed by \( n \) RC integrators with identical time constants \( \tau \).  
The general case is defined by the following transfer function:

\begin{equation}
  H(s) = \left( \frac{s \tau}{1 + s \tau} \right) \left( \frac{1}{1 + s \tau} \right)^n.
\end{equation}

\noindent
The time-domain response to a unit voltage step is:

\begin{equation}
  v_{\text{out}}(t) = \frac{1}{n!} \left( \frac{t}{\tau} \right)^n \exp\left( -\frac{t}{\tau} \right),
\end{equation}

\noindent
with peaking time \( t_p = n \tau \). The peak amplitude is:

\begin{equation}
  v_{\text{out,max}} = \frac{n^n}{n!} e^{-n}.
\end{equation}

%   Shaper circuit implementation
\subsection{Shaper circuit implementation}

\begin{table}[tbp]
  \centering
  {\renewcommand{\arraystretch}{1.1}
  \setlength{\tabcolsep}{8pt}
  \begin{tabular}{ccccc}
    \toprule
    \multirow{2}{*}{$n$} &
    $t_p$ & $C_1$ & $C_2$ & $C_p$ \\
    & (\si{\nano\second}) & (\si{\femto\farad}) & (\si{\femto\farad}) & (\si{\pico\farad}) \\
    \midrule
    1 & 200  & 48.9  & 30   & 5.022 \\
    2 & 400  & 97.8  & 60   & 10.043 \\
    3 & 600  & 146.7 & 90   & 15.065 \\
    4 & 800  & 195.6 & 120  & 20.087 \\
    5 & 1000 & 244.5 & 150  & 25.108 \\
    6 & 1200 & 293.4 & 180  & 30.130 \\
    7 & 1400 & 342.3 & 210  & 35.152 \\
    8 & 1600 & 391.2 & 240  & 40.174 \\
    \bottomrule
  \end{tabular}
  }
  \caption{Calculated values of the input capacitance \( C_1 \), feedback capacitance \( C_2 \), and pole-setting capacitance \( C_p \) corresponding to eight programmable peaking times ranging from \SIrange{0.2}{1.6}{\micro\second}. The values are obtained using the design equations with a capacitance ratio \( \alpha = 1.63 \) and scaling factor \( \gamma = 167.4 \).}
  \label{tab:ch4-shaper_capacitors}
\end{table}

\par
The shaping architecture implemented in this work adopts the \( \text{CR-(RC)}^n \) topology in its simplest form, with \( n = 1 \), corresponding to a first-order CR-RC filter.
The simplified schematic diagram of the shaping stage is shown in~\Cref{fig:ch4-shaper_diagram}.
This configuration provides a unipolar semi-Gaussian response with a single time constant and is well suited for low-complexity integration in a multichannel front-end.  

\par
The choice of the lowest filter order was motivated by a comparative evaluation with the second-order CR-RC\textsuperscript{2} architecture implemented in the \SI{180}{\nano\meter} version of the readout channel.  
The evaluation was performed through schematic-level simulations at \SI{-40}{\celsius}, using ideal shaping stages and the actual CSA with Krummenacher feedback, in combination with a detector capacitance of \SI{40}{\pico\farad}.
Simulation results reported in~\Cref{fig:ch4-shaper_CR-RC_CR-RC2} show that the improvement in energy resolution obtained by increasing the order is marginal: the CR-RC filter achieves the best energy resolution of \SI{3.01}{\kilo\electronvolt} at \SI{0.6}{\micro\second}, compared to \SI{2.88}{\kilo\electronvolt} at \SI{0.8}{\micro\second} for the CR-RC\textsuperscript{2} filter.  
This corresponds to a relative improvement of only \SI{4.26}{\percent}, which does not justify the additional complexity, power consumption and area required for implementing an extra integrator.  

\par
The shaping time is programmable and is set by selecting appropriate capacitance values through a 3-bit digital control signal, \texttt{SHAP\_MODE}.  
The input capacitance \( C_1 \), feedback capacitance \( C_2 \), and pole-setting capacitance \( C_p \) in the forward gain stage are each implemented as arrays of eight unit capacitors, which are selectively enabled by a \hbox{binary-to-thermometric} decoder.  
These capacitances are increased proportionally according to a relationship detailed later in the chapter, enabling control of the shaping time—defined as the time at which the output transient reaches its peak, and corresponding to the inverse of the filter's peak frequency.

\begin{figure}[t!]
    \centering
    \includegraphics[width=0.9\textwidth]{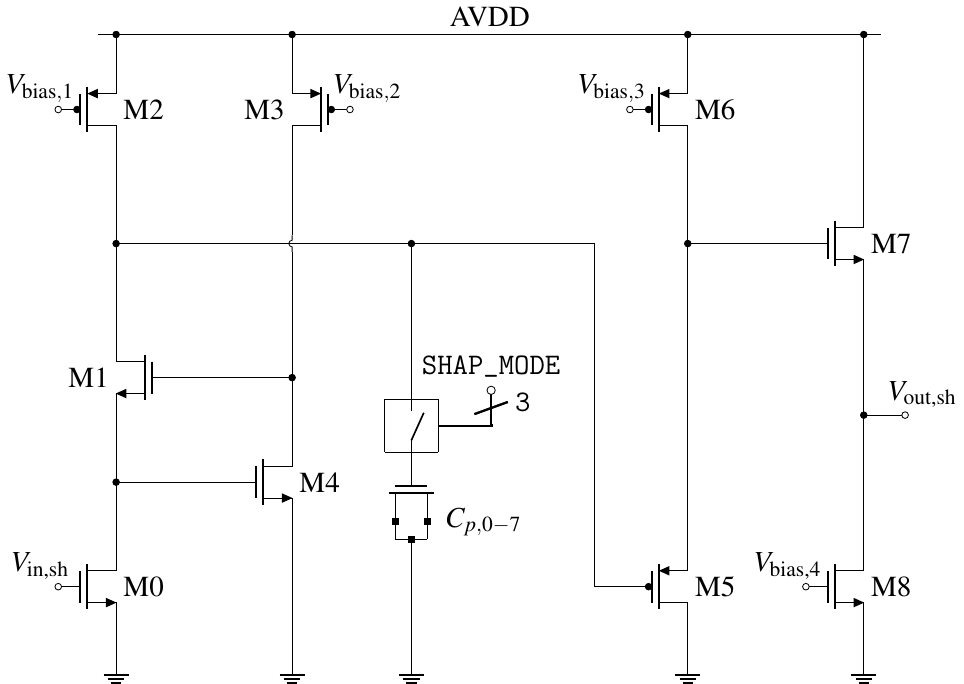}
    \caption{Schematic of the forward gain stage implemented in the shaper. The circuit consists of a common-source amplifier with an NMOS input transistor (M0) and local feedback to increase the impedance seen at the output node. The dominant pole is set by a programmable capacitance \( C_p \), implemented as an array of NMOS transistors and controlled via the 3-bit digital signal \texttt{SHAP\_MODE}. The output stage comprises a PMOS source follower (M5--M6) followed by an NMOS source follower (M7--M8), providing low output impedance. The bias network is omitted for clarity.}
    \label{fig:ch4-shaper_gain_stage}
\end{figure}

\par
The circuit topology incorporates, in its feedback path, a transconductance element with transconductance \( G_{f,\text{sh}} \), which discharges the feedback capacitor \( C_2 \) and establishes a DC output baseline of \( 2V_{\text{ref,sh}} \) (\SI{900}{\milli\volt}, with \( V_{\text{ref,sh}} = \SI{450}{\milli\volt} \)).
From this point on, the input and output voltages of the shaping stage will be denoted as \( V_{\text{in,sh}} \) and \( V_{\text{out,sh}} \), respectively.

\par
In order to derive the shaping stage transfer function, as well as the parameters relevant to the calculation of the input, feedback, and pole-setting capacitances, the peaking time, and the feedback transconductance, the forward gain stage can be modeled as a single-pole amplifier with the following transfer function:

\begin{equation}
    H_{\text{fgs}}(s) = \frac{A_{\text{sh}}}{1 + s \tau_0},
\end{equation}
\noindent
where \( A_{\text{sh}} = g_{m,0} R_{\text{out}} \) is the low-frequency gain and \( \tau_0 = R_{\text{out}} C_p \) is the dominant pole time constant.
The capacitance \( C_p \) is used to tune the bandwidth of the stage and is defined as a scaled multiple of the seed capacitance \( C_0 \).
Specifically, the following relation holds:

\begin{equation}
  C_p = n \gamma C_0,
\end{equation}

\noindent
where \( C_0 = \SI{30}{\femto\farad} \), and \( \gamma = 167.4 \) is a scaling coefficient derived later in the chapter.
The shaping filter transfer function is given by:

\begin{equation}
  \frac{V_{\text{out,sh}}}{V_{\text{in,sh}}}(s) = -\frac{A_{\text{sh}}}{\tau_0} \frac{C_1}{C_1 + C_2} \frac{s}{s^2 + 2\xi_0 \omega_f s+ \omega_0^2}.
\end{equation}

\begin{table}[t!]
    \centering
    \renewcommand{\arraystretch}{1.1}
    \setlength{\tabcolsep}{8pt}
    \begin{tabular}{ccccccc}
        \toprule
        \multirow{2}{*}{Device} & \multirow{2}{*}{Type} & $L$ & $W_\text{tot}$ & $W_\text{finger}$ & Fingers \\
                                &                        & (\si{\micro\meter}) & (\si{\micro\meter}) & (\si{\micro\meter}) & \# \\
        \midrule
        M0 & NMOS, LVT & 0.2 & 30   & 5.0  & 6  \\
        M1 & NMOS, LVT & 0.1 & 35   & 5.0  & 7  \\
        M2 & PMOS, LVT & 3.0 & 2    & 2.0  & 1  \\
        M3 & PMOS, LVT & 1.0 & 0.3  & 0.3  & 1  \\
        M4 & NMOS, LVT & 0.1 & 75   & 5.0  & 15 \\
        M5 & PMOS, LVT & 1.0 & 100  & 5.0  & 20 \\
        M6 & PMOS, LVT & 0.5 & 2    & 2.0  & 1  \\
        M8 & NMOS, LVT & 1.0 & 0.7  & 0.7  & 1  \\
        M9 & NMOS, LVT & 0.2 & 50   & 5.0  & 10 \\
        \bottomrule
    \end{tabular}
    \caption{Geometrical parameters of all transistors used in the forward gain stage of the shaper, as shown in~\Cref{fig:ch4-shaper_gain_stage}. The table reports the channel length \( L \), total width \( W_\text{tot} \), width per finger \( W_\text{finger} \), and number of fingers for each device.}
    \label{tab:ch4-shaper_gain_stage_devices}
\end{table}

\noindent
The damping factor and natural frequency are defined as:

\begin{equation}
  2\xi_0 \omega_f = \frac{C_1 + C_2 + G_{f,\text{sh}}\tau_0 + A_{\text{sh}} C_2}{\tau_0 (C_1 + C_2)} \approx \frac{A_{\text{sh}}}{\tau_0} \frac{C_2}{C_1 + C_2},
  \label{eq:twoxi}
\end{equation}

\begin{equation}
  \omega_f^2 = \frac{G_{f,\text{sh}} (1 + A_{\text{sh}})}{\tau_0 (C_1 + C_2)} \approx \frac{A_{\text{sh}}}{\tau_0} \frac{G_{f,\text{sh}}}{C_1 + C_2},
  \label{eq:omega}
\end{equation}

\noindent
where \( G_{f,\text{sh}} \) is the transconductance of the feedback stage.
Assuming \( A_{\text{sh}} \gg 1 \), the damping factor becomes:

\begin{equation}
  \xi_0 = \frac{1}{2} \frac{C_2}{\sqrt{C_1 + C_2}} \sqrt{\frac{A_{\text{sh}}}{\tau_0 G_{f,\text{sh}}}}.
\end{equation}

\noindent
The condition for critical damping (\( \xi_0 = 1 \)) leads to the requirement on the transconductance of the feedback element:

\begin{equation}
  G_{f,\text{sh}} = \frac{1}{4} \frac{A_{\text{sh}}}{\tau_0} \frac{C_2^2}{C_1 + C_2}.
  \label{eq:gm}
\end{equation}

\noindent
In this case, the shaping function reduces to:

\begin{equation}
  H_{\text{CR-RC}}(s) = \frac{s\tau}{(1+\tau s)^2},
\end{equation}

\noindent
with time constant \( \tau_f = \omega_f^{-1} \), which also defines the peaking time:

\begin{equation}
  t_p = \tau_f = 2 \frac{\tau_0}{A_{\text{sh}}} \left(1 + \frac{C_1}{C_2}\right).
  \label{eq:tp}
\end{equation}

\begin{figure}[t!]
    \centering
    \includegraphics[width=0.75\textwidth]{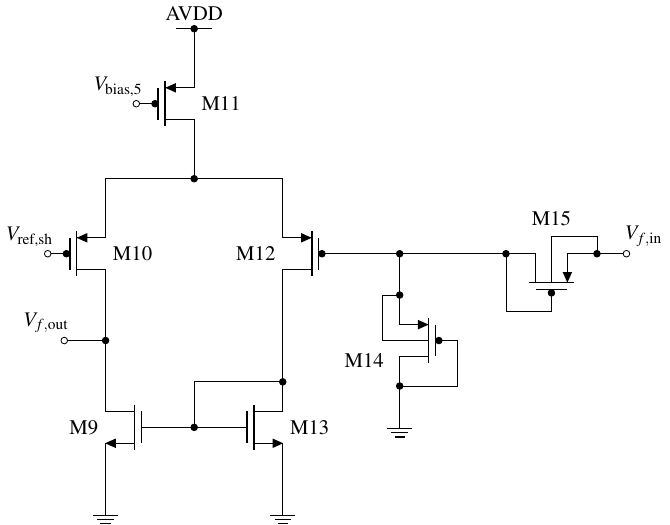}
    \caption{Schematic of the transconductor stage used in the shaper feedback path. The circuit includes a voltage divider at the input, implemented using identical PMOS devices M14 and M15 ($2/\SI{0.5}{\micro\meter/\micro\meter}$), to scale the input voltage and prevent the differential pair from entering a fully unbalanced configuration. The transconductor sets the DC baseline voltage of \SI{900}{\milli\volt} at the output of the shaper via the reference input \( V_{\text{ref,sh}} \).}
    \label{fig:ch4-shaper_transconductor}
\end{figure}

\noindent
Using the expressions for \( A_{\text{sh}} \) and \( \tau_0 \), the peaking time can be rewritten as:

\begin{equation}
  t_p = 2 \frac{C_p}{g_{m,0}} \left(1 + \frac{C_1}{C_2} \right).
\end{equation}

\noindent
The capacitors \( C_1 \) and \( C_2 \) are defined as multiples of the seed capacitance \( C_0 \), scaled according to:

\begin{align}
  C_2 &= n C_0, \\
  C_1 &= \alpha C_2,
\end{align}

\noindent
where \( \alpha \) defines the input-to-feedback capacitance ratio.
Assuming an input dynamic range \( \Delta V_{\text{in,sh}} = \SI{500}{\milli\volt} \) and a required output range \( \Delta V_{\text{out,sh}} = \SI{600}{\milli\volt} \), the required shaping gain is \( k_{\text{sh}} = \Delta V_{\text{out,sh}} / \Delta V_{\text{in,sh}} = 1.2 \).
Since the CR-RC response has an exponential attenuation factor \( 1/e \) at the peak, the gain must be corrected accordingly:

\begin{equation}
  \frac{C_1}{C_2} = \alpha = k_f \frac{e}{2} \approx 1.63.
\end{equation}

\noindent
The pole-setting capacitance \( C_p \) can also be expressed using \( C_2 \) and \( \alpha \), yielding:

\begin{equation}
  C_p = n \gamma C_0 = \frac{g_{m,0}}{2(1 + \alpha)} C_2.
\end{equation}

\noindent
Finally, the nominal value of the feedback transconductance is set from the seed parameters as:

\begin{equation}
  G_{f,\text{sh}} = \frac{C_0}{2 t_{p,0}} = \SI{75}{\nano\siemens},
\end{equation}

\noindent
with \( t_{p,0} = \SI{200}{\nano\second} \) as the reference peaking time for \( n = 1 \).  
Table~\ref{tab:ch4-shaper_capacitors} summarizes the values of the input capacitance \( C_1 \), scaled by a factor \( \alpha \) relative to the feedback capacitance \( C_2 \), and the pole-setting capacitance \( C_p \), scaled by a factor \( \gamma \).  
Among these, \( C_p \) is the largest and, due to area constraints, has been implemented as an array of MOS transistors.  
The input and feedback capacitors, on the other hand, are implemented as arrays of MIM capacitors.

\begin{table}[t!]
    \centering
    \renewcommand{\arraystretch}{1.1}
    \setlength{\tabcolsep}{8pt}
    \begin{tabular}{ccccc}
        \toprule
        \multirow{2}{*}{Device} & \multirow{2}{*}{Type} & $L$ & $W_\text{tot}$ & Fingers \\
                                &                        & (\si{\micro\meter}) & (\si{\micro\meter}) & \# \\
        \midrule
        M9  & NMOS, LVT & 1.0  & 0.2 & 1 \\
        M10 & PMOS, LVT & 12.0 & 0.2 & 1 \\
        M11 & PMOS, LVT & 1.0  & 1.0 & 1 \\
        M12 & PMOS, LVT & 12.0 & 0.2 & 1 \\
        M13 & NMOS, LVT & 1.0  & 0.2 & 1 \\
        M14 & PMOS, SVT & 0.5  & 2.0 & 1 \\
        M15 & PMOS, SVT & 0.5  & 2.0 & 1 \\
        \bottomrule
    \end{tabular}
    \caption{Geometrical parameters of the transistors used in the transconductor stage, as shown in~\Cref{fig:ch4-shaper_transconductor}. The table reports the channel length \( L \), total width \( W_\text{tot} \), and number of fingers for each device.}
    \label{tab:transconductor_device_sizes}
\end{table}

% Forward gain stage
\subsection{Forward gain stage}

\begin{table}[t!]
    \centering
    \renewcommand{\arraystretch}{1.1}
    \setlength{\tabcolsep}{5pt}
    \begin{tabular}{ccc|cccccccc}
        \toprule
        $b_{\text{in},2}$ & $b_{\text{in},1}$ & $b_{\text{in},0}$ &
        $b_{\text{out},7}$ & $b_{\text{out},6}$ & $b_{\text{out},5}$ &
        $b_{\text{out},4}$ & $b_{\text{out},3}$ & $b_{\text{out},2}$ & $b_{\text{out},1}$ &
        $b_{\text{out},0}$ \\
        \midrule
        0 & 0 & 0 & 0 & 0 & 0 & 0 & 0 & 0 & 0 & 1 \\
        0 & 0 & 1 & 0 & 0 & 0 & 0 & 0 & 0 & 1 & 1 \\
        0 & 1 & 0 & 0 & 0 & 0 & 0 & 0 & 1 & 1 & 1 \\
        0 & 1 & 1 & 0 & 0 & 0 & 0 & 1 & 1 & 1 & 1 \\
        1 & 0 & 0 & 0 & 0 & 0 & 1 & 1 & 1 & 1 & 1 \\
        1 & 0 & 1 & 0 & 0 & 1 & 1 & 1 & 1 & 1 & 1 \\
        1 & 1 & 0 & 0 & 1 & 1 & 1 & 1 & 1 & 1 & 1 \\
        1 & 1 & 1 & 1 & 1 & 1 & 1 & 1 & 1 & 1 & 1 \\
        \bottomrule
    \end{tabular}
    \caption{Truth table of the 3-to-8 thermometric decoder with fixed high output \( b_{\text{out},0} \). Inputs \( b_{\text{in},2} \), \( b_{\text{in},1} \), and \( b_{\text{in},0} \) are ordered from MSB to LSB. Outputs are ordered from \( b_{\text{out},7} \) (MSB) to \( b_{\text{out},1} \).}
    \label{tab:ch4-thermo_decoder_table}
\end{table}

\begin{table}[t!]
    \centering
    \renewcommand{\arraystretch}{1.2}
    \setlength{\tabcolsep}{6pt}
    \begin{tabular}{c|l}
        \toprule
        Output & Minimized Boolean expression \\
        \midrule
        $b_{\text{out},0}$ & $1$ (tied to DVDD) \\
        $b_{\text{out},1}$ & $b_{\text{in},2} \lor b_{\text{in},1} \lor b_{\text{in},0}$ \\
        $b_{\text{out},2}$ & $b_{\text{in},2} \lor b_{\text{in},1}$ \\
        $b_{\text{out},3}$ & $b_{\text{in},2} \lor \left( b_{\text{in},1} \land b_{\text{in},0} \right)$ \\
        $b_{\text{out},4}$ & $b_{\text{in},2}$ \\
        $b_{\text{out},5}$ & $(b_{\text{in},2} \land b_{\text{in},1}) \lor (b_{\text{in},2} \land b_{\text{in},0})$ \\
        $b_{\text{out},6}$ & $b_{\text{in},2} \land b_{\text{in},1}$ \\
        $b_{\text{out},7}$ & $b_{\text{in},2} \land b_{\text{in},1} \land b_{\text{in},0}$ \\
        \bottomrule
    \end{tabular}
    \caption{Minimized Boolean expressions for the 3-to-8 thermometric decoder. Inputs are $b_{\text{in},2}$ (MSB), $b_{\text{in},1}$, and $b_{\text{in},0}$ (LSB), representing the 3-bit binary input. Outputs $b_{\text{out},0}$ to $b_{\text{out},7}$ implement a thermometer code, where each output is logic high if the binary input is greater than or equal to its index. The output $b_{\text{out},0}$ is fixed high via DVDD.}
    \label{tab:thermo_decoder_min_exprs}
\end{table}

\par
The forward gain stage used in the shaping circuit is shown in~\Cref{fig:ch4-shaper_gain_stage}, and the geometrical parameters of each device are listed in~\Cref{tab:ch4-shaper_gain_stage_devices}, including the channel length \( L \), total width \( W_{\text{tot}} \), width per finger \( w_{\text{finger}} \), and number of fingers.

\par
The input stage is implemented as a common-source amplifier with an NMOS input transistor (M0) and local feedback to increase the impedance seen at the output node.
Schematic-level simulations at \SI{-40}{\celsius} in the typical process corner report a DC \hbox{open-loop} gain of approximately \SI{35}{\decibel}.
The dominant pole frequency is set by a switchable capacitor array implementing the pole-setting capacitance \( C_p \), which is controlled by a 3-bit digital signal \texttt{SHAP\_MODE} provided by the on-chip shift register.

\par
The output stage consists of a PMOS source follower (M5--M6) followed by an NMOS source follower (M7--M8), forming a level-shifting buffer with low output impedance.
This configuration supports the required output dynamic range of approximately \SI{600}{\milli\volt} and a baseline voltage of \SI{900}{\milli\volt}.
In the typical process corner, the output stage achieves a DC gain of \SI{-0.31}{\decibel}.

\begin{figure}[t!]
    \centering
    \includegraphics[width=0.7\textwidth]{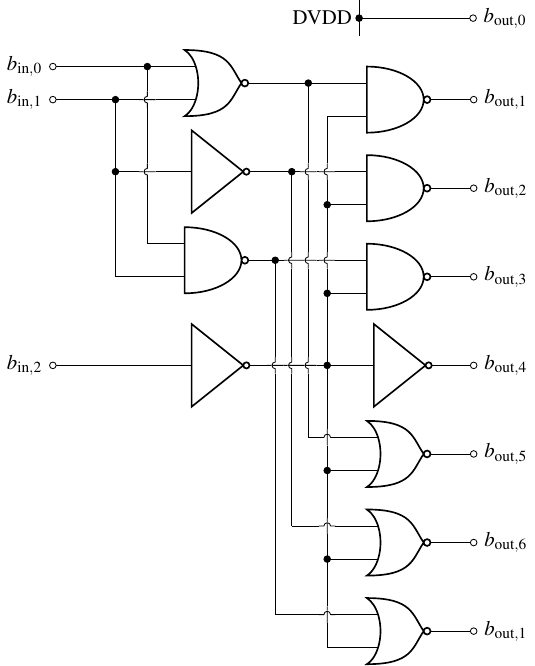}
    \caption{Logic diagram of the 3-bit binary-to-thermometric decoder used to control the shaping time selection. The circuit is implemented exclusively using NOT, NOR, and NAND gates. The 3-bit binary input $b_{\text{in}}$[2:0] is decoded into an 8-bit thermometric output $b_{\text{out}}$[7:0], with $b_{\text{out},0}$ permanently tied to logic high via DVDD.}
    \label{fig:ch4-BTM_decoder_schematic}
\end{figure}

\begin{figure}[t!]
    \centering
    \includegraphics[width=\textwidth]{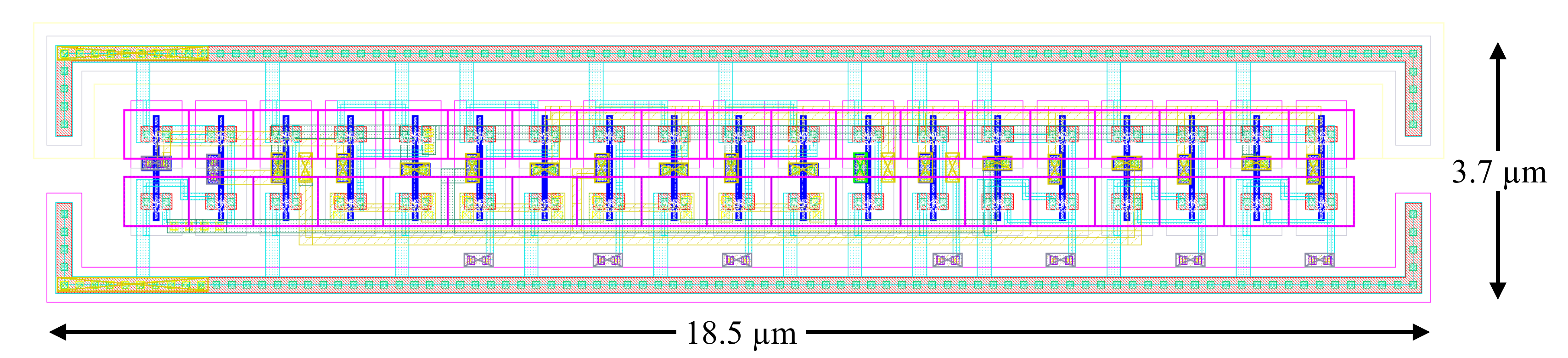}
    \caption{Layout of the 3-to-8 thermometric decoder implementing the logic defined in~\Cref{tab:thermo_decoder_min_exprs}. The layout excludes power routing for clarity and occupies an area of \SI{18.5}{\micro\meter}~$\times$~\SI{3.7}{\micro\meter}.}
    \label{fig:ch4-BTM_decoder_layout}
\end{figure}

%   Feedback transconductor
\subsection{Feedback transconductor}

\par
The transconductor stage shown in~\Cref{fig:ch4-shaper_transconductor} is implemented in the feedback path of the shaping stage to fulfill two critical functions: charge restoration in the feedback capacitor \( C_2 \) and setting of the DC output baseline to twice the reference voltage \( V_{\text{ref,sh}} \).
In this implementation, the baseline is defined at \SI{900}{\milli\volt}, while the output voltage swing must accommodate signals up to \SI{600}{\milli\volt} for the maximum input energy of \SI{100}{\mega\electronvolt}.

\par
To support this dynamic range without driving the differential pair into a fully unbalanced state, the input of the transconductor includes a voltage divider formed by two identical SVT PMOS transistors, M14 and M15, with dimensions \( W/L = \SI{2.0}{\micro\meter}/\SI{0.5}{\micro\meter} \).
This divider effectively doubles the reference voltage \(V_{\text{ref,sh}} = \SI{450}{\milli\volt}\), resulting in an effective DC baseline voltage \(V_{f,\text{in}}\) of about \SI{900}{\milli\volt}, while the gate of M12 remains biased at \SI{450}{\milli\volt} in steady-state conditions.
As a result, the differential pair (M10 and M12), implemented with long-channel PMOS devices (\SI{12}{\micro\meter}), preserves its symmetry across the full output dynamic range.

\par
This approach enables the design to meet the required low transconductance value of approximately \SI{75}{\nano\siemens}, which is critical to achieving the desired shaping stage peaking times.
The differential pair is biased with a current of \SI{10}{\nano\ampere}, provided by a dedicated bias network implemented in the periphery of the ASIC.
The device dimensions for all transistors used in the transconductor stage are reported in~\Cref{tab:transconductor_device_sizes}.

%   Binary-to-thermometric decoder
\subsection{Binary-to-thermometric decoder}

\par
The 3-bit binary-to-thermometric (BTM) decoder shown in~\Cref{fig:ch4-BTM_decoder_schematic} is used to select the shaping time within the signal processing chain. It converts a 3-bit binary input \( b_{\text{in}}[2{:}0] \) into an 8-bit thermometer code \( b_{\text{out}}[7{:}0] \), where each successive output bit is asserted high when the binary input exceeds or equals its corresponding index. The input bits \( b_{\text{in},2:0} \) are provided by the slow control shift register implemented in the periphery of the ASIC, which programs the internal configuration bits of the chip.

\par
The output \( b_{\text{out},0} \) is permanently tied to logic high via DVDD to ensure that the smallest peaking time configuration (\( t_{p,0} = \SI{0.2}{\micro\second} \)) always includes at least one active capacitor. This guarantees that all capacitor sets, including the one corresponding to the shortest peaking time, see the same effective on-resistance. The associated pass gate for the smallest capacitance is always enabled to match the resistance seen by the other pass-gated capacitor sets.

\par
The thermometer decoder drives pass gates implemented with minimum-length transistors to minimize their on-resistance. These pass gates progressively connect increasing values of capacitance in the input, feedback, and pole-setting capacitor arrays, enabling programmable shaping times from the shortest to the longest configuration, as defined by the output states in~\Cref{tab:ch4-thermo_decoder_table}.

\par
The decoder is implemented using standard digital logic gates (NAND, NOR, and NOT), as defined by the minimized Boolean expressions in~\Cref{tab:thermo_decoder_min_exprs}. The layout of the circuit, shown in~\Cref{fig:ch4-BTM_decoder_layout}, follows best practices for symmetry and routing regularity, and occupies an area of \SI{18.5}{\micro\meter}~$\times$~\SI{3.7}{\micro\meter} (excluding power routing).

\begin{figure}[t!]
    \centering
    \captionsetup[subfigure]{justification=justified,singlelinecheck=false}
    \begin{subfigure}{0.486\textwidth}
        \centering
        \includegraphics[width=\linewidth]{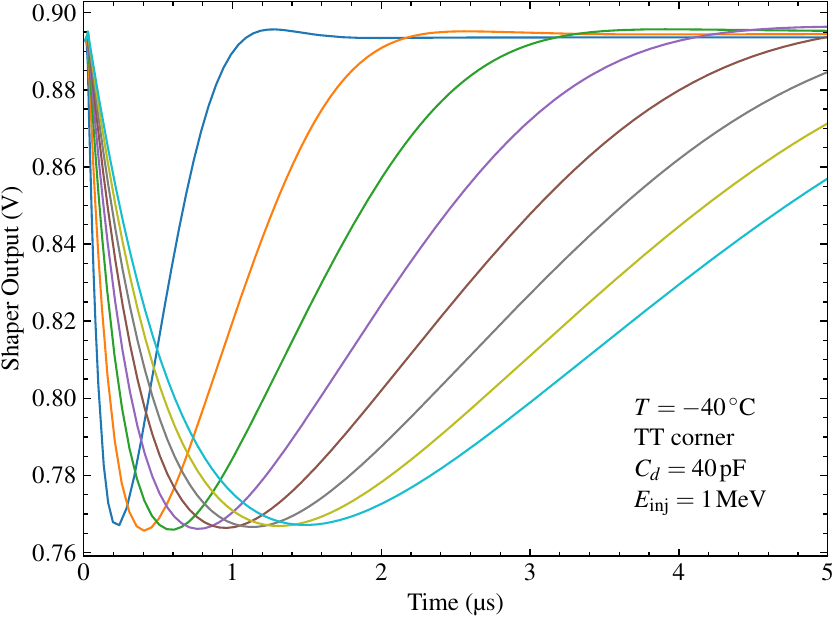}
        \caption{Transient simulation of the shaper output for \SI{1}{\mega\electronvolt} input energy across all eight peaking time settings.}
        \label{fig:ch4-CR_RC_transient_allpts}
    \end{subfigure}
    \hfill
    \begin{subfigure}{0.494\textwidth}
        \centering
        \includegraphics[width=\linewidth]{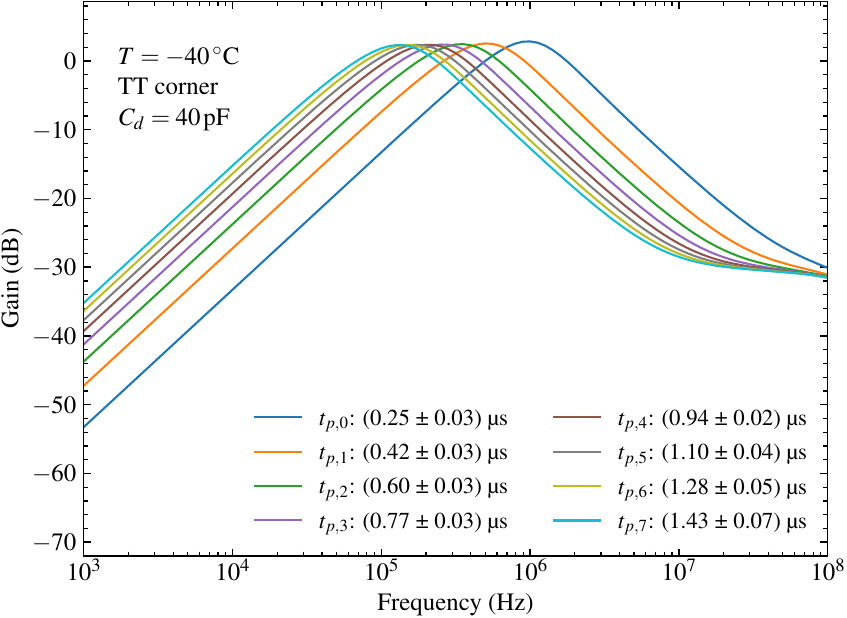}
        \caption{Simulated frequency response of the shaping stage for all eight programmable peaking time settings.}
        \label{fig:ch4-CR_RC_ac_allpts}
    \end{subfigure}
    \caption{Transient and frequency-domain response of the CR-RC shaper across all peaking time settings. Simulations were performed at \SI{-40}{\celsius} in the typical process corner with a detector capacitance of \SI{40}{\pico\farad}.}
    \label{fig:ch4-CR_RC_transient_ac_allpts}
\end{figure}

%   Performance evaluation
\subsection{Performance evaluation}\label{sec:ch5-shaper_noise_performance}

\par
The performance of the CR-RC shaper has been evaluated through schematic-level and post-layout simulations to verify its functionality across all programmable shaping configurations and operating conditions.

\par
\Cref{fig:ch4-CR_RC_transient_ac_allpts} shows both the transient and frequency-domain response of the shaping stage for all eight nominal peaking time settings, with an input signal corresponding to an energy deposition of \SI{1}{\mega\electronvolt}. The shaping times are controlled by the BTM decoder and nominally span from \( t_{p,0} = \SI{0.2}{\micro\second} \) to \( t_{p,7} = \SI{1.6}{\micro\second} \).

\par
In the transient simulation plot in~\Cref{fig:ch4-CR_RC_transient_allpts}, the output waveform peak amplitude remains nearly constant across all peaking time settings.
This is consistent with the simulated frequency response shown in~\Cref{fig:ch4-CR_RC_ac_allpts}, which confirms that the shaping filter maintains uniform gain at the peak frequency while shifting the passband toward lower frequencies as the peaking time increases.
The frequency response exhibits the expected band-pass behavior, with the passband progressively shifting toward lower frequencies as the peaking time increases.

\begin{figure}[t!]
    \centering
    \captionsetup[subfigure]{justification=justified,singlelinecheck=false}
    \begin{subfigure}{0.482\textwidth}
        \centering
        \includegraphics[width=\linewidth]{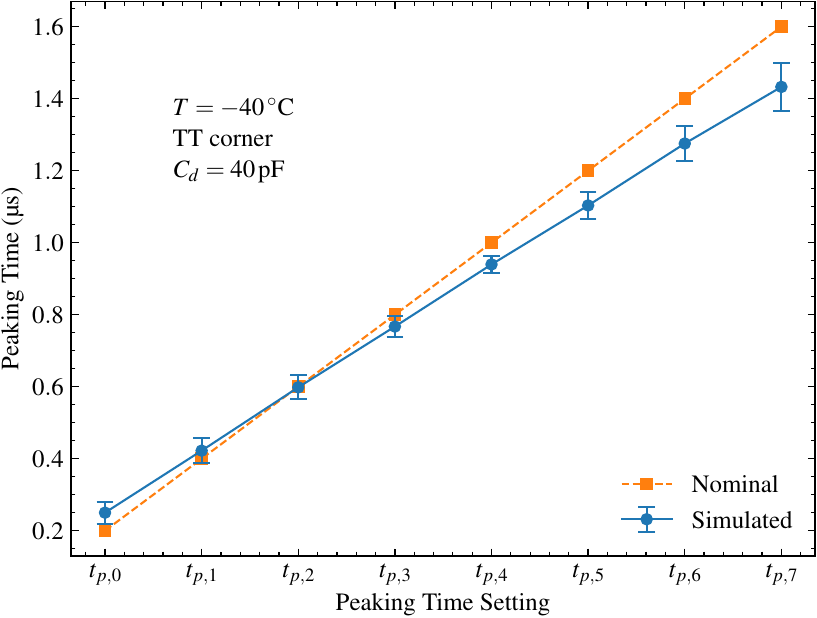}
        \caption{Simulated versus nominal peaking times for the eight programmable shaping configurations (\(t_{p,0}\)–\(t_{p,7}\)) in the typical process corner.}
        \label{fig:ch4-shaper_peaking_times}
    \end{subfigure}
    \hfill
    \begin{subfigure}{0.498\textwidth}
        \centering
        \includegraphics[width=\linewidth]{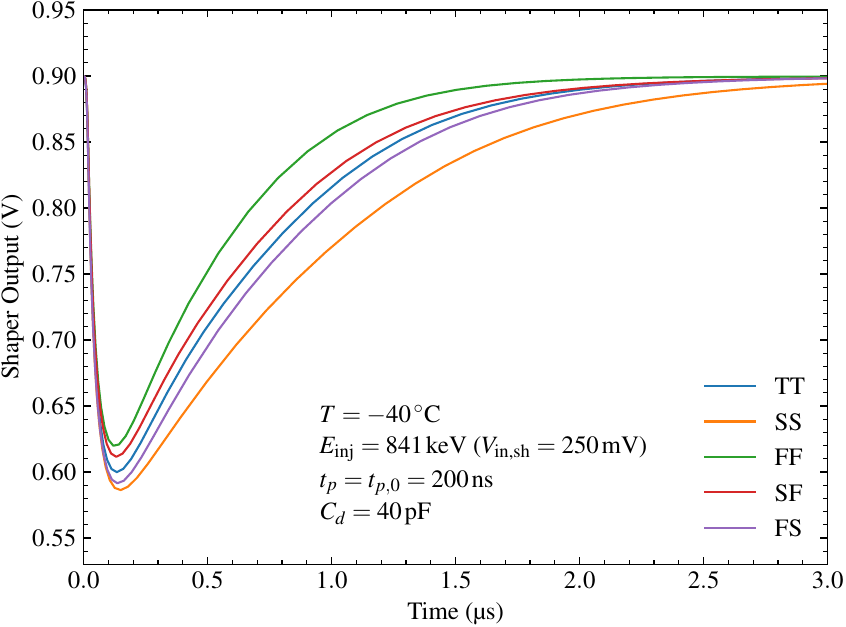}
        \caption{Transient shaper output for an incoming energy of \SI{841}{\kilo\electronvolt} (\SI{250}{\milli\volt} input) at a peaking time of \SI{0.2}{\micro\second}, simulated across five process corners.}
        \label{fig:ch4-shaper_transient_corners}
    \end{subfigure}
    \caption{Shaper output response and peaking-time characterization from \hbox{schematic-level} simulations at \SI{-40}{\celsius}. (\subref{fig:ch4-shaper_peaking_times}) Comparison of average simulated and nominal peaking times across shaping configurations, with error bars showing the variation across the full input dynamic range. (\subref{fig:ch4-shaper_transient_corners}) Output waveforms for five process corners under fixed input and shaping conditions.}
    \label{fig:ch4-shaper_transient_peaking_times}
\end{figure}

\begin{table}[t!]
    \centering
    \renewcommand{\arraystretch}{1.1}
    \setlength{\tabcolsep}{6pt}
    \begin{tabular}{cccccccc}
        \toprule
        \multirow{2}{*}{Peaking Time} & Nominal & Simulated & \multicolumn{2}{c}{Variation} & \multicolumn{2}{c}{Deviation} & Gain \\
        \cmidrule(lr){4-5} \cmidrule(lr){6-7}
        & (\si{\micro\second}) & (\si{\micro\second}) & (\si{\nano\second}) & (\si{\percent}) & (\si{\nano\second}) & (\si{\percent}) & (\si{\decibel}) \\
        \midrule
        $t_{p,0}$ & 0.20 & 0.25 & 31 & 12.25 & 50  & +25.04 & 2.84 \\
        $t_{p,1}$ & 0.40 & 0.42 & 34 & 8.03  & 22  & +5.59  & 2.55 \\
        $t_{p,2}$ & 0.60 & 0.60 & 33 & 5.56  & 2   & $-$0.28 & 2.44 \\
        $t_{p,3}$ & 0.80 & 0.77 & 29 & 3.76  & 33  & $-$4.11 & 2.39 \\
        $t_{p,4}$ & 1.00 & 0.94 & 24 & 2.60  & 60  & $-$5.35 & 2.35 \\
        $t_{p,5}$ & 1.20 & 1.10 & 37 & 3.39  & 100 & $-$8.06 & 2.35 \\
        $t_{p,6}$ & 1.40 & 1.23 & 50 & 3.89  & 170 & $-$8.90 & 2.32 \\
        $t_{p,7}$ & 1.60 & 1.43 & 67 & 4.68  & 170 & $-$10.43 & 2.30 \\
        \bottomrule
    \end{tabular}
    \caption{Nominal and simulated peaking times for the eight programmable shaping configurations \( t_{p,0} \) to \( t_{p,7} \). Reported values include the nominal target, average simulated value, variation across the input dynamic energy range, deviation from the nominal, and corresponding closed-loop gain. Simulations were performed at \SI{-40}{\celsius} in the typical process corner.}
    \label{tab:sim_peaking_stats}
\end{table}

\par
An artifact is observed above \SI{10}{\mega\hertz}, most prominently at the longest peaking time settings.
This behavior is attributed to gate leakage currents associated with the MOS implementation of the pole-setting capacitor \( C_p \), whose contribution becomes significant when the largest capacitance combinations are enabled.
Despite this effect, the shaper maintains acceptable performance across all configurations.
The eight programmable peaking time settings were implemented in this prototype ASIC to enable a systematic study of the noise performance of the readout channel under varying detector capacitance, leakage current, and temperature.
In the final design, a reduced set of shaping configurations will likely be adopted, since only the peaking time that yields the best energy resolution under flight conditions will be used during the experiment.
For this reason, the presence of non-ideal behavior for such long peaking times is not expected to impact the overall system performance.

\par
The comparison between simulated and nominal peaking times is reported in~\Cref{fig:ch4-shaper_peaking_times} and quantified in~\Cref{tab:sim_peaking_stats}. The measured peaking times closely track the nominal targets with a maximum deviation of \SI{170}{\nano\second} (\SI{-10.43}{\percent}) at the two longest configurations.
The peaking time variation across the full input dynamic energy range from \SI{10}{\kilo\electronvolt} to \SI{100}{\mega\electronvolt} is minimal for all configurations beyond \( t_{p,0} \), remaining below \SI{5}{\percent} in most cases and reaching a maximum of \SI{8.03}{\percent} for \( t_{p,1} \).
As expected, the highest relative variation of \SI{12.25}{\percent} occurs at \( t_{p,0} \), which corresponds to the shortest peaking time setting.
The closed-loop gain, measured as the peak value of the frequency-domain transfer function of the shaping stage and reported in~\Cref{tab:sim_peaking_stats}, decreases slightly as the peaking time increases, ranging from \SI{2.84}{\decibel} at \( t_{p,0} \) to \SI{2.30}{\decibel} at \( t_{p,7} \), corresponding to a relative reduction of approximately \SI{19}{\percent}.

\begin{figure}[tbp]
    \centering
    \includegraphics[width=0.8\textwidth]{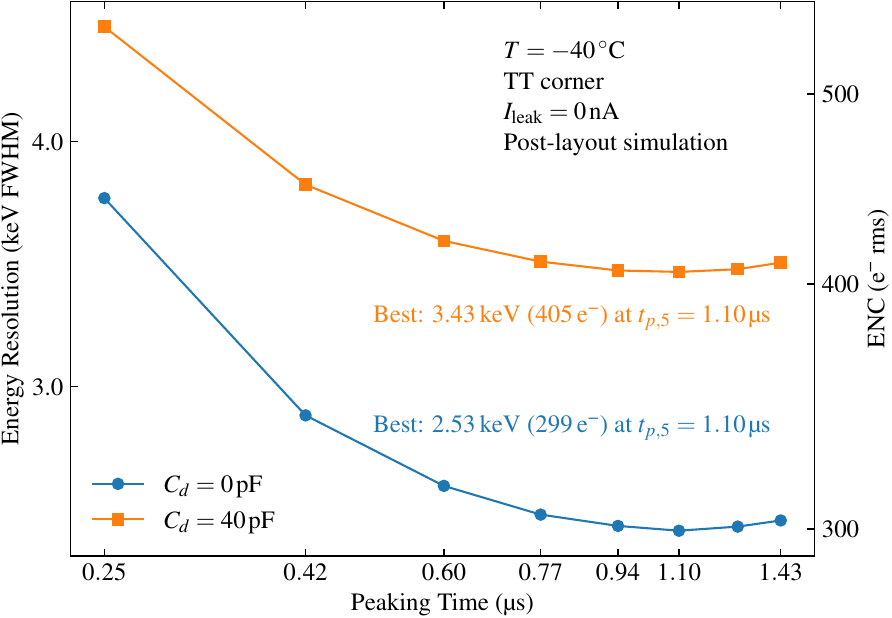}
    \caption{Energy resolution and ENC of the complete readout channel with the \hbox{CR-RC} shaper as a function of measured peaking times, evaluated with and without detector capacitance. Post-layout simulations assume the typical process corner, $T = \SI{-40}{\celsius}$, and $I_\text{leak} = \SI{0}{\nano\ampere}$. The best result without detector capacitance is \SI{2.53}{\kilo\electronvolt} (\SI{299}{e^{\text{-}}}) at $t_p = t_{p,5} = \SI{1.10}{\micro\second}$, increasing to \SI{3.43}{\kilo\electronvolt} (\SI{405}{e^{\text{-}}}) with a \SI{40}{\pico\farad} detector, corresponding to a \SI{36}{\percent} degradation.}
    \label{fig:ch4-shaper_noise}
\end{figure}

\par
Robustness against process variation was verified by performing simulations of the shaper in five process corners (TT, SS, FF, SF, FS).
Here, only the transient response is reported, as shown in~\Cref{fig:ch4-shaper_transient_corners}.
The results demonstrate consistent shaping behavior and timing across all corners, including the mixed slow-fast (SF) and fast-slow (FS) process corners.
Simulations were performed at \SI{-40}{\celsius} with a peaking time setting of \SI{0.2}{\micro\second} and an input signal corresponding to an incoming energy of \SI{841}{\kilo\electronvolt} (\SI{\sim250}{\milli\volt} input amplitude).

\par
The noise performance of the full readout channel up to the shaper was evaluated through post-layout simulations including detector capacitance. As shown in~\Cref{fig:ch4-shaper_noise}, the best energy resolution of \SI{2.53}{\kilo\electronvolt} is achieved at \( t_{p,5} = \SI{1.10}{\micro\second} \) in the absence of detector capacitance, corresponding to an equivalent noise charge (ENC) of \SI{299}{e^{\text{-}}}.
When a detector capacitance of \SI{40}{\pico\farad} is added, the energy resolution degrades to \SI{3.43}{\kilo\electronvolt}, with a corresponding ENC of \SI{405}{e^{\text{-}}}, reflecting a \SI{36}{\percent} increase in noise.

%   Shaping stage layout implementation
\subsection{Shaping stage layout implementation}

\begin{figure}[tbp]
    \centering
    \includegraphics[width=0.8\textwidth]{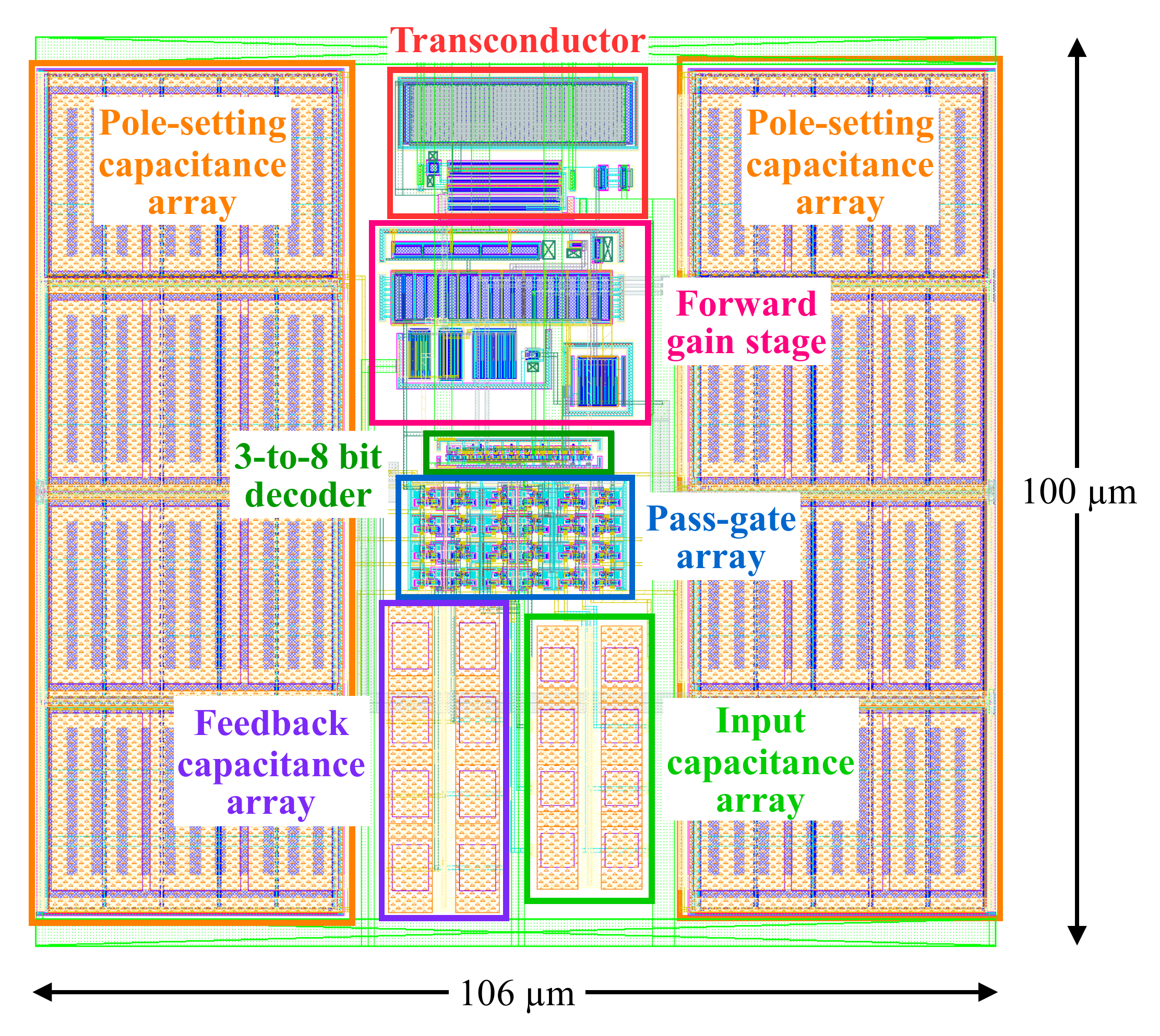}
    \caption{Complete layout of the CR-RC shaper, including all functional blocks and their relative placement. The layout integrates the input and feedback capacitance arrays, transconductor, forward gain stage, pole-setting capacitance arrays, pass-gate array, and the 3-to-8 bit thermometric decoder. The total area is \SI{106}{\micro\meter}~$\times$~\SI{100}{\micro\meter}.}
    \label{fig:ch4-shaper_layout}
\end{figure}

\par
The complete layout of the CR-RC shaper is shown in~\Cref{fig:ch4-shaper_layout}.  
It includes all functional blocks required by the shaping stage: the input and feedback capacitance arrays, the transconductor, the forward gain stage, the pole-setting capacitance arrays, the pass-gate array, and the 3-to-8 bit thermometric decoder.  
The total layout area is \SI{106}{\micro\meter}~$\times$~\SI{100}{\micro\meter}, corresponding to \SI{10600}{\micro\meter\squared}.

\par
Using the same design strategy adopted for the Krummenacher feedback element in the CSA, the layout of the shaping stage includes two \SI{4}{\micro\meter}-wide metal 7 tracks for the analog \( V_{\text{DD}} \) (AVDD) supply and analog ground, matching the corresponding tracks of the Krummenacher block.
This, in combination with the total layout height of \SI{100}{\micro\meter}, reflects the modular design approach adopted for the channel blocks described earlier.
To minimize the total length of the input and output signal paths of the shaper, these have been implemented using metal 4 tracks placed between the MOS capacitance units at the layout edges.
This choice helps reduce the distributed resistance of the routing and minimizes track length.

\begin{figure}[t!]
    \centering
    \includegraphics[width=0.99\textwidth]{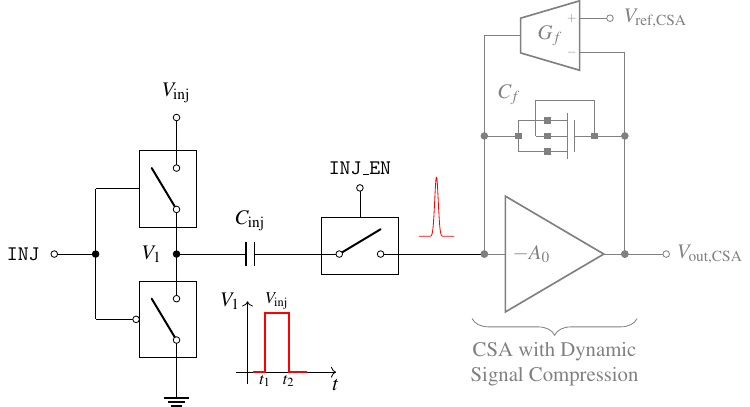}
    \caption{Conceptual diagram of the charge injection circuit. A voltage step \( V_\text{inj} \) is applied at \( V_1 \) across the \SI{3.7}{\pico\farad} injection capacitor \( C_\text{inj} \) by toggling two switches in opposite phase. The resulting charge is delivered to the CSA input. The injection is enabled by the digital signal \texttt{INJ\_EN}. \( V_\text{inj} \) is set by an external discrete 16-bit DAC over a \SI{1.2}{\volt} range, covering the full input dynamic range from \SI{10}{\kilo\electronvolt} to \SI{100}{\mega\electronvolt}.}
    \label{fig:ch4-injection_circuit_diagram}
\end{figure}

\par
The layout has been organized such that the forward gain stage and the transconductor---representing the core of the shaping function---are placed adjacent to each other near the center of the layout.
The auxiliary blocks are positioned toward the edges and include the most area-consuming components: the capacitors.
The input capacitance \( C_1 \) and feedback capacitance \( C_2 \) are implemented as MIM capacitors, while the pole-setting capacitance \( C_p \) is realized using a MOS capacitor array.
As described in the previous sections, the progressive activation of the capacitor arrays with increasing peaking time is controlled by the 3-to-8 bit thermometric decoder, which is placed in close proximity to the pass-gate array at the layout level. A total of 24 pass-gates are used, grouped into three banks of eight, one for each of the three capacitance types. These are controlled via the corresponding thermometric bits \( b_{\text{out}}[7{:}0] \).

%-------------------------------------------------------------------------------
%   Charge injection circuit for calibration
%-------------------------------------------------------------------------------
\section{Charge injection circuit for calibration}\label{sec:ch4-injection_circuit}

\par
Each readout channel implemented in the ASIC includes a charge injection circuit for calibration purposes.  
Its role is to inject a well-defined charge at the CSA input, reproducing the effect of charge collection on the Si(Li) strip.
The circuit is designed to cover the entire expected input dynamic energy range, spanning from \SI{10}{\kilo\electronvolt} to \SI{100}{\mega\electronvolt}.  
This calibration circuit must provide the best possible linearity in the charge delivered to the channel input, ensuring a reliable reference for calibrating the channel's energy response.

\par
\Cref{fig:ch4-injection_circuit_diagram} presents the conceptual diagram of the charge injection circuit.  
It can be modeled simply as an injection capacitance \( C_{\text{inj}} \) of known value, charged to a reference voltage \( V_{\text{inj}} \) at node \( V_1 \) by toggling two switches in opposite phase.  
The result is a voltage step of duration \( t_2 - t_1 \), corresponding to a known amount of charge and thus to a specific energy value within the \SI{10}{\kilo\electronvolt}--\SI{100}{\mega\electronvolt} input range.

\begin{figure}[t!]
    \centering
    \includegraphics[width=0.99\textwidth]{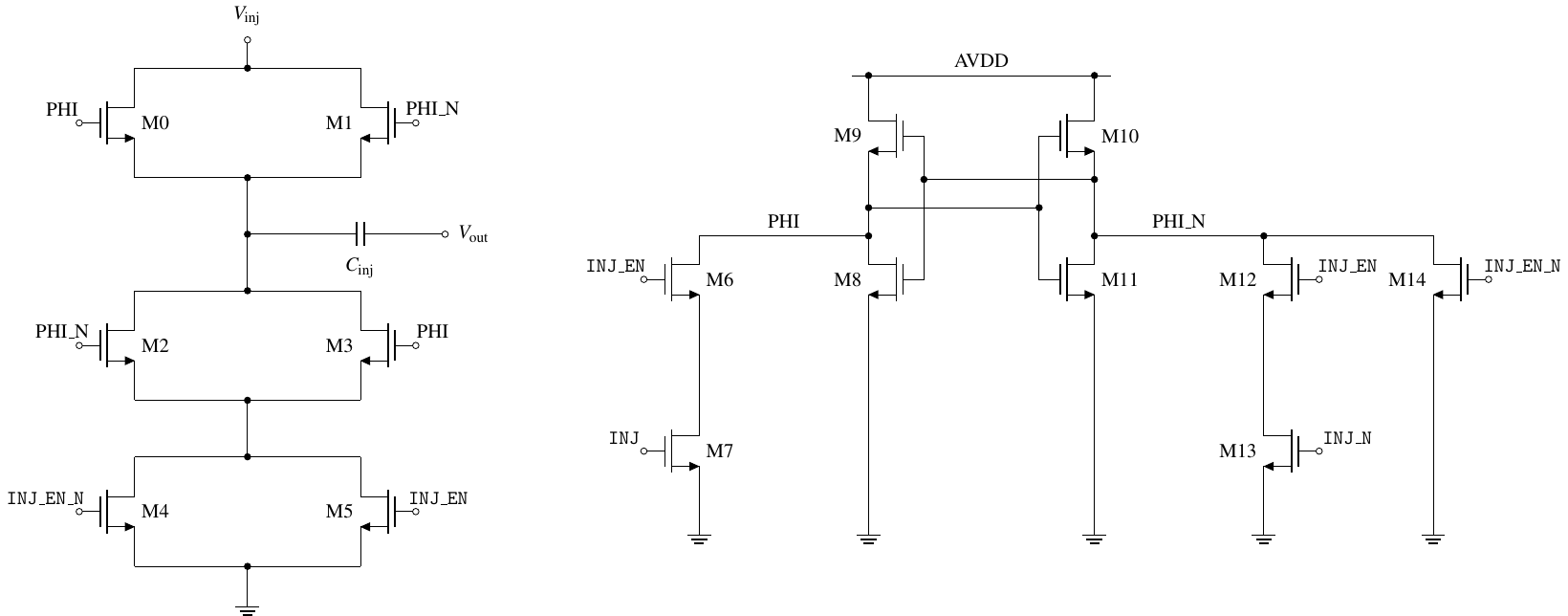}
    \caption{Complete schematic of the charge injection circuit (excluding the inverters generating \texttt{INJ\_EN\_N} and \texttt{INJ\_N}). On the left, the injection block uses NMOS switches to apply the voltage step \( V_\text{inj} \) across the \SI{3.7}{\pico\farad} MOM injection capacitor \( C_\text{inj} \). On the right, the phase generator produces the control signals \texttt{PHI} and \texttt{PHI\_N}, which drive the switches to generate the voltage step across \( C_\text{inj} \) when \texttt{INJ} is toggled and \texttt{INJ\_EN} is high.}
    \label{fig:ch4-injection_circuit_schematic}
\end{figure}

\par
The limiting factors in the design of this circuit are the precision in the implementation of the injection capacitance and the resolution of the reference voltage \( V_{\text{inj}} \).  
The latter limits the overall energy resolution, i.e., the minimum energy step that can be delivered to the CSA input.  
To address the first limiting factor, the injection capacitance has been implemented as a metal-oxide-metal (MOM) capacitor, using metal layers 1 through 7.  
This choice, instead of using a MIM or MOS capacitor, ensures the best precision in the capacitance value, at the expense of a lower capacitance density, as MOM capacitors typically provide less capacitance per unit area than MIM.
Given the relatively small value of \SI{3.7}{\pico\farad}, the design prioritizes precision over area occupation.

\par
The injection control signal \texttt{INJ} is provided by the slow control digital section of the ASIC, in combination with the channel-specific enable signal \texttt{INJ\_EN}.  
The latter activates the injection for the corresponding channel and allows the readout channel to be completely decoupled from the injection circuit during normal operation.

\par
The definition of the injection capacitance value \( C_{\text{inj}} = \SI{3.7}{\pico\farad} \) arises from considerations regarding the expected energy resolution of the channel, which in turn determines the required resolution of \( V_{\text{inj}} \).  
The injection reference voltage is provided externally by a precision DAC mounted on the front-end board.  
To ensure that the injected energy step remains below the target energy resolution of the readout channel, a 16-bit precision DAC (Texas Instruments DAC8541) was adopted.

\begin{figure}[tbp]
    \centering
    \includegraphics[width=0.75\textwidth]{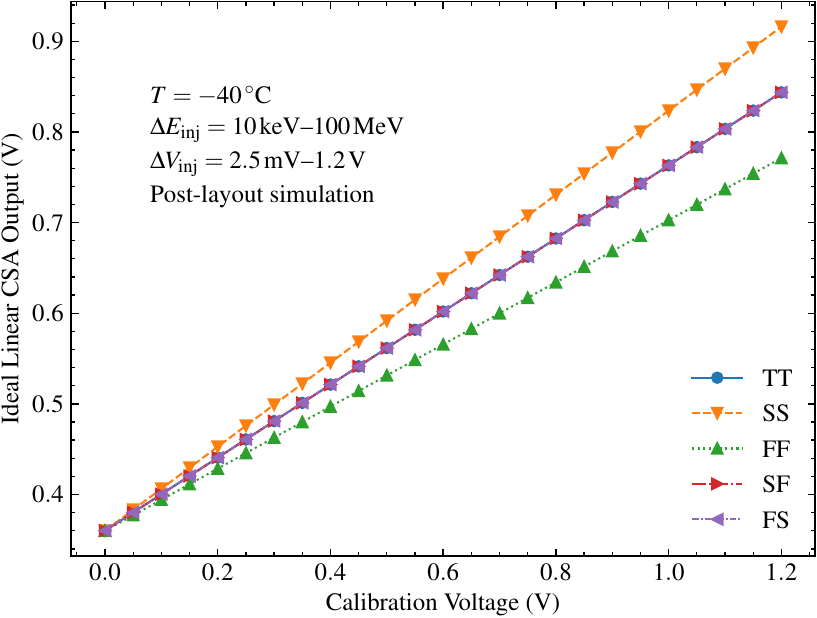}
    \caption{Post-layout simulation of the charge injection circuit at \SI{-40}{\celsius}, over the full input voltage range (\SI{2.5}{\milli\volt}--\SI{1.2}{\volt}), corresponding to \SI{10}{\kilo\electronvolt}--\SI{100}{\mega\electronvolt}. The output of an ideal linear CSA is shown, used to evaluate the linearity of the injected charge. Linearity is maintained across all process corners.}
    \label{fig:ch4-injection_circuit_performance}
\end{figure}

\par
Considering a \SI{1.2}{\volt} calibration voltage range, the minimum voltage step is:

\begin{equation}
    V_{\text{LSB}} = \frac{\SI{1.2}{\volt}}{2^{16}} \approx \SI{18.31}{\micro\volt}.
\end{equation}

\noindent
The corresponding charge increment is:

\begin{equation}
    Q_{\text{LSB}} = C_{\text{inj}} V_{\text{LSB}} \approx \SI{67.9}{\atto\coulomb}.
\end{equation}

\noindent
The charge-to-energy conversion, which yields the minimum injectable energy step, is calculated as:

\begin{equation}
    E_{\text{LSB}} = E_{\text{inj,min}} = \frac{\epsilon}{1000} \frac{Q_{\text{LSB}}}{e^{\text{-}}} \approx \SI{1.53}{\kilo\electronvolt},
\end{equation}

\noindent
where \( \epsilon \) is the ionization energy of silicon (\SI{3.6}{\electronvolt} per electron-hole pair).
It follows that the maximum injectable energy is:

\begin{equation}
    E_{\text{inj,max}} = E_{\text{LSB}} \cdot 2^{16} \approx \SI{100.04}{\mega\electronvolt},
\end{equation}

\noindent
thus covering the full expected input energy range.

\begin{figure}[tbp]
    \centering
    \includegraphics[width=0.99\textwidth]{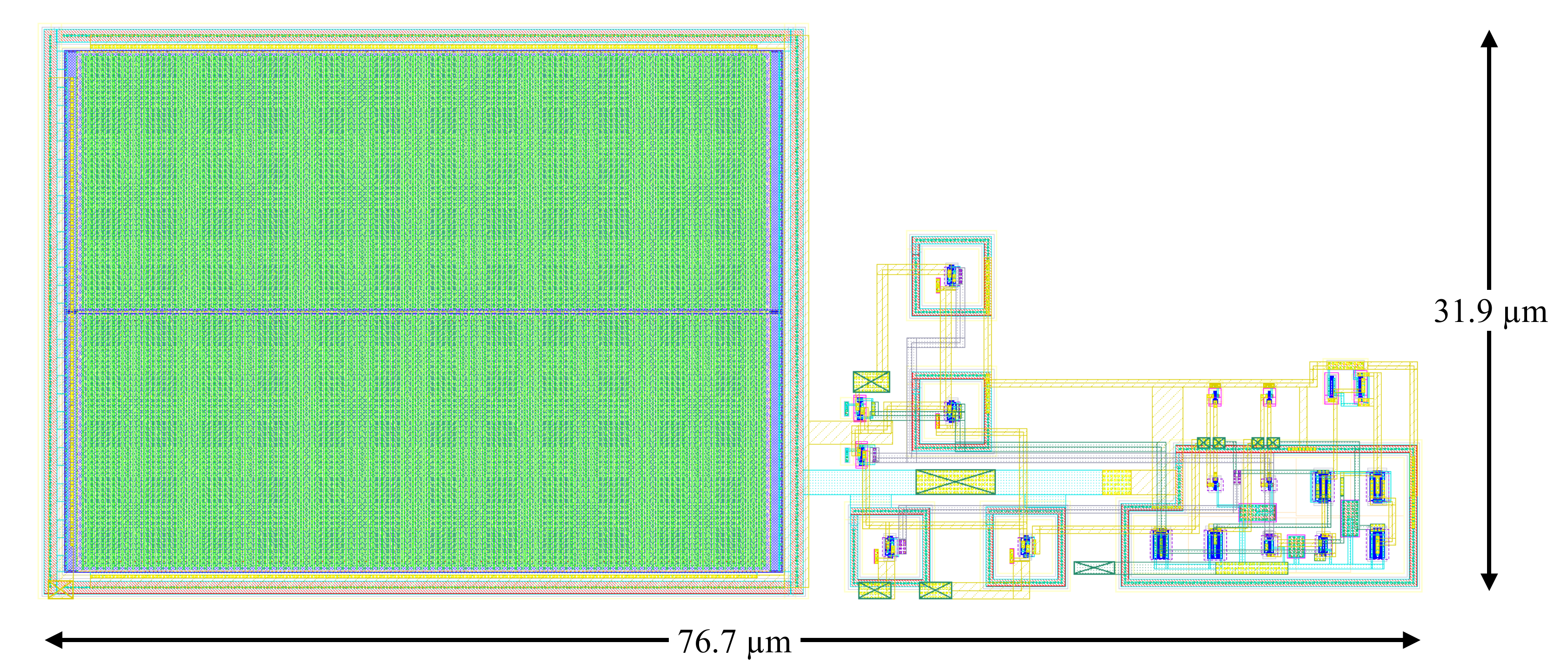}
    \caption{Layout of the charge injection circuit. The total area occupation is \SI{76.7}{\micro\meter}~\(\times\)~\SI{31.9}{\micro\meter}. All active devices are implemented in deep-\textit{n}-well to prevent substrate coupling with adjacent channels. The \SI{3.7}{\pico\farad} injection capacitor is realized as an array of two \SI{1.85}{\pico\farad} metal-oxide-metal (MOM) capacitors using metal layers 1 through 7.}
    \label{fig:ch4-injection_circuit_layout}
\end{figure}

\par
The schematic-level implementation of the charge injection circuit is shown in~\Cref{fig:ch4-injection_circuit_schematic}.  
On the left, the switch array is depicted, where each switch is implemented as a pair of minimum-length NMOS transistors to minimize the on-resistance.  
The M0--M1 and M2--M3 structures implement the two switches, driven by the control signals PHI and PHI\_N, which are generated by the phase generator circuit shown on the right.

\par
Decoupling of the charge injection circuit from the CSA input is achieved through the NMOS switches M4--M5, which are controlled by \texttt{INJ\_EN} and its complementary signal \texttt{INJ\_EN\_N}.  
These control signals are provided by the slow control digital section of the ASIC via the 12-bit shift register, described later in this chapter.

\par
On the right, the schematic of the phase generator circuit is shown.  
This block is responsible for generating the PHI and PHI\_N signals that control the MOS switches, thereby producing the input calibration voltage step of amplitude \( V_{\text{inj}} \).  
These signals are activated when \texttt{INJ} transitions from low to high and \texttt{INJ\_EN} is asserted.

\par
To evaluate the performance of the calibration circuit, post-layout simulations were carried out using an ideal CSA with a linear response.  
The CSA was modeled as an ideal \hbox{voltage-controlled} voltage source with a fixed feedback capacitor and an ideal \hbox{voltage-controlled} current source emulating the transconductance gain stage.  
This setup allowed the assessment of the intrinsic linearity of the charge injection circuit, independently of non-idealities in the front-end.

\par
\Cref{fig:ch4-injection_circuit_performance} shows the simulated response of the ideal CSA combined with the designed injection circuit across all process corners at \SI{-40}{\celsius}, covering the full input energy range from \SI{10}{\kilo\electronvolt} to \SI{100}{\mega\electronvolt}.  
The results exhibit excellent linearity throughout the range, with a maximum nonlinearity of \SI{0.21}{\percent} observed in the FF corner, as evaluated using the INL formula in~\eqref{eq:ch4-index_non_linearity}.
The simulation swept the entire calibration voltage range \( V_{\text{inj}} \) from \SI{2.5}{\milli\volt} to \SI{1.2}{\volt} in steps of approximately \SI{18.31}{\micro\volt}, corresponding to the resolution of the injection DAC.  
The plot presents a downsampled dataset comprising 24 injection steps, representing a voltage increment of roughly \SI{50}{\milli\volt}.

\begin{figure}[tbp]
    \centering
    \includegraphics[width=0.99\textwidth]{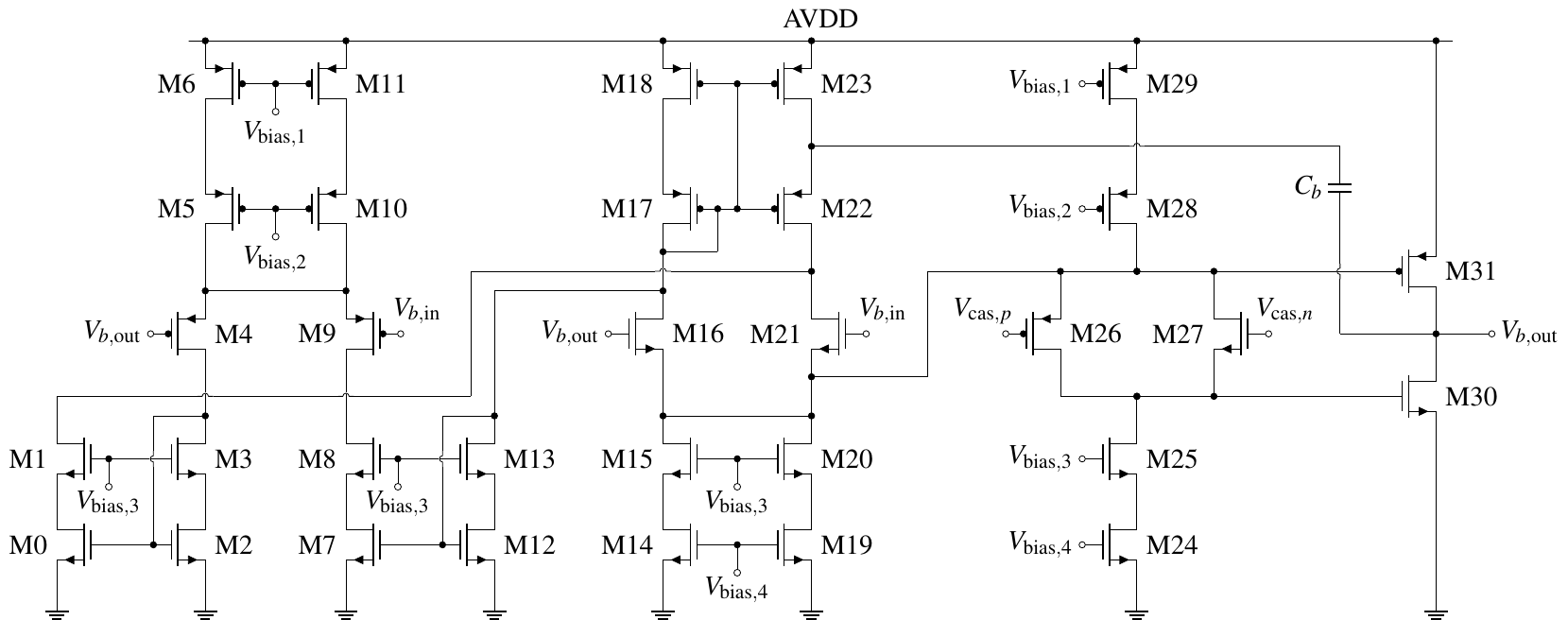}
    \caption{Schematic of the analog output buffer used at the output of either the CSA or the shaper stage to drive the dedicated output pad and interface with the external analog chain (e.g., oscilloscope) during ASIC testing. The capacitor \( C_b = \SI{340}{\femto\farad} \), implemented as a MOM capacitor, is connected to the output node to enhance stability.}
    \label{fig:ch4-output_driver_schematic}
\end{figure}

\par
The layout of the complete charge injection block is shown in~\Cref{fig:ch4-injection_circuit_layout}, occupying a total area of \( \SI{76.7}{\micro\meter} \times \SI{31.9}{\micro\meter} \).
All NMOS devices are implemented in deep-\textit{n}-well to prevent substrate coupling with adjacent channels.
The injection capacitance \( C_{\text{inj}} \) is realized as an array of two MOM capacitors, each with a value of \SI{1.85}{\pico\farad}, implemented using metal layers 1 through 7, resulting in a total capacitance of \SI{3.7}{\pico\farad}.
An asymmetric layout has been adopted, with the MOS switches and the phase generator placed in the bottom-right corner and the injection capacitance on the left.
This configuration minimizes the track length between the injection circuit and the CSA input while reserving space for the Krummenacher circuit in the top-right corner.
The modular design ensures optimal area utilization, fitting within the \SI{100}{\micro\meter} channel height and minimizing longitudinal footprint, thereby enabling the shortest possible readout chain.

%-------------------------------------------------------------------------------
%   Analog output buffer
%-------------------------------------------------------------------------------
\section{Analog output buffer}\label{ch4:sec-analog_outoput_buffer}

\par
Since the \antareslogo chip is a fully analog prototype of the next-generation ASIC for the GAPS experiment, testing is performed entirely in the analog domain.  
In this prototype, the readout chain is implemented up to the shaping stage only, with no on-chip digitization.
An external data acquisition system, described in~\Cref{chapter5}, will be used, with digitization performed via an oscilloscope.
To support this setup, output buffers are placed at the outputs of the selected CSA and shaper stages to drive the oscilloscope inputs, which typically present an input capacitance of \SIrange{10}{20}{\pico\farad}.

\begin{figure}[tbp]
    \centering
    \captionsetup[subfigure]{justification=justified,singlelinecheck=false}
    \begin{subfigure}{0.49\textwidth}
        \centering
        \includegraphics[width=\linewidth]{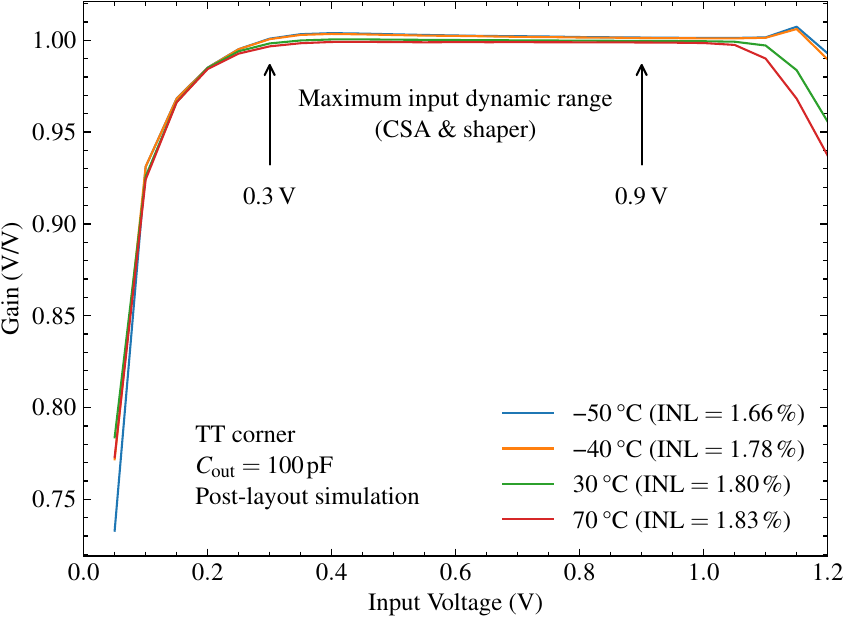}
        \caption{Gain across temperature at TT corner (\SI{-50}{\celsius} to \SI{70}{\celsius}). Post-layout simulation with \SI{100}{\pico\farad} load. Good linearity is maintained in the \SI{0.3}{\volt}–\SI{0.9}{\volt} input range.}
        \label{fig:ch4-buffer_performance_temps}
    \end{subfigure}
    \hfill
    \begin{subfigure}{0.49\textwidth}
        \centering
        \includegraphics[width=\linewidth]{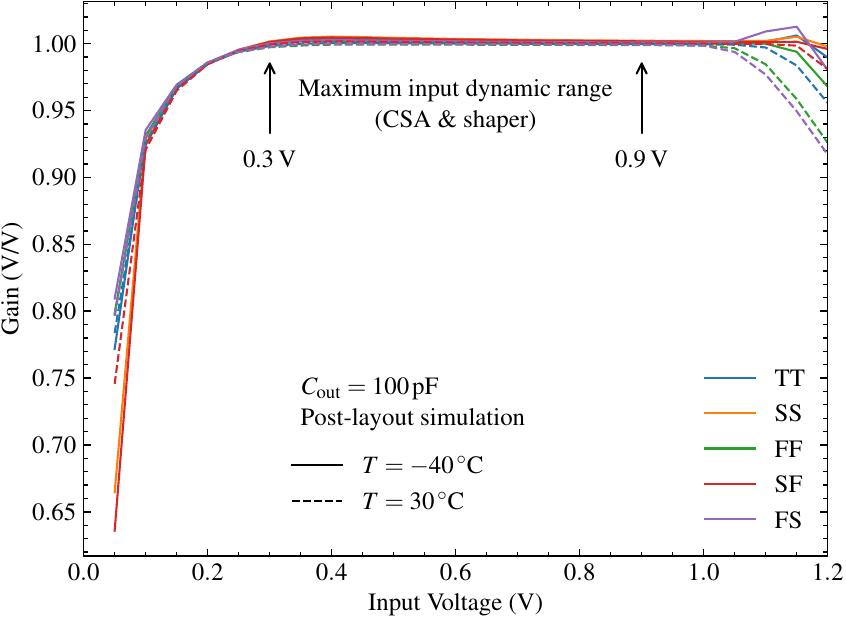}
        \caption{Gain of the analog output driver across process corners at \SI{-40}{\celsius} and \SI{30}{\celsius}. \hbox{Post-layout} simulations confirm consistent linearity in the \SI{0.3}{\volt}–\SI{0.9}{\volt} range.}
        \label{fig:ch4-buffer_performance_corners}
    \end{subfigure}
    \caption{Post-layout performance of the analog output driver used at the output of both CSA and shaper in the 8-channel ASIC. Simulations include a \SI{100}{\pico\farad} capacitive load to model oscilloscope probing conditions. The driver ensures unity gain and good linearity across temperature and process variations within the expected dynamic range.}
    \label{fig:ch4-buffer_performance}
\end{figure}

\par
In order to handle the capacitive loading of the measurement setup, the output buffer was designed to provide a flat response with unitary gain across both the positive voltage swing of the CSA, approximately spanning \SIrange{360}{900}{\milli\volt}, and the negative swing of the shaper outputs, ranging from \SIrange{900}{300}{\milli\volt}.  
\Cref{fig:ch4-output_driver_schematic} shows the schematic of the analog output buffer, which implements a Class~AB stage based on complementary source followers with dynamic gate biasing via a differential pair and current mirrors.  
The biasing network ensures low static power consumption while allowing the output stage to deliver sufficient drive capability for capacitive loads~\cite{bakerCMOSCircuitDesign2010}.  
A MOM capacitor of \SI{340}{\femto\farad} is connected to the output node to improve stability.  
A total of eight outputs---six from the CSA stages and two from the shaper stages---are then routed to dedicated output pads for readout.

\par
The open-loop differential gain of the amplifier can be approximated as the product of the gain of the rail-to-rail input stage and that of the push--pull buffer. 
Under small-signal conditions, the low-frequency gain can be expressed as

\begin{equation}
    A_{d,\mathrm{OL}} \simeq 2\,g_{m,\mathrm{in}}\,R_{1}\,g_{m,\mathrm{out}}\,R_{2},
\end{equation}

\noindent
where $g_{m,\mathrm{in}} = g_{m,\mathrm{N,in}} + g_{m,\mathrm{P,in}}$ denotes the effective transconductance of the combined NMOS and PMOS differential pairs, $R_{1}$ is the equivalent output resistance at the intermediate node $V_{b,\mathrm{in}}$, $g_{m,\mathrm{out}} = g_{m,30} + g_{m,31}$ is the transconductance of the push--pull output stage, and $R_{2} = r_{o,30} \parallel r_{o,31} \parallel R_\text{L}$ is the resistance seen at the output node $V_{b,\mathrm{out}}$.

\par
The buffer bias current is supplied by a two-terminal floating current source, as described later in~\Cref{chapter5}. 
Alternatively, during automated testing the same current can be generated on the test board using a programmable digital potentiometer.  
For the nominal bias current of \SI{5}{\milli\ampere}, the digital potentiometer is set to approximately \SI{21}{\ohm}, which results in a voltage drop of about \SI{142}{\milli\volt}.
For direct noise characterization, an additional test buffer---identical in topology---is included in the left-side driver array and equipped with dedicated input and output pads.  
This configuration enables evaluation of the analog buffer's noise performance as a function of the bias current, with the aim of quantifying its contribution to the overall noise of the readout channel.  
Such analysis is particularly relevant in the context of this prototype and supports future reuse of the output stage as a known IP block in ASICs fabricated in the \SI{65}{\nano\meter} process, benefiting from a measured and verified noise profile.

\begin{figure}[tbp]
    \centering
    \includegraphics[width=0.99\textwidth]{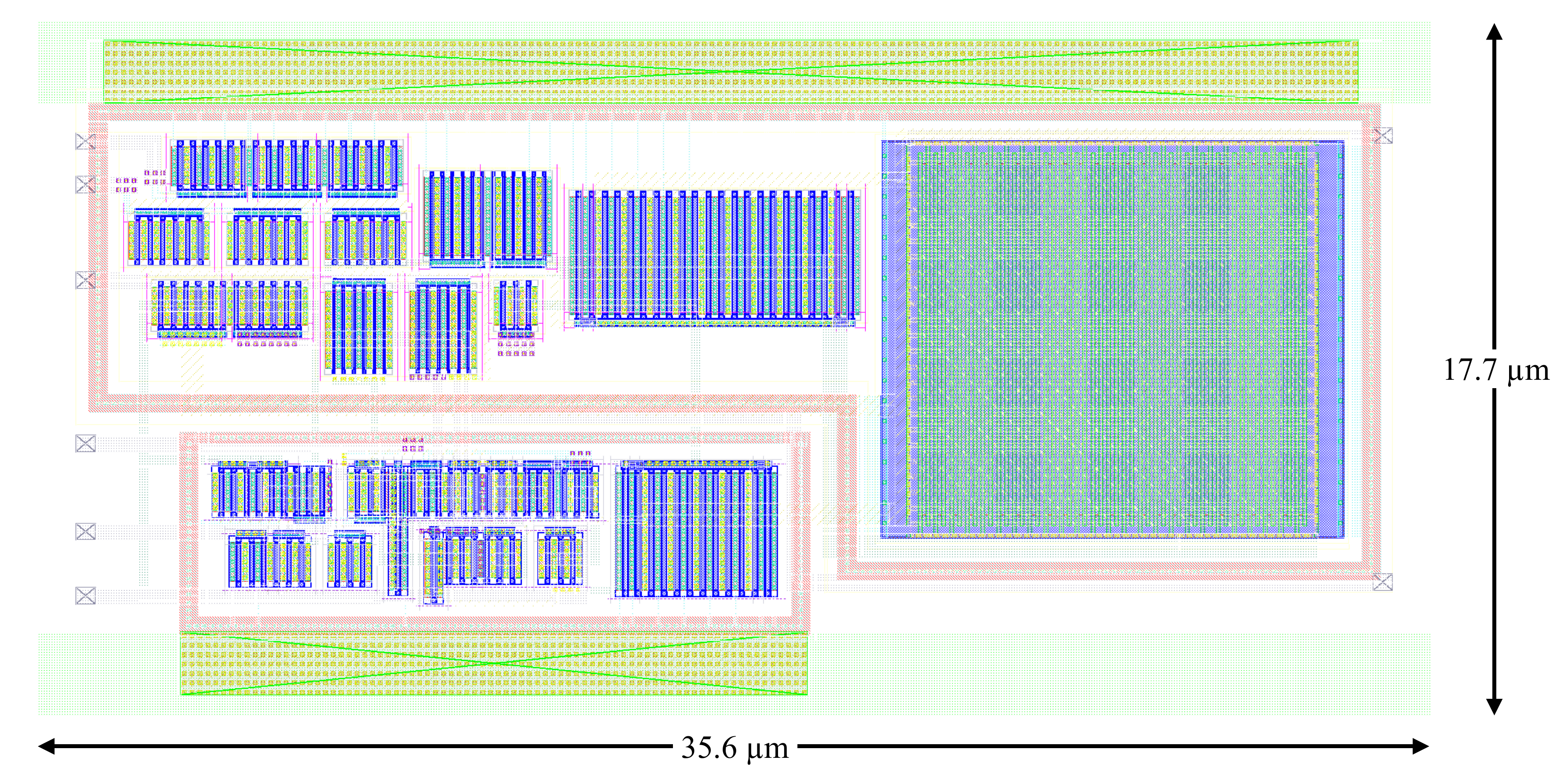}
    \caption{Layout of the analog output driver. The total area is \SI{35.6}{\micro\meter}~\(\times\)~\SI{17.7}{\micro\meter}. The capacitor \( C_b \) is implemented as a \SI{340}{\femto\farad} MOM capacitor. Analog \( V_{\text{DD}} \) and ground are routed through \SI{2}{\micro\meter}-tall metal 7 tracks to enable modular integration of multiple mirrored drivers. References and bias signals (not shown) are supplied from the left.}
    \label{fig:ch4-output_driver_layout}
\end{figure}

\par
\Cref{fig:ch4-buffer_performance} reports the post-layout simulated gain of the analog output buffer under varying process corners and temperatures, assuming a \SI{100}{\pico\farad} load to model the \hbox{worst-case} scenario associated with oscilloscope probe capacitance.  
\Cref{fig:ch4-buffer_performance_temps} shows the buffer gain across the full \SI{0}{\volt} to \SI{1.2}{\volt} input range at the typical process corner (TT) for temperatures from \SI{-50}{\celsius} to \SI{70}{\celsius}.  
Within the nominal dynamic range of the CSA and shaper stages---approximately \SIrange{0.3}{0.9}{\volt}---the output buffer maintains a linear response, with a maximum INL of \SI{1.83}{\percent} at \SI{70}{\celsius}.  
\Cref{fig:ch4-buffer_performance_corners} shows the gain variation across process corners (TT, SS, FF, SF, FS) at both \SI{-40}{\celsius} and \SI{30}{\celsius}.  
In all simulated corners, the buffer exhibits consistent linearity over the target dynamic range, confirming robust operation under both temperature and process variations.

\par
The complete layout of the analog output buffer is shown in~\Cref{fig:ch4-output_driver_layout}.
The total area occupation is \( \SI{35.6}{\micro\meter} \times \SI{17.7}{\micro\meter} \), corresponding to approximately \( \SI{360}{\micro\meter\squared} \).
The MOM capacitor \( C_b \) of \SI{340}{\femto\farad} was implemented using metal layers 1 through 7.

% -------------------------------------------------------------------------------
%   Si(Li) detector leakage current emulator
% -------------------------------------------------------------------------------
\section{Si(Li) detector leakage current emulator}\label{sec:ch4-leakage_current_emulator}

\begin{figure}[t!]
    \centering
    \includegraphics[width=0.85\textwidth]{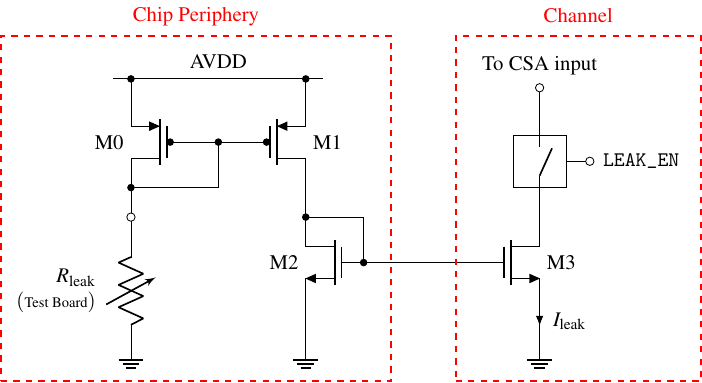}
    \caption{Schematic of the leakage current emulator. It consists of a PMOS current mirror (M0--M1) with a 50:2 ratio in the periphery, biased through the external resistor \(R_{\text{leak}}\), and an NMOS mirror in which M2 is implemented in the periphery and M3 is replicated in each channel with a 50:1 ratio to generate the programmable leakage current.}
    \label{fig:ch4-leakage_current_emulator_schematic}
\end{figure}

\begin{figure}[t!]
    \centering
    \includegraphics[width=0.75\textwidth]{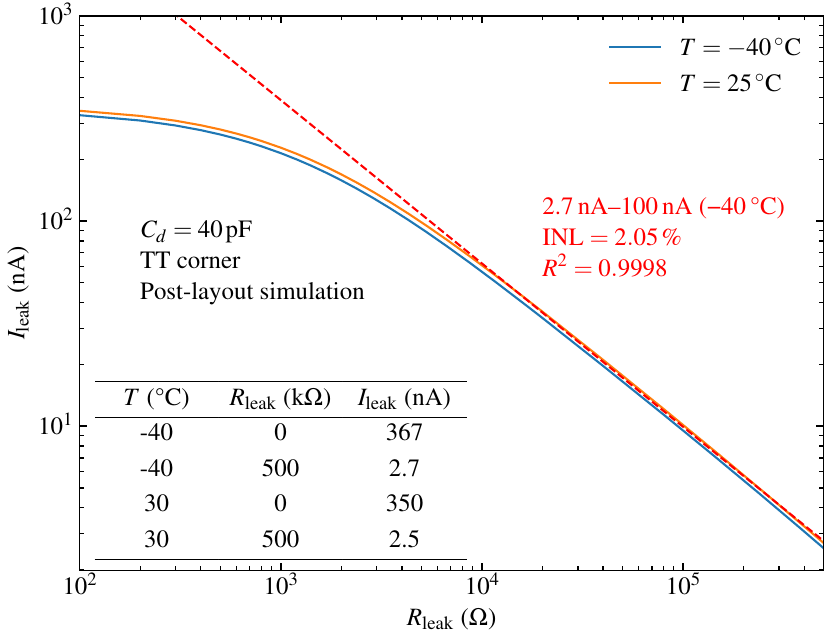}
    \caption{Post-layout simulation of the leakage current generated by the leakage current emulator as a function of the control resistance \( R_{\text{leak}} \), with \( C_d = \SI{40}{\pico\farad} \) in the typical process corner. Results are shown for \( \SI{-40}{\celsius} \) and \( \SI{25}{\celsius} \). At \( \SI{-40}{\celsius} \), the output current ranges from \SI{2.7}{\nano\ampere} to \SI{367}{\nano\ampere}, exhibiting linear behavior in the range from \SI{2.7}{\nano\ampere} to \SI{100}{\nano\ampere}, with an INL of \SI{2.05}{\percent} and an \( R^2 \) of 0.9998.}
    \label{fig:ch4-leakage_current_emulator}
\end{figure}

\par
Four of the eight channels implemented in the ASIC include a functional block designed to generate a programmable current at the input of the CSA to emulate the Si(Li) strip detector leakage current.  
The circuit is intended to reproduce the nominal leakage conditions expected during flight at \SI{-40}{\celsius}, typically between \SI{2.5}{\nano\ampere} and \SI{10}{\nano\ampere}~\cite{xiaoLargescaleDetectorTesting2023}.  
At the same temperature, it can also generate currents up to about \SI{100}{\nano\ampere}, while higher currents are mainly used for room-temperature characterization and worst-case studies.

\par
\Cref{fig:ch4-leakage_current_emulator_schematic} shows the leakage current emulator circuit.  
In the periphery, M0–M1 form a PMOS current mirror with a 50:2 ratio, with M0 biased through the external resistor \(R_{\text{leak}}\), implemented on the test board as a digitally controlled variable resistor.  
M2 provides the corresponding NMOS device in the periphery, and M3 is the channel-local NMOS device that mirrors this branch with a 50:1 ratio when \texttt{LEAK\_EN} is asserted.
To support the leakage current compensation capability of the newly implemented Krummenacher feedback network---capable of compensating up to \SI{200}{\nano\ampere} during \hbox{on-ground} testing at room temperature---the emulator circuit can supply up to approximately \SI{367}{\nano\ampere} at \SI{-40}{\celsius}, decreasing to approximately \SI{350}{\nano\ampere} at \SI{30}{\celsius}.  
This allows testing of the CSA compensation network beyond the maximum expected leakage current at room temperature, enabling the study of its behavior under worst-case conditions.

\begin{figure}[t!]
    \centering
    \includegraphics[width=\textwidth]{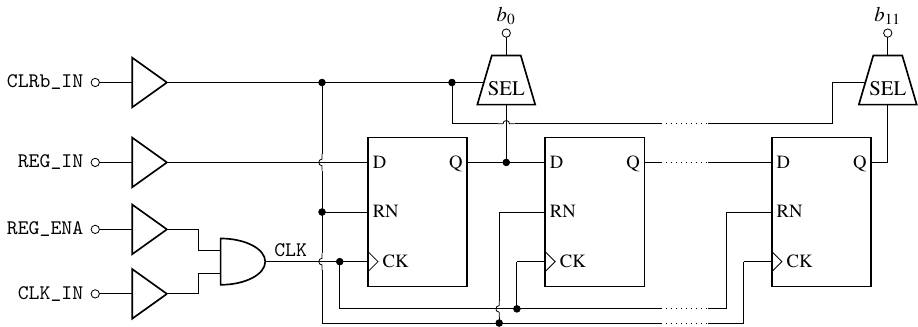}
    \caption{Diagram of the 12-bit slow-control shift register. The configuration word is shifted through a cascade of D flip-flops, each featuring a clock input, an asynchronous clear, and a registered output. The signals \texttt{REG\_IN}, \texttt{REG\_ENA}, \texttt{CLK\_IN}, and \texttt{CLRb\_IN} drive the register chain from the left, while the 12 stored bits ($b_{0}$--$b_{11}$) are available at the outputs.}
    \label{fig:ch4-shift_register_diagram}
\end{figure}

\par
To regulate the detector leakage current supplied at the input of the readout channel, each channel implementing this feature includes a current mirror that mirrors a current controlled by an external resistor \( R_{\text{leak}} \).  
\Cref{fig:ch4-leakage_current_emulator} shows the leakage current generated by the block as \( R_{\text{leak}} \) is varied from \SI{0}{\ohm} to \SI{500}{\kilo\ohm}.  
The output current spans from \SI{2.7}{\nano\ampere} to \SI{367}{\nano\ampere} at \SI{-40}{\celsius}, with a linear region between \SI{2.7}{\nano\ampere} and \SI{100}{\nano\ampere}, characterized by an INL of \SI{2.05}{\percent} and an \( R^2 \) of 0.9998, indicating good linearity in the lowest portion of the operating range.  
At \SI{30}{\celsius}, the current varies from \SI{2.5}{\nano\ampere} to \SI{350}{\nano\ampere}, following a similar monotonic behavior.  
The difference in current at the lowest value between the two temperatures is approximately \SI{7.4}{\percent}.  
This range was selected to ensure that the generated current could cover both the nominal low-current leakage levels of \SI{2.5}{\nano\ampere} and the extended operating region up to and beyond the \SI{200}{\nano\ampere} requirement.

%-------------------------------------------------------------------------------
%   Shift register for slow control
%-------------------------------------------------------------------------------
\section{Shift register for slow control}\label{sec:ch4-shift_register}

\par
In order to provide the necessary configuration bits to the charge injection circuit for calibration, the Si(Li) detector leakage current emulator and the shaping stage peaking time, a basic slow control system has been implemented using a 12-bit shift register.
The latter has been implemented as a cascade of 12 positive-edge-triggered D flip-flops.
Each flip-flop is built using a master-slave architecture based on two transmission gates controlled by complementary clock phases and two AND gates.
The first transmission gate samples the input data when the clock is high and propagates it to the intermediate node.
The second transmission gate transfers the stored value to the output when the clock is low.
The AND gates combine the data path with an asynchronous active-low reset signal to initialize or clear the register.
Data are shifted serially through the register on each rising clock edge, with the final state latched at the output of the last flip-flop.
The overall structure of the 12-bit shift register is shown in \Cref{fig:ch4-shift_register_diagram}.

\par
The input signals to the shift register are buffered via digital buffers implemented as a cascade of inverters to ensure logic level restoration and sufficient drive capability for the internal control lines.
The injection trigger signal \texttt{INJ} is provided independently via a dedicated pad, bypassing the shift register, and is buffered through a separate digital buffer.

\begin{figure}[t!]
    \centering
    \includegraphics[width=\textwidth]{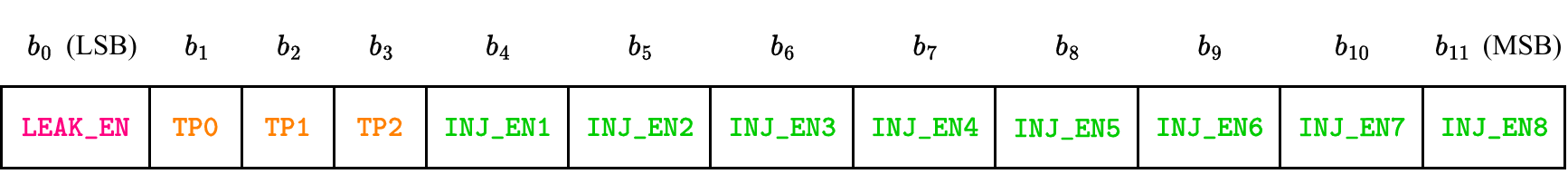}
    \caption{Bit assignment of the 12-bit configuration word shifted serially into the register via \texttt{REG\_IN}, mapping $b_{0}$ to \texttt{LEAK\_EN}, $b_{1}$--$b_{3}$ to the peaking-time selection (\hbox{\texttt{TP0}--\texttt{TP2}}), and $b_{4}$--$b_{11}$ to the eight injection-enable signals (\texttt{INJ\_EN1}--\texttt{INJ\_EN8}).}
    \label{fig:ch4-shift_register_bits}
\end{figure}

\begin{table}[t!]
    \centering
    \renewcommand{\arraystretch}{1.1}
    \setlength{\tabcolsep}{5pt}
    \begin{tabular}{cll}
        \toprule
        Type & Signal / Bit & Function \\ 
        \midrule
        \multirow{4}{*}{Input}
            & \texttt{CLRb\_IN}   & Asynchronous active-low reset \\   
            & \texttt{REG\_ENA}   & Enables data loading during the 12 clock cycles \\  
            & \texttt{CLK\_IN}    & Shift clock (data sampled on rising edge) \\   
            & \texttt{REG\_IN}    & Serial data input (LSB\,→\,MSB) for configuration word \\  
        \midrule
        \multirow{3}{*}{Output}
            & \texttt{LEAK\_EN}    & Enables or disables the leakage current generator \\   
            & \texttt{TP0--TP2}    & Select the shaping amplifier peaking time \\  
            & \texttt{INJ\_EN1--8} & Enable the injection switches for each channel \\  
        \bottomrule
    \end{tabular}
    \caption{Input signals and programmable output bits of the 12-bit shift register used for slow control.}
    \label{tab:ch4-shift_register_configuration}
\end{table}

\begin{figure}[t!]
    \centering
    \includegraphics[width=0.9\textwidth]{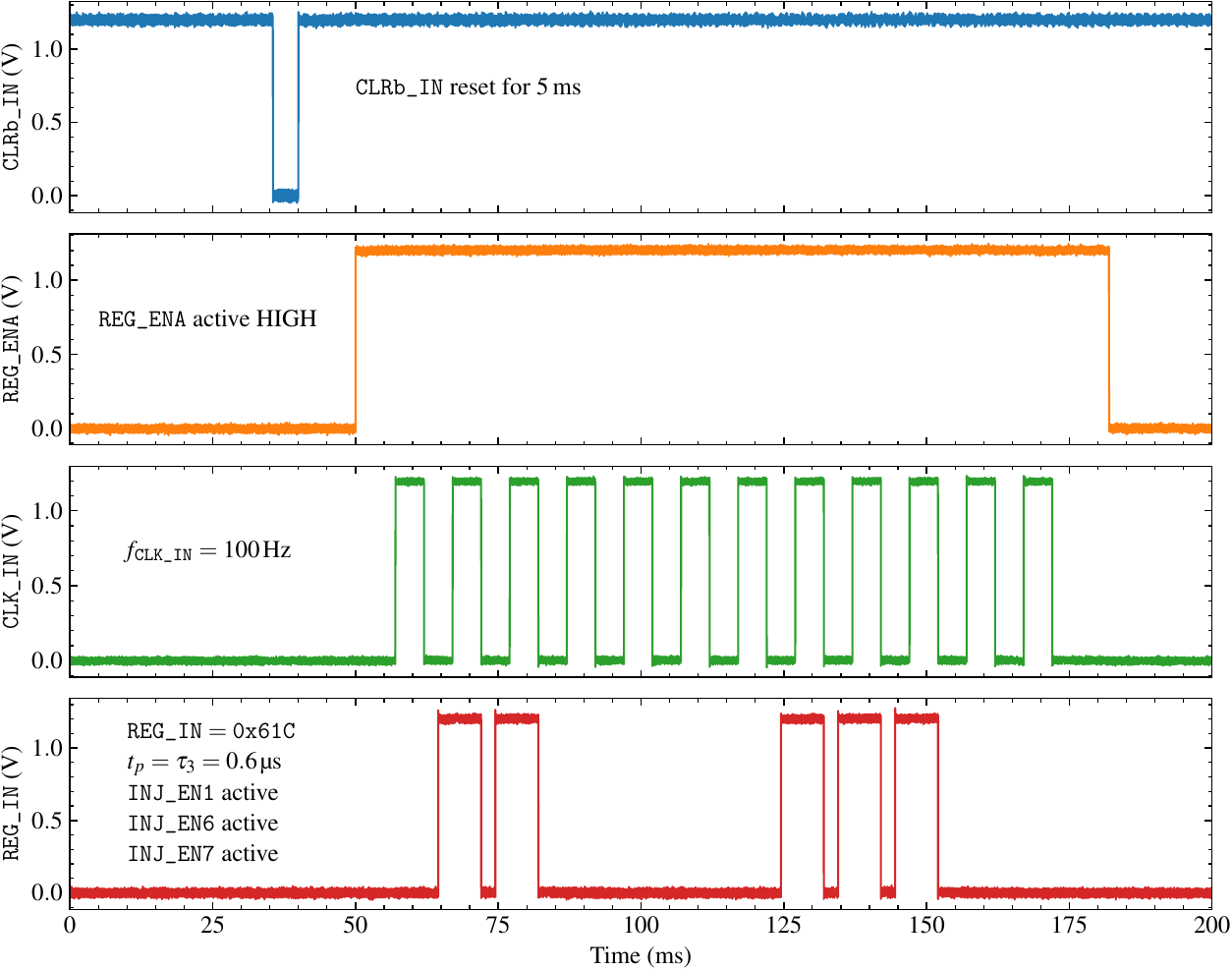}
    \caption{Measured waveforms of the four digital signals that control the configuration of the 12-bit shift register. The asynchronous clear signal \texttt{CLRb\_IN} is active LOW and is pulsed for \SI{5}{\milli\second} prior to configuration to reset the register. \texttt{REG\_ENA} is active HIGH and defines the write window, enabling the loading of data only during the 12 clock cycles. The \texttt{CLK\_IN} signal provides the synchronization clock at \SI{100}{\hertz}, with data sampled on each rising edge. The serial data input \texttt{REG\_IN} carries the 12-bit configuration word, which in this example corresponds to the hexadecimal value \texttt{0x61C}, enabling injection on channels 1, 6, and 7 and setting the peaking time to $\tau_3$ (\SI{0.6}{\micro\second}).}
    \label{fig:ch4-shift_register_signals}
\end{figure}

\begin{figure}[t!]
    \centering
    \includegraphics[width=0.69\textwidth]{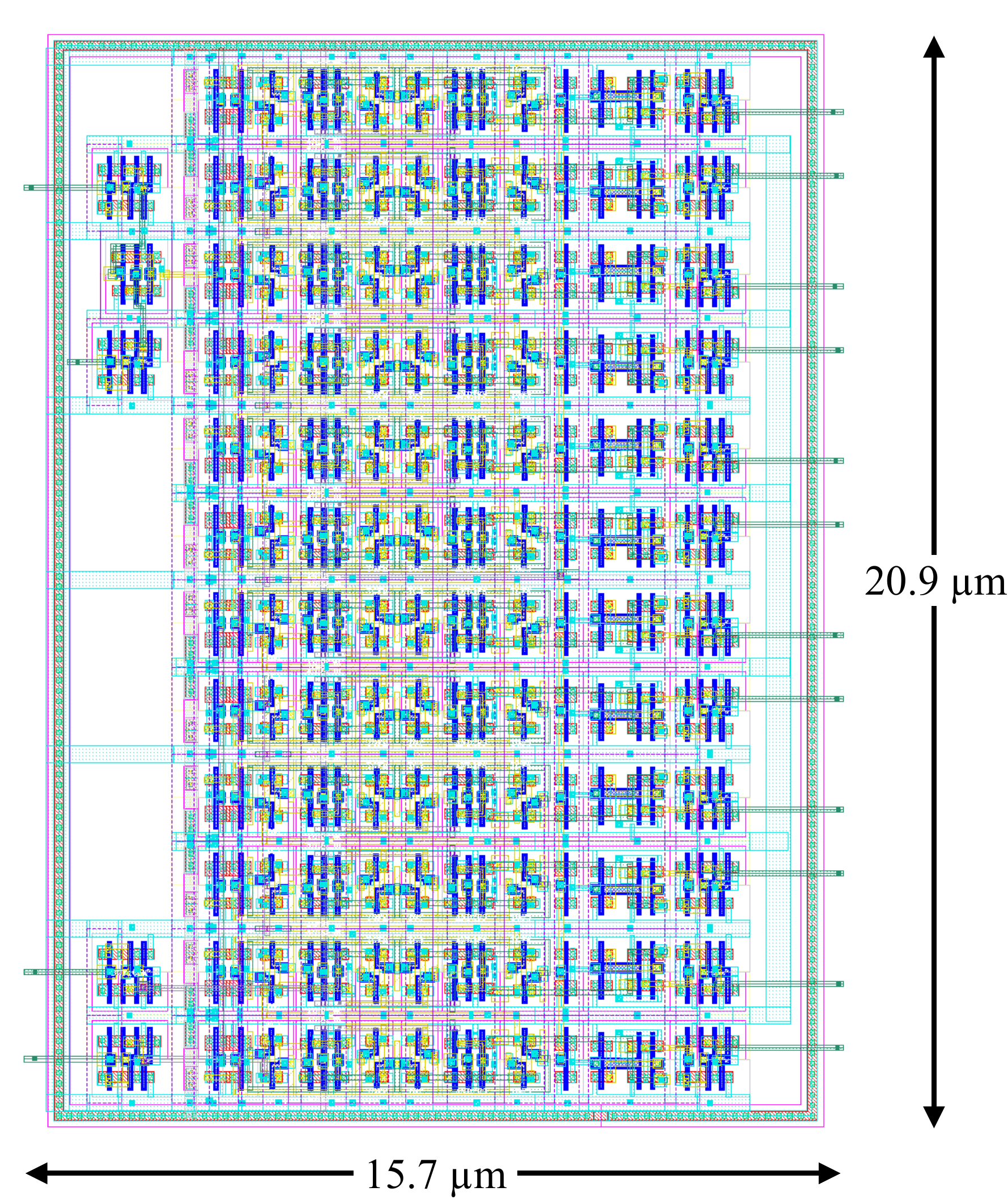}
    \caption{Layout of the 12-bit shift register used for slow control. The total area is \SI{15.7}{\micro\meter}~\(\times\)~\SI{20.9}{\micro\meter}. On the left, the input signals are provided (from top to bottom): \texttt{REG\_ENA} (write enable), \texttt{CLK\_IN} (clock), \texttt{CLRb\_IN} (clear), and \texttt{REG\_IN} (serial data input). On the right, the 12 output bits are routed: \texttt{LEAK\_EN} (leakage current generator enable), \texttt{TP0}--\texttt{TP2} (3-bit peaking time setting), and \texttt{INJ\_EN1}--\texttt{INJ\_EN8} (injection enable signals for the eight channels).}
    \label{fig:ch4-shift_register_layout}
\end{figure}

\par
Four input signals control the shift register: \texttt{REG\_ENA} (write enable), \texttt{CLK\_IN} (clock), \texttt{CLRb\_IN} (asynchronous clear), and \texttt{REG\_IN} (serial data input).
An example of the shift register configuration is presented in~\Cref{fig:ch4-shift_register_signals}, which shows the measured waveforms of these four signals.  
The asynchronous clear signal \texttt{CLRb\_IN} is active LOW and is pulsed for \SI{5}{\milli\second} prior to configuration to reset the register.  
The register enable \texttt{REG\_ENA} is active HIGH and defines the write window, enabling data loading only during the 12 clock cycles required for transmitting the 12-bit configuration word.  
The clock signal \texttt{CLK\_IN} operates at a frequency of \SI{100}{\hertz}, with data sampled on each rising edge.  
The serial data input \texttt{REG\_IN} carries the configuration word, transmitted from LSB to MSB in synchrony with the clock.  

\par
To ensure the correct bit value is present at the time of sampling, each bit is set on \texttt{REG\_IN} approximately one-quarter period before the corresponding rising clock edge, compensating for transmission delays and setup time.  
In the specific example shown, the configuration word is \texttt{0x61C}, which enables injection on channels 1, 6, and 7, sets the peaking time to \( \tau_3 \), and disables the leakage current generator.
The 12 output bits configure various blocks within the readout channel: \texttt{LEAK\_EN} enables the leakage current generator, \texttt{TP0} to \texttt{TP2} set the shaping amplifier peaking time, and \texttt{INJ\_EN1} to \texttt{INJ\_EN8} control the injection switches for the eight charge injection circuits.
The sequence of bits shifted serially into the register via \texttt{REG\_IN} and their assignment to the control signals is illustrated in \Cref{fig:ch4-shift_register_bits}. 
The role of the input control signals and the output configuration bits is summarized in \Cref{tab:ch4-shift_register_configuration}.

\begin{figure}[t!]
    \centering
    \includegraphics[width=\textwidth]{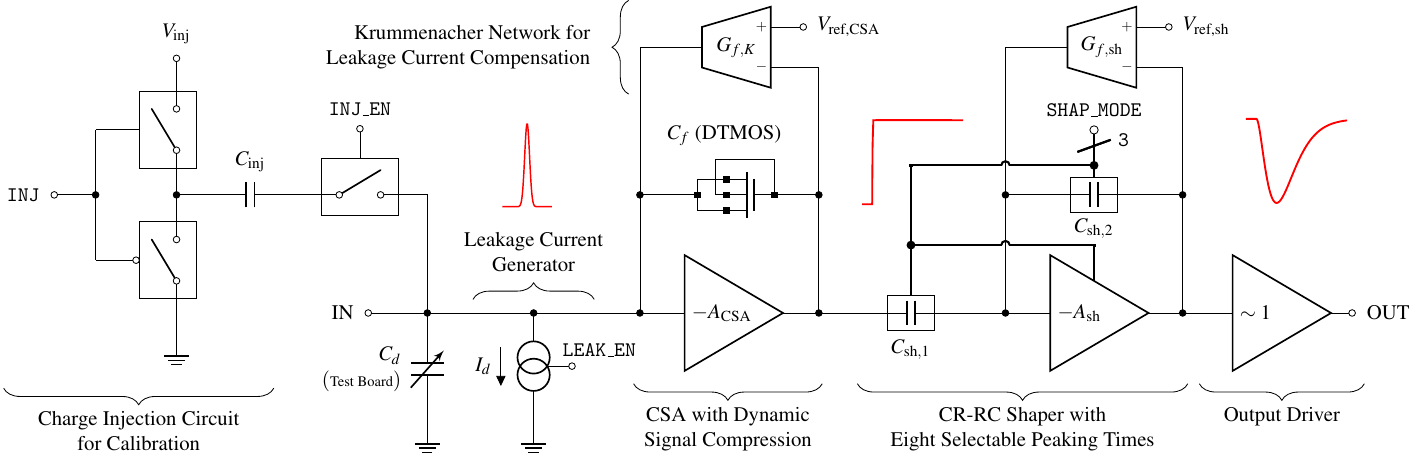}
    \caption{Complete functional diagram of the \antareslogo prototype readout channel, featuring a CSA with $n$-type DTMOS feedback, followed by a CR-RC shaper with eight selectable peaking times. The architecture also includes a charge injection circuit for calibration, a leakage current generator to emulate the Si(Li) strip detector leakage, and a Krummenacher network for charge restoration and leakage compensation.}
    \label{fig:ch4-complete_channel_diagram}
\end{figure}

\par
\Cref{fig:ch4-shift_register_layout} shows the layout capture of the shift register.
It occupies a total area of \( \SI{15.7}{\micro\meter} \times \SI{20.9}{\micro\meter} \) (\( \SI{\sim 328}{\micro\meter\squared} \)).
The input signals \texttt{REG\_ENA}, \texttt{CLK\_IN}, \texttt{CLRb\_IN}, and \texttt{REG\_IN} enter from the left, while the 12-bit control signals exit on the right and are subsequently routed to the corresponding blocks within the ASIC.
In the default configuration---i.e., with no inputs applied to the shift register---the eight charge injection enable signals \texttt{INJ\_EN1} to \texttt{INJ\_EN8} are all set high, the leakage current generator (\texttt{LEAK\_EN}) is active, and the shaping stage peaking time is set to \( \tau_5 \), corresponding to the nominal \SI{1}{\micro\second} (\SI{1.10}{\micro\second} from simulation), which was selected based on the optimal noise performance discussed in~\Cref{sec:ch5-shaper_noise_performance}.

%-------------------------------------------------------------------------------
%   Complete channel and ASIC layout
%-------------------------------------------------------------------------------
\section{Complete channel and ASIC layout} \label{sec:ch4-complete_channel_layout}

\par
\begin{figure}[t!]
    \centering
    \includegraphics[width=\textwidth]{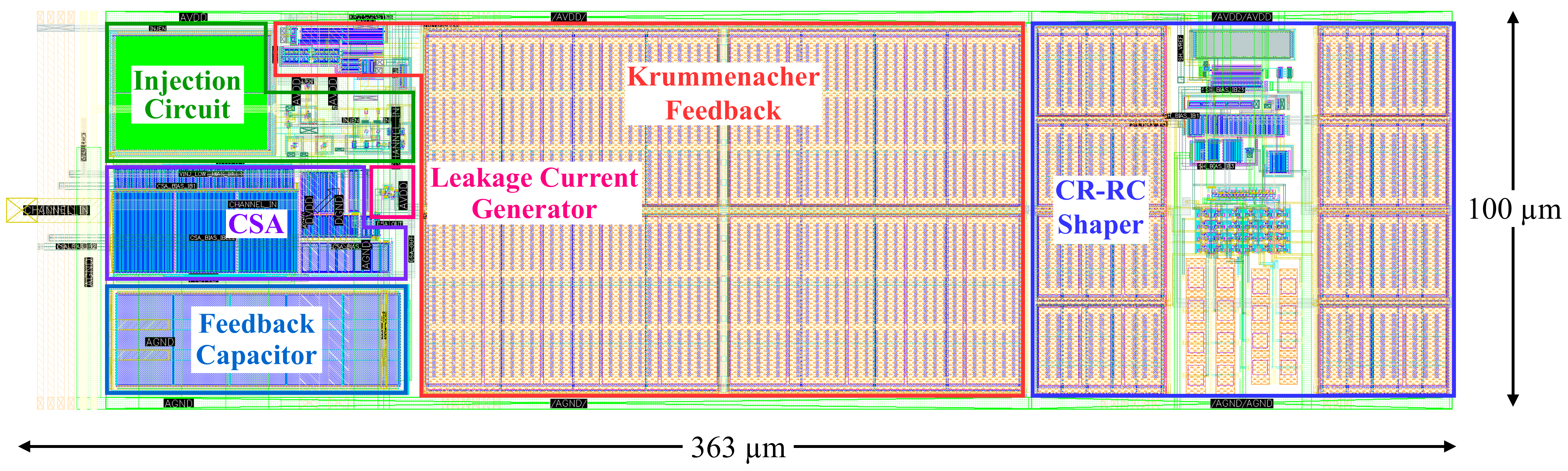}
    \caption{Layout view of the complete \antareslogo readout channel. The layout includes the charge injection circuit, leakage current generator, CSA with dynamic signal compression ($n$-type DTMOS feedback variant), Krummenacher network and CR-RC shaper. It fits within the \SI{100}{\micro\meter} channel height constraint, with a total width of \SI{363}{\micro\meter}.}
    \label{fig:ch4-complete_channel_layout}
\end{figure}

\Cref{fig:ch4-complete_channel_diagram} shows the complete readout channel diagram with all its functional blocks.  
The specific implementation depicted corresponds to the channel featuring an $n$-type DTMOS feedback capacitor in the CSA, with the shaper output selected for readout and thus routed to the corresponding ASIC pad.  
As a result, the analog output driver discussed in~\Cref{ch4:sec-analog_outoput_buffer} is placed at the output of the shaper in this configuration.  
The complete layout view of this implementation is shown in~\Cref{fig:ch4-complete_channel_layout}.

\par
\Cref{tab:ch4-antares4_channels_info} summarizes the characteristics of all eight analog channels implemented in the ASIC, indicating for each of them the type of CSA feedback device, the presence of the leakage current generator and the selected output stage (CSA or shaper).  
The selected combinations of CSA feedback device, output, and leakage current emulator have been chosen to enable a direct comparison between the reference $n$-type DTMOS and its conventional MOS variant, both with and without the leakage current emulator, and at both the CSA and shaper output levels.  
The $p$-type MOS and the associated DTMOS topology have been included for comparison, but without leakage current emulator and at CSA level only, as these configurations were not found to be optimal, though still valuable for reference.

\begin{table}[t!]
    \centering
    \renewcommand{\arraystretch}{1.1}
    \setlength{\tabcolsep}{5pt}
    \begin{tabular}{cccc}
        \toprule
        Channel & CSA feedback device & Leakage current generator & Output \\
        \midrule
        1 & $n$-type MOS   & \xmark & CSA    \\
        2 & $n$-type DTMOS & \xmark & CSA    \\
        3 & $n$-type MOS   & \cmark & Shaper \\
        4 & $n$-type DTMOS & \cmark & Shaper \\
        5 & $p$-type MOS   & \xmark & CSA    \\
        6 & $p$-type DTMOS & \xmark & CSA    \\
        7 & $n$-type MOS   & \cmark & CSA    \\
        8 & $n$-type DTMOS & \cmark & CSA    \\
        \bottomrule
    \end{tabular}
    \caption{Summary of the key characteristics of the eight analog channels implemented in the ASIC. For each channel, the type of MOS transistor used to implement the CSA feedback capacitance, the presence of the leakage current emulator circuit, and the output tap point (CSA or shaper) are indicated.}
    \label{tab:ch4-antares4_channels_info}
\end{table}

\par
The specific channel implementation shown in the diagram is representative of the general structure of the readout channel, which, in its various configurations, may include the leakage current generator and an output buffer placed either at the shaper output (as in this case) or at the CSA output.  
All readout channels are based on a CSA that implements dynamic signal compression through a nonlinear capacitor realized using a MOS device.  
This capacitor may take the form of a conventional MOS capacitance, with the bulk tied to ground (NMOS) or to \( V_{\text{DD}} \) (PMOS), or a DTMOS capacitance, in which the bulk is shorted to the gate, resulting in a reduced threshold voltage as \( V_{\text{GS},f} \) increases.  
In the CSA feedback path, all channels incorporate a Krummenacher network, which performs both charge restoration in the feedback capacitor \( C_f \) and detector leakage current compensation, maintaining good performance over the entire expected leakage range up to \SI{200}{\nano\ampere} at room temperature for on-ground calibration.  
The CSA is followed by a CR-RC shaping stage designed to optimize the signal-to-noise ratio.  
This stage provides eight selectable peaking times, ranging from \SI{0.2}{\micro\second} to \SI{1.6}{\micro\second}.
Slow control is implemented through a 12-bit shift register, described in~\Cref{sec:ch4-shift_register}, which allows configuration of the channel's adjustable blocks, including the leakage current emulator, the charge injection circuit, and the peaking time selection of the shaping stage.

\par
\begin{figure}[t!]
    \centering
    \includegraphics[width=0.93\textwidth]{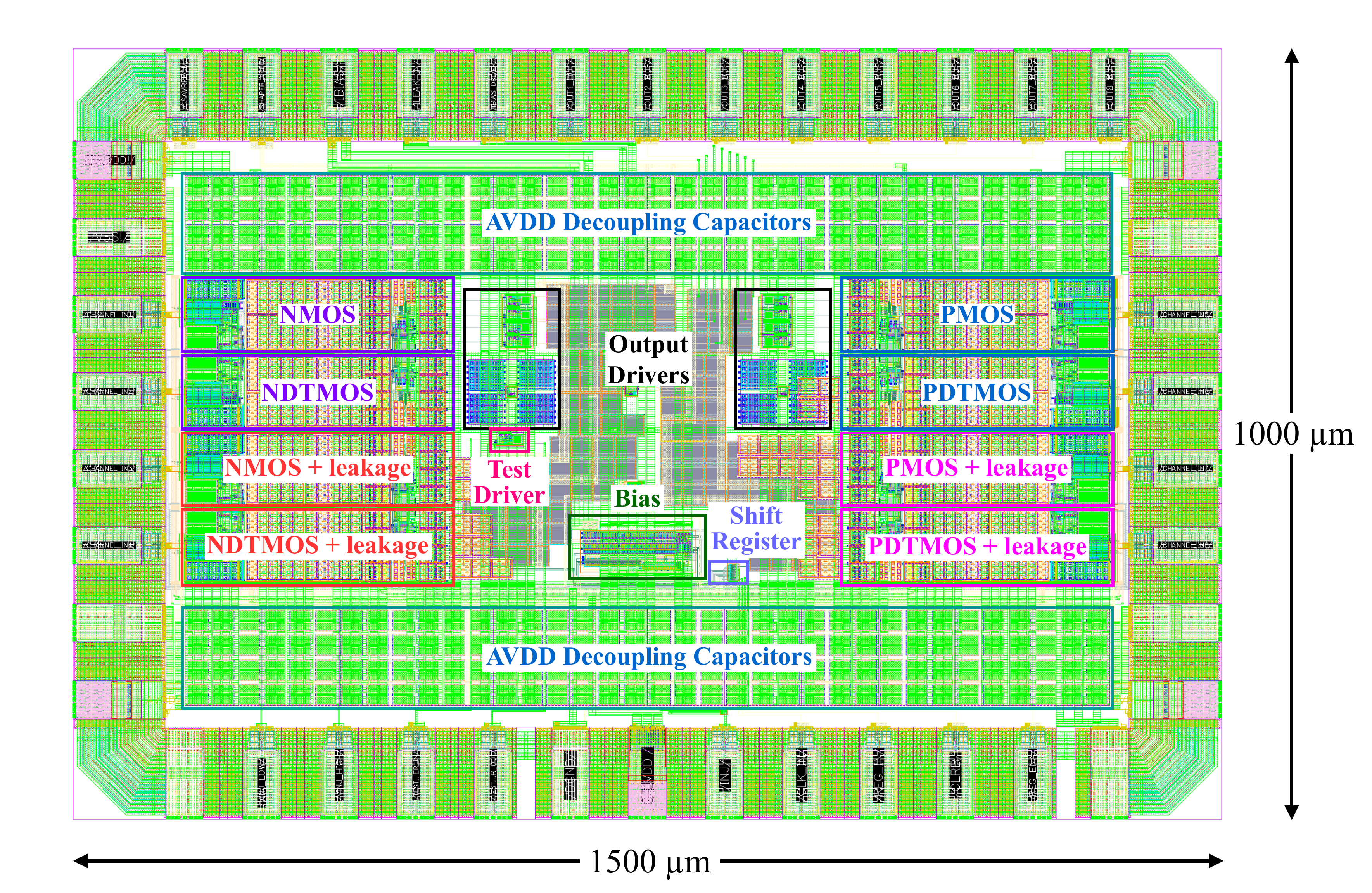}
    \caption{Layout view of the complete \antareslogo prototype ASIC. The chip integrates eight readout channels, each labeled with its specific CSA feedback capacitor type. Four channels include the leakage current generator. The layout also features eight output buffers (CSA or shaper outputs), a bias generation block, and a shift register for configuration. At the top and bottom edges, MOM capacitor arrays provide approximately \SI{870}{\pico\farad} of AVDD filtering. Additional MOM arrays of approximately \SI{80}{\pico\farad} are included for filtering the CSA baseline and shaper voltage references. Dummy metal fill is omitted in this layout view.}
    \label{fig:ch4-complete_asic_layout}
\end{figure}

The final ASIC layout, shown in~\Cref{fig:ch4-complete_asic_layout}, includes a total of eight readout channels, four of which integrate the leakage current emulator circuit described in~\Cref{sec:ch4-leakage_current_emulator}.  
The eight channels implement all four configurations of the feedback capacitance: standard MOS and DTMOS types, each realized using both NMOS and PMOS transistors.  
Each configuration is implemented in two versions, one with the leakage current generator and one without.

\par
The overall ASIC occupies an area of \( \SI{1500}{\micro\meter} \times \SI{1000}{\micro\meter} \), corresponding to a total area of \( \SI{1.5}{\milli\meter\squared} \), and was submitted in a \SI{65}{\nano\meter} CMOS technology using the MiniASIC \hbox{multi-project} format, which provides a fixed padframe and shared reticle for prototype designs.
According to the foundry design rules for MiniASIC submissions, the horizontal dimension must exceed \SI{1000}{\micro\meter}, while the vertical dimension must be set to one of the allowed values: \SI{1000}{\micro\meter}, \SI{1500}{\micro\meter}, or \SI{2000}{\micro\meter}.  
The selected dimensions of \hbox{\( \SI{1500}{\micro\meter} \times \SI{1000}{\micro\meter} \)} comply with these constraints and were chosen to accommodate a padring with 42 pads, required to read out all analog signals during ASIC testing. In particular, the vertical dimension was minimized to reduce the distance between the channel input pads and the corresponding CSA input, thereby minimizing routing length and parasitic effects.

\par
The layout placement of the readout channels and associated functional blocks follows this criterion: the eight readout channels are placed vertically, centered with respect to the ASIC area limits, and horizontally aligned with the vertical pads.
The peripheral circuits are arranged in the central region of the ASIC. The output drivers are placed symmetrically at the center to ensure the minimum track distance from either the CSA or the shaper output of the corresponding left or right channels.
The buffer voltage references and bias currents are generated by a dedicated bias block, centrally located between the two 4-buffer arrays.
The additional test buffer, included for noise and overall performance characterization, is placed below the left buffer array.

\begin{figure}[t!]
    \centering
    \includegraphics[width=0.9\textwidth]{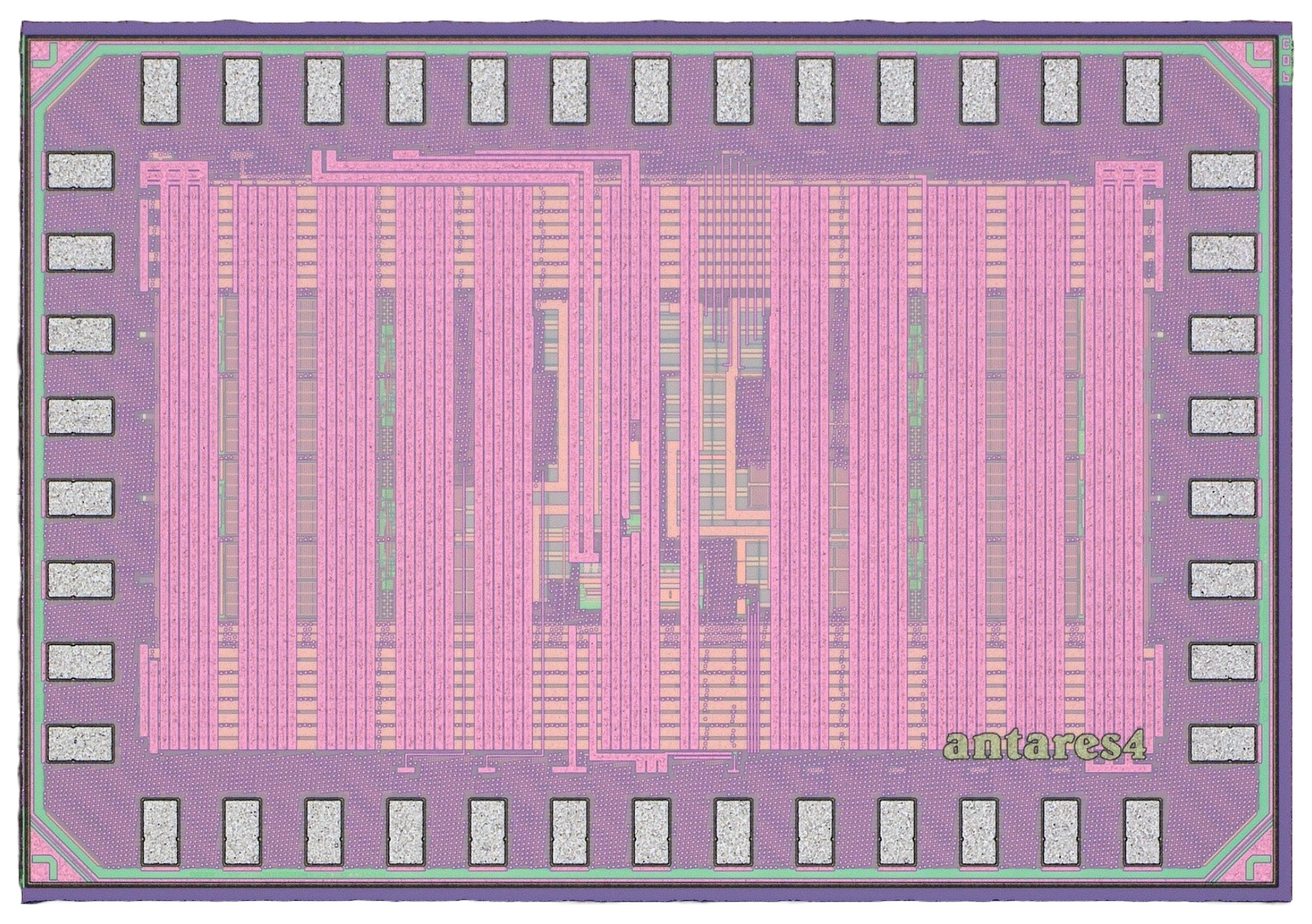}
    \caption{Photograph of the \antareslogo ASIC die (\texttt{antares4\_001}), acquired using a Keyence VHX-X1 digital microscope. The die measures \( \SI{1.5}{\milli\meter} \times \SI{1.0}{\milli\meter} \).}
    \label{fig:ch4-asic_picture_nobg}
\end{figure}

\par
The peripheral bias block, located at the center of the ASIC, generates all necessary bias voltages for both the CSA and shaping stage. The main bias current, mirrored across all circuits inside the readout channels, has a nominal value of \SI{5}{\milli\ampere} at \SI{-40}{\celsius} and is supplied externally through a dedicated pad in the top section of the ASIC. This current is generated by a two-terminal \SI{200}{\milli\ampere} programmable floating current source (Analog Devices LT3092) mounted on the test board.
The 12-bit shift register for slow control is placed in the lower central section of the ASIC, near the lower-left part of the padring, where the corresponding digital signals are provided. The analog power supply (AVDD and AGND) is connected to the four corner pads of the ASIC, each tied to an array of MOM capacitors designed in a modular pattern of \SI{26}{\pico\farad} units, filling all available space in the top and bottom parts of the layout and implementing a total decoupling capacitance of approximately \SI{870}{\pico\farad}. The digital power supply (DVDD and DGND), the CSA reference voltage \( V_{\text{ref,CSA}} \), and the shaper reference voltage \( V_{\text{ref,sh}} \) are also decoupled by MOM capacitors, each providing approximately \SI{80}{\pico\farad}, and are located in the central part of the layout.

\begin{figure}[t!]
    \centering
    \captionsetup[subfigure]{justification=justified,singlelinecheck=false}
    \begin{subfigure}{0.49\textwidth}
        \centering
        \includegraphics[width=.8\linewidth]{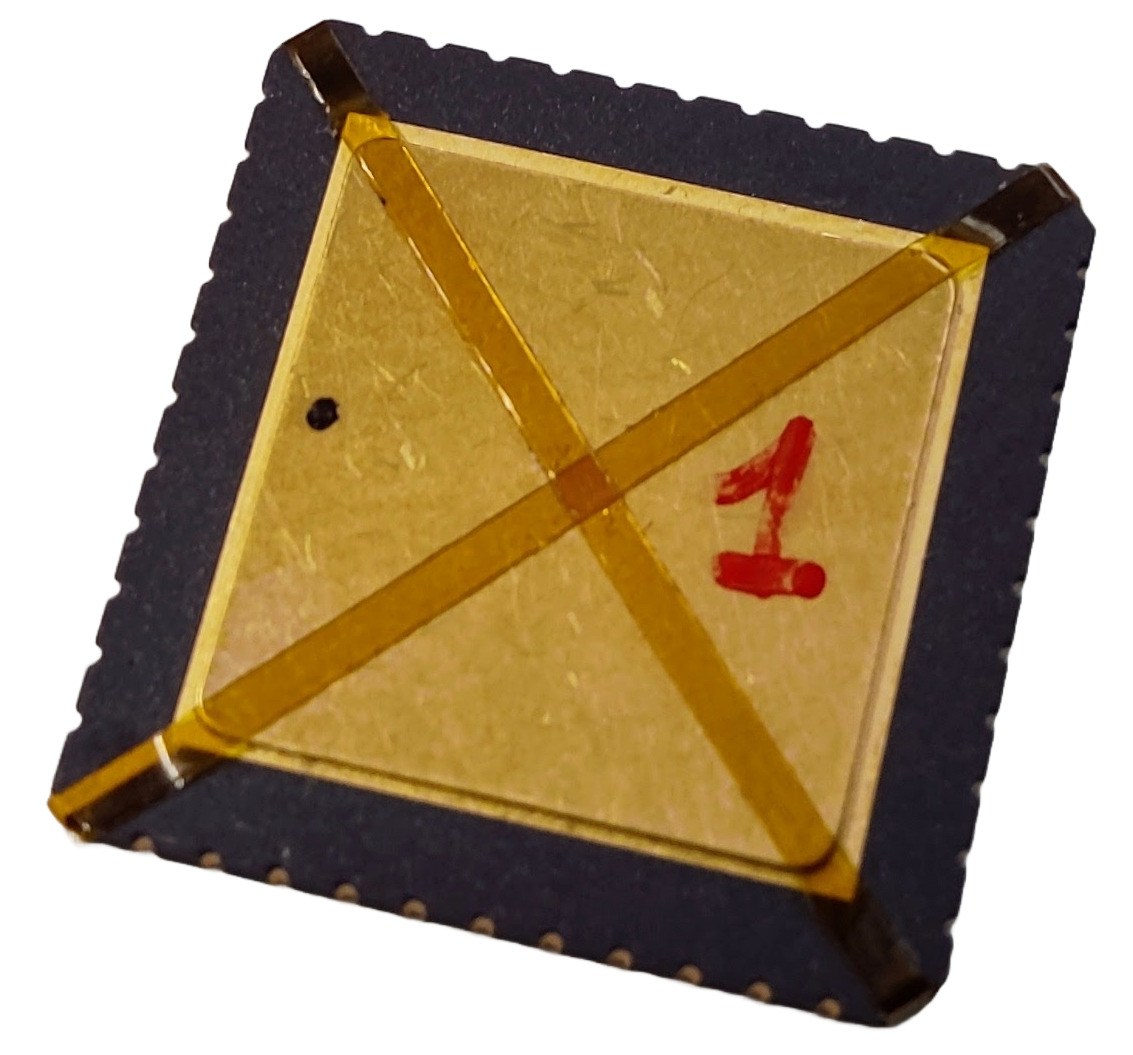}
        \caption{External view of the CLCC44 package of the \texttt{antares4\_001} ASIC with the lid closed, showing the protective metal lid and package enclosure.}
        \label{fig:ch4-asic_package_close}
    \end{subfigure}
    \hfill
    \begin{subfigure}{0.49\textwidth}
        \centering
        \includegraphics[width=.8\linewidth]{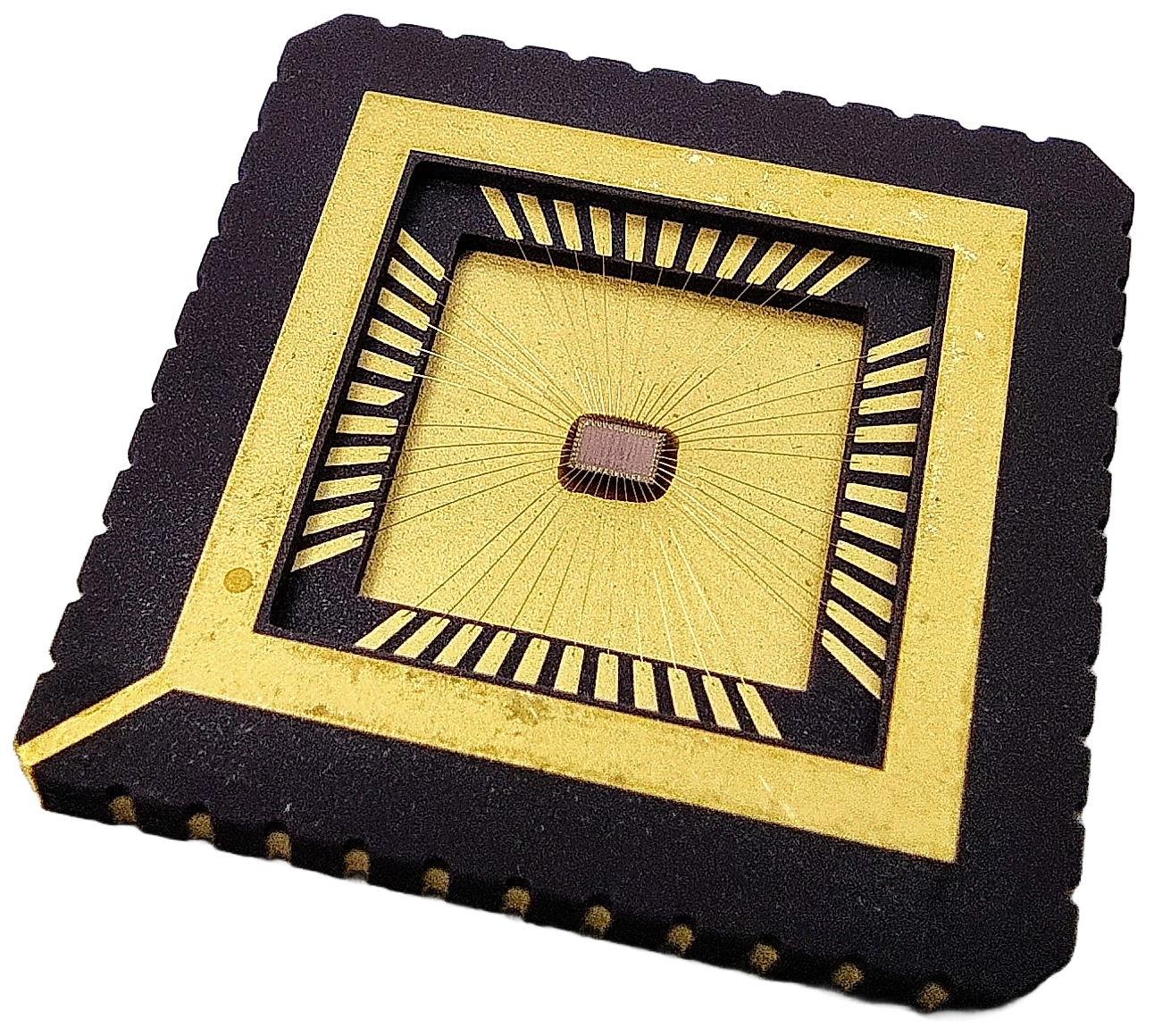}
        \caption{Internal view of the same package with the lid removed, revealing the ASIC die glued to the cavity and connected to the pads through wire bonds.}
        \label{fig:ch4-asic_package_open}
    \end{subfigure}
    \caption{Photographs of the \antareslogo ASIC (\texttt{antares4\_001}) in its CLCC44 package. (\subref{fig:ch4-asic_package_close}) External view with the package closed. (\subref{fig:ch4-asic_package_open}) Internal view with the die visible and bonded to the pins.}
    \label{fig:ch4-asic_package}
\end{figure}

\par
The MiniASIC submission was completed on March 3, 2025, with ASICs delivered in late July 2025. 
A total of \num{100} ASICs have been produced, of which \num{10} have been installed in a ceramic leadless chip carrier with \num{44} pins (CLCC44) package.
\Cref{fig:ch4-asic_picture_nobg} shows a photograph of the \antareslogo ASIC die, acquired using a Keyence VHX-X1 digital microscope, while 
\Cref{fig:ch4-asic_package} shows the \texttt{antares4\_001} device in its CLCC44 package, in both closed (\Cref{fig:ch4-asic_package_close}) and opened (\Cref{fig:ch4-asic_package_open}) views.

\par
The CLCC44 is a square ceramic package with \num{44} I/O pads (\num{11} per side) and a protective metal lid. 
For this chip, the lid was made removable and fixed with Kapton tape, in order to allow inspection and evaluation of the internal die and bonding.
The package body measures approximately \SI{17.8}{\milli\meter} per side, with a pad pitch of \SI{1.27}{\milli\meter}. 
This package was selected based on the ASIC's overall footprint, the required number of I/O connections, and its compatibility with the S-LCC-00-044-B socket used on the test board. 
This socket allows for repeated insertion and removal of ASICs, enabling the testing of multiple devices without the need for permanent installation on carrier boards, which are less suitable for quick-swapping during characterization.

%% file: maintext/chapters/chapter5.tex
\chapter{\texorpdfstring
  {Characterization of the \includegraphics[height=1.5ex]{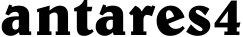} readout ASIC}
  {Characterization of the ANTARES4 readout ASIC}
}\label{chapter5}

\par
This chapter presents the characterization of the \antareslogo ASIC, summarizing the key results obtained during the initial test campaign conducted using a custom-designed test board.
This first assessment of the ASIC's main performance metrics is essential to verify correct functionality and to compare the measured behavior against the simulation results discussed in~\Cref{chapter4}. % previously

\par
The chapter is organized as follows: the first section describes the test setup, detailing all the instrumentation used for measurements at both room temperature and \SI{-40}{\celsius}, the latter achieved using a climate chamber.
Next, the custom test board is introduced, along with the supporting electronics designed to generate the control signals, bias currents, and reference voltages required for proper ASIC operation.
Finally, the results of the characterization measurements are discussed and compared with the corresponding simulation data.
These results include the transient response at the output of the CSA and shaper stages, the energy-to-voltage transcharacteristic, analyzed with a focus on gain, linearity, and dynamic signal compression metrics, the effect of shaping time on the signal response, and the impact of leakage current on both transient response and transcharacteristic, followed by an initial noise assessment.

%-------------------------------------------------------------------------------
%   Test setup
%-------------------------------------------------------------------------------
\section{Test setup} \label{sec:ch5-test_setup}

\par
\begin{figure}[tbp]
    \centering
    \includegraphics[width=\textwidth]{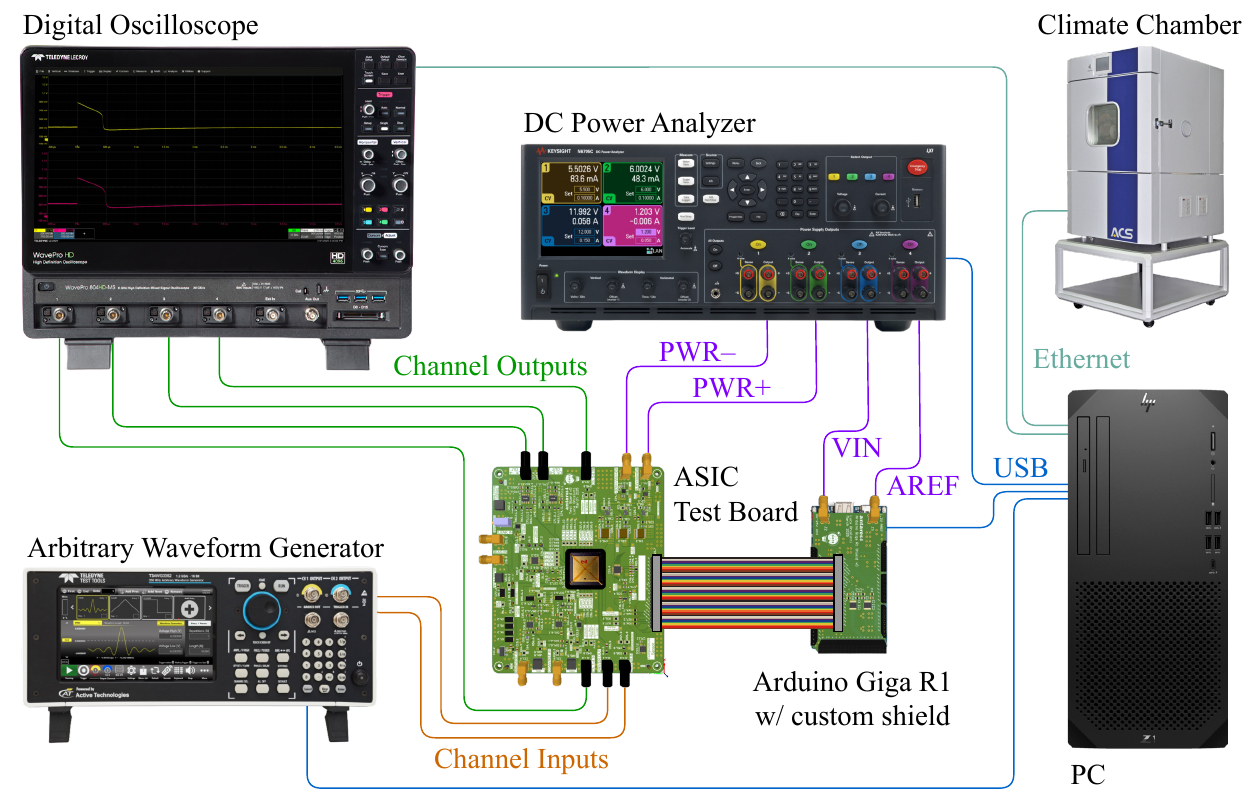}
    \caption{Block diagram of the test setup used for the initial characterization of the \antareslogo{} ASIC. The setup is centered around a custom-designed test board controlled by an Arduino Giga R1 equipped with a dedicated shield for generating the required control signals. A DC power analyzer provides power to both the board and the shield. The ASIC outputs are monitored using a digital oscilloscope. To enable \hbox{temperature-controlled} measurements, including operation at the target flight temperature of \SI{-40}{\celsius}, the entire setup is placed inside a climate chamber. A PC running a dedicated Python script manages the setup and automates data acquisition.}
    \label{fig:ch5-antares4_test_setup}
\end{figure}

In order to perform the characterization of the chip, a test setup was built to validate the performance metrics of the electronics and ensure consistency with specifications.
The setup is based on a custom-designed test board equipped with a socket that allows for quick and easy swapping between different ASICs under test.
The test board provides the required analog and digital power supplies, calibration voltages, and the necessary voltage and current biases, as well as the control signals for ASIC configuration.
An overview of the setup is shown in the block diagram of \Cref{fig:ch5-antares4_test_setup}.

\par
The digital control signals are delivered via a custom-designed shield board developed for the Arduino Giga R1 Wi-Fi, a development board based on the STM32H7 microcontroller, featuring dual-core Arm Cortex-M7 and Cortex-M4 32-bit processors.
In this setup, the Arduino acts as a digital signal processor (DSP), managing the digital signals required for configuring the ASIC's slow control interface and the on-chip injection circuit.
This choice was made considering the complexity of the test setup and the limited number of digital signals to be managed, which, for this prototype ASIC, are restricted to controlling the on-chip leakage current emulator, the charge injection circuit for calibration, and shaper peaking time selection.
The connection between the Arduino shield and the test board is established using a \SI{50}{\centi\meter}-long, 40-wire flat ribbon cable.

\par
Since the prototype chip lacks on-chip analog-to-digital conversion, signal acquisition at the output of both the CSA and shaper blocks---within the eight readout channels and the test buffer---is carried out using a Teledyne LeCroy WavePro 804HD digital oscilloscope via LEMO connectors on the test board and coaxial cables to the oscilloscope.
The PC, running Pop!\_OS 22.04 LTS, serves as the main control unit of the test setup, sending commands to the Arduino Giga, acquiring data from the oscilloscope, and managing the benchtop power supply, waveform generator, and climate chamber.

\par
The PC runs a dedicated Python test program~\cite{ghislottiLucaghisloANTARES4ASICTestBench2025} that controls all the instruments in the setup as well as the climate chamber.
Power is supplied by a Keysight N6705C DC power analyzer, a modular \SI{600}{\watt}, \SI{4}{channel} programmable power supply with logging capabilities.
An arbitrary waveform generator, the Teledyne LeCroy T3AWG3352 (2 channels, \SI{350}{\mega\hertz}, \SI{16}{\bit}), is included in the setup to provide additional charge calibration capability in combination with the on-chip charge injection circuit. 
However, this instrument was not employed during the characterization campaign reported in this thesis.

\par
\begin{figure}[tbp]
    \centering
    \captionsetup[subfigure]{justification=justified,singlelinecheck=false}
    \begin{subfigure}{0.49\textwidth}
        \centering
        \includegraphics[width=\linewidth]{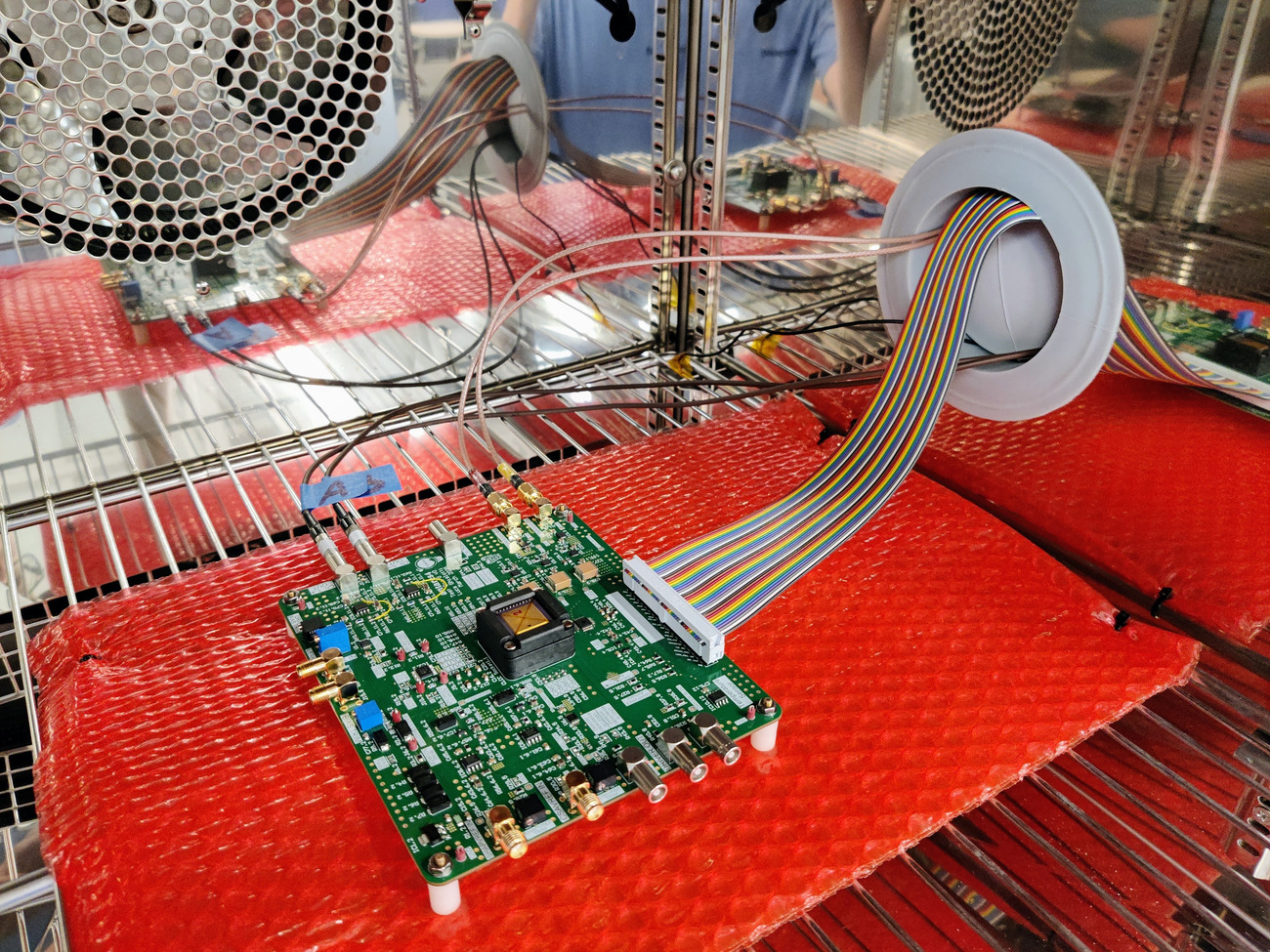}
        \caption{ASIC test board inside the climate chamber, with coaxial cables routed via LEMO and SMA connectors and the \hbox{flat-ribbon} cable connected to the Arduino Giga shield.}
        \label{fig:ch5-test_setup-board}
    \end{subfigure}
    \hfill
    \begin{subfigure}{0.49\textwidth}
        \centering
        \includegraphics[width=\linewidth]{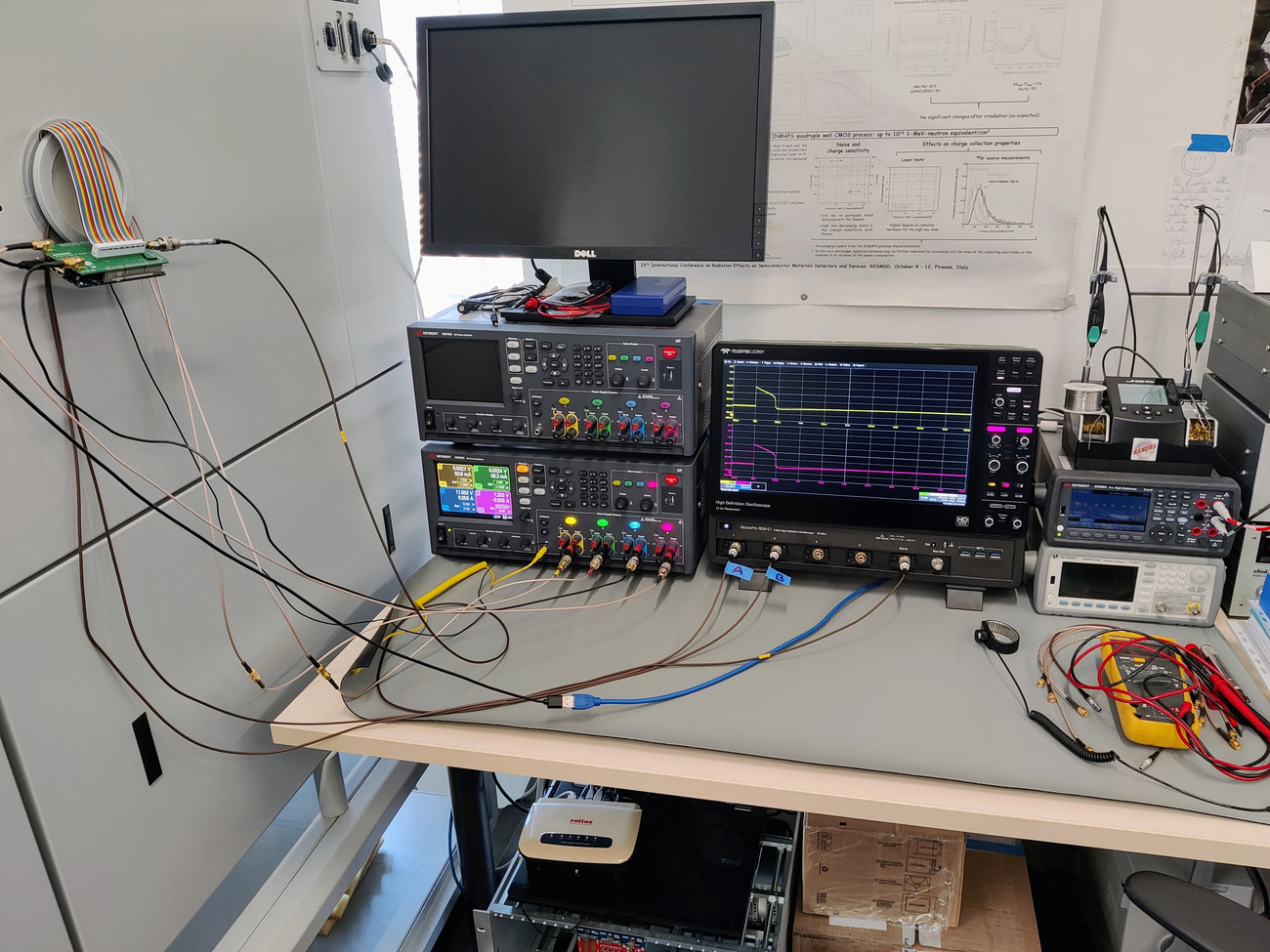}
        \caption{Measurement instruments with the DC power analyzer on the left and the oscilloscope on the right, with the Arduino Giga and its custom shield placed outside the chamber.}
        \label{fig:ch5-test_setup-instruments}
    \end{subfigure}
    \caption{Photographs of the laboratory test setup used for the characterization of the \antareslogo ASIC, showing (\subref{fig:ch5-test_setup-board}) the test board inside the climate chamber and (\subref{fig:ch5-test_setup-instruments}) the external instrumentation.}
    \label{fig:ch5-test_setup}
\end{figure}

The setup can be operated both at room temperature and inside a climate chamber (ACS DY110), which supports a temperature range of \SIrange{-40}{180}{\celsius}, enabling \hbox{temperature-dependent} characterization and operation of the chip at the target flight temperature of \SI{-40}{\celsius}.
The climate chamber can be controlled directly by the testbench through a custom Python driver, which allows setting temperature and humidity setpoints, turning the chamber on and off, and reading the measured temperature and humidity in real time.
This enables fully automated testing of the chip under varying thermal conditions.
Photographs of the setup are provided in \Cref{fig:ch5-test_setup}, showing the test board inside the climate chamber (\Cref{fig:ch5-test_setup-board}) and the external instrumentation (\Cref{fig:ch5-test_setup-instruments}).

%-------------------------------------------------------------------------------
%   ASIC test board
%-------------------------------------------------------------------------------
\section{ASIC test board}

\par
The ASIC test PCB was designed using a 4-layer stack-up fabricated on standard FR-4 material and has dimensions of \SI{148}{\milli\meter}~\(\times\)~\SI{124}{\milli\meter}.
A plastic spring-action socket (Emulation Technology S-LCC-00-044-B) is mounted at the center of the board to accommodate the CLCC44 package.
The choice of using a socket instead of a daughterboard-motherboard solution was driven by the need to quickly swap ASICs during repeated characterization cycles.
Socket-based replacement avoids the mechanical wear and potential failure associated with male-female pin connectors, which are prone to degradation over time and may lead to premature contact issues.
Using one dedicated test board per ASIC was also ruled out due to the cost and complexity of replicating the full board design for each device.
A possible drawback of the socket solution, however, is the risk of unreliable contact, which requires careful attention during chip insertion.

\par
All components selected for the design have a temperature rating of at least \SIrange{-40}{70}{\celsius} or better.
All passive components were selected in the 0805 package.
General purpose and decoupling capacitors employ X7R dielectric, while resistors were chosen with a tolerance not exceeding \SI{5}{\percent} and a temperature coefficient (TC) no greater than \SI{100}{ppm/\celsius}.
Critical components, in terms of precision and temperature stability, were selected with the highest available accuracy and thermal performance, including C0G (NP0) dielectric capacitors~\cite{keithleyinstrumentsinc.LowLevelMeasurements2016}.

\begin{figure}[tbp]
    \centering
    \includegraphics[width=.85\textwidth]{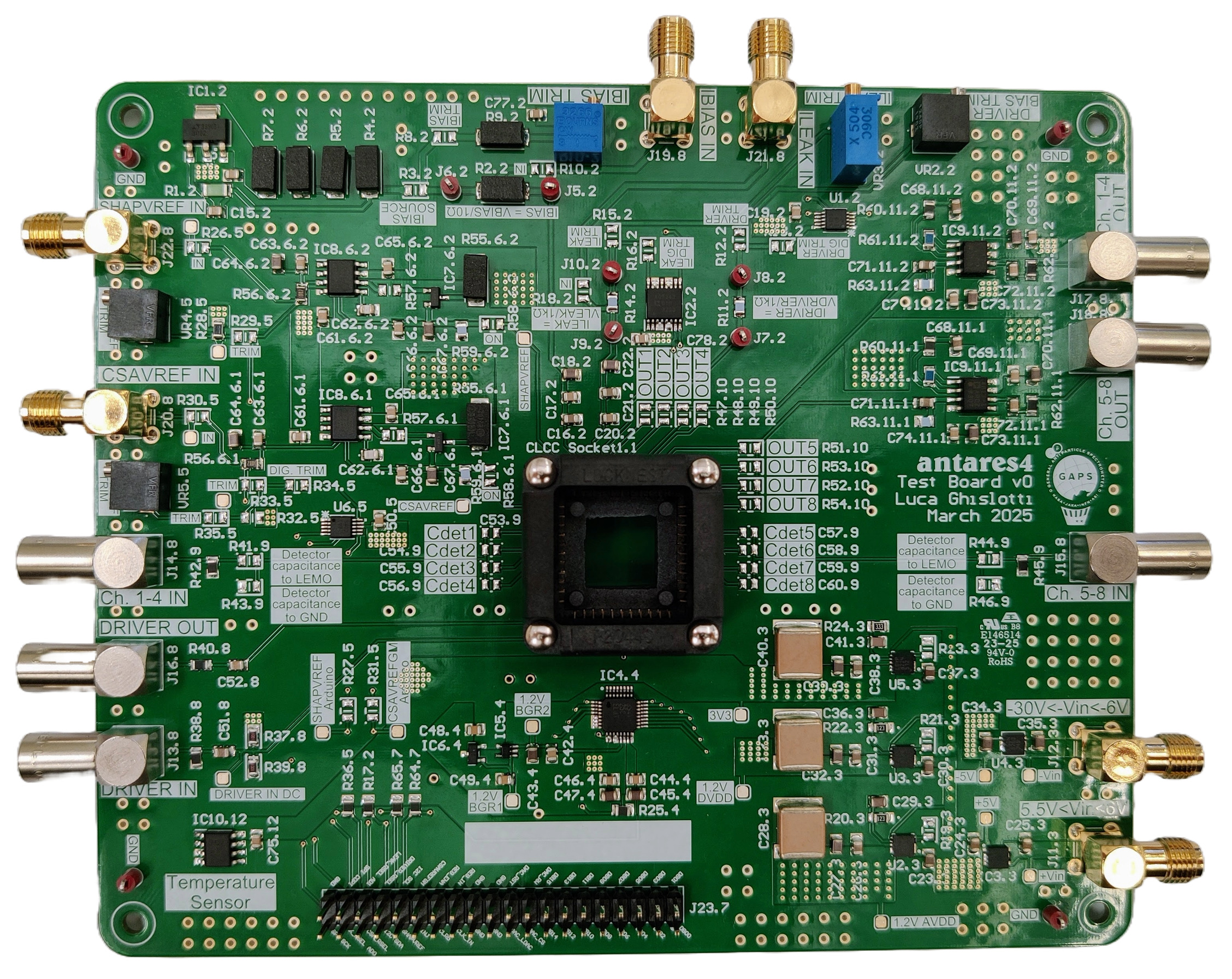}
    \caption{Photograph of the \antareslogo ASIC test board. The CLCC44 socket is centrally located, surrounded by supporting circuitry. The analog and digital power rails (\SI{\pm5}{\volt}, \SI{3.3}{\volt}, \SI{1.2}{\volt} AVDD and DVDD) are generated in the bottom right section. A 40-pin header at the bottom edge allows connection to the Arduino shield via a flat ribbon cable. Voltage references and current bias circuits are located in the top left, while the top right section houses the output buffers, each connected to dedicated LEMO connectors for signal readout.}
    \label{fig:ch5-antares4_test_board}
\end{figure}

\par
To monitor the temperature during thermal characterization, the board includes a \hbox{16-bit} digital temperature sensor (Analog Devices ADT7410) with an accuracy of \SI{0.5}{\celsius} and an operating range from \SI{-55}{\celsius} to \SI{150}{\celsius}. 
This enables direct measurement of the temperature inside the climate chamber, allowing correlation between the measured data and the actual thermal conditions of the device under test. 
The use of a dedicated on-board temperature sensor provides the most accurate estimation of the ASIC temperature, since relying on the temperature measured by the chamber probe may result in discrepancies caused by thermal gradients, air circulation effects, and the limited heat transfer between the chamber air and the board.

%   Power supply
\subsection{Power supply}

\par
The power supply section is based on a dual \SI{\pm5}{\volt} supply, implemented using two \hbox{high-PSRR}, low-noise, low-dropout regulators (LDOs).
The \SI{+5}{\volt} rail is provided by the Texas Instruments TPS7A21-Q1, a \SI{500}{\milli\ampere} LDO with a PSSR of \SI{91}{\decibel}, while the \SI{-5}{\volt} rail is supplied by the Analog Devices LT3015, a \SI{1.5}{\ampere} high-PSRR LDO.
The input voltages (\SIrange{5.5}{6}{\volt} for the \SI{+5}{\volt} rail and \SIrange{-30}{-6}{\volt} for the \SI{-5}{\volt} rail) are delivered via dedicated SMA connectors.
Test points are provided for all relevant voltage references and power rails, with ground points implemented using Keystone through-hole terminals to facilitate probing with multimeter leads during testing procedures.

\par
In cascade to the regulated \SI{5}{\volt} supply---also used, together with the \SI{-5}{\volt} rail, to power components such as the operational amplifiers implementing the output buffers---the \SI{+5}{\volt} line is further used to derive the \SI{3.3}{\volt}, \SI{1.2}{\volt} analog (AVDD), and \SI{1.2}{\volt} digital (DVDD) supplies.
These are generated using three Analog Devices LT3045 \hbox{ultralow-noise}, ultrahigh-PSRR linear regulators (\SI{20}{\volt}, \SI{500}{\milli\ampere}), each configured with precision resistors on the \texttt{SET} pin to produce the required output voltage.
The \SI{3.3}{\volt} supply powers components that operate at this lower voltage, such as the DAC providing the reference voltage \( V_{\text{inj}} \) for the charge injection circuit used in calibration.
The \SI{1.2}{\volt} analog and digital supplies (AVDD and DVDD) directly power the ASIC.
All LDO outputs have been decoupled with arrays of capacitors, with values specified according to the recommendations provided in the respective component datasheets.
In addition, \SI{1}{\micro\farad} and \SI{0.1}{\micro\farad} decoupling capacitors were placed on the bottom side of the board, close to the socket pins of AVDD, DVDD, $V_{\text{ref,CSA}}$, and $V_{\text{ref,sh}}$.

%   Voltage and current references
\subsection{Voltage and current references}

\par
The voltage and current references include the CSA baseline voltage \( V_{\text{ref,CSA}} = \SI{360}{\milli\volt} \), the shaper baseline voltage \( V_{\text{ref,sh}} = \SI{450}{\milli\volt} \), and the \SI{5}{\milli\ampere} master bias current, from which all on-chip bias currents are internally derived by the bias block located in the periphery of the chip.
Additional references include the bias current for the leakage current emulator and the bias current for the on-chip analog output buffers.

\par
A redundant design approach was adopted for the implementation of these references: for each voltage or current reference, at least two independent sources are available.
This redundancy ensures that, in the event of a malfunction or failure in one path, operation can be promptly restored by switching to an alternative source.
In several cases, one of the available options is a programmable reference voltage or current, which can be dynamically adjusted via the Arduino and its custom shield, allowing automated variation of the reference during test procedures.

\par
For the CSA output baseline voltage \( V_{\text{ref,CSA}} \), four options have been implemented. The primary option consists of a buffered voltage reference, configured as a voltage follower using a high-speed, low-noise operational amplifier (Analog Devices AD829). The op-amp operates in unity-gain mode and includes an active load implemented with an NPN bipolar junction transistor (Onsemi MMBT3904LT1G) in a common-emitter configuration, biased in the active region to enhance the output drive capability.

\par
The input voltage to the follower can be manually adjusted around the nominal \SI{360}{\milli\volt} reference using a voltage divider, selectable between an ultra-high precision \SI{10}{\kilo\ohm} trimming potentiometer (Vishay Accutrim 1240, \SI{0.1}{\percent} tolerance, \SI{10}{ppm/\celsius} TC) and a \SI{10}{\kilo\ohm} digital potentiometer (Maxim MAX5403) controlled via SPI by the Arduino. The latter enables remote tuning of the CSA baseline during automated testing inside the climate chamber.

\begin{figure}[tbp]
    \centering
    \captionsetup[subfigure]{justification=justified,singlelinecheck=false}
    \begin{subfigure}{0.43\textwidth}
        \centering
        \includegraphics[width=\linewidth]{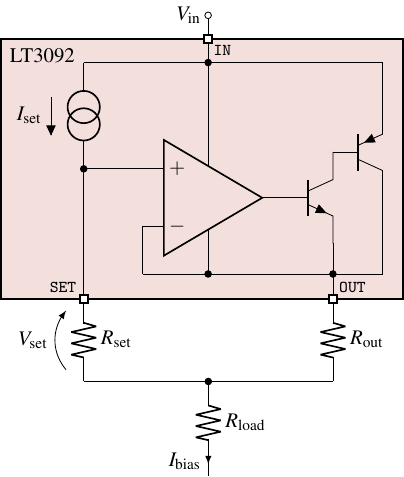}
        \caption{Functional schematic of the LT3092 programmable current source. The output current is set by the external resistors $R_{\text{set}}$ and $R_{\text{out}}$ based on the reference current \( I_{\text{set}} = \SI{10}{\micro\ampere} \).}
        \label{fig:ch5-LT3092_current_source_diagram}
    \end{subfigure}
    \hfill
    \begin{subfigure}{0.55\textwidth}
        \centering
        \includegraphics[width=\linewidth]{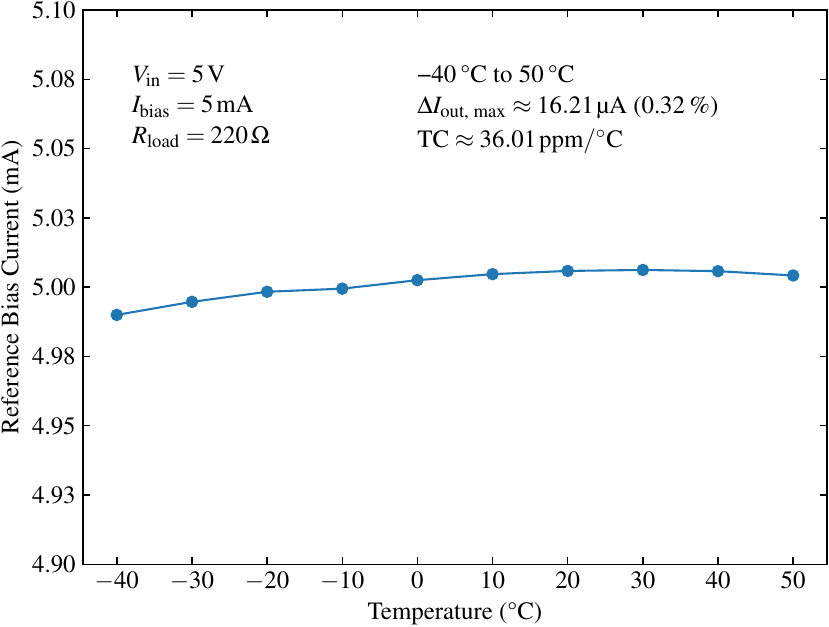}
        \vspace{0.4em}
        \caption{Measured \( I_\text{bias} \) current as a function of temperature, showing a maximum variation of approximately \SI{16.2}{\micro\ampere} ($\SI{\sim 0.32}{\percent}$) over the range from \SI{-40}{\celsius} to \SI{50}{\celsius}. The extracted temperature coefficient is \SI{36.01}{ppm\per\celsius}.}
        \label{fig:ch5-LT3092_current_source_plot}
    \end{subfigure}
    \caption{Analog Devices LT3092-based programmable current source implementation and validation. The schematic (\subref{fig:ch5-LT3092_current_source_diagram}) illustrates the internal structure and external resistor configuration, while (\subref{fig:ch5-LT3092_current_source_plot}) presents the results of the dedicated temperature characterization.}
    \label{fig:ch5-LT3092_current_source}
\end{figure}

\par
The shaper baseline voltage \( V_{\text{ref,sh}} \) is implemented with the same architecture, excluding the digital potentiometer option.
For both references, an SMA input connector is provided on the test board to allow external sourcing of the voltage during debugging or initial startup.
Additionally, both can be supplied via dedicated lines from the Arduino shield, using the two onboard 12-bit DACs of the STM32H7 microcontroller.
Selection among the available options is made by installing the appropriate \SI{0}{\ohm} jumper (Vishay high reliability thick film resistor, \SI{0.5}{\percent} tolerance) on the test board, ensuring that only one reference is active at a time.

\par  
The current references are implemented following a similar approach.
The main bias current is generated by a two-terminal programmable current source (Analog Devices LT3092), configured to deliver \SI{5}{\milli\ampere}.
Given the critical importance of this block in terms of current precision and temperature stability, it was validated prior to integration through SPICE simulations and dedicated prototype measurements, including a preliminary temperature characterization. The output current of the LT3092 is determined according to the following equations from the datasheet~\cite{analogdevicesAnalogDevicesLT3092}:

\begin{align}
    V_{\text{set}} &= I_{\text{set}} R_{\text{set}}, \\
    I_{\text{out}} &= \frac{V_{\text{set}}}{R_{\text{out}}} + I_{\text{set}}. \label{eq:ch5-LT3092_Iout}
\end{align}

\noindent
For a target output current of \SI{5}{\milli\ampere}, the resistor values were selected accordingly, based on the internal reference current \( I_{\text{set}} = \SI{10}{\micro\ampere} \). In line with datasheet recommendations, \( R_{\text{set}} \) was chosen as \SI{20}{\kilo\ohm}, and \( R_{\text{out}} \) was calculated from~\eqref{eq:ch5-LT3092_Iout} to be \SI{40}{\ohm}. Both resistors were implemented as high-precision metal film resistors with \SI{0.01}{\percent} tolerance and a low temperature coefficient of \SI{0.01}{ppm\per\celsius}, to ensure optimal thermal performance.

\begin{figure}[tbp]
    \centering
    \includegraphics[width=.75\textwidth]{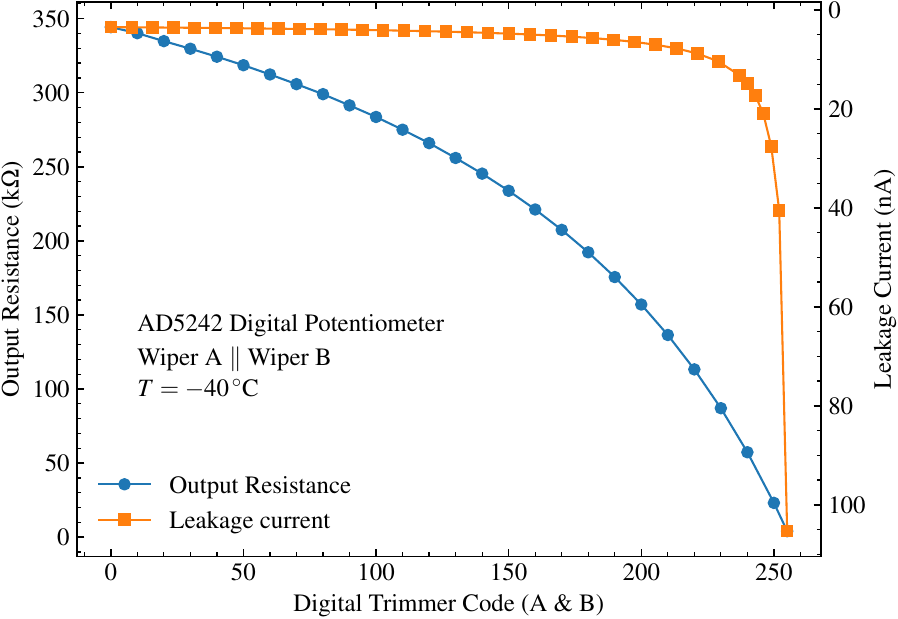}
    \caption{Output resistance of the AD5242 I\textsuperscript{2}C digital potentiometer as a function of the digital code applied simultaneously to both internal \SI{1}{\mega\ohm} channels connected in parallel. The corresponding leakage current is also reported, spanning a range from approximately \SI{2.5}{\nano\ampere} to \SI{105}{\nano\ampere} at \SI{-40}{\celsius}.}
    \label{fig:ch5-leakage_current_trimmer_resistance_vs_code}
\end{figure}

\par
For prototype testing of the current source, a load resistor ($R_\text{load}$ in~\Cref{fig:ch5-LT3092_current_source_diagram}) of \SI{220}{\ohm} was used to emulate the effective resistance seen at the drain of the input bias NMOS on the chip.  
It should be noted that this initial characterization was performed using resistors with a tolerance of \SI{5}{\percent} and a temperature coefficient of \SI{20}{ppm\per\celsius}.  
Therefore, the results presented in~\Cref{fig:ch5-LT3092_current_source_plot} represent a worst-case scenario and are expected to improve in the final test board implementation.  
The plot shows the current source behavior across temperature, confirming a maximum output variation of \SI{16.21}{\micro\ampere} ($\SI{\sim 0.32}{\percent}$) and a temperature coefficient of $\SI{\sim 36.01}{ppm\per\celsius}$.

\par
The input leakage current, mirrored across the four channels implementing the leakage current emulator, is generated using an I\textsuperscript{2}C-controlled 8-bit digital potentiometer (Analog Devices AD5241).  
This configuration enables remote adjustment during temperature characterization of the CSA Krummenacher feedback compensation mechanism within the climate chamber.
The bias current for the eight output buffers and the test buffer is also supplied via a digital potentiometer, allowing for noise performance evaluation of the output stage at \SI{-40}{\celsius} under varying bias conditions. 

\par
A particularly delicate aspect was the precise setting of the leakage current, as discussed in \Cref{sec:ch4-leakage_current_emulator} and illustrated in~\Cref{fig:ch4-leakage_current_emulator}, where post-layout simulations highlighted a non-linear dependence on the resistance value.
To address this, an initial characterization of the output resistance of the I\textsuperscript{2}C AD5242 digital potentiometer was performed at \SI{-40}{\celsius}, sweeping all 8-bit codes from 0 to 255 with both internal \SI{1}{\mega\ohm} potentiometers connected in parallel and set to the same code, as shown in~\Cref{fig:ch5-leakage_current_trimmer_resistance_vs_code}.

\par
This calibration was integrated into a dedicated function of the Python control software, which automatically determined the appropriate wiper code corresponding to the resistance required to generate the desired leakage current, as inferred from the post-layout simulation results.
This software-based approach, combining the measured resistance of the digital trimmer with the simulated leakage transfer function, makes the non-linear relationship between leakage current and bias resistance effectively transparent during testing.
It enables fine resolution adjustment of the leakage current in the range from $\sim$\SI{2.5}{\nano\ampere} to $\sim$\SI{100}{\nano\ampere} at the operating temperature of \SI{-40}{\celsius}, while the study of the compensation capabilities at room temperature for higher leakage values up to $\sim$\SI{300}{\nano\ampere} is delegated to the manually adjustable Bourns PV36 25-turn \SI{500}{\kilo\ohm} precision potentiometer with a temperature coefficient of \SI{100}{ppm/\celsius}, thereby avoiding the need for remote adjustment when the climate chamber is in operation.

\par
For all three current references---main bias, leakage, and buffer bias---external alternatives are also provided for flexibility and debugging, including SMA connectors for current injection and manually adjustable trimmers (Vishay Accutrim 1240 for the main and buffer bias, and the aforementioned Bourns PV36 for the leakage current bias).

%   Readout channel I/O interfaces
\subsection{Readout channel I/O interfaces}

\par
The outputs of each channel---available at both the CSA and shaper stages---are already provided with on-chip drivers capable of directly driving the inputs of a digital storage oscilloscope.  
Nevertheless, additional output buffers were included on the test board, implemented with low-noise, \SI{1}{\giga\hertz} bandwidth Analog Devices ADA4817 operational amplifiers.  
These external stages, configured for a voltage gain of 5, ensure increased signal amplitude for measurements at low injected charges, guaranteeing improved flexibility during laboratory testing.
In the characterization campaign reported in this thesis, however, the test board output buffers were bypassed in order to directly evaluate the driving capability of the on-chip output stages.

\par
To reduce the number of connectors, inputs and outputs are organized into two groups of four channels, with channel selection within each group enabled by soldering the corresponding \SI{0}{\ohm} jumper.
The board also allows direct access to the channel inputs, bypassing the on-chip charge injection circuit, which enables preliminary testing of the ASIC with actual Si(Li) detectors.

\par  
Eight dedicated 0805 footprints are included on the board for capacitors that emulate the detector capacitance, nominally \SI{40}{\pico\farad}, though this value can be adjusted to study the noise performance as a function of the detector capacitance.
Each capacitor can be installed in two configurations: either with one terminal connected to the CSA input and the other to ground---mimicking a standard detector connection with on-chip charge injection---or with one terminal connected to the input LEMO connector via a jumper, allowing for external signal injection.
This latter configuration serves two purposes: it enables direct interfacing of the CSA input with a real detector (by replacing the capacitor with a jumper) and allows the use of the capacitor as an injection capacitor when driven by the external Teledyne LeCroy T3AWG3352 arbitrary waveform generator.

%   DAC for charge injection
\subsection{DAC for charge injection}

\begin{figure}[tbp]
    \centering
    \includegraphics[width=.75\textwidth]{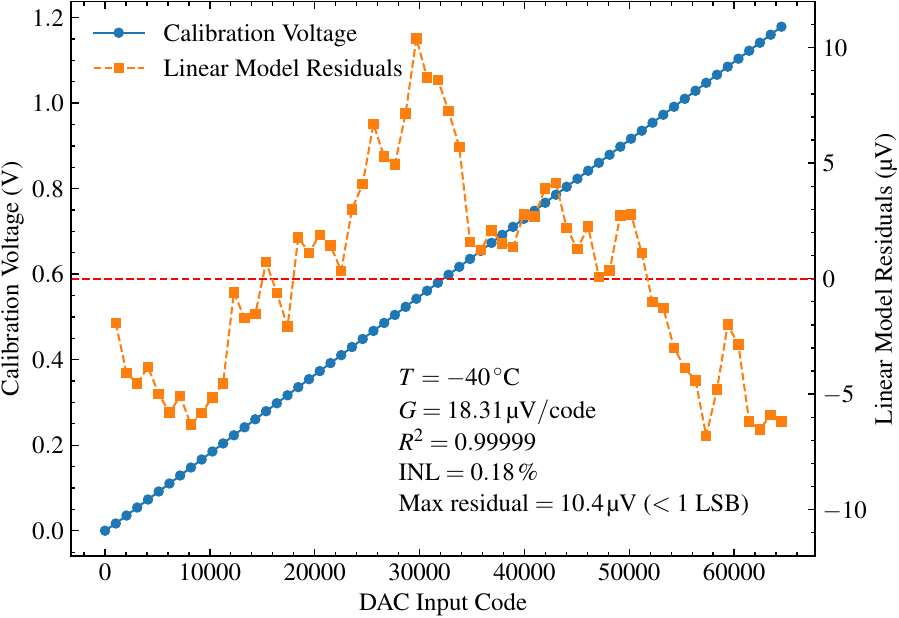}
    \caption{Measured calibration voltage provided by the DAC8541 DAC as a function of the 16-bit input code at \SI{-40}{\celsius}. The primary $y$-axis shows the output voltage, while the secondary $y$-axis displays the residuals of the linear fit. The extracted gain is \SI{18.31}{\micro\volt\per\text{code}}, an INL of \SI{0.18}{\percent} (\( R^2 = 0.99999 \)), and a maximum residual of \SI{10.4}{\micro\volt}.}
    \label{fig:ch5-dac_injection_vs_code}
\end{figure}

\par
As discussed in~\Cref{sec:ch4-injection_circuit}, one of the limiting factors in accurately reconstructing the input--output charge-to-voltage transcharacteristic of the readout channel is the resolution of the injection reference voltage \( V_{\text{inj}} \), which must be provided with a resolution of at least \SI{18.31}{\micro\volt} to achieve a minimum energy resolution of approximately \SI{1.53}{\kilo\electronvolt}.
This requirement can be met by employing a 16-bit DAC operating over a \SI{1.2}{\volt} dynamic range.

\par
The selected DAC is the Texas Instruments DAC8541, one of the few commercially available devices capable of delivering a true rail-to-rail output with an upper voltage limit externally constrained to \SI{1.2}{\volt}.
This constraint is achieved by supplying the DAC with a high-precision \SI{1.2}{\volt} bandgap voltage reference (BGR), the Analog Devices ADR280, which features a high PSRR of \SI{80}{\decibel} at \SI{220}{\hertz} and a temperature coefficient of \SI{40}{ppm\per\celsius}.
To ensure that the BGR is capable of driving the DAC reference input, its output is buffered using a low-noise, wide-bandwidth operational amplifier, the Analog Devices AD8541. 

\par
The DAC configuration signals and the 16-bit input word are delivered through dedicated pins on the shield and are managed by the Arduino.
The input word is transferred in parallel format.
To ensure compatibility with the Arduino's \SI{3.3}{\volt} logic levels and avoid the need for level shifting, the digital section of the DAC is powered at \SI{3.3}{\volt}.

\par
To validate the performance of the DAC-based injection calibration chain, the output voltage was measured as a function of the input code at \SI{-40}{\celsius} using a Keysight 34461A 6\textonehalf-digit digital multimeter.  
The results are shown in~\Cref{fig:ch5-dac_injection_vs_code}.  
The measured transfer curve exhibits excellent linearity over the full 16-bit DAC range (0--65535). 
The DAC was sampled with unit step size, while only one point every 1024 steps is shown in the plot for clarity. 
The extracted gain is \SI{18.31}{\micro\volt\per\text{code}}, the coefficient of determination is \( R^2 = 0.99999 \), and the INL is \SI{0.18}{\percent} relative to the full-scale voltage swing.  
The residuals with respect to a linear fit remain below a maximum of \SI{10.4}{\micro\volt}, corresponding to less than one LSB (\SI{18.31}{\micro\volt}), thus confirming the suitability of the selected DAC for providing the injection reference voltage \( V_{\text{inj}} \).

%-------------------------------------------------------------------------------
%   Arduino Giga R1 shield board
%-------------------------------------------------------------------------------
\section{Arduino Giga R1 interface shield board}

\begin{figure}[tbp]
    \centering
    \includegraphics[width=.85\textwidth]{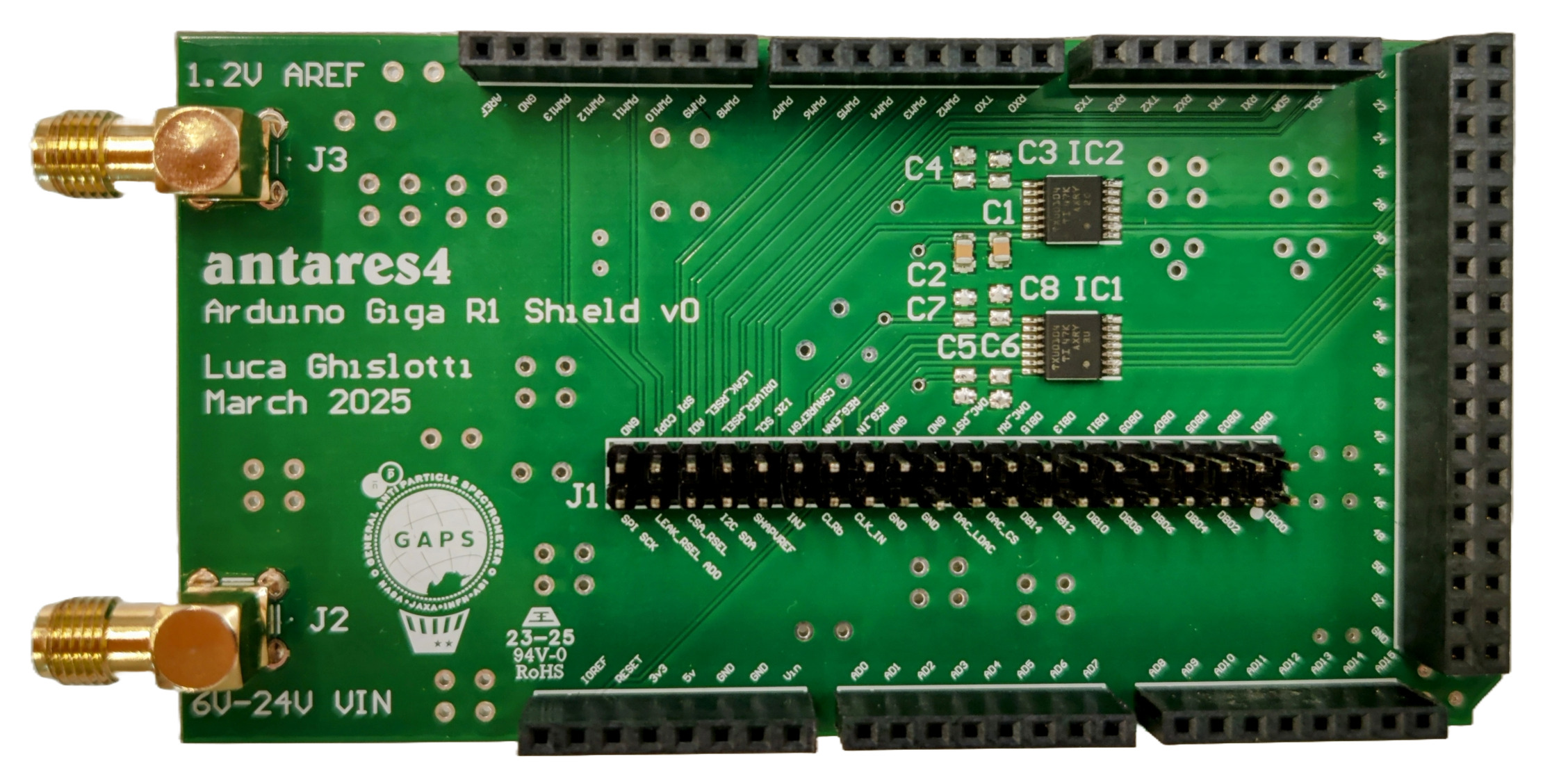}
    \caption{Photograph of the \antareslogo Arduino Giga R1 shield. The board is designed to interface directly with the Arduino Giga R1 Wi-Fi as a stackable shield and enables the interconnection of the digital signals required for the configuration of the ASIC's slow control. It connects to the Arduino via Samtec through-hole headers, and to the ASIC test board through a 40-pin Amphenol through-hole header using a 40-wire flat-ribbon cable.}
    \label{fig:ch5-arduino_giga_shield}
\end{figure}

\par
An Arduino Giga R1 Wi-Fi was selected as the DSP for the test setup.
It provides the digital control signals required for the ASIC slow control, as well as the configuration signals for the DAC that generates the calibration voltage \( V_{\text{inj}} \) for the on-chip charge injection circuit. 
In addition, the Arduino controls the auxiliary reference voltages \( V_{\text{ref,CSA}} \) and \( V_{\text{ref,sh}} \), which can be remotely adjusted during temperature calibration in the climate chamber, and manages all the digital potentiometers via SPI or I\textsuperscript{2}C interfaces.

\par
A custom interface shield board was designed to provide the Arduino main board with the required connectors and supporting electronics to interface with the ASIC test board.
The layout is based on the standard footprint of the Arduino Mega and Giga families, with dimensions of \SI{101.6}{\milli\meter} in width and \SI{53.4}{\milli\meter} in height.
A picture of the board is provided in~\Cref{fig:ch5-arduino_giga_shield}.
The shield was implemented as a two-layer PCB in standard FR-4 material, following the open-source mechanical and electrical standards for this class of Arduino boards~\cite{ghislottiLucaghisloArduinoMegaGigaShield2025}.

\par
The signals associated with the ASIC slow control (\texttt{REG\_ENA}, \texttt{CLK\_IN}, \texttt{CLRb}, and \texttt{REG\_IN}) as well as the injection trigger signal \texttt{INJ} are level-shifted from the \SI{3.3}{\volt} logic level of the Arduino to the \SI{1.2}{\volt} logic level required by the ASIC using two 4-bit level shifters (Texas Instruments TXU0104).
Connection to the Arduino's main headers is achieved using Amphenol 8-pin and 36-pin through-hole headers with a \SI{2.54}{\milli\meter} pitch.
The interface with the ASIC test board is realized using a \SI{50}{\centi\meter}-long 40-wire flat-ribbon cable terminated by Amphenol 40-pin male headers on both the shield and the test board.

%-------------------------------------------------------------------------------
%   Measurement results
%-------------------------------------------------------------------------------
\section{Measurement results}

\par
This section presents the results of the initial characterization of the \antareslogo ASIC. 
The measurements were carried out in the climate chamber using the setup described in~\Cref{sec:ch5-test_setup}, with the test board placed inside the chamber and the Arduino Giga, equipped with its dedicated shield, located outside. 
The test campaign began with standalone measurements of the ASIC test board, performed both at room temperature (\SI{25}{\celsius}) and at \SI{-40}{\celsius}, in order to validate the performance of the supporting electronics, including the power supplies, voltage and current references, and the DAC used for charge injection. 
The device under test was ASIC \texttt{antares4\_002} of the ten packaged prototypes fabricated.

\par
The characterization was conducted by varying the temperature from room conditions down to \SI{-40}{\celsius}, and by evaluating the performance of the prototype channels with and without the \SI{40}{\pico\farad} detector capacitance, as well as under different leakage current levels provided by the on-chip emulator introduced in~\Cref{sec:ch4-leakage_current_emulator}. 
Owing to the prototype implementation described in~\Cref{sec:ch4-complete_channel_layout}, only a subset of CSA feedback devices, output configurations, and leakage current emulator options were available (\Cref{tab:ch4-antares4_channels_info}). 
The characterization was therefore performed by exploiting the accessible configurations of CSA and shaper outputs together with the leakage current generators.

\par
In all the following measurements, the on-chip output buffers were set to a bias resistance of \SI{6}{\kilo\ohm} through the Maxim MAX5403 8-bit, \SI{10}{\kilo\ohm}, 3-wire SPI digital potentiometer.
The configuration used in each measurement is reported in the corresponding plot, including CSA feedback configuration, temperature, detector capacitance, and leakage current.

\par
This section is organized as follows. 
First, the ASIC and test board power consumption are presented. 
This is followed by the evaluation of the CSA and shaper transient response, transcharacteristics, noise performance, and corresponding performance metrics, compared with post-layout simulation results when available.
Finally, results obtained with the \SI{40}{\pico\farad} detector capacitance and with leakage currents ranging from \SI{2.5}{\nano\ampere} to \SI{300}{\nano\ampere} are reported.

%   ASIC power consumption
\subsection{ASIC power consumption}

\par
An initial evaluation of the ASIC power consumption was carried out by measuring the total current drawn by the test board at room temperature (\SI{25}{\celsius}) and at \SI{-40}{\celsius} inside the climate chamber. 
The measurement was performed on the current drawn from the \SI{5}{\volt} rail, which provides the main supply for the ASIC and the supporting electronics, from which the \SI{3.3}{\volt} and \SI{1.2}{\volt} supplies (AVDD and DVDD) are derived. 

\begin{figure}[tbp]
    \centering
    \includegraphics[width=.75\textwidth]{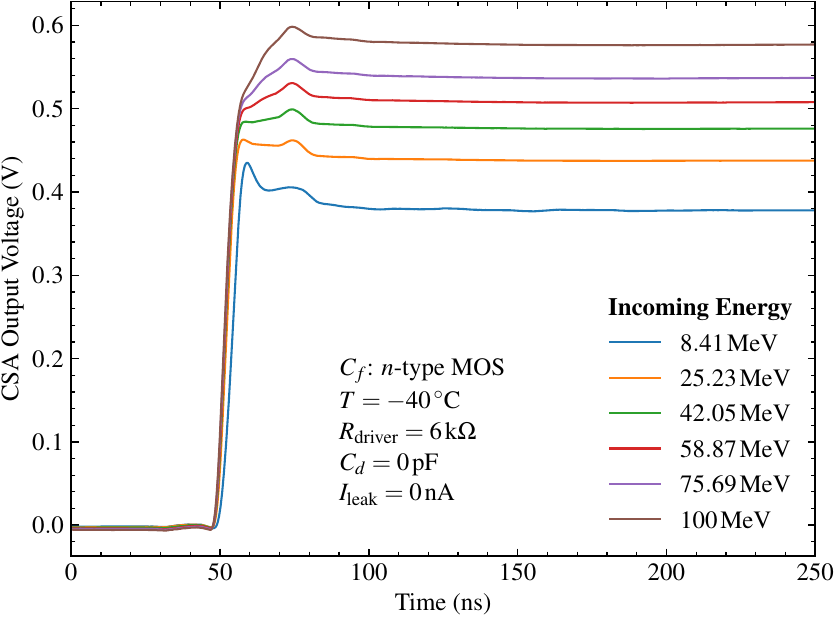}
    \caption{CSA output waveforms as a function of incoming energy, considering the first \SI{200}{\nano\second} window to highlight the rising edge. Results are shown for the $n$-type MOS feedback device at $T=\SI{-40}{\celsius}$, $R_{\text{driver}}=\SI{6}{\kilo\ohm}$, $C_d=\SI{0}{\pico\farad}$, and $I_{\text{leak}}=\SI{0}{\nano\ampere}$.}
    \label{fig:ch5-meas-CSA_rise_buffer}
\end{figure} 

\par
At room temperature, the current consumption of the test board alone was measured to be \SI{56.2}{\milli\ampere}, decreasing to \SI{45.4}{\milli\ampere} at \SI{-40}{\celsius}, corresponding to a reduction of approximately \SI{19}{\percent}. 
When repeating the measurement with the ASIC placed in the socket, the ASIC current consumption was obtained by subtracting the board-only contribution from the total current, yielding \SI{31.3}{\milli\ampere} at room temperature and \SI{26.6}{\milli\ampere} at \SI{-40}{\celsius}, i.e. a decrease of about \SI{15}{\percent}. 
From these values, a conservative estimate of the per-channel power consumption, including both the analog (AVDD) and digital (DVDD) supplies, is $\SI{\sim4.69}{\milli\watt\per ch}$ at room temperature and $\SI{\sim3.99}{\milli\watt\per ch}$ at \SI{-40}{\celsius}, corresponding to a $\SI{\sim12}{\percent}$ increase with respect to the simulated $\SI{\sim3.50}{\milli\watt\per ch}$ at \SI{-40}{\celsius} in the typical process corner. 

\par
For comparison, the previous \SI{180}{\nano\meter} implementation achieved a per-channel power consumption of about \SI{8.3}{\milli\watt} at \SI{-40}{\celsius}, including the complete readout chain with self-trigger circuitry (active differentiation and zero-crossing discriminator) and a shared 11-bit SAR ADC. 
Although the prototype ASIC presented here implements only the analog front-end up to the shaper stage, the measured value of \SI{3.99}{\milli\watt\per ch} still indicates a reduction of approximately \SI{52}{\percent} and leaves ample margin to accommodate the additional circuit blocks required in the final flight design.

\begin{figure}[t!]
    \centering
    \captionsetup[subfigure]{justification=justified,singlelinecheck=false}
    \begin{subfigure}{0.49\textwidth}
        \centering
        \includegraphics[width=\linewidth]{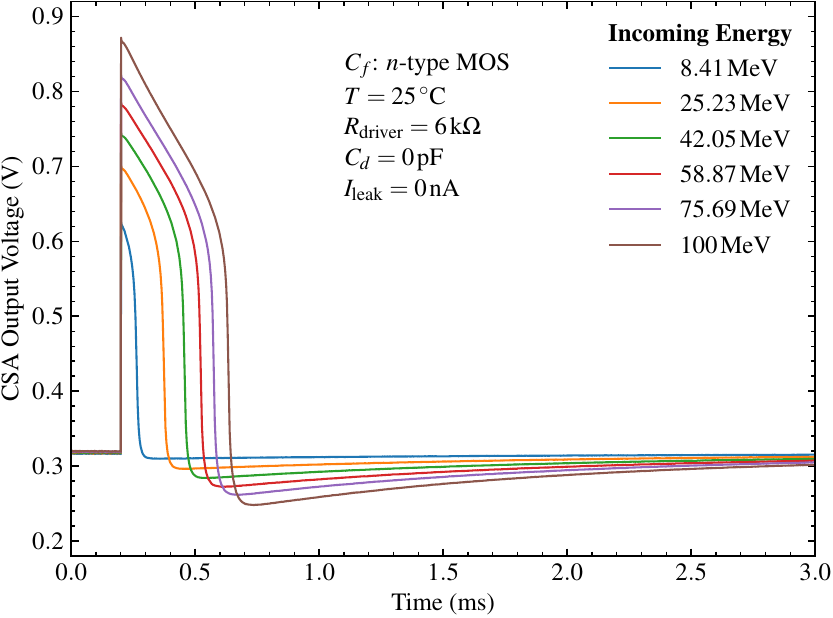}
        \caption{CSA output waveforms for the $n$-type MOS feedback device.}
        \label{fig:ch5-meas-CSA1_NMOS_30C}
    \end{subfigure}
    \hfill
    \begin{subfigure}{0.49\textwidth}
        \centering
        \includegraphics[width=\linewidth]{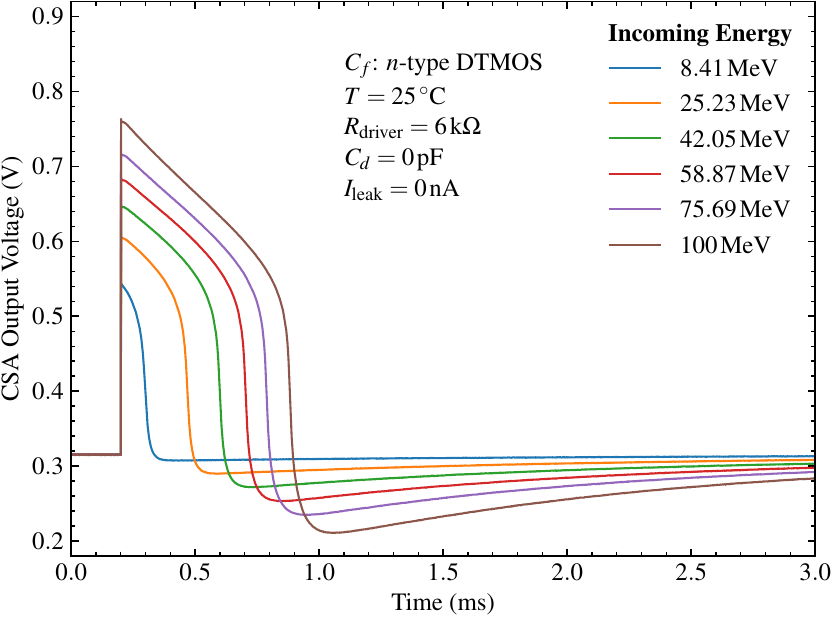}
        \caption{CSA output waveforms for the $n$-type DTMOS feedback device.}
        \label{fig:ch5-meas-CSA2_NDTMOS_30C}
    \end{subfigure}
    \caption{CSA output waveforms as a function of incoming energy at $T=\SI{25}{\celsius}$ for (\subref{fig:ch5-meas-CSA1_NMOS_30C}) the $n$-type MOS and (\subref{fig:ch5-meas-CSA2_NDTMOS_30C}) the $n$-type DTMOS feedback devices, with $C_d=\SI{0}{\pico\farad}$, $R_{\text{driver}}=\SI{6}{\kilo\ohm}$, and $I_{\text{leak}}=\SI{0}{\nano\ampere}$.}
    \label{fig:ch5-meas-CSA_NMOS_NDTMOS_30C}
\end{figure}

\begin{figure}[t!]
    \centering
    \captionsetup[subfigure]{justification=justified,singlelinecheck=false}
    \begin{subfigure}{0.49\textwidth}
        \centering
        \includegraphics[width=\linewidth]{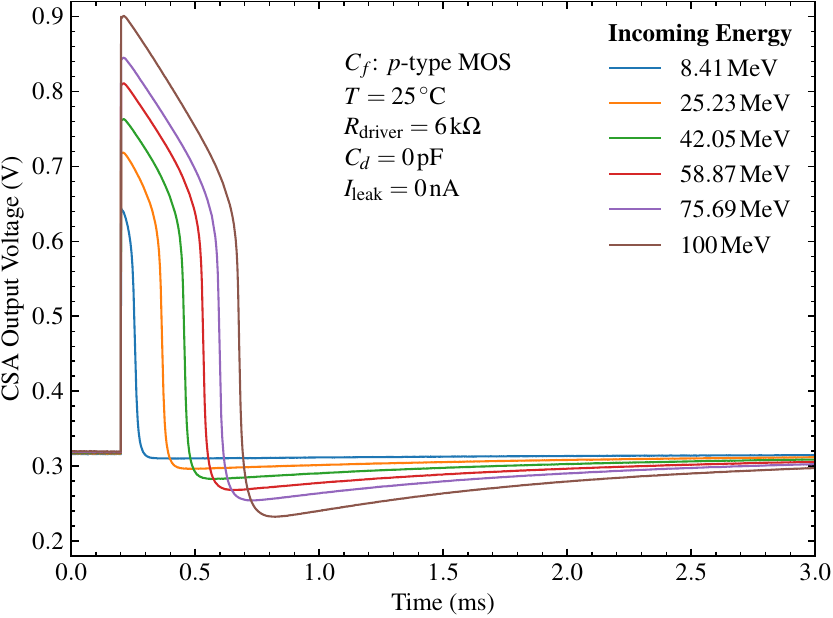}
        \caption{CSA output waveforms for the $p$-type MOS feedback device.}
        \label{fig:ch5-meas-CSA5_PMOS_30C}
    \end{subfigure}
    \hfill
    \begin{subfigure}{0.49\textwidth}
        \centering
        \includegraphics[width=\linewidth]{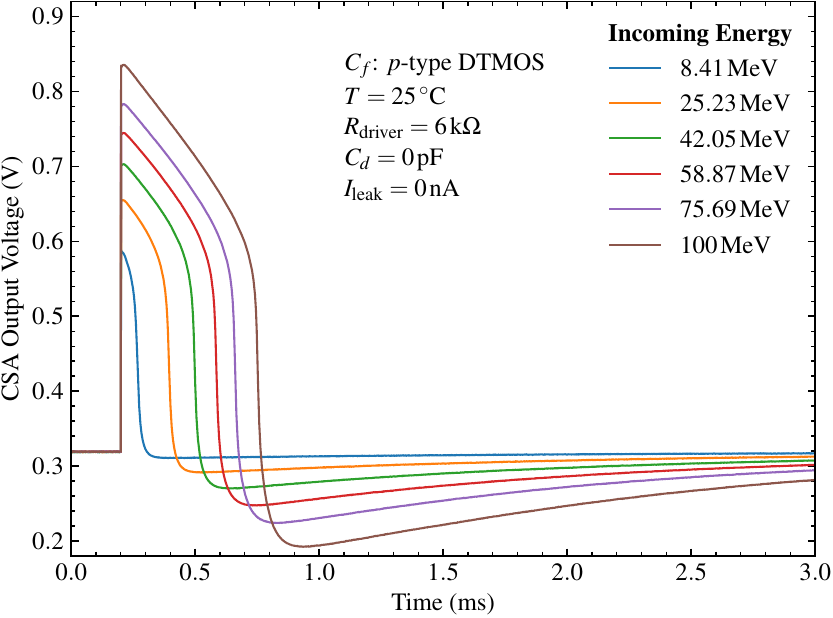}
        \caption{CSA output waveforms for the $p$-type DTMOS feedback device.}
        \label{fig:ch5-meas-CSA6_PDTMOS_30C}
    \end{subfigure}
    \caption{CSA output waveforms as a function of incoming energy at $T=\SI{25}{\celsius}$ for (\subref{fig:ch5-meas-CSA5_PMOS_30C}) the $p$-type MOS and (\subref{fig:ch5-meas-CSA6_PDTMOS_30C}) the $p$-type DTMOS feedback devices, with $C_d=\SI{0}{\pico\farad}$, $R_{\text{driver}}=\SI{6}{\kilo\ohm}$, and $I_{\text{leak}}=\SI{0}{\nano\ampere}$.}
    \label{fig:ch5-meas-CSA_PMOS_PDTMOS_30C}
\end{figure}

%   CSA transient response
\subsection{CSA transient response} \label{sec:ch5-CSA_transient_response}

\par
The CSA transient response was evaluated at the outputs of channels 1, 2, 5, and 6, corresponding respectively to the $n$- and $p$-type MOS and DTMOS CSA feedback devices. 
The characterization of the transient response as a function of the leakage current for the $n$-type DTMOS feedback device was carried out on channel 8.

\begin{figure}[t!]
    \centering
    \captionsetup[subfigure]{justification=justified,singlelinecheck=false}
    \begin{subfigure}{0.49\textwidth}
        \centering
        \includegraphics[width=\linewidth]{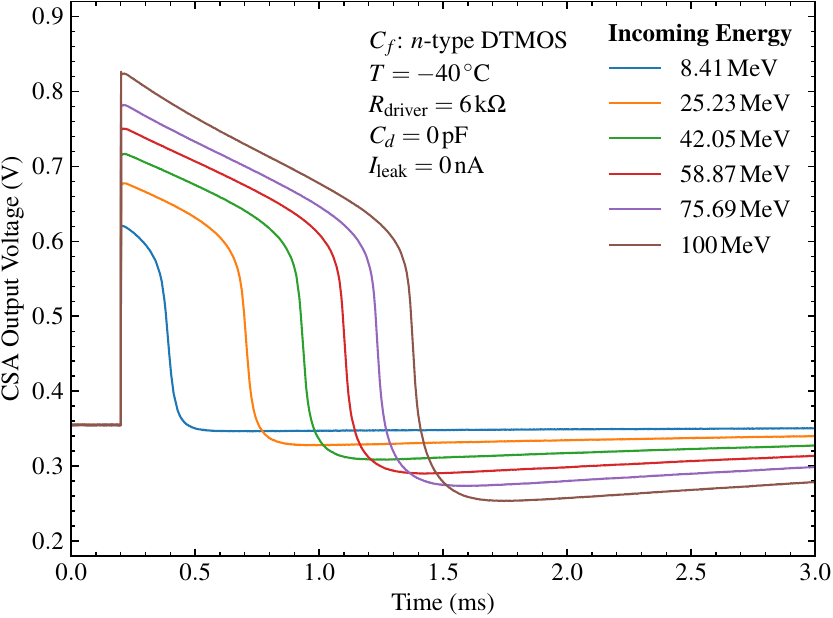}
        \caption{CSA output waveforms for the $n$-type DTMOS feedback device with $C_d=\SI{0}{\pico\farad}$.}
        \label{fig:ch5-meas-CSA2_NDTMOS_m40C_Cd0}
    \end{subfigure}
    \hfill
    \begin{subfigure}{0.49\textwidth}
        \centering
        \includegraphics[width=\linewidth]{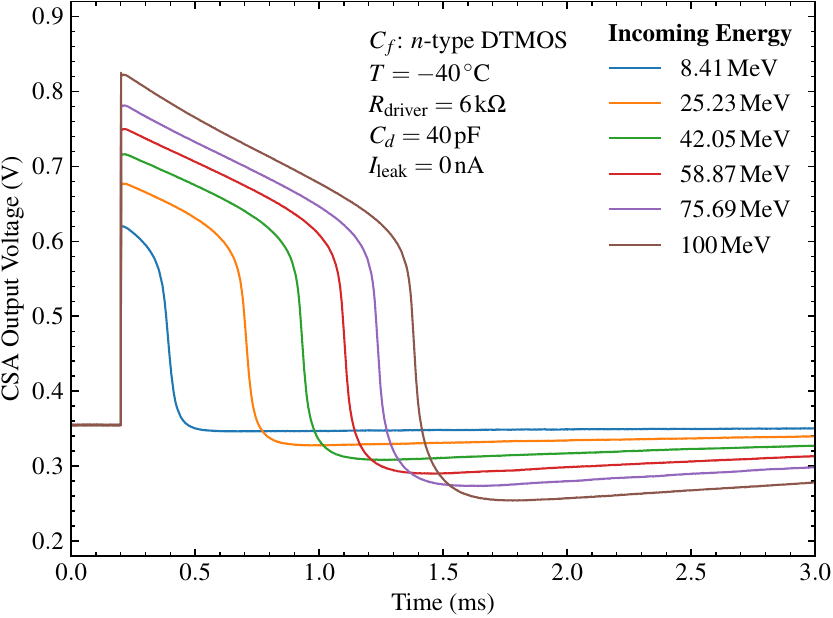}
        \caption{CSA output waveform for the $n$-type DTMOS feedback device with $C_d=\SI{40}{\pico\farad}$.}
        \label{fig:ch5-meas-CSA2_NDTMOS_m40C_Cd40}
    \end{subfigure}
    \caption{CSA output waveforms for the $n$-type DTMOS feedback device at \hbox{$T=\SI{-40}{\celsius}$} with $R_{\text{driver}}=\SI{6}{\kilo\ohm}$ and $I_{\text{leak}}=\SI{0}{\nano\ampere}$, comparing the cases with detector capacitances $C_d=\SI{0}{\pico\farad}$ (\subref{fig:ch5-meas-CSA2_NDTMOS_m40C_Cd0}) and $C_d=\SI{40}{\pico\farad}$ (\subref{fig:ch5-meas-CSA2_NDTMOS_m40C_Cd40}). Incoming energy is varied from \SI{8.41}{\mega\electronvolt} to \SI{100}{\mega\electronvolt}.}
    \label{fig:ch5-meas-CSA2_NDTMOS_m40C}
\end{figure}

\begin{figure}[t!]
    \centering
    \captionsetup[subfigure]{justification=justified,singlelinecheck=false}
    \begin{subfigure}{0.49\textwidth}
        \centering
        \includegraphics[width=\linewidth]{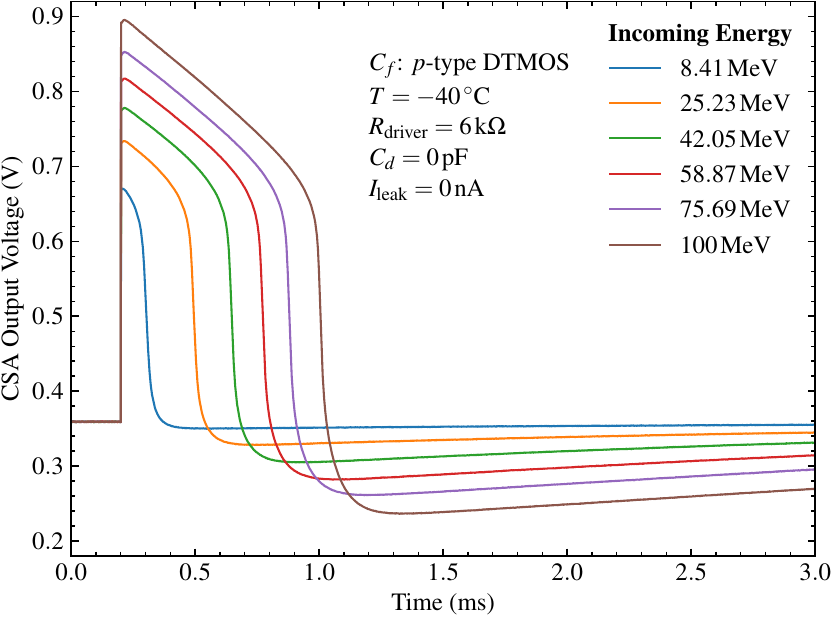}
        \caption{CSA output waveforms for the $p$-type DTMOS feedback device with $C_d=\SI{0}{\pico\farad}$.}
        \label{fig:ch5-meas-CSA6_PDTMOS_m40C_Cd0}
    \end{subfigure}
    \hfill
    \begin{subfigure}{0.49\textwidth}
        \centering
        \includegraphics[width=\linewidth]{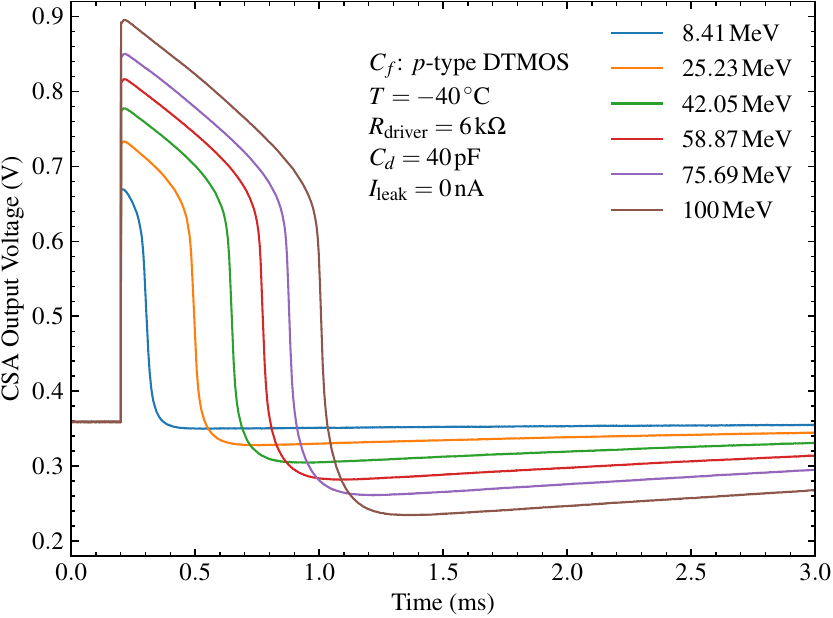}
        \caption{CSA output waveforms for the $p$-type DTMOS feedback device with $C_d=\SI{40}{\pico\farad}$.}
        \label{fig:ch5-meas-CSA6_PDTMOS_m40C_Cd40}
    \end{subfigure}
    \caption{CSA output waveforms for the $p$-type DTMOS feedback device at $T=\SI{-40}{\celsius}$ with $R_{\text{driver}}=\SI{6}{\kilo\ohm}$ and $I_{\text{leak}}=\SI{0}{\nano\ampere}$, comparing the cases with detector capacitances $C_d=\SI{0}{\pico\farad}$ (\subref{fig:ch5-meas-CSA6_PDTMOS_m40C_Cd0}) and $C_d=\SI{40}{\pico\farad}$ (\subref{fig:ch5-meas-CSA6_PDTMOS_m40C_Cd40}). Incoming energy is varied from \SI{8.41}{\mega\electronvolt} to \SI{100}{\mega\electronvolt}.}
    \label{fig:ch5-meas-CSA6_PDTMOS_m40C}
\end{figure}

\begin{table}[t!]
    \centering
    \renewcommand{\arraystretch}{1.1}
    \setlength{\tabcolsep}{5pt}
    \begin{tabular}{
        c
        c
        *{2}{S[table-format=4.0]}
        *{2}{S[table-format=5.0]}
    }
        \toprule
        \multirow{2}{*}{\vspace{-0.8em}\shortstack{CSA Feedback\\Device}} &
        \multirow{2}{*}{\vspace{-0.8em}\shortstack{Incoming Energy\\(\si{\mega\electronvolt})}} &
        \multicolumn{2}{c}{Discharge Time (\si{\micro\second})} &
        \multicolumn{2}{c}{Return to Baseline (\si{\micro\second})} \\
        \cmidrule(lr){3-4}\cmidrule(lr){5-6}
        & &
        \multicolumn{1}{c}{$C_d=\SI{0}{\pico\farad}$} &
        \multicolumn{1}{c}{$C_d=\SI{40}{\pico\farad}$} &
        \multicolumn{1}{c}{$C_d=\SI{0}{\pico\farad}$} &
        \multicolumn{1}{c}{$C_d=\SI{40}{\pico\farad}$} \\
        \midrule
        \multirow{6}{*}{$n$-type DTMOS}
            &  8.41  &  269 &  268 &    <1 &    <1 \\
            & 25.23  &  555 &  556 &  3279 &  3304 \\
            & 42.05  &  773 &  772 &  5465 &  5507 \\
            & 58.87  &  937 &  937 &  7651 &  7709 \\
            & 75.69  & 1066 & 1067 &  9837 &  9912 \\
            & 100.00 & 1205 & 1210 & 12996 & 13095 \\
        \midrule
        \multirow{6}{*}{$p$-type DTMOS}
            &  8.41  &  170 &  171 &    <1 &    <1 \\
            & 25.23  &  338 &  339 &  3123 &  3159 \\
            & 42.05  &  482 &  482 &  5204 &  5265 \\
            & 58.87  &  602 &  602 &  7286 &  7371 \\
            & 75.69  &  706 &  706 &  9368 &  9477 \\
            & 100.00 &  829 &  831 & 12376 & 12521 \\
        \bottomrule
    \end{tabular}
    \caption{Discharge and return-to-baseline times for the two dynamic threshold MOSFET CSA feedback devices ($n$-type DTMOS and $p$-type DTMOS) with detector capacitances $C_d=\SI{0}{\pico\farad}$ and $C_d=\SI{40}{\pico\farad}$ as a function of incoming energy, for $T=\SI{-40}{\celsius}$, $R_{\text{driver}}=\SI{6}{\kilo\ohm}$, and $I_{\text{leak}}=\SI{0}{\nano\ampere}$.}
    \label{tab:ch5-meas-discharge_baseline_comparison}
\end{table}

\par
The first step in this study was to verify the rise time of the CSA output voltage. 
For simplicity, the case of the $n$-type MOS device at \SI{-40}{\celsius} with $C_d=\SI{0}{\pico\farad}$ is reported in~\Cref{fig:ch5-meas-CSA_rise_buffer}, where the incoming energy was varied from the minimum of \SI{8.41}{\mega\electronvolt} to the maximum expected energy of \SI{100}{\mega\electronvolt}. 
As expected, the rise time decreases with the charge released in the detector, from a maximum of \SI{11.95}{\nano\second} at low deposited energy to \SI{5.39}{\nano\second} at the maximum deposited energy.
These results are consistent with the expected behavior of this type of feedback device under the reported conditions, and confirm the fast-rising nature of the implemented CSA block.

\par
The key aspect in evaluating the CSA transient response is the analysis of the discharge time and the return-to-baseline time for the four CSA feedback configurations as a function of temperature, detector capacitance, and leakage current.  
In this context, the \hbox{return-to-baseline} time is defined as the interval required for the output to settle within \SI{5}{\percent} of the CSA baseline.  
\Cref{fig:ch5-meas-CSA_NMOS_NDTMOS_30C} and~\Cref{fig:ch5-meas-CSA_PMOS_PDTMOS_30C} show this analysis performed at room temperature, without detector capacitance or leakage current, for all four feedback devices.  
Although these conditions are not representative of nominal operation, they provide a useful step in verifying the correct functionality of the CSA at room temperature, which is required to validate the channel response during ASIC integration and pre-flight testing.

\par
Under these conditions, the response follows the expected behavior: for an input energy of \SI{100}{\mega\electronvolt}, the discharge time is \SI{442}{\micro\second} for the $n$-type MOS device (\Cref{fig:ch5-meas-CSA1_NMOS_30C}), increasing to \SI{693}{\micro\second} for the dynamic-threshold variant of the same polarity (\Cref{fig:ch5-meas-CSA2_NDTMOS_30C}). 
The corresponding return-to-baseline times, measured after the undershoot, are \SI{3437}{\micro\second} and \SI{4619}{\micro\second}, respectively.

\par
For the $p$-type MOS device, the discharge and return-to-baseline times are \SI{491}{\micro\second} and \SI{3789}{\micro\second} (\Cref{fig:ch5-meas-CSA5_PMOS_30C}), while for the $p$-type DTMOS they increase to \SI{564}{\micro\second} and \SI{4714}{\micro\second} respectevely (\Cref{fig:ch5-meas-CSA6_PDTMOS_30C}). 
Overall, the dynamic-threshold feedback configurations (NDTMOS and PDTMOS) exhibit systematically longer discharge and return-to-baseline times compared to the corresponding MOS configurations, as expected from their feedback implementation.

\par
The same evaluation was carried out at the nominal flight temperature, comparing the two dynamic-threshold MOSFET feedback configurations, $n$-type (\Cref{fig:ch5-meas-CSA2_NDTMOS_m40C}) and $p$-type (\Cref{fig:ch5-meas-CSA6_PDTMOS_m40C}), with and without detector capacitance for $I_{\text{leak}}=\SI{0}{\nano\ampere}$. 
A comprehensive comparison of the discharge and return-to-baseline times is reported in~\Cref{tab:ch5-meas-discharge_baseline_comparison}, where the two configurations are analyzed side by side under both detector capacitance conditions.

\par
Both $n$- and $p$-type DTMOS feedback configurations show discharge and \hbox{return-to-baseline} times increasing monotonically with the deposited energy, spanning from a few hundred microseconds at low energy to values above \SI{1}{\milli\second} for the discharge time and up to \SI{13}{\milli\second} for the \hbox{return-to-baseline} at the maximum input energy of \SI{100}{\mega\electronvolt}. 
The effect of the detector capacitance is found to be negligible, with differences below \SI{2}{\percent} when comparing the cases of $C_d=\SI{0}{\pico\farad}$ and $C_d=\SI{40}{\pico\farad}$ for both configurations. 
This indicates that the transient response is primarily determined by the feedback topology rather than the detector capacitance in the investigated operating range. 

\begin{figure}[tbp]
    \centering
    \includegraphics[width=.75\textwidth]{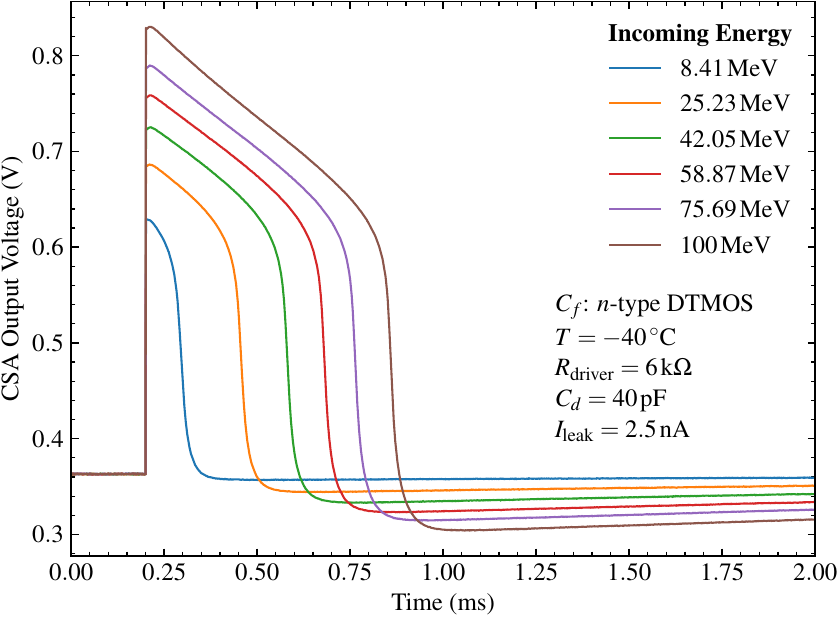}
    \caption{Most representative CSA transient response, measured for the $n$-type DTMOS feedback device at $T=\SI{-40}{\celsius}$, $R_{\text{driver}}=\SI{6}{\kilo\ohm}$, $C_d=\SI{40}{\pico\farad}$, and $I_{\text{leak}}=\SI{2.5}{\nano\ampere}$, the nominal leakage current expected during operation. The incoming energy is varied from \SI{8.41}{\mega\electronvolt} to \SI{100}{\mega\electronvolt}.}
    \label{fig:ch5-meas-CSA8_NDTMOS_Ileak2p5nA}
\end{figure}

\par
A consistent difference is observed between the two configurations: the $n$-type DTMOS exhibits discharge times that are on average $\SI{\sim45}{\percent}$ longer than those of the $p$-type, while the return-to-baseline is $\SI{\sim5}{\percent}$ longer at the highest energies. 
This behavior, fully consistent with expectations, reflects the slower recovery dynamics of the $n$-type DTMOS feedback implementation, highlighting its stronger influence on both discharge and baseline restoration compared to the $p$-type counterpart.

\par
The most representative case is reported in~\Cref{fig:ch5-meas-CSA8_NDTMOS_Ileak2p5nA}, where the CSA transient response for the $n$-type DTMOS feedback device is shown at the nominal leakage current of $I_{\text{leak}}=\SI{2.5}{\nano\ampere}$, with $T=\SI{-40}{\celsius}$ and $C_d=\SI{40}{\pico\farad}$. 
In these conditions, the discharge time increases monotonically with the deposited energy, ranging from \SI{154}{\micro\second} at \SI{8.41}{\mega\electronvolt} up to \SI{682}{\micro\second} at \SI{100}{\mega\electronvolt}, while the return-to-baseline evolves from negligible values at low energy to approximately \SI{7.9}{\milli\second} at the maximum input charge. 
The effect of the leakage current is thus to reduce the overall times, guaranteeing that even for the feedback device exhibiting the slowest dynamics the complete recovery at the maximum input energy of \SI{100}{\mega\electronvolt} remains well within the timing constraints required to sustain a maximum event rate of \SI{100}{\hertz}, as specified by the experiment requirements. 
This analysis was carried out for the $n$-type DTMOS device, which represents the most relevant case. It consistently exhibited the longest discharge and recovery times, and also demonstrated in simulations the best overall performance in terms of channel transcharacteristics, including low- and high-energy gain, kink behavior, and compression factor.

\begin{figure}[t!]
    \centering
    \captionsetup[subfigure]{justification=justified,singlelinecheck=false}
    \begin{subfigure}{0.49\textwidth}
        \centering
        \includegraphics[width=\linewidth]{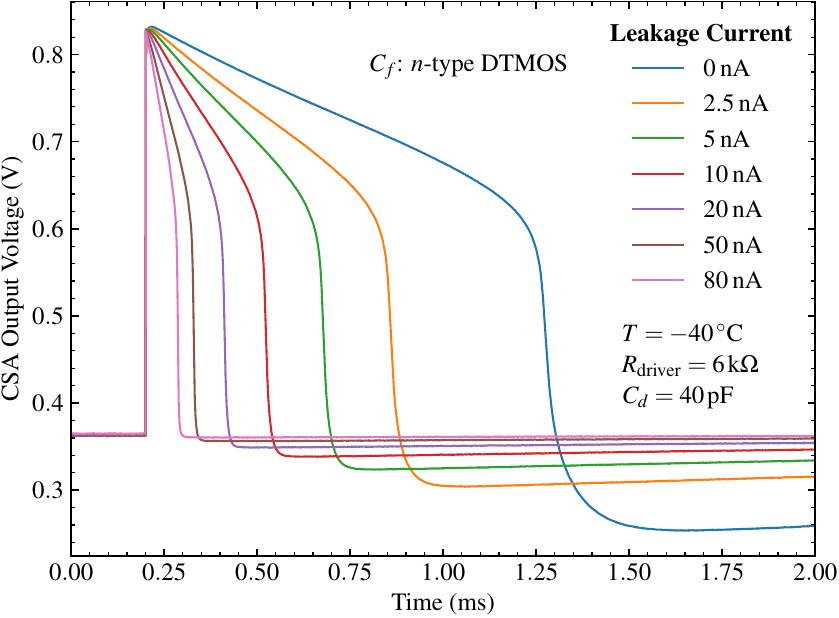}
        \caption{CSA output waveforms at $T=\SI{-40}{\celsius}$, varying the leakage current from 0 to \SI{80}{\nano\ampere}, above the nominal value expected during flight operation.}
        \label{fig:ch5-meas-CSA8_NDTMOS_m40C_Ileak}
    \end{subfigure}
    \hfill
    \begin{subfigure}{0.49\textwidth}
        \centering
        \includegraphics[width=\linewidth]{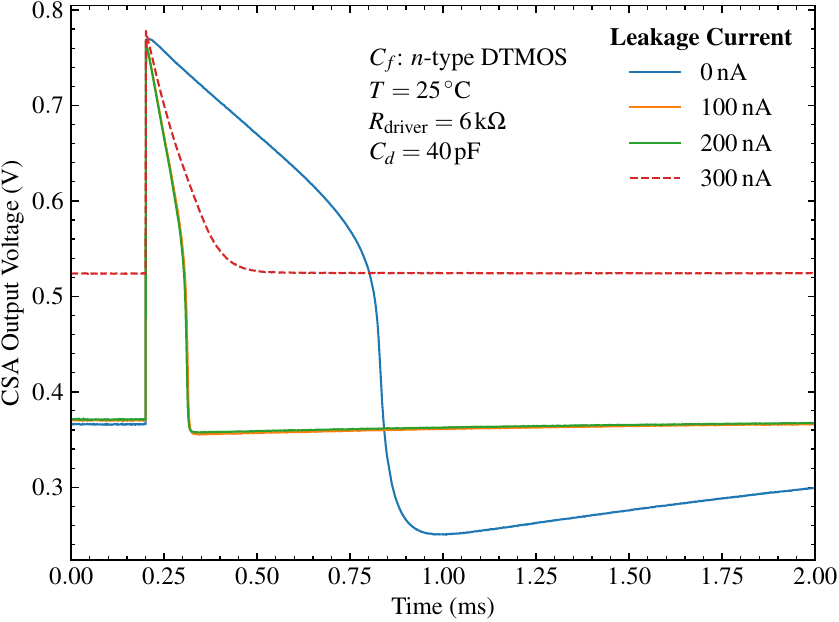}
        \caption{CSA output waveforms at $T=\SI{25}{\celsius}$, varying the leakage current from 0 to \SI{300}{\nano\ampere} to probe the maximum compensation capability of the Krummenacher feedback.}
        \label{fig:ch5-meas-CSA8_NDTMOS_30C_Ileak}
    \end{subfigure}
    \caption{Effect of leakage current on the CSA transient response for the $n$-type DTMOS feedback device at $R_{\text{driver}}=\SI{6}{\kilo\ohm}$ and $C_d=\SI{40}{\pico\farad}$. (\subref{fig:ch5-meas-CSA8_NDTMOS_m40C_Ileak}) At $T=\SI{-40}{\celsius}$, leakage current was swept from 0 to \SI{80}{\nano\ampere}, extending above the nominal value of \SI{2.5}{\nano\ampere} expected during flight. (\subref{fig:ch5-meas-CSA8_NDTMOS_30C_Ileak}) At room temperature, the leakage current was swept up to \SI{300}{\nano\ampere} to simulate on-ground operation and verify the Krummenacher feedback limit, with correct response guaranteed up to \SI{200}{\nano\ampere}.}
    \label{fig:ch5-meas-CSA8_NDTMOS_Ileak}
\end{figure}

\par
The effect of the leakage current above the nominal \SI{2.5}{\nano\ampere} case was investigated in two scenarios, as reported in~\Cref{fig:ch5-meas-CSA8_NDTMOS_Ileak}. 
The first, presented in~\Cref{fig:ch5-meas-CSA8_NDTMOS_m40C_Ileak}, examines the CSA behavior at the expected flight conditions (\SI{-40}{\celsius}), varying the leakage current up to \SI{80}{\nano\ampere}, well above the maximum value foreseen in flight. 
The second, shown in~\Cref{fig:ch5-meas-CSA8_NDTMOS_30C_Ileak}, explores the compensation limits of the Krummenacher feedback at room temperature, with leakage currents up to \SI{300}{\nano\ampere}. 
This configuration, combining a detector capacitance of \SI{40}{\pico\farad} with room temperature, represents the most demanding operating condition and provides a worst-case benchmark for assessing the ultimate compensation capability of the circuit during on-ground calibration.

\par
At the expected flight temperature of \SI{-40}{\celsius}, the effect of increasing the leakage current up to \SI{80}{\nano\ampere} is modest in terms of signal shape. 
The baseline shows only a small upward drift of about \SI{0.3}{\percent} between 0 and \SI{80}{\nano\ampere}. 
Since the signal amplitude is defined as the peak value relative to the baseline, the same proportional variation is reflected in the extracted amplitude, which therefore remains practically constant across the full leakage current range. 
The most pronounced impact is instead observed on the discharge dynamics: the discharge time decreases steadily from \SI{1103}{\micro\second} at $I_{\text{leak}}=\SI{0}{\nano\ampere}$ to below \SI{100}{\micro\second} at \SI{80}{\nano\ampere}, while the return-to-baseline time is progressively shortened and eventually suppressed at high leakage values.

\par
At room temperature, sweeping the leakage current from \SI{0}{\nano\ampere} to \SI{200}{\nano\ampere} produces only modest baseline and amplitude variations, remaining within the compensation range specified for on-ground calibration.
The baseline increases by about \SI{1.13}{\percent} at \SI{100}{\nano\ampere} and \SI{1.38}{\percent} at \SI{200}{\nano\ampere} with respect to \SI{0}{\nano\ampere}, while the amplitude decreases by \SI{0.93}{\percent} and \SI{1.16}{\percent}, respectively---indicating practically stable gain across the compensated range.
In the same conditions, the discharge time shortens from \SI{640}{\micro\second} at $I_{\text{leak}}=\SI{0}{\nano\ampere}$ to \SI{117}{\micro\second} at \SI{100}{\nano\ampere} and \SI{115}{\micro\second} at \SI{200}{\nano\ampere}, while the return-to-baseline time improves from \SI{444}{\micro\second} to \SI{357}{\micro\second}.
At \SI{300}{\nano\ampere}$\,$---included only to illustrate operation beyond the specified compensation limit---the baseline exhibits a large drift of about \SI{43}{\percent}, the amplitude drops by roughly \SI{36}{\percent}, and the transient dynamics deteriorate, confirming that \SI{200}{\nano\ampere} represents the practical upper limit of the compensation capability of the Krummenacher feedback network, in agreement with the simulations reported in~\Cref{sec:ch4-krummenacher_feedback}.

%   Shaper transient response
\subsection{Shaper transient response}

\par
The shaper transient response was evaluated at the output of channels~3 and~4, the only two in the prototype chip that provide direct access to the shaping stage output. 
As reported in~\Cref{tab:ch4-antares4_channels_info}, these channels implement as CSA feedback devices the $n$-type MOS and the $n$-type DTMOS, respectively, and are also equipped with the detector leakage current emulator feature. 
For this reason, the evaluation of the shaper performance was carried out both with and without detector capacitance, and under different leakage current conditions. 
In addition, a complete temperature scan from \SI{-40}{\celsius} to \SI{40}{\celsius} was performed to assess the shaping performance across temperature variations. 
From the shaper transient response analysis, the peaking time was determined as the instant corresponding to the minimum output voltage, i.e. the maximum excursion below the baseline.

\begin{figure}[t!]
    \centering
    \captionsetup[subfigure]{justification=justified,singlelinecheck=false}
    \begin{subfigure}{0.49\textwidth}
        \centering
        \includegraphics[width=\linewidth]{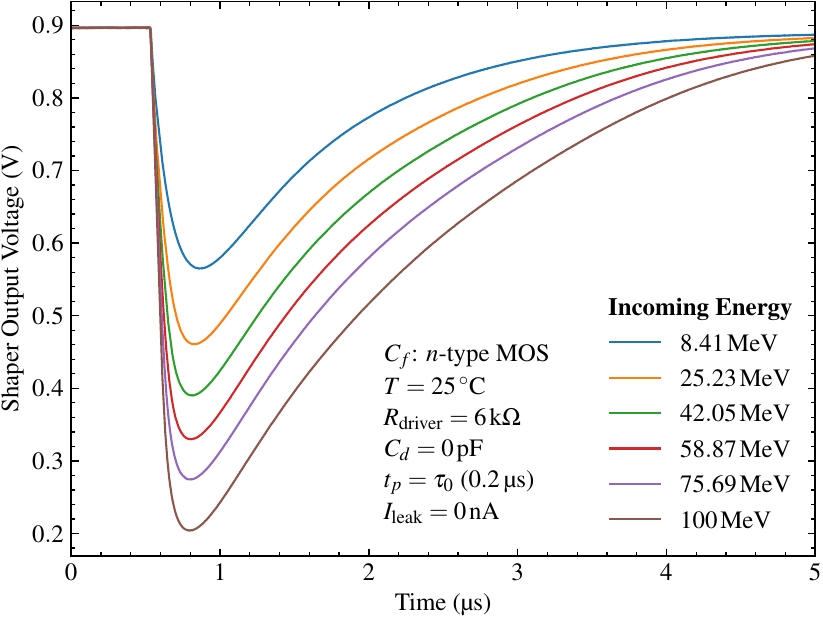}
        \caption{Shaper output for the $n$-type MOS CSA feedback device.}
        \label{fig:ch5-shaper_NMOS_30C}
    \end{subfigure}
    \hfill
    \begin{subfigure}{0.49\textwidth}
        \centering
        \includegraphics[width=\linewidth]{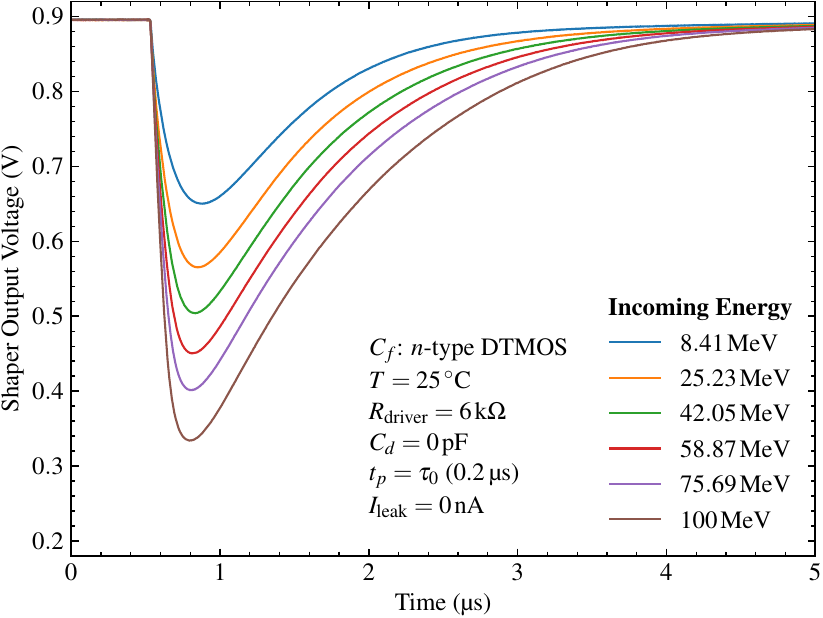}
        \caption{Shaper output for the $n$-type DTMOS CSA feedback device.}
        \label{fig:ch5-shaper_NDTMOS_30C}
    \end{subfigure}
    \caption{Shaper output waveforms measured at $T=\SI{25}{\celsius}$, $R_{\text{driver}}=\SI{6}{\kilo\ohm}$, $C_d=\SI{0}{\pico\farad}$, $t_p=\tau_0=\SI{0.2}{\micro\second}$, and $I_{\text{leak}}=\SI{0}{\nano\ampere}$. The incoming energy is varied between \SI{8.41}{\mega\electronvolt} and \SI{100}{\mega\electronvolt}, comparing the cases of the $n$-type MOS (\subref{fig:ch5-shaper_NMOS_30C}) and $n$-type DTMOS (\subref{fig:ch5-shaper_NDTMOS_30C}) CSA feedback devices.}
    \label{fig:ch5-shaper_MOS_vs_NDTMOS_30C}
\end{figure}

\begin{figure}[t!]
    \centering
    \captionsetup[subfigure]{justification=justified,singlelinecheck=false}
    \begin{subfigure}{0.484\textwidth}
        \centering
        \includegraphics[width=\linewidth]{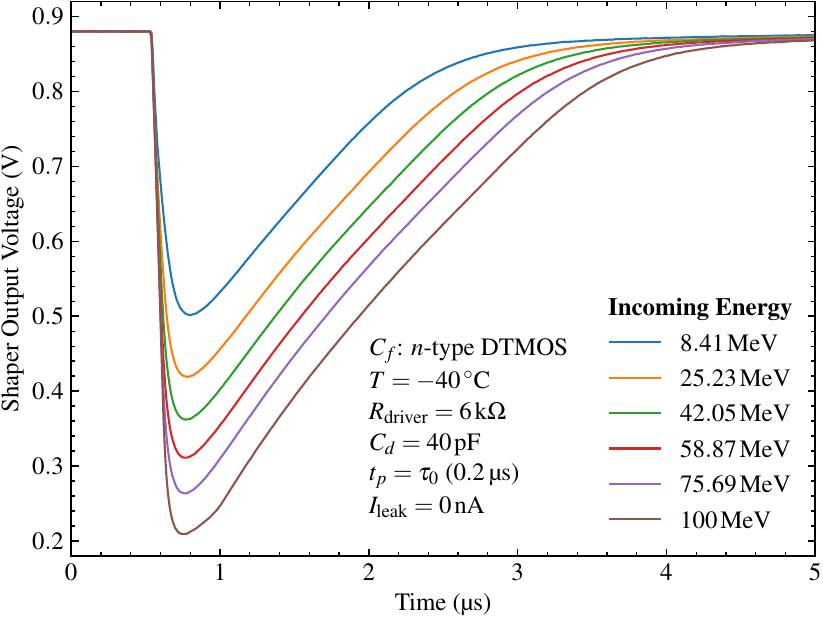}
        \caption{Output waveforms as a function of input energy between \SI{8.41}{\mega\electronvolt} and \SI{100}{\mega\electronvolt} at peaking time $t_{p,0}$.}
        \label{fig:ch5-shaper-NDTMOS-m40C-energy}
    \end{subfigure}
    \hfill
    \begin{subfigure}{0.496\textwidth}
        \centering
        \includegraphics[width=\linewidth]{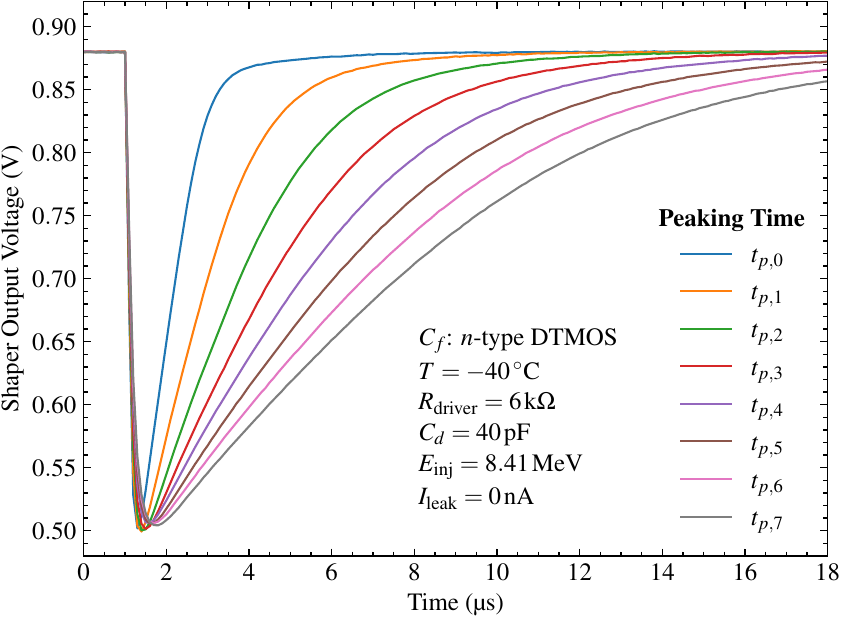}
        \caption{Output waveforms for a fixed input energy of \SI{8.41}{\mega\electronvolt} for different values of the peaking time $t_{p,0}$--$t_{p,7}$.}
        \label{fig:ch5-shaper-NDTMOS-m40C-tp}
    \end{subfigure}
    \caption{Transient response of the shaper output for the $n$-type DTMOS CSA feedback device at $T=\SI{-40}{\celsius}$ and $C_d=\SI{40}{\pico\farad}$. (\subref{fig:ch5-shaper-NDTMOS-m40C-energy}) Dependence on input energy at peaking time $t_{p,0}$. (\subref{fig:ch5-shaper-NDTMOS-m40C-tp}) Dependence on peaking time for a fixed input energy of \SI{8.41}{\mega\electronvolt}.}
    \label{fig:ch5-shaper-NDTMOS-m40C}
\end{figure}

\begin{figure}[t!]
    \centering
    \captionsetup[subfigure]{justification=justified,singlelinecheck=false}
    \begin{subfigure}{0.49\textwidth}
        \centering
        \includegraphics[width=\linewidth]{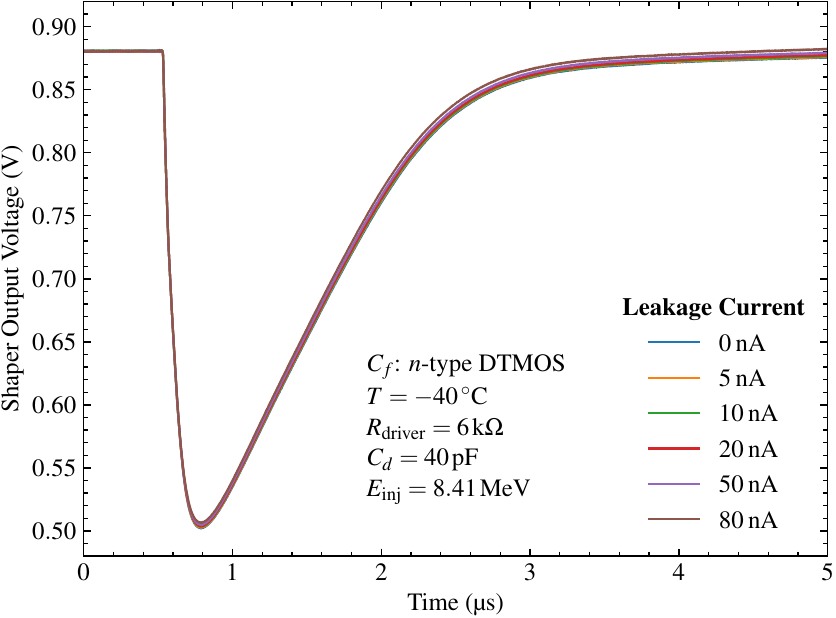}
        \caption{Shaper output for an input energy of \SI{8.41}{\mega\electronvolt} at $T=\SI{-40}{\celsius}$ and $C_d=\SI{40}{\pico\farad}$, for different detector leakage current values between \SI{0}{\nano\ampere} and \SI{80}{\nano\ampere}.}
        \label{fig:ch5-shaper-NDTMOS-Ileak}
    \end{subfigure}
    \hfill
    \begin{subfigure}{0.49\textwidth}
        \centering
        \includegraphics[width=\linewidth]{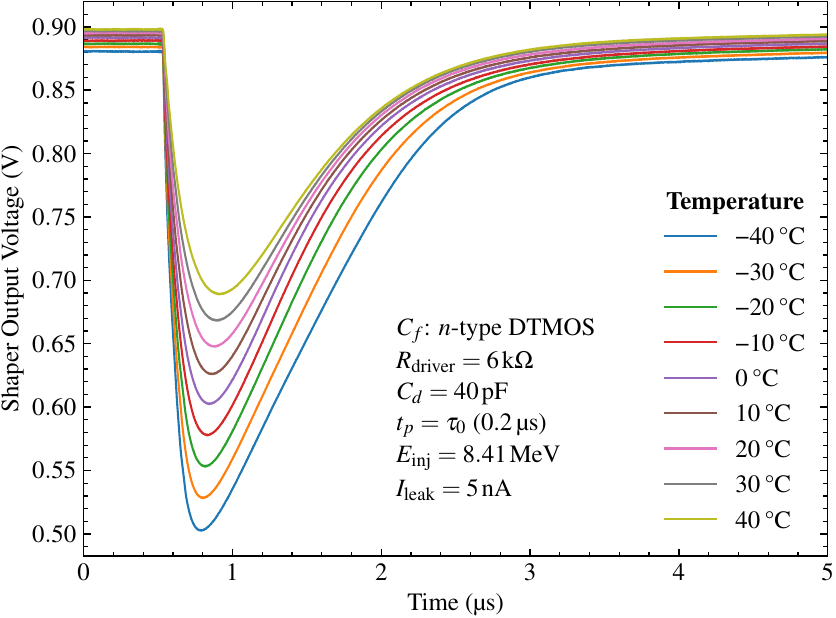}
        \caption{Shaper output for the same fixed input energy and $C_d=\SI{40}{\pico\farad}$, with leakage current $I_{\text{leak}}=\SI{5}{\nano\ampere}$, at different temperatures between \SI{-40}{\celsius} and \SI{+40}{\celsius}.}
        \label{fig:ch5-shaper-NDTMOS-temp}
    \end{subfigure}
    \caption{Shaper transient response for the $n$-type DTMOS CSA feedback device with $C_d=\SI{40}{\pico\farad}$, $R_{\text{driver}}=\SI{6}{\kilo\ohm}$. (\subref{fig:ch5-shaper-NDTMOS-Ileak}) Effect of leakage current at $T=\SI{-40}{\celsius}$. (\subref{fig:ch5-shaper-NDTMOS-temp}) Effect of temperature at $I_{\text{leak}}=\SI{5}{\nano\ampere}$.}
    \label{fig:ch5-shaper-NDTMOS-Ileak-temp}
\end{figure}

\par
The first evaluation was carried out for both configurations at room temperature, with no leakage current or detector capacitance, as shown in~\Cref{fig:ch5-shaper_MOS_vs_NDTMOS_30C}.
The measurement illustrates, for both feedback types, the shaper output waveforms as the input energy is varied from \SI{8.41}{\mega\electronvolt} to \SI{100}{\mega\electronvolt}, using the shortest selectable peaking time $t_{p,0}$, corresponding to a nominal value of \SI{200}{\nano\second}.
For the $n$-type MOS configuration, reported in~\Cref{fig:ch5-shaper_NMOS_30C}, the measured peaking times range from \SI{0.33}{\micro\second} at the lowest input energy down to \SI{0.26}{\micro\second} at the highest.
This corresponds to a reduction of about \SI{21}{\percent} across the energy range, with most of the variation occurring below \SI{60}{\mega\electronvolt}, while above this value the peaking time stabilizes around \SI{0.26}{\micro\second} with less than \SI{4}{\percent} fluctuation.
Compared to the nominal value of \SI{0.2}{\micro\second}, the measured peaking times are consistently longer by \SI{30}{\percent}--\SI{65}{\percent}, showing that the circuit provides a peaking time longer than expected, though with only weak dependence on the input energy in the high-energy range.

\par
For the $n$-type DTMOS case, shown in~\Cref{fig:ch5-shaper_NDTMOS_30C}, the measured peaking times decrease from \SI{0.35}{\micro\second} at \SI{8.41}{\mega\electronvolt} to \SI{0.27}{\micro\second} at \SI{100}{\mega\electronvolt}, corresponding to a variation of about \SI{23}{\percent} across the full energy range. 
Compared to the nominal value of \SI{0.2}{\micro\second}, the circuit provides peaking times consistently longer by \SI{35}{\percent} to \SI{75}{\percent}. 
The dependence on the input energy is more pronounced at low energies, while for $E_{\text{inj}} \geq \SI{75.69}{\mega\electronvolt}$ the peaking time stabilizes around \SI{0.27}{\micro\second}.

\par
It should be noted that these measurements do not represent the conditions expected during flight operation. 
The observed variation of the peaking time with respect to the nominal value does not pose a critical issue, as long as the range remains sufficient to select the optimal shaping constant for achieving the best energy resolution. 
In practice, only a single peaking time will be used during the experiment, while the availability of multiple shaping constants is mainly intended for laboratory characterization to study the noise behavior of the channel.

\par
A second evaluation of the shaper transient response was carried out for the $n$-type DTMOS feedback device at the expected flight temperature of \SI{-40}{\celsius}, with a detector capacitance of \SI{40}{\pico\farad}, as reported in~\Cref{fig:ch5-shaper-NDTMOS-m40C}. 
For this configuration, the shaper response shown in~\Cref{fig:ch5-shaper-NDTMOS-m40C-energy} was obtained by varying the input energy from \SI{8.41}{\mega\electronvolt} to \SI{100}{\mega\electronvolt}.
The measured peaking times decrease from \SI{0.25}{\micro\second} at the lowest energy (about \SI{25}{\percent} longer than the nominal \SI{0.20}{\micro\second}) to \SI{0.20}{\micro\second} at the highest energy, fully matching the nominal value. 
The overall variation remains within \SI{0.05}{\micro\second}, confirming that the shaper dynamics are only weakly dependent on the input energy and remain stable across the full range.

\par
The corresponding evaluation with a fixed input energy of \SI{8.41}{\mega\electronvolt} for varying peaking times $t_{p,0}$--$t_{p,7}$ is shown in~\Cref{fig:ch5-shaper-NDTMOS-m40C-tp}, with the corresponding simulation study at the typical process corner reported in~\Cref{fig:ch4-CR_RC_transient_ac_allpts} and discussed in \Cref{sec:ch5-shaper_noise_performance}.
The measured peaking times increase from \SI{0.25}{\micro\second} at $t_{p,0}$ to \SI{0.70}{\micro\second} at $t_{p,7}$, following a nearly linear progression across the selectable range. 
Although the absolute values are shorter than the design expectation of \SI{0.2}{\micro\second} to \SI{1.6}{\micro\second}, the relative scaling is preserved, confirming that the peaking-time programmability is correctly implemented. 
From the point of view of the experiment this does not represent a major issue, although the reduced absolute values may limit the flexibility for laboratory characterization of noise performance. 
Nevertheless, the implemented range still provides sufficient coverage for optimizing noise conditions under flight operation, while highlighting an aspect to be improved in the final ASIC design.

\par
The effect of the detector leakage current on the shaper output is shown in~\Cref{fig:ch5-shaper-NDTMOS-Ileak}, for a fixed input energy of \SI{8.41}{\mega\electronvolt} at $T=\SI{-40}{\celsius}$ and $C_d=\SI{40}{\pico\farad}$, varying the leakage current between \SI{0}{\nano\ampere} and \SI{80}{\nano\ampere}. 
As expected, the shaper response is essentially unaffected, since the leakage current acts primarily at the CSA level. 
The measured baselines remain stable within \SI{0.03}{\percent}, and the peak amplitudes show a variation below \SI{1.2}{\milli\volt} (\SI{0.5}{\percent}) across the full range. 
The peaking time is also preserved, with only a marginal change from \SI{0.25}{\micro\second} to \SI{0.26}{\micro\second}. 
These results confirm that, even when the leakage current is pushed well above the levels expected in flight conditions, the shaping stage performance remains unaffected in terms of baseline stability, signal amplitude, and timing. 

\begin{figure}[t!]
    \centering
    \captionsetup[subfigure]{justification=justified,singlelinecheck=false}
    \begin{subfigure}{0.49\textwidth}
        \centering
        \includegraphics[width=\linewidth]{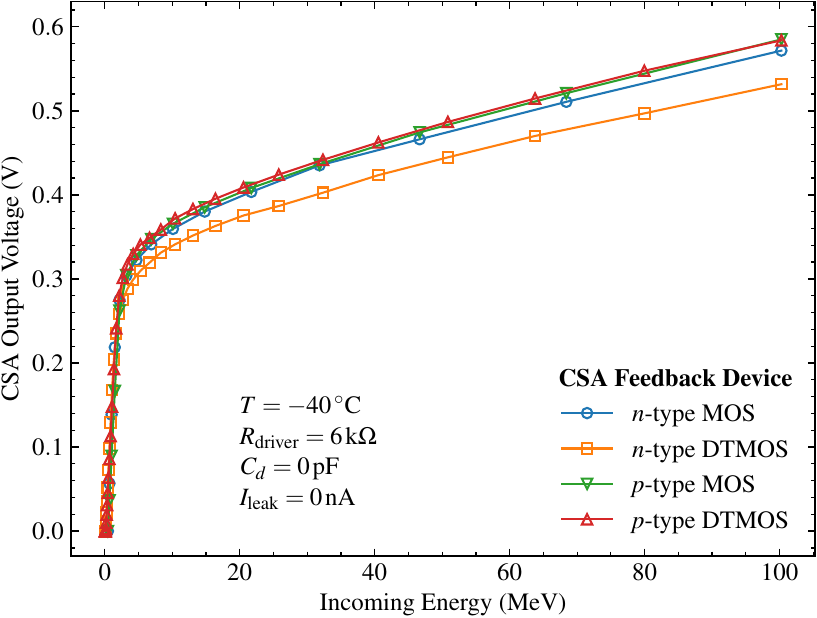}
        \caption{Case without detector capacitance for the four CSA feedback devices: \(n\)-type MOS and DTMOS, \(p\)-type MOS and DTMOS.}
        \label{fig:ch5-csa-transchar-0pf}
    \end{subfigure}
    \hfill
    \begin{subfigure}{0.49\textwidth}
        \centering
        \includegraphics[width=\linewidth]{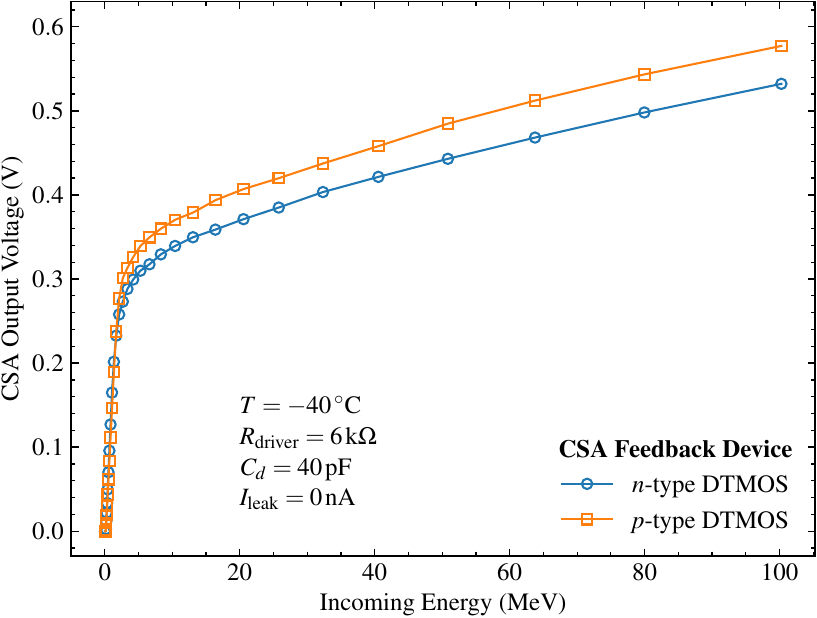}
        \caption{Case with detector capacitance \hbox{(\(C_d=\SI{40}{\pico\farad}\))} for the two DTMOS feedback configurations: \(n\)-type and \(p\)-type.}
        \label{fig:ch5-csa-transchar-40pf}
    \end{subfigure}
    \caption{CSA transcharacteristics at \(T=\SI{-40}{\celsius}\), with \(R_{\text{driver}}=\SI{6}{\kilo\ohm}\), \(I_{\text{leak}}=\SI{0}{\nano\ampere}\). (\subref{fig:ch5-csa-transchar-0pf}) Case without detector capacitance for the four feedback configurations (\(n\)-type MOS, \(n\)-type DTMOS, \(p\)-type MOS, \(p\)-type DTMOS). (\subref{fig:ch5-csa-transchar-40pf}) Case with detector capacitance \hbox{(\(C_d=\SI{40}{\pico\farad}\))} for the two DTMOS feedback configurations (\(n\)-type and \(p\)-type).}
    \label{fig:ch5-csa-transchar}
\end{figure}

\par
The effect of temperature on the shaper output was studied at a fixed input energy of \SI{8.41}{\mega\electronvolt} and leakage current of \SI{5}{\nano\ampere}, as reported in~\Cref{fig:ch5-shaper-NDTMOS-temp}. 
The baseline shows a gradual upward drift of about \SI{18}{\milli\volt} (\SI{2}{\percent}) across the full temperature span, while the peak amplitude decreases proportionally.
The peaking time exhibits an increase from \SI{0.25}{\micro\second} at \SI{-40}{\celsius} to \SI{0.36}{\micro\second} at \SI{+40}{\celsius}, corresponding to a relative variation of approximately \SI{45}{\percent}. 
This analysis therefore provides a verification of the correct functionality of the shaping stage across a wide temperature range. While proper operation must be guaranteed only under flight conditions, such measurements remain valuable to assess the channel behavior during laboratory characterization and pre-flight testing.

%   Channel input--output transcharacteristic
\subsection{Channel input--output transcharacteristic}

\par
In this section, the channel input--output transcharacteristics are presented, measured at the output of the CSA stage for all four feedback configurations (\(n\)-type MOS, \(n\)-type DTMOS, \(p\)-type MOS, and \(p\)-type DTMOS), and at the output of the shaper for the \(n\)-type DTMOS feedback topology. 
The transcharacteristics were obtained by recording the CSA output voltage as a function of the input energy, extracting both the baseline and the peak amplitude for each input value. 
For the shaper, the output voltage was evaluated at the effective peaking time corresponding to the specific configuration under measurement.

\begin{table}[t!]
    \centering
    \renewcommand{\arraystretch}{1.1}
    \setlength{\tabcolsep}{7pt}
    \begin{tabular}{@{}ccccccccc@{}}
        \toprule
        \multirow{2}{*}{\vspace{-0.3em}\shortstack{CSA Feedback\\Device}} & $C_d$ & $G_{\text{le}}$ & INL$_\text{le}$ & $G_{\text{he}}$ & INL$_\text{he}$ & $k$ & Kink & $\Delta V_{\text{out}}$ \\
        & (\si{\pico\farad}) & (\si{\micro\volt/\kilo\electronvolt}) & (\%) & (\si{\micro\volt/\kilo\electronvolt}) & (\%) & & (\si{\kilo\electronvolt}) & (\si{\milli\volt}) \\
        \midrule
        $n$-type MOS    & 0  & 259 & 0.15 & 2.03 & 4.59 & 127 & 1918 & 653 \\
        \textcolor{green!60!black}{$n$-type DTMOS}  & \textcolor{green!60!black}{0}  & \textcolor{green!60!black}{187} & \textcolor{green!60!black}{0.18} & \textcolor{green!60!black}{1.91} & \textcolor{green!60!black}{4.99} & \textcolor{green!60!black}{98} & \textcolor{green!60!black}{2023} & \textcolor{green!60!black}{532} \\
        $p$-type MOS    & 0  & 168 & 0.29 & 2.18 & 4.06 &  77 & 2711 & 639 \\
        $p$-type DTMOS  & 0  & 162 & 0.08 & 2.14 & 4.33 &  76 & 2502 & 584 \\
        \midrule
        \textcolor{green!60!black}{$n$-type DTMOS}  & \textcolor{green!60!black}{40} & \textcolor{green!60!black}{180} & \textcolor{green!60!black}{1.40} & \textcolor{green!60!black}{1.94} & \textcolor{green!60!black}{5.12} & \textcolor{green!60!black}{93} & \textcolor{green!60!black}{2081} & \textcolor{green!60!black}{532} \\
        $p$-type DTMOS  & 40 & 156 & 0.50 & 2.11 & 5.01 &  74 & 2576 & 577 \\
        \bottomrule
    \end{tabular}
    \caption{Performance parameters extracted from the CSA input--output transcharacteristics for $n$- and $p$-type MOS and DTMOS feedback devices, evaluated at \SI{-40}{\celsius} with $R_{\text{driver}}=\SI{6}{\kilo\ohm}$, $I_{\text{leak}}=\SI{0}{\nano\ampere}$. The table reports the low- and high-energy gain ($G_{\text{le}}$, $G_{\text{he}}$), index of nonlinearity (INL), compression factor $k$, kink, and output voltage swing $\Delta V_{\text{out}}$ for two values of detector capacitance ($C_d=\SI{0}{\pico\farad}$ and $C_d=\SI{40}{\pico\farad}$).}
    \label{tab:ch5-csa-transchar-summary}
\end{table}

%   Transcharacteristic evaluated at the CSA output
\subsubsection{Transcharacteristic evaluated at the CSA output}

\begin{figure}[t!]
    \centering
    \captionsetup[subfigure]{justification=justified,singlelinecheck=false}
    \begin{subfigure}{0.49\textwidth}
        \centering
        \includegraphics[width=\linewidth]{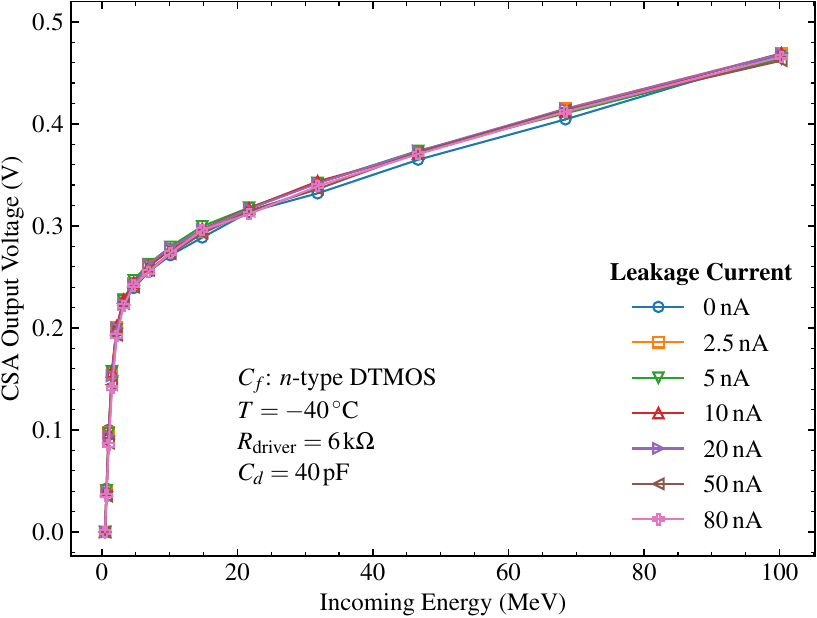}
        \caption{Variation of the leakage current between \SI{0}{\nano\ampere} and \SI{80}{\nano\ampere} at \(T=\SI{-40}{\celsius}\).}
        \label{fig:ch5-csa-transchar-Ileak-m40C}
    \end{subfigure}
    \hfill
    \begin{subfigure}{0.49\textwidth}
        \centering
        \includegraphics[width=\linewidth]{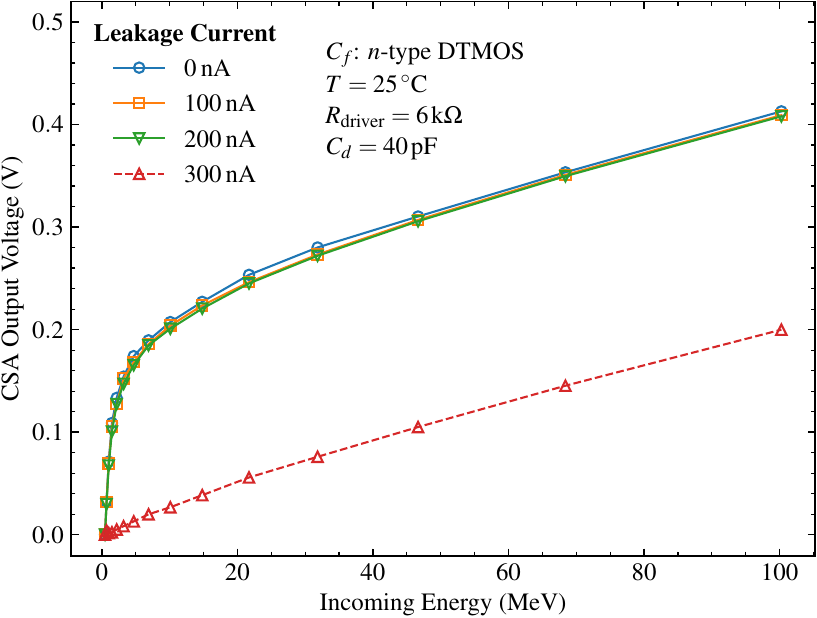}
        \caption{Evaluation of the compensation capability up to \SI{300}{\nano\ampere} at \(T=\SI{25}{\celsius}\).}
        \label{fig:ch5-csa-transchar-Ileak-25C}
    \end{subfigure}
    \caption{Effect of detector leakage current on the CSA transcharacteristics for the \(n\)-type DTMOS feedback device, with \(C_d=\SI{40}{\pico\farad}\), \(R_{\text{driver}}=\SI{6}{\kilo\ohm}\). (\subref{fig:ch5-csa-transchar-Ileak-m40C}) Analysis at \(-40^{\circ}\text{C}\) with leakage current varied from \SI{0}{\nano\ampere} to \SI{80}{\nano\ampere}. (\subref{fig:ch5-csa-transchar-Ileak-25C}) Evaluation at room temperature (\SI{25}{\celsius}) demonstrating the compensation capability of the Krummenacher feedback circuit up to \SI{200}{\nano\ampere}.}
    \label{fig:ch5-csa-transchar-Ileak}
\end{figure}

\par
The first evaluation of the CSA transcharacteristic was performed to compare the four feedback configurations without detector capacitance at \SI{-40}{\celsius}, as shown in~\Cref{fig:ch5-csa-transchar-0pf}. 
Moreover, the performance evaluation with the detector capacitance of \SI{40}{\pico\farad} was carried out for the two DTMOS configurations, as reported in~\Cref{fig:ch5-csa-transchar-40pf}. 
A comprehensive summary of the performance parameters extracted from the transcharacteristics is provided in~\Cref{tab:ch5-csa-transchar-summary}, which includes the low-energy gain $G_{\text{le}}$, the high-energy gain $G_{\text{he}}$, the index of nonlinearity INL, the compression factor $k$, the kink energy, and the output voltage swing $\Delta V_{\text{out}}$, for both cases with and without detector capacitance.

\par
The comparison between the different feedback devices highlights the superior performance of the $n$-type DTMOS configuration. 
At $C_d=\SI{0}{\pico\farad}$ the $n$-type DTMOS exhibits a low-energy gain about \SI{15}{\percent} higher than the $p$-type counterpart, together with a compression factor nearly \SI{30}{\percent} larger. 
The slightly lower high-energy gain (about \SI{10}{\percent} smaller) is a positive feature, as it reflects the stronger dynamic compression that extends the usable range without requiring a larger output swing. 
By contrast, the high-energy gain, although still valuable, is less critical, while the maximization of the low-energy gain remains the primary objective, as it directly limits the achievable energy resolution in the 10--\SI{100}{\kilo\electronvolt} x-ray detection range.

\par
The swing of the $n$-type DTMOS is reduced by \SI{9}{\percent}, which is advantageous as it relaxes the headroom requirements. 
At $C_d=\SI{40}{\pico\farad}$ the trends are confirmed, with the $n$-type DTMOS retaining a \SI{15}{\percent} advantage in low-energy gain and a \SI{25}{\percent} higher compression factor, while maintaining practically unchanged output swing. 
The index of nonlinearity at high energy is comparable between the two polarities, whereas at low energy the $p$-type device shows a slightly smaller value, though this marginal benefit does not outweigh the overall advantages of the $n$-type. 

\begin{table}[t!]
    \centering
    \renewcommand{\arraystretch}{1.1}
    \setlength{\tabcolsep}{7pt}
    \begin{tabular}{@{}ccccccccc@{}}
        \toprule
        \multicolumn{1}{c}{$I_{\text{leak}}$} &
        \multicolumn{1}{c}{$T$} &
        \multicolumn{1}{c}{$G_{\text{le}}$} &
        \multicolumn{1}{c}{INL$_\text{le}$} &
        \multicolumn{1}{c}{$G_{\text{he}}$} &
        \multicolumn{1}{c}{INL$_\text{he}$} &
        \multicolumn{1}{c}{$k$} &
        \multicolumn{1}{c}{Kink} &
        \multicolumn{1}{c}{$\Delta V_{\text{out}}$} \\
        \multicolumn{1}{c}{(\si{\nano\ampere})} &
        \multicolumn{1}{c}{(\si{\celsius})} &
        \multicolumn{1}{c}{(\si{\micro\volt/\kilo\electronvolt})} &
        \multicolumn{1}{c}{(\%)} &
        \multicolumn{1}{c}{(\si{\micro\volt/\kilo\electronvolt})} &
        \multicolumn{1}{c}{(\%)} &
        & 
        \multicolumn{1}{c}{(\si{\kilo\electronvolt})} &
        \multicolumn{1}{c}{(\si{\milli\volt})} \\
        \midrule
        \textbf{0}   & \textbf{-40} & \textbf{220} & \textbf{1.29} & \textbf{1.96} & \textbf{1.45} & \textbf{112} & \textbf{1724} & \textbf{535} \\
        2.5 & -40 & 186 & 0.36 & 1.89 & 3.58 &  99 & 2013 & 529 \\
        5   & -40 & 185 & 1.04 & 1.79 & 4.51 & 104 & 2046 & 525 \\
        10  & -40 & 183 & 0.27 & 1.86 & 3.14 &  98 & 2047 & 528 \\
        20  & -40 & 180 & 1.90 & 1.88 & 4.17 &  96 & 2069 & 526 \\
        50  & -40 & 173 & 0.87 & 1.86 & 4.91 &  93 & 2123 & 520 \\
        80  & -40 & 166 & 1.13 & 1.88 & 4.98 &  88 & 2184 & 519 \\
        \midrule
        \textbf{0}   & \textbf{25} & \textbf{180} & \textbf{2.40} & \textbf{1.96} & \textbf{3.21} & \textbf{92} & \textbf{1701} & \textbf{462} \\
        100 & 25 & 147 & 0.82 & 2.01 & 4.16 &  73 & 1933 & 450 \\
        200 & 25 & 148 & 0.99 & 2.01 & 4.12 &  73 & 1921 & 449 \\
        \bottomrule
    \end{tabular}
    \caption{Performance parameters extracted from the CSA input--output transcharacteristics of the $n$-type DTMOS feedback device with $C_d=\SI{40}{\pico\farad}$ under different leakage currents at $T=\SI{-40}{\celsius}$ and $T=\SI{25}{\celsius}$. The table reports the low- and high-energy gain ($G_{\text{le}}$, $G_{\text{he}}$), INL, compression factor $k$, kink energy, and output swing $\Delta V_{\text{out}}$.}
    \label{tab:ch5-csa-transchar-Ileak-summary}
\end{table}

\par
The $n$-type DTMOS achieves the most favorable combination of high gain, strong compression, reduced output swing, and acceptable linearity, which makes it the preferred feedback solution for the CSA.
The observed behavior is in line with the expected characteristics and with the considerations already discussed for the $n$-type DTMOS feedback device in~\Cref{chapter4}.
Overall, although the standard NMOS feedback device may appear advantageous if evaluated solely in terms of low-energy gain and compression factor, the $n$-type DTMOS achieves a more favorable overall trade-off.  
In addition to providing sufficiently high low-energy gain and strong compression, it also reduces the output swing and relaxes headroom requirements, mitigating the risk of saturation and preserving correct CSA functionality across varying operating conditions, including temperature, detector capacitance, and leakage current.
This combination makes it the preferred feedback solution for the CSA, as it extends the usable dynamic range without compromising the energy resolution in the low-energy x-ray band.

\par
It should be noted that these transcharacteristics and evaluations have been performed at the output of the CSA, where the overall gain appears smaller since the shaping stage introduces additional amplification.
This approach is dictated by the architecture of the prototype chip, in which all four CSA feedback devices can be compared only at the CSA level, while the shaper output is available for the $n$-type MOS and DTMOS configurations only.

\par
Following from the analysis of the CSA transient response under different leakage current conditions reported in~\Cref{sec:ch5-CSA_transient_response}, the effect of the leakage current on the CSA transcharacteristics was evaluated for the $n$-type DTMOS feedback device.
The results of \Cref{fig:ch5-csa-transchar-Ileak} and the corresponding summary in \Cref{tab:ch5-csa-transchar-Ileak-summary} provide a comprehensive evaluation of the effect of detector leakage current on the CSA transcharacteristic for the $n$-type DTMOS feedback device.
At $T=\SI{-40}{\celsius}$, the transcharacteristic remains stable up to $I_{\text{leak}}=\SI{80}{\nano\ampere}$, well above the maximum leakage current expected under flight conditions, thus validating the robustness of the proposed Krummenacher feedback scheme.
As the leakage current increases from \SI{0}{\nano\ampere} to \SI{80}{\nano\ampere}, the low-energy gain decreases by approximately \SI{25}{\percent}, the compression factor reduces from about 112 to 88, and the output swing decreases by only \SI{3}{\percent}, while the high-energy gain and the kink position remain substantially unchanged.

\begin{figure}[t!]
    \centering
    \captionsetup[subfigure]{justification=justified,singlelinecheck=false}
    \begin{subfigure}{0.49\textwidth}
        \centering
        \includegraphics[width=\linewidth]{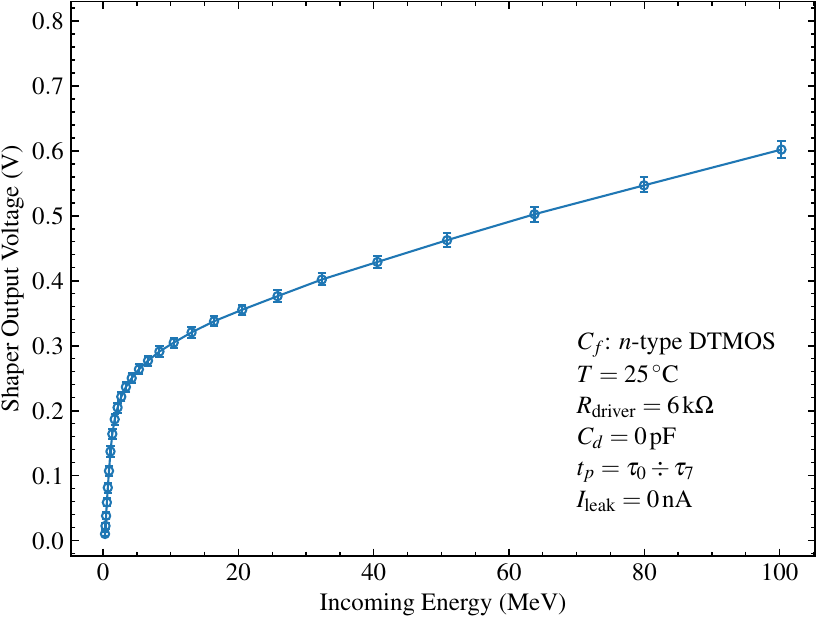}
        \caption{Evaluation of the shaper transcharacteristics at \(T=\SI{25}{\celsius}\) with $C_d=\SI{0}{\pico\farad}$ and $I_{\text{leak}}=\SI{0}{\nano\ampere}$, varying the peaking time between $\tau_0$ and $\tau_7$.}
        \label{fig:ch5-shaper-transchar-25C}
    \end{subfigure}
    \hfill
    \begin{subfigure}{0.49\textwidth}
        \centering
        \includegraphics[width=\linewidth]{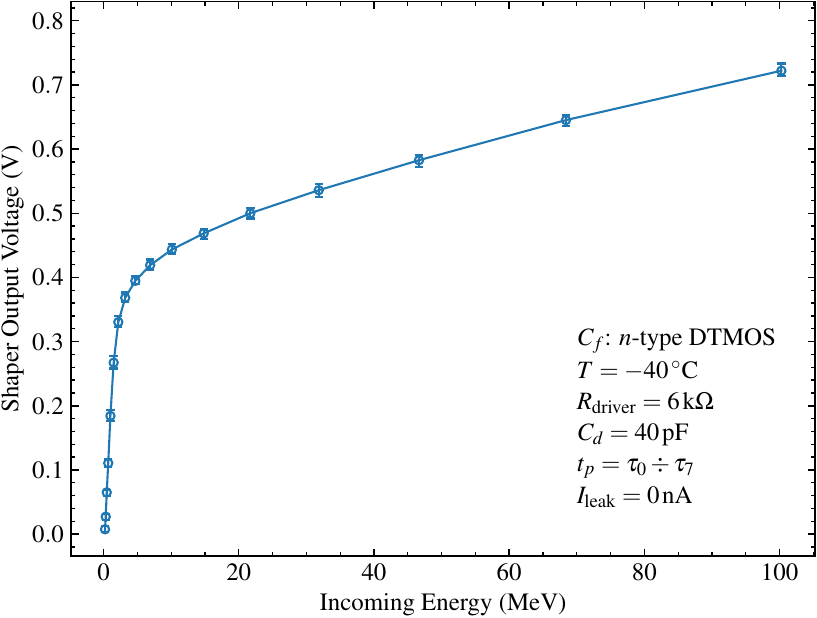}
        \caption{Evaluation of the shaper transcharacteristics at \(T=\SI{-40}{\celsius}\) with $C_d=\SI{40}{\pico\farad}$ and $I_{\text{leak}}=\SI{0}{\nano\ampere}$, for peaking times between $\tau_0$ and $\tau_7$.}
        \label{fig:ch5-shaper-transchar-m40C}
    \end{subfigure}
    \caption{Input--output transcharacteristics of the shaping stage for the $n$-type DTMOS feedback configuration, obtained by sweeping the peaking time setting from $\tau_0$ to $\tau_7$. The results are reported for two conditions: (\subref{fig:ch5-shaper-transchar-25C}) at room temperature without detector capacitance and (\subref{fig:ch5-shaper-transchar-m40C}) at \SI{-40}{\celsius} with a detector capacitance of \SI{40}{\pico\farad}.}
    \label{fig:ch5-shaper-transchar-tau}
\end{figure}

\begin{table}[t!]
    \centering
    \renewcommand{\arraystretch}{1}
    \setlength{\tabcolsep}{6pt}
    \begin{tabular}{@{}cccccccccc@{}}
        \toprule
        \multirow{2}{*}{\vspace{0em}\shortstack{$t_p$}} &
        \multicolumn{1}{c}{$T$} & \multicolumn{1}{c}{$C_d$} &
        \multicolumn{1}{c}{$G_{\text{le}}$} & \multicolumn{1}{c}{INL$_\text{le}$} &
        \multicolumn{1}{c}{$G_{\text{he}}$} & \multicolumn{1}{c}{INL$_\text{he}$} &
        \multicolumn{1}{c}{$k$} & \multicolumn{1}{c}{Kink} &
        \multicolumn{1}{c}{$\Delta V_{\text{out}}$} \\
        & \multicolumn{1}{c}{(\si{\celsius})} & \multicolumn{1}{c}{(\si{\pico\farad})} &
        \multicolumn{1}{c}{(\si{\micro\volt/\kilo\electronvolt})} & \multicolumn{1}{c}{(\%)} &
        \multicolumn{1}{c}{(\si{\micro\volt/\kilo\electronvolt})} & \multicolumn{1}{c}{(\%)} &
        \multicolumn{1}{c}{} & \multicolumn{1}{c}{(\si{\kilo\electronvolt})} &
        \multicolumn{1}{c}{(\si{\milli\volt})} \\
        \midrule
        \textbf{\boldmath$\tau_0$} & \textbf{25}  & \textbf{0}  & \textbf{185} & \textbf{1.22} & \textbf{3.02} & \textbf{4.34} &  \textbf{61} & \textbf{1915} & \textbf{611} \\
        $\tau_1$ & 25  & 0  & 186 & 1.09 & 3.07 & 3.19 &  61 & 1920 & 616 \\
        $\tau_2$ & 25  & 0  & 184 & 0.54 & 2.97 & 4.44 &  62 & 1974 & 614 \\
        $\tau_3$ & 25  & 0  & 180 & 0.33 & 2.96 & 3.90 &  61 & 1988 & 607 \\
        $\tau_4$ & 25  & 0  & 174 & 1.08 & 2.96 & 4.25 &  59 & 2015 & 598 \\
        $\tau_5$ & 25  & 0  & 172 & 0.68 & 2.99 & 3.45 &  57 & 2013 & 600 \\
        $\tau_6$ & 25  & 0  & 168 & 0.63 & 2.93 & 4.81 &  57 & 2070 & 594 \\
        $\tau_7$ & 25  & 0  & 168 & 1.03 & 3.01 & 3.50 &  56 & 2053 & 600 \\
        \midrule
        \textbf{\boldmath$\tau_0$} & \textbf{-40} & \textbf{40} & \textbf{245} & \textbf{1.91} & \textbf{2.80} & \textbf{3.93} &  \textbf{88} & \textbf{2076} & \textbf{728} \\
        $\tau_1$ & -40 & 40 & 249 & 2.53 & 2.79 & 3.60 &  89 & 2059 & 734 \\
        $\tau_2$ & -40 & 40 & 247 & 3.15 & 2.80 & 3.93 &  88 & 2067 & 734 \\
        $\tau_3$ & -40 & 40 & 244 & 2.81 & 2.70 & 5.34 &  90 & 2062 & 725 \\
        $\tau_4$ & -40 & 40 & 243 & 2.95 & 2.70 & 4.90 &  90 & 2075 & 719 \\
        $\tau_5$ & -40 & 40 & 239 & 2.72 & 2.72 & 5.05 &  88 & 2066 & 723 \\
        $\tau_6$ & -40 & 40 & 241 & 3.00 & 2.82 & 3.28 &  85 & 2107 & 722 \\
        $\tau_7$ & -40 & 40 & 242 & 3.08 & 2.79 & 3.44 &  87 & 2090 & 724 \\
        \bottomrule
    \end{tabular}
    \caption{Performance parameters extracted from the input--output transcharacteristics evaluated at the shaper output for the $n$-type DTMOS feedback device as a function of peaking time setting ($\tau_0$--$\tau_7$). Measurements were performed at $T=\SI{25}{\celsius}$ with $C_d=\SI{0}{\pico\farad}$ and at $T=\SI{-40}{\celsius}$ with $C_d=\SI{40}{\pico\farad}$, both with $I_{\text{leak}}=\SI{0}{\nano\ampere}$. Reported parameters are the low- and high-energy gain ($G_{\text{le}}$, $G_{\text{he}}$), index of nonlinearity (INL), compression factor $k$, kink energy, and output swing $\Delta V_{\text{out}}$.}
    \label{tab:ch5-shaper-transchar-summary}
\end{table}

\par
At room temperature (\SI{25}{\celsius}), the Krummenacher feedback circuit demonstrates compensation capability up to about \SI{200}{\nano\ampere}, as anticipated from the transient analysis, with the transcharacteristic becoming strongly degraded at \SI{300}{\nano\ampere}.
Within the compensated range, the low-energy gain is reduced by about \SI{20}{\percent}, while the high-energy gain, kink energy, and output swing remain relatively stable.
These results confirm that the $n$-type DTMOS device maintains acceptable performance under realistic leakage current conditions, with the expected compensation limit around \SI{200}{\nano\ampere}, thereby ensuring correct CSA operation even during on-ground calibration activities at room temperature.

%   Transcharacteristic evaluated at the shaper output
\subsubsection{Transcharacteristic evaluated at the shaper output}

\par
Following the analysis of the CSA transcharacteristics, the evaluation was extended to the output of the shaping stage, focusing on the $n$-type DTMOS feedback device. 
While at the CSA level the study addressed the impact of detector leakage current both at nominal temperature and at room temperature, the investigation at the shaper output was instead devoted to assessing the influence of the peaking time setting and the variation of the operating temperature.

\par
As a first step in the validation of the shaper transcharacteristic, a comparison was performed between measurements at room temperature (\SI{25}{\celsius}) without detector capacitance, shown in~\Cref{fig:ch5-shaper-transchar-25C}, and at \SI{-40}{\celsius} with a detector capacitance of \SI{40}{\pico\farad}, reported in~\Cref{fig:ch5-shaper-transchar-m40C}, while varying the peaking time setting from $\tau_0$ to $\tau_7$.
In both plots, the error bars indicate the maximum variation of the shaper output voltage at a given input energy across the different peaking time settings.

\par
The comparison between the transcharacteristics at the shaper output is reported in~\Cref{tab:ch5-shaper-transchar-summary} and shows that, while the overall shape is preserved, the low-temperature configuration with detector capacitance leads to a clear increase in low-energy gain (up to \SI{30}{\percent} higher) and a stronger compression factor, at the expense of a modest reduction in high-energy gain.  
At \SI{25}{\celsius} with $C_d=\SI{0}{\pico\farad}$ (\Cref{fig:ch5-shaper-transchar-25C}) the low-energy gain remains in the range \SI{170}{\micro\volt/\kilo\electronvolt}--\SI{185}{\micro\volt/\kilo\electronvolt}, while the output swing is about \SI{600}{\milli\volt}.

\par
In contrast, at \SI{-40}{\celsius} with $C_d=\SI{40}{\pico\farad}$ (\Cref{fig:ch5-shaper-transchar-m40C}) the low-energy gain increases by nearly \SI{30}{\percent}, reaching \SI{239}{\micro\volt/\kilo\electronvolt}--\SI{249}{\micro\volt/\kilo\electronvolt}, and the compression factor almost doubles, with values close to 90. 
The high-energy gain decreases slightly, from about \SI{3.0}{\micro\volt/\kilo\electronvolt} at room temperature to \SI{2.7}{\micro\volt/\kilo\electronvolt} at low temperature, while the output swing increases to approximately \SI{720}{\milli\volt}. 
The index of nonlinearity remains contained below \SI{3}{\percent} at low energy and \SI{5.3}{\percent} at high energy in both cases, confirming good linearity of the shaping stage under varying temperature and detector capacitance. 
Notably, the behavior observed at \SI{-40}{\celsius} with $C_d=\SI{40}{\pico\farad}$ is consistent with the simulation results discussed in~\Cref{sec:ch4-feedback_device_dimensioning}, although that earlier study was carried out assuming an ideal shaping stage.

\begin{figure}[tbp]
    \centering
    \includegraphics[width=.75\textwidth]{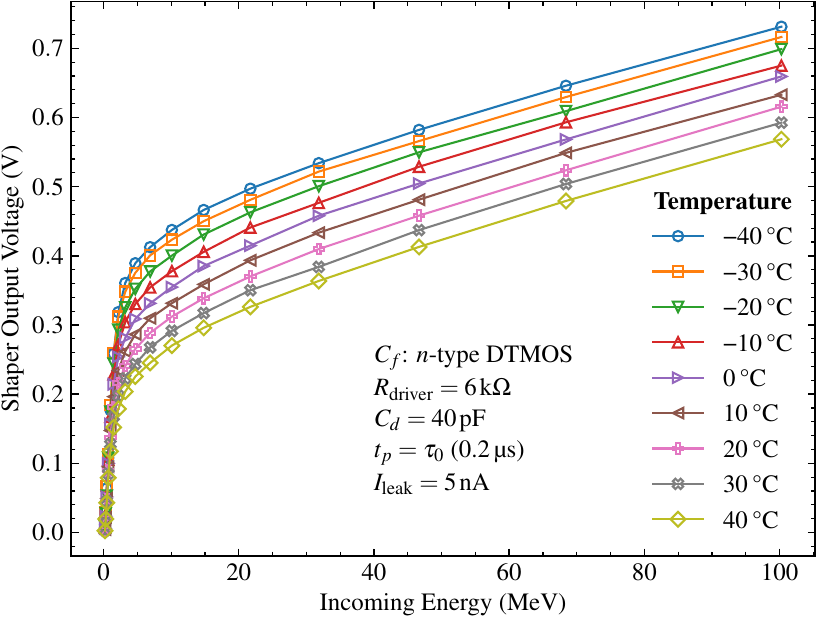}
    \caption{Input--output transcharacteristics at the shaper output for the $n$-type DTMOS feedback device with $C_d=\SI{40}{\pico\farad}$, $R_{\text{driver}}=\SI{6}{\kilo\ohm}$, $t_p=\tau_0$ (\SI{0.2}{\micro\second}), and $I_{\text{leak}}=\SI{5}{\nano\ampere}$. The curves show the effect of temperature variation between \SI{-40}{\celsius} and \SI{40}{\celsius}.}
    \label{fig:ch5-shaper-transchar-temp}
\end{figure}

\par
Nevertheless, the most relevant performance figures, namely the low-energy gain and the compression factor, are consistent with expectations. 
In particular, the gain in the \hbox{low-energy} range remains very close to the required \SI{250}{\micro\volt/\kilo\electronvolt} under the expected operating conditions ($T=\SI{-40}{\celsius}$ and $C_d=\SI{40}{\pico\farad}$), while the compression factor stays consistently above 80. 
These results confirm the validity of the design choices adopted for the $n$-type DTMOS feedback device, and indicate that this feedback topology is the most suitable solution for the CSA in the flight ASIC implementation.

\par
The variation of the transcharacteristic parameters with the peaking time setting indicates only moderate changes across $\tau_0$--$\tau_7$.  
At \SI{25}{\celsius}, the low-energy gain decreases by about \SI{9}{\percent} with respect to $\tau_0$, while the compression factor reduces by \SI{10}{\percent}.
The output swing remains nearly constant around \SI{600}{\milli\volt}.  
At \SI{-40}{\celsius} with $C_d=\SI{40}{\pico\farad}$, the low-energy gain changes by less than \SI{4}{\percent} relative to $\tau_0$, while the compression factor and output swing stay stable within \SI{5}{\percent} of the shortest peaking time setting.

\begin{table}[t!]
    \centering
    \renewcommand{\arraystretch}{1.1}
    \setlength{\tabcolsep}{6pt}
    \begin{tabular}{@{}cccccccc@{}}
        \toprule
        \multicolumn{1}{c}{$T$} &
        \multicolumn{1}{c}{$G_{\text{le}}$} & \multicolumn{1}{c}{INL$_\text{le}$} &
        \multicolumn{1}{c}{$G_{\text{he}}$} & \multicolumn{1}{c}{INL$_\text{he}$} &
        \multicolumn{1}{c}{$k$} & \multicolumn{1}{c}{Kink} &
        \multicolumn{1}{c}{$\Delta V_{\text{out}}$} \\
        (\si{\celsius}) &
        (\si{\micro\volt/\kilo\electronvolt}) & (\%) &
        (\si{\micro\volt/\kilo\electronvolt}) & (\%) &
        & (\si{\kilo\electronvolt}) & (\si{\milli\volt}) \\
        \midrule
        \textbf{-40} & \textbf{229} & \textbf{2.08} & \textbf{2.74} & \textbf{5.29} & \textbf{83} & \textbf{2196} & \textbf{716} \\
        -30 & 245 & 1.79 & 2.89 & 3.57 & 85 & 1977 & 716 \\
        -20 & 173 & 4.20 & 2.89 & 3.56 & 60 & 2575 & 699 \\
        -10 & 181 & 3.13 & 2.92 & 4.08 & 62 & 2367 & 675 \\
         0  & 188 & 1.79 & 2.98 & 3.93 & 63 & 2169 & 659 \\
        10  & 196 & 1.75 & 3.02 & 3.82 & 65 & 1986 & 641 \\
        20  & 198 & 2.23 & 2.98 & 3.88 & 66 & 1835 & 629 \\
        30  & 203 & 2.00 & 3.02 & 3.97 & 67 & 1661 & 614 \\
        40  & 206 & 1.83 & 3.01 & 4.09 & 68 & 1500 & 597 \\
        \bottomrule
    \end{tabular}
    \caption{Performance parameters extracted from the input--output transcharacteristics evaluated at the shaper output for the $n$-type DTMOS feedback device with $C_d=\SI{40}{\pico\farad}$, $I_{\text{leak}}=\SI{5}{\nano\ampere}$ and $t_p=\tau_0$ (\SI{0.2}{\micro\second}) as a function of temperature. Reported parameters are the low- and high-energy gain ($G_{\text{le}}$, $G_{\text{he}}$), index of nonlinearity (INL), compression factor $k$, kink energy, and output voltage swing $\Delta V_{\text{out}}$.}
    \label{tab:ch5-shaper-transchar-temp}
\end{table}

\par
Finally, the effect of temperature variation between \SI{-40}{\celsius} and \SI{40}{\celsius} on the shaper transcharacteristic was evaluated, as shown in~\Cref{fig:ch5-shaper-transchar-temp}, with the corresponding performance parameters summarized in~\Cref{tab:ch5-shaper-transchar-temp}.  
At \SI{-40}{\celsius}, a comparison between the two reference cases with $I_{\text{leak}}=\SI{0}{\nano\ampere}$ and $I_{\text{leak}}=\SI{5}{\nano\ampere}$ shows that the low-energy gain decreases by about \SI{6.5}{\percent} (245 to \SI{229}{\micro\volt/\kilo\electronvolt}), while the high-energy gain decreases slightly by \SI{2.1}{\percent} (2.80 to \SI{2.74}{\micro\volt/\kilo\electronvolt}).
The compression factor $k$ is reduced by \SI{5.7}{\percent}, accompanied by a kink energy shift of \SI{5.8}{\percent} (2076 to \SI{2196}{\kilo\electronvolt}).
The nonlinearity indices increase marginally from \SI{1.91}{\percent} to \SI{2.08}{\percent} (low-energy) and more noticeably from \SI{3.93}{\percent} to \SI{5.29}{\percent} (high-energy).
Despite these variations, the overall response remains consistent, indicating that the leakage current has only a limited effect on the shaper transcharacteristic under these conditions.

\par
The low-energy gain decreases by about \SI{10}{\percent} between \SI{-40}{\celsius} and \SI{40}{\celsius}, while the high-energy gain increases by roughly \SI{10}{\percent}.  
The compression factor $k$ decreases systematically with temperature (from 83 to 68, about \SI{-18}{\percent}), accompanied by a reduction of the output swing by roughly \SI{17}{\percent}.  
The kink energy shifts progressively to lower values with increasing temperature, consistent with enhanced compression at lower operating points, while the nonlinearity indices remain contained below \SI{5.3}{\percent}, confirming great linearity in both the high and especially the low energy range.  

%   Pedestal and preliminary noise assessment
\subsection{Pedestal and preliminary noise assessment}

\par
In this section, a preliminary evaluation of the noise performance of the channel at the shaper output is presented, focusing on the $n$-type DTMOS feedback device. 
The noise analysis is based on the measurement of the pedestal, defined as the output voltage at the shaper output in the absence of an input signal, and evaluated in terms of its full width at half maximum (FWHM), consistently with the methodology adopted in the previous chapters for both measurements (\Cref{chapter2} and \Cref{chapter3}) and simulations (\Cref{chapter4}). 
The energy resolution and the ENC are then obtained by dividing the standard deviation of the pedestal distribution by the channel low-energy gain measured under the same conditions.

%   Setup and conditions for noise measurements
\subsubsection{Setup and conditions for noise measurements}

\par
An important disclaimer for correctly interpreting the results presented in this section concerns the setup adopted for noise measurements. 
The test setup presented in~\Cref{sec:ch5-test_setup} and used throughout all measurements up to this point was not specifically optimized to achieve the best possible noise performance, but was instead designed to enable a flexible and comprehensive evaluation of the overall behavior of the prototype ASIC.
This choice inevitably sacrifices some details relevant to achieving optimal noise conditions in favor of a more automated and versatile measurement approach.

\par
For instance, the ASIC test board integrates several auxiliary components, such as digital potentiometers and a temperature sensor, which are controlled by the Arduino Giga R1 board through the dedicated interface shield via I\textsuperscript{2}C and SPI buses. 
In addition, the use of a socket to house the ASIC, while convenient for testing, is not optimal for noise measurements, as the spring-loaded contacts may introduce parasitic capacitances and inductances, as well as additional noise pickup. 
Although these elements can slightly degrade the measured noise performance, their inclusion was necessary to ensure automated bias control, environmental monitoring, and fast chip swapping, thereby enabling a systematic and reproducible characterization of the ASIC.

\par
The possibility of adjusting the CSA and shaper reference voltages with the 12-bit DACs of the STM32H7 microcontroller onboard the Arduino requires a constant connection between the Arduino and the ASIC test board, which may itself influence the measured noise performance. 
In principle, the best condition for minimizing external noise coupling would be to use the onboard manually adjustable trimmers to set the two references, thereby avoiding the continuous use of the Arduino and the associated \SI{50}{\centi\meter} ribbon cable once the ASIC configuration has been completed. 
This approach was not adopted for the first characterization campaign, but it will be implemented in future dedicated noise evaluation measurements.

\begin{figure}[t!]
    \centering
    \includegraphics[width=.75\textwidth]{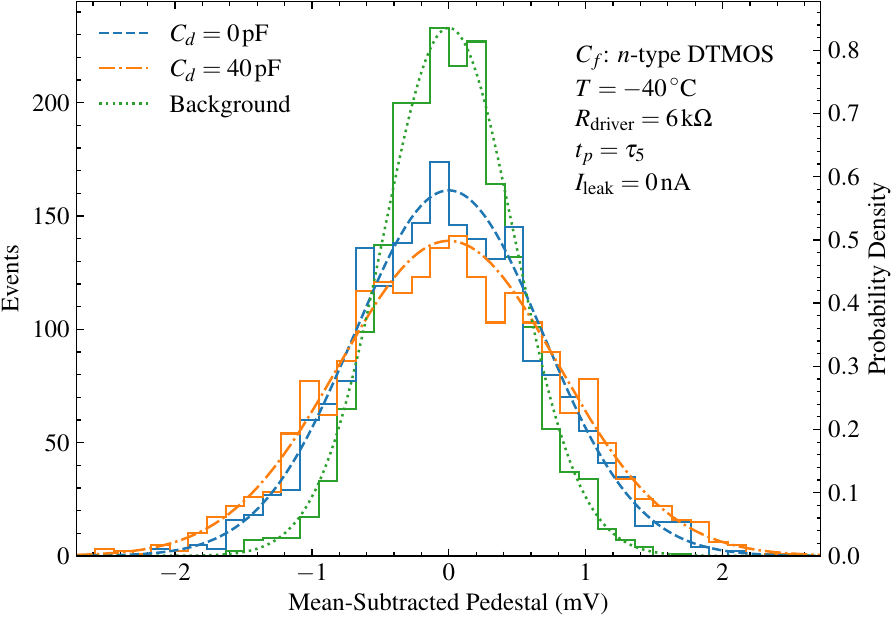}
    \caption{Mean-subtracted pedestal distributions at the shaper output for the $n$-type DTMOS CSA feedback device, comparing $C_d=\SI{0}{\pico\farad}$ and $C_d=\SI{40}{\pico\farad}$ at $T=\SI{-40}{\celsius}$, $I_{\text{leak}}=\SI{0}{\nano\ampere}$, $R_{\text{driver}}=\SI{6}{\kilo\ohm}$, and $t_p=\tau_5$. The background, measured under the same conditions using the test board without the ASIC, is also shown. Histograms are plotted as event counts (left $y$-axis), while Gaussian fits are reported as probability densities (right $y$-axis).}
    \label{fig:ch5-NDTMOS_pedestal}
\end{figure}

\par
Similarly, the use of relatively long coaxial cables to route the supply voltages and the channel outputs through the climate chamber toward the laboratory instruments (mainly the power supply and the digital oscilloscope) is not optimal for noise performance. 
Finally, no measures were taken to shield the chip and the test board from external noise sources, either due to electromagnetic interference from nearby instrumentation in the laboratory or from the climate chamber itself. 
While the chamber is usually switched off during noise measurements to eliminate any contribution from its electronics, in this first assessment it was kept on to avoid the additional time required to bring it back to the operating temperature of \SI{-40}{\celsius}.

\par
Another key consideration concerns the noise contribution of the on-chip output buffers. 
This aspect is twofold: first, the intrinsic noise of the buffer stage is not negligible and should be accounted for in the overall noise budget, ideally through dedicated measurements using the test buffer included on the chip for this purpose. 
Second, it should be noted that these buffers were implemented in the prototype to allow direct readout of the analog outputs of the CSA and shaper, a feature not required in the final flight ASIC. 
Moreover, throughout this first measurement campaign, the buffer bias current was set to the minimum value allowed by the digital potentiometer ($R_{\text{bias}}=\SI{6}{\kilo\ohm}$ instead of the nominal $R_{\text{bias}}=\SI{5}{\kilo\ohm}$), a condition that is not optimal for minimizing their noise contribution.

%   Noise measurement results
\subsubsection{Noise measurement results}

\par
The pedestal measurement has been carried out at the output of the shaper for the \hbox{$n$-type} DTMOS feedback device using an automated testing procedure, that performs the acquisition of the voltage at the output of the shaper for a fixed number of events using the Teledyne LeCroy WavePro 804HD remotely controlled via the Python test script and saving the acquired data to a database for analysis.
In the same conditions, the channel input--output transcharacteristic evaluated at the output of the shaper is also acquired, allowing the low-energy gain to be determined.

\begin{table}[t!]
    \centering
    \renewcommand{\arraystretch}{1.2}
    \setlength{\tabcolsep}{7pt}
    \begin{tabular}{cccccc}
        \toprule
        \multicolumn{1}{c}{$C_d$} & 
        \multicolumn{1}{c}{$G_{\text{le}}$} & 
        \multicolumn{1}{c}{$\sigma_{\text{ped}}$} & 
        \multicolumn{1}{c}{$\sqrt{\sigma_{\text{ped}}^2 - \sigma_{\text{bg}}^2}$} & 
        \multicolumn{1}{c}{$E_{\mathrm{res}}$} & 
        \multicolumn{1}{c}{ENC} \\
        (\si{\pico\farad}) & 
        (\si{\micro\volt/\kilo\electronvolt}) & 
        (\si{\micro\volt}) & 
        (\si{\micro\volt}) & 
        (\si{\kilo\electronvolt} FWHM) & 
        ($\si{e^-}$) \\
        \midrule
         0  & 277 & 690 & 497 & 4.22 & 499 \\
        40  & 239 & 800 & 641 & 6.30 & 746 \\
        \bottomrule
    \end{tabular}
    \caption{Pedestal standard deviation and ENC at the shaper output for the $n$-type DTMOS CSA feedback device, measured at $T=\SI{-40}{\celsius}$, $t_p=\tau_5$, $R_{\text{driver}}=\SI{6}{\kilo\ohm}$, and $I_{\text{leak}}=\SI{0}{\nano\ampere}$. The table reports the low-energy gain $G_{\text{le}}$, the measured pedestal standard deviation $\sigma_{\text{ped}}$, the value corrected for the background contribution ($\sigma_{\text{bg}}=\SI{478}{\micro\volt}$), the corresponding energy resolution $E_{\mathrm{res}}$, and the ENC expressed in electrons, for $C_d=\SI{0}{\pico\farad}$ and $C_d=\SI{40}{\pico\farad}$.}
    \label{tab:ch5-pedestal-noise}
\end{table}

\par
\Cref{fig:ch5-NDTMOS_pedestal} shows the results of the pedestal measurements performed at $T=\SI{-40}{\celsius}$, both in the absence and in the presence of detector capacitance, with $I_{\text{leak}}=\SI{0}{\nano\ampere}$. 
The background pedestal acquisition was carried out using the test board without the ASIC, under the same conditions as the ASIC measurements, in order to account for the noise contribution of the measurement setup itself.
The standard deviation of the measured voltage was found to be $\sigma_{\text{bg}}=\SI{478}{\micro\volt}$.
This includes the intrinsic contribution of the test board, the Arduino Giga board with its dedicated shield connected outside the climate chamber, as well as the noise introduced by the oscilloscope readout circuitry. 
In the plot, the 2000 events corresponding to each of the three pedestal distributions are shown after subtraction of their respective mean values, to allow a fair comparison of the distribution widths. 
Superimposed on the histograms, the corresponding Gaussian fits are reported and used to extract the standard deviation of each distribution, which is then employed for the evaluation of the energy resolution and ENC.

\par
\Cref{tab:ch5-pedestal-noise} summarizes the results of the analysis, including the low-energy gain $G_{\text{le}}$ extracted from the low-energy portion of the transcharacteristic evaluated at the shaper output, and the measured pedestal standard deviation $\sigma_{\text{ped}}$ for both $C_d=\SI{0}{\pico\farad}$ and \hbox{$C_d=\SI{40}{\pico\farad}$}. 
To account for the background contribution and isolate the pedestal standard deviation associated solely with the ASIC, the corrected value is obtained as $\sqrt{\sigma_{\text{ped}}^2 - \sigma_{\text{bg}}^2}$.

\par
The energy resolutions reported in \Cref{tab:ch5-pedestal-noise}, although higher than the values obtained from post-layout simulations presented in \Cref{sec:ch5-shaper_noise_performance}, provide a first experimental estimate of the achievable noise performance. 
As already discussed in the previous section, these preliminary measurements indicate figures that approach the simulated trends and remain reasonably close to the reference requirement of \SI{4}{\kilo\electronvolt}. 
It should be emphasized that this is a prototype ASIC developed in the context of R\&D on novel CSA feedback solutions based on dynamic-threshold MOS devices, as well as on an improved Krummenacher network for leakage current compensation. 
As such, the chip is less suited for dedicated noise measurements, but nevertheless allows meaningful indications on the intrinsic performance of the proposed design topologies to be extracted.

%-------------------------------------------------------------------------------
%   Comparative performance analysis
%-------------------------------------------------------------------------------
\section{Comparative performance analysis}

\begin{table}[t!]
    \centering
    {\fontsize{10}{12}\selectfont
    \renewcommand{\arraystretch}{1.15}
    \setlength{\tabcolsep}{2.6pt}
    \begin{tabular}{lllll}
        \toprule
        Parameter &
        ANTARES4 &
        SLIDER32~\cite{manghisoni32channelsReadoutASIC2023} &
        Nam et al.~\cite{namCMOSReadoutASIC2026} &
        Gao et al.~\cite{gaoDesignCharacterizationLowNoise2018} \\
        \midrule
        Application                                             & Astrophysics & Astrophysics & General purpose & Dosimetry \\
        Process                                                 & CMOS \SI{65}{\nano\meter} & CMOS \SI{180}{\nano\meter} & CMOS \SI{180}{\nano\meter} & CMOS \SI{180}{\nano\meter} \\
        Supply voltage (\si{\volt})                             & 1.2 & 1.8 & 3.3 & 3.3 \\
        Operating temperature (\si{\celsius})                   & $-40$ & $-40$ & Room & Room \\
        Detector type                                           & Si(Li) strip & Si(Li) strip & Si & CdZnTe/Si \\
        Detector capacitance (\si{\pico\farad})                 & 38 & 38 & 3 & 20 \\
        Input dynamic range (\si{\kilo\electronvolt})           & \SI{10}{\kilo\electronvolt}--\SI{100}{\mega\electronvolt} & \SI{10}{\kilo\electronvolt}--\SI{100}{\mega\electronvolt}      & $\le81$ & $\le203$ (CdZnTe) \\
        High/low gain modes                                     & Yes & Yes & No & No \\
        CSA feedback capacitor                                  & N/P (DT)MOS & PMOS & Linear $C_f$ & Linear $C_f$ \\
        CSA gain H/L (\si{\micro\volt\per\kilo\electronvolt})   & 250/2 & 250/3 & $\sim1241$ & $\sim890$ \\
        CSA output dyn. range (\si{\volt})                      & 0.5 & 0.4 & -- & 1 \\
        ENC (\si{e^-})                                          & 746 @ $C_d=\SI{40}{\pico\farad}$ & 480 @ $C_d=\SI{40}{\pico\farad}$ & 169 & 112 \\
        Energy resolution (\si{\kilo\electronvolt})             & 6.3 @ $\le\SI{100}{\kilo\electronvolt}$ & 4.0 @ $\le\SI{100}{\kilo\electronvolt}$ & 1.5 @ \SI{30.8}{\kilo\electronvolt} & 3.27 @ \SI{59.5}{\kilo\electronvolt} \\
        Shaping stage                                           & CR-RC & CR-RC$^2$ & CR-RC & CR-RC \\
        \# Peaking times                                        & 8 (\SI{0.2}{\micro\second} step) & 8 (\SI{0.2}{\micro\second} step) & 1 & 1 \\
        Peaking time(s) (\si{\micro\second})                    & 0.2--1.6 & 0.25--1.6 & 20 & 1.3 \\
        \# channels                                             & 8 & 32 & 2 & 4 \\
        Channel area (\si{\milli\meter\squared})                & 0.04\;\textsuperscript{(*)} & 0.15 & 0.27 & 5.25 (die) \\
        Power consumption (\si{\milli\watt/ch})                 & 3.99\;\textsuperscript{(*)} & 8.3 & 16.2 & 2 \\
        Charge restoration technique                            & Krummenacher & Krummenacher & Pseudo-resistive & Resistive \\
        Max leakage current (\si{\nano\ampere})                 & 200 & 50 & -- & 10 \\
        Leakage current emulator                                & Yes ($\le\SI{300}{\nano\ampere}$) & No & -- & -- \\
        Channel output stage                                    & Shaper & ADC & Shaper & Sallen-Key filter \\                                              
        \bottomrule
    \end{tabular}}
    \caption{Comparison of the architectural features and measured performance parameters of ANTARES4, SLIDER32, and two representative readout ASICs reported in the literature, covering technology, detector type, gain configuration, noise, shaping, and power consumption. Performance metrics for ANTARES4 and SLIDER32 are obtained at $T=\SI{-40}{\celsius}$ with $C_d=\SI{40}{\pico\farad}$. \textsuperscript{(*)}\;The reported channel area and power consumption for ANTARES4 refer to the readout chain up to the shaper stage only.}
    \label{tab:ch5-asic_comparison}
\end{table}

\par
This section presents a comparative analysis of the main characteristics and performance metrics of the first-generation readout ASIC designed in a \SI{180}{\nano\meter} CMOS technology for the initial flight, SLIDER32, and the newly developed ANTARES4.
The comparison also includes two additional ASICs, both implemented in \SI{180}{\nano\meter} CMOS technology and employed for the readout of cadmium-zinc-telluride (CZT)~\cite{gaoDesignCharacterizationLowNoise2018} and silicon strip detectors~\cite{namCMOSReadoutASIC2026}, respectively, in order to provide a broader perspective and to benchmark the proposed design against representative state-of-the-art solutions for applications similar to the one discussed in this work.

\par
\Cref{tab:ch5-asic_comparison} summarizes the main architectural features and measured performance parameters of the ANTARES4 front-end ASIC in comparison with the first-generation SLIDER32 design. The table provides a direct side-by-side comparison of the two implementations under equivalent operating conditions, allowing the impact of the design choices introduced in ANTARES4 to be evaluated with respect to technology scaling, power consumption, noise performance, dynamic range, and channel-level integration.

\par
The most relevant comparison emerges from the analysis of the performance metrics of the newly developed ANTARES4 front-end with respect to the first-generation SLIDER32 readout ASIC.
The migration to the more scaled \SI{65}{\nano\meter} technology node enabled a reduction of the supply voltage from \SI{1.8}{\volt} to \SI{1.2}{\volt}, with direct implications for channel-level power consumption.
In the current configuration, considering the readout chain up to the shaping stage, the measured power dissipation of ANTARES4 is \SI{3.99}{\milli\watt} at an operating temperature of \SI{-40}{\celsius}, corresponding to a reduction of approximately \SI{52}{\percent} with respect to the first-generation design.
Although this figure does not represent the complete channel configuration, it provides a clear indication of the available power margin for the integration of the remaining channel blocks, while remaining compliant with the \SI{10}{\milli\watt\per channel} limit imposed by the experiment.

\par
The channel area occupation has been reduced from \SI{0.15}{\milli\meter\squared} to \SI{0.04}{\milli\meter\squared}, corresponding to a reduction of approximately \SI{73}{\percent}.
This comparison should be interpreted considering that the current ANTARES4 channel implementation does not yet include part of the self-trigger and discriminator circuitry.
Nevertheless, the higher integration density offered by the scaled technology node enables a substantial reduction of the overall channel footprint, opening the possibility of increased channel density for future tracker architectures employing higher strip-count detectors or alternative layout strategies.

\par
Significant effort was devoted to the design of the CSA feedback capacitor, including the investigation of multiple topologies based on both $n$- and $p$-type devices, implemented in standard MOS and DTMOS configurations.
Among the explored solutions, the dynamic-threshold configuration proved particularly effective in ensuring compliance with the required output dynamic range of \SI{0.5}{\volt} at the CSA level and \SI{0.6}{\volt} at the shaper output.
At the same time, this approach preserves the target energy-to-voltage conversion gain of \SI{250}{\micro\volt\per\kilo\electronvolt} in the low-energy spectroscopy range between \SIrange{10}{100}{\kilo\electronvolt}, while providing a reduced gain of approximately \SI{2}{\micro\volt\per\kilo\electronvolt} in the particle-tracking regime spanning \SIrange{25}{100}{\mega\electronvolt}.

\par
As discussed in~\Cref{sec:ch3-fdt_calibration}, a dedicated calibration procedure of the channel \hbox{input--output} transcharacteristic was required to enable accurate reconstruction of the deposited energy from the ADC output.
This procedure proved particularly challenging during \hbox{on-ground} calibration using cosmic muons, since the expected energy deposition of vertically traversing muons at ground level is approximately \SI{842}{\kilo\electronvolt}, which lies in the transition region between the low-energy and particle-tracking regimes.
To simplify the calibration process and avoid the need for complex fitting strategies, a key design objective of the \SI{65}{\nano\meter} implementation was to shift the location of this transition above \SI{1.5}{\mega\electronvolt}.
This objective was successfully achieved at both the CSA and shaper levels, as summarized in~\Cref{tab:ch5-csa-transchar-summary} and~\Cref{tab:ch5-shaper-transchar-summary}, respectively.

\par
A further significant improvement concerns the compensation of the Si(Li) detector leakage current, which has been extended from the \SI{50}{\nano\ampere} limit of the \SI{180}{\nano\meter} implementation to \SI{200}{\nano\ampere} at room temperature in the newly developed readout channel.
This enhancement is particularly beneficial during on-ground calibration and integration activities, where higher operating temperatures were observed during the preparation phase of the first flight.
The increased compensation capability reduces the likelihood of channel saturation in Si(Li) strips subject to large temperature gradients and elevated leakage currents, enabling more robust and extensive calibration procedures to be performed on ground.
In addition, ANTARES4 introduces an integrated leakage current generator at the CSA input, allowing direct verification of the compensation capability during early-stage chip characterization.

\par
As a final remark, a preliminary noise performance study, although showing promising results, does not yet meet the noise performance requirements imposed by the experiment, nor does it fully match the performance achieved by the first-generation design.
For this reason, dedicated noise characterization activities will be carried out in future to provide a more accurate assessment of the actual noise performance of the newly developed readout channel.
These studies will enable a systematic comparison of the different CSA feedback capacitor configurations in terms of noise behavior and will allow the evaluation of potential differences between the standard MOS implementation and the novel DTMOS approach adopted in this work.

%-------------------------------------------------------------------------------
%   Lessons learned and future prospects
%-------------------------------------------------------------------------------
\section{Lessons learned and future prospects}

\par
The first characterization campaign on the \antareslogo ASIC provided a comprehensive initial assessment of the channel performance up to the shaper stage, highlighting both the strengths and the aspects for improvement, while confirming the effectiveness of the novel circuit topologies implemented in the prototype chip, namely the dynamic-threshold MOSFET feedback device in the CSA and the improved Krummenacher network for leakage current compensation.

\par
The transient response measurements of the CSA confirmed the expected behavior in terms of discharge dynamics and baseline recovery.
The circuit demonstrated a robust response to varying detector leakage currents, both at room temperature and at the operational temperature, for values up to \SI{200}{\nano\ampere}, well above the requirement and the capability of the current \SI{180}{\nano\meter} implementation.
This robustness is particularly useful for detector characterization during the on-ground calibration campaign of the instrument.

\par
The transient analysis of the shaper further confirmed the expected behavior as a function of input energy, showing consistent peaking times for a given setting across the full energy range.
A point to be considered for the future flight ASIC design is the implementation of the complete set of peaking times, from the shortest to the longest values compatible with the expected dynamic range and timing spacing.
Although the experiment itself makes use of only a single peaking time during flight, this extended functionality would increase the flexibility of laboratory characterization, enabling noise performance optimization and supporting extended noise studies under different operating conditions or varying detector capacitance.

\par
The CSA and shaper transcharacteristic measurements confirmed the expected behavior in both the low- and high-energy regions, with excellent gain performance, especially at low energies under realistic operating conditions with detector capacitance.
In the $n$-type DTMOS configuration, the low-energy gain reached the required \SI{250}{\micro\volt/\kilo\electronvolt} and exhibited good linearity, while the compression factor remained consistently above 80, as characteristic of this feedback topology.
The kink energy was found to be well above \SI{2}{\mega\electronvolt}, thereby extending the linear region beyond $\SI{\sim800}{\kilo\electronvolt}$.
This represents a significant improvement compared to the \SI{180}{\nano\meter} implementation, where the curvature region overlapped with the energy of cosmic muons used for on-ground calibration, complicating energy reconstruction and requiring specific calibration strategies, as discussed in~\Cref{chapter3}.
Overall, the DTMOS configuration has proven highly promising, both in terms of absolute low-energy gain and in combination with the reduced output dynamic range compared to the non-DTMOS counterparts.
In particular, the $n$-type implementation emerged as the most effective, as expected.

\par
From the noise perspective, this first evaluation provided an initial picture of the performance.
Although the results are reasonable in absolute terms, they did not fully meet the expectations derived from simulations.
Future efforts should therefore focus on improving the test setup for dedicated noise measurements and on a systematic evaluation of all noise sources, including the contributions from the on-chip buffers and from the external readout chain on the test board and in the test setup.

\par
Finally, the evaluation of the full set of ten packaged chips will be crucial to building a statistical picture of the overall performance of the \antareslogo ASIC, providing guidance for the design of the final flight version.
In this respect, the test setup developed for this campaign has proven to meet all characterization requirements while offering a high degree of automation and flexibility.
This allows the complete set of measurements to be performed on a single ASIC in a fully automated manner, without human intervention inside the climate chamber, even when varying the temperature conditions.

%% file: maintext/conclusions.tex
\chapter*{Conclusions}

\par
The work presented in this thesis has addressed two main objectives: the validation of the flight electronics based on the SLIDER32 readout ASIC, implemented in \SI{180}{\nano\meter} CMOS technology and selected for the first GAPS flight scheduled for December 2025, and the design and characterization of the \antareslogo prototype ASIC in \SI{65}{\nano\meter} CMOS technology, conceived as an upgrade of the tracker readout system for the second flight planned from 2026 onwards.
These two complementary activities have allowed to validate the GAPS tracker readout electronics ahead of the first Antarctic flight, while laying the foundations for the next generation Si(Li) readout ASIC, proving the effectiveness of novel solutions aimed at improving the performance of the current implementation and extending the capabilities of the readout front-end.

\par
An important aspect of this work is that it encompasses the complete chain of validation of the GAPS tracker readout electronics. 
Beginning with the single-ASIC characterization on a dedicated test board, presented in the author's master thesis~\cite{ghislottiCharacterisationReadoutElectronics2022} and published in~\cite{manghisoni32channelsReadoutASIC2023}, the activity progressed to the module-level validation discussed in~\Cref{chapter2} and was subsequently extended to the performance evaluation of the fully integrated tracker presented in~\Cref{chapter3}. 
In addition to confirming the readiness of the current implementation for the first flight, this multi-level validation also provided the foundation for the development of the next-generation readout system, realized through the design and characterization of the \antareslogo prototype ASIC in \SI{65}{\nano\meter} CMOS technology.

\par
The first part of the work focused on the laboratory characterization of a complete Si(Li) tracker module instrumented with the SLIDER32 ASIC, selected for the first flight.
Dedicated test procedures verified that the integrated readout electronics satisfied the stringent requirements of the experiment, especially in terms of gain, linearity, dynamic range, and energy resolution.
In addition, acquisitions performed with an Americium-241 source and with on-ground cosmic muons validated the correct operation of the electronics deployed in a flight-ready tracker module configuration, representing the first successful measurements with both a radioactive source and charged particles as a clear benchmark of system functionality.

\par
This effort was extended to the fully assembled tracker during the integration and testing campaign, which represented a crucial step toward flight readiness.
In this context, a set of calibration and verification procedures for the Si(Li) detector readout electronics was developed and successfully implemented in the instrument flight software.
These methodologies enabled the first acquisition of x-rays from a Cadmium-109 source with the complete tracker as well as measurements with cosmic muons at ground level.
They now form the basis for the calibration workflow of the entire instrument, and have supported both the Antarctic deployment in December 2024 and the preparation for the first scientific flight scheduled for December 2025.

\par
The experience gained during this extensive validation campaign highlighted the opportunity to develop an upgraded ASIC aimed at overcoming specific limitations of the current implementation and at extending the operational capabilities of the tracker.
This led to the design of the \antareslogo chip, comprising eight prototype analog channels up to the shaping stage and conceived as a test platform for novel circuit topologies.
Among these, the adoption of the dynamic threshold MOSFET for the implementation of the CSA feedback capacitance introduced, for the first time in a charge-sensitive amplifier, a technique typically employed in digital electronics.
The prototype has demonstrated the capability to achieve the required dynamic range, gain and linearity across both the x-ray energy domain (10--\SI{100}{\kilo\electronvolt}) and the charged-particle tracking regime (up to \SI{100}{\mega\electronvolt}), validating the effectiveness of this novel solution.

\par
Another critical improvement concerned the detector leakage current compensation.
While limited to approximately \SI{50}{\nano\ampere} in the previous \SI{180}{\nano\meter} implementation, the newly improved feedback scheme achieved compensation up to \SI{200}{\nano\ampere} at room temperature and was also tested at leakage currents exceeding \SI{300}{\nano\ampere}, although compensation is no longer achieved in this regime.
This enhancement substantially broadens the calibration capabilities of the readout system, especially during ground-based testing at room temperature, and represents a significant step forward with respect to the current ASIC.

\par
The laboratory characterization of the chip, performed with a dedicated test setup, confirmed its correct operation across the expected temperature range and validated the expected behavior of the novel feedback solutions.
Measurements of the CSA and shaper transcharacteristics demonstrated compliance with the requirements imposed by the experiment in terms of dynamic range, compression factor, and linearity.
In addition, the auxiliary circuit blocks were successfully validated, namely the charge injection circuit for calibration, the leakage current emulator, the on-chip output buffers, and the digital control logic comprising the 12-bit shift register and the binary-to-thermometric decoder for peaking time selection.
These results provide a solid experimental basis for the integration of the validated blocks into the next-generation flight ASIC.

\par
Overall, the activities carried out in this thesis have had a direct impact on the progress of the GAPS experiment.
On one side, the characterization of the SLIDER32 ASIC supported the tracker integration, testing, and validation during the assembly phases in preparation for the first Antarctic flight.
On the other, the design and laboratory validation of the \antareslogo chip extended and improved the functionality of several blocks of the current ASIC, introducing innovative circuit topologies and addressing the requirements that emerged for the second flight.

\par
Equally important, this development has laid the foundations for the upgrade of the readout electronics in a more advanced and scaled technology node, capable of accommodating possible redesigns of the tracker geometry and detector layout.
The reduced footprint and power consumption of the channel, enabled by the \SI{65}{\nano\meter} technology, provide the flexibility to integrate a larger number of channels while sustaining the stringent power budget of the experiment.
The results of this work contributed to the successful preparation of the GAPS instrument for its first Antarctic flight and established a solid foundation for the next generation of the Si(Li) tracker readout electronics in preparation for future missions.

%% file: maintext/appendix/appendixA.tex
\chapter{Contribution to the GAPS integration and first Antarctic campaign} \label{appendix_gaps_collaboration}

\par
\textit{This appendix is written in the first person to reflect the author's direct involvement in the integration, testing, and deployment phases of the GAPS experiment.}

%-------------------------------------------------------------------------------
%   Instrument integration and testing (I\&T)
%-------------------------------------------------------------------------------
\section{Instrument integration and testing (I\&T)}

\par
\noindent
In October 2022, I joined the GAPS collaboration, contributing to the assembly, validation, and testing of the Si(Li) tracker readout electronics.
I visited the Space Sciences Laboratory (SSL) at the University of California, Berkeley, in December 2022 and March 2023, where I worked on the integration of the tracker modules and the associated readout electronics.
During this phase, individual tracker modules, previously tested under laboratory conditions as described in \Cref{chapter2}, were assembled into rows and planes to construct the full tracker system.

\par
\begin{figure}[t!]
    \centering
    \captionsetup[subfigure]{justification=justified,singlelinecheck=false}
    \begin{subfigure}{0.49\textwidth}
        \centering
        \includegraphics[width=\linewidth]{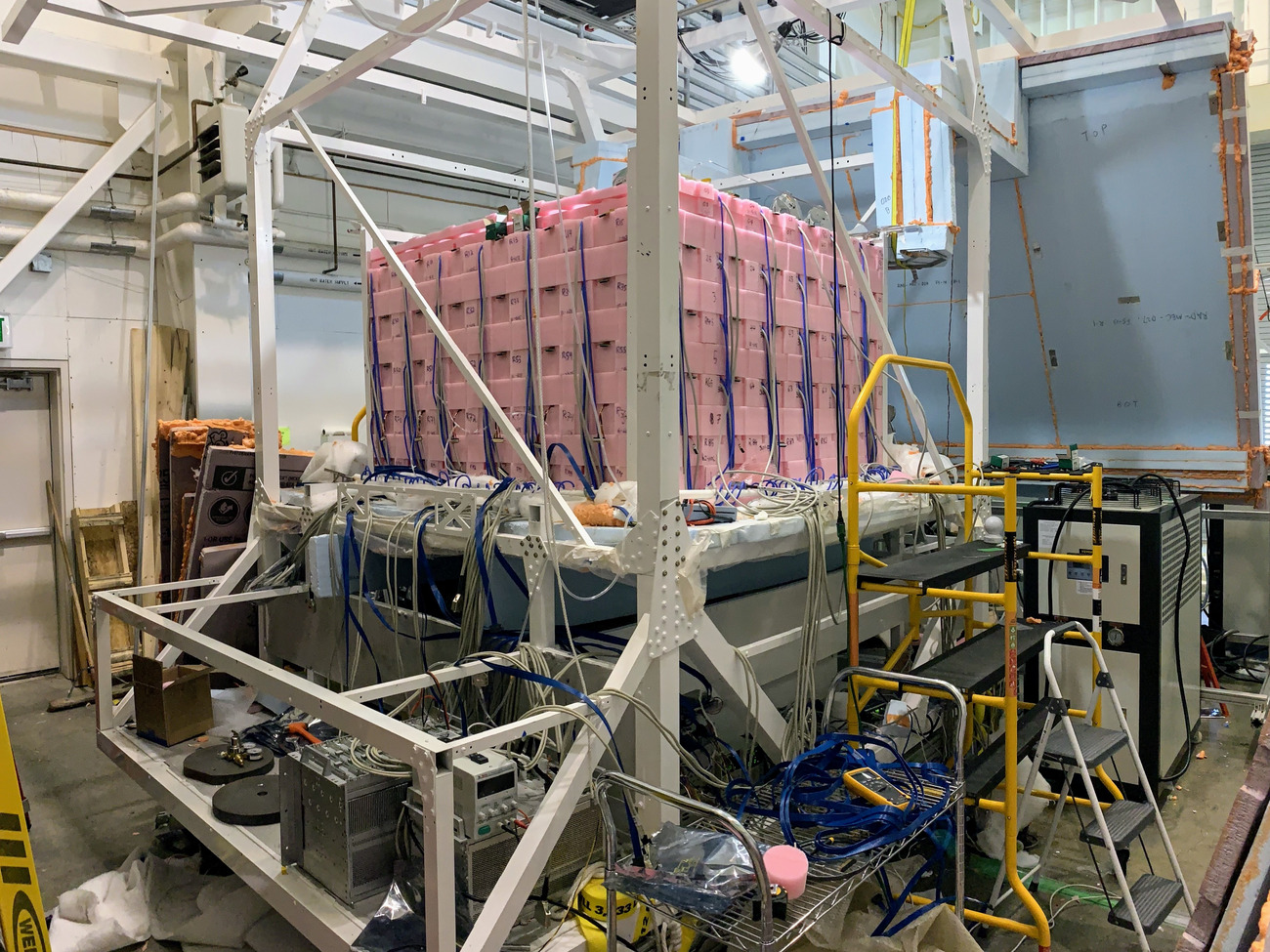}
        \caption{Integration of the Si(Li) tracker at the Space Sciences Laboratory (SSL) in Berkeley, December 2022.}
        \label{fig:app-B_gaps_tracker_berkeley}
    \end{subfigure}
    \hfill
    \begin{subfigure}{0.49\textwidth}
        \centering
        \includegraphics[width=\linewidth]{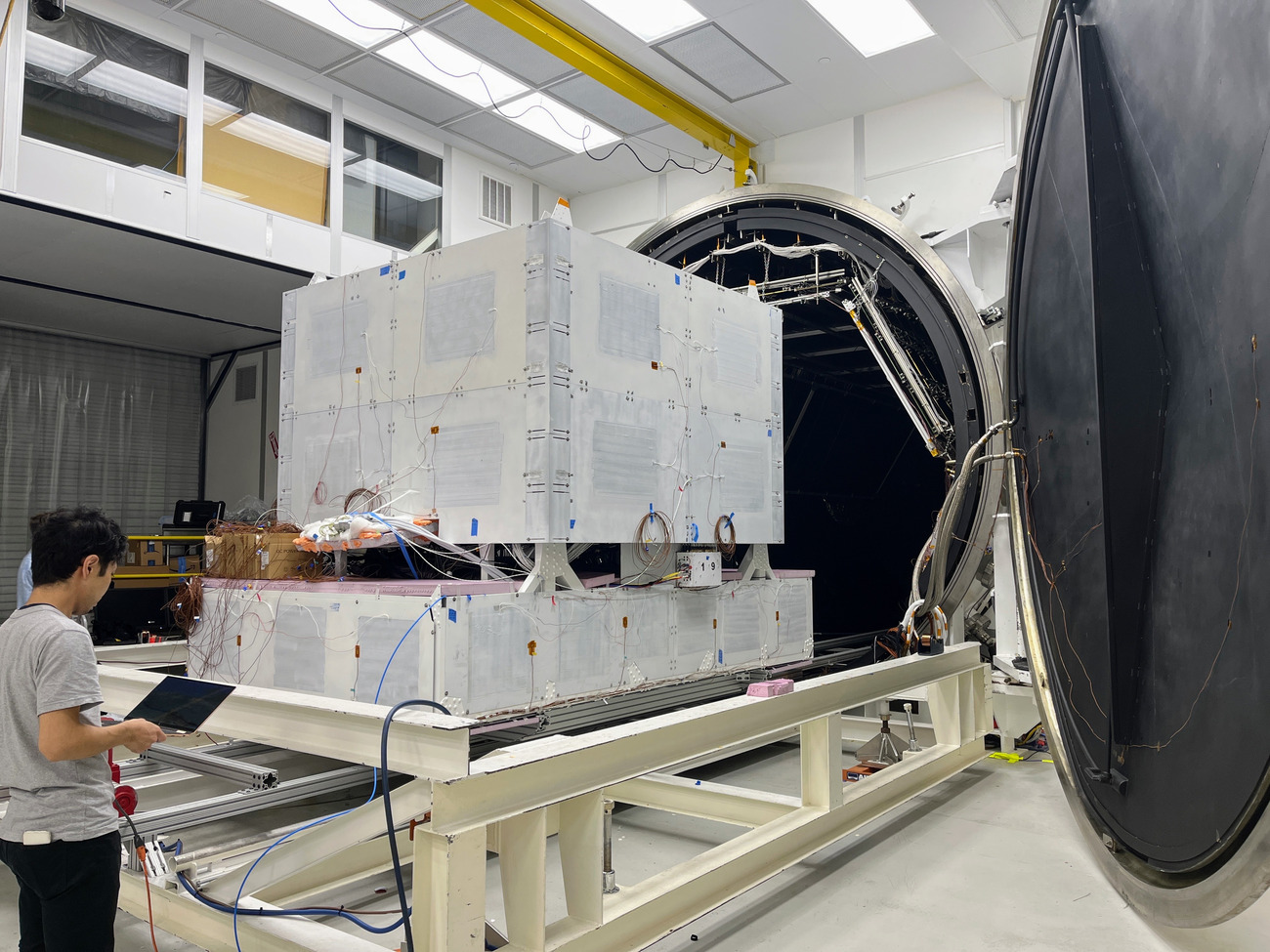}
        \caption{Thermal vacuum (TVAC) testing of the full GAPS instrument at National Technical Systems (NTS), Los Angeles.}
        \label{fig:app-B_gaps_nts_los_angeles}
    \end{subfigure}
    \caption{Two key milestones in the U.S.-based integration and testing (I\&T) campaign of the GAPS experiment: (\subref{fig:app-B_gaps_tracker_berkeley}) tracker integration at SSL and (\subref{fig:app-B_gaps_nts_los_angeles}) full-system TVAC testing at NTS.}
    \label{fig:app-B_berkeley_los_angeles_integration}
\end{figure}

As shown in \Cref{fig:app-B_gaps_tracker_berkeley}, the tracker was assembled from the bottom up, with rows of modules stacked on custom-cut Styrofoam blocks that housed the Si(Li) detectors.
A preliminary validation of the readout electronics was performed on each integrated row by powering on the readout ASICs and verifying its functionality.
Dedicated procedures were developed to validate the entire readout chain of the Si(Li) tracker, culminating in the first successful detection of x-rays from a Cadmium-109 source and cosmic muons at ground level using the fully instrumented tracker, as detailed in \Cref{chapter3}.

\par
Following successful integration and ambient-pressure testing at SSL---where the system was cooled using the ground cooling system (GCS)---the next objective was to validate the full GAPS instrument under flight-like conditions.
To that end, I participated in a one-month-long thermal vacuum (TVAC) campaign at the National Technical Systems (NTS) facility in Los Angeles, California, where the complete GAPS payload was tested under vacuum and cooled to the target flight temperature of \SI{-40}{\celsius}.
As shown in \Cref{fig:app-B_gaps_nts_los_angeles}, the instrument was placed inside a thermal vacuum chamber and operated to assess its thermal performance.
The test validated the tracker and time-of-flight (ToF) electronics under vacuum conditions, as well as the effectiveness of the tracker's thermal management system.

\begin{figure}[t!]
    \centering
    \captionsetup[subfigure]{justification=justified,singlelinecheck=false}
    \begin{subfigure}{0.49\textwidth}
        \centering
        \includegraphics[width=\linewidth]{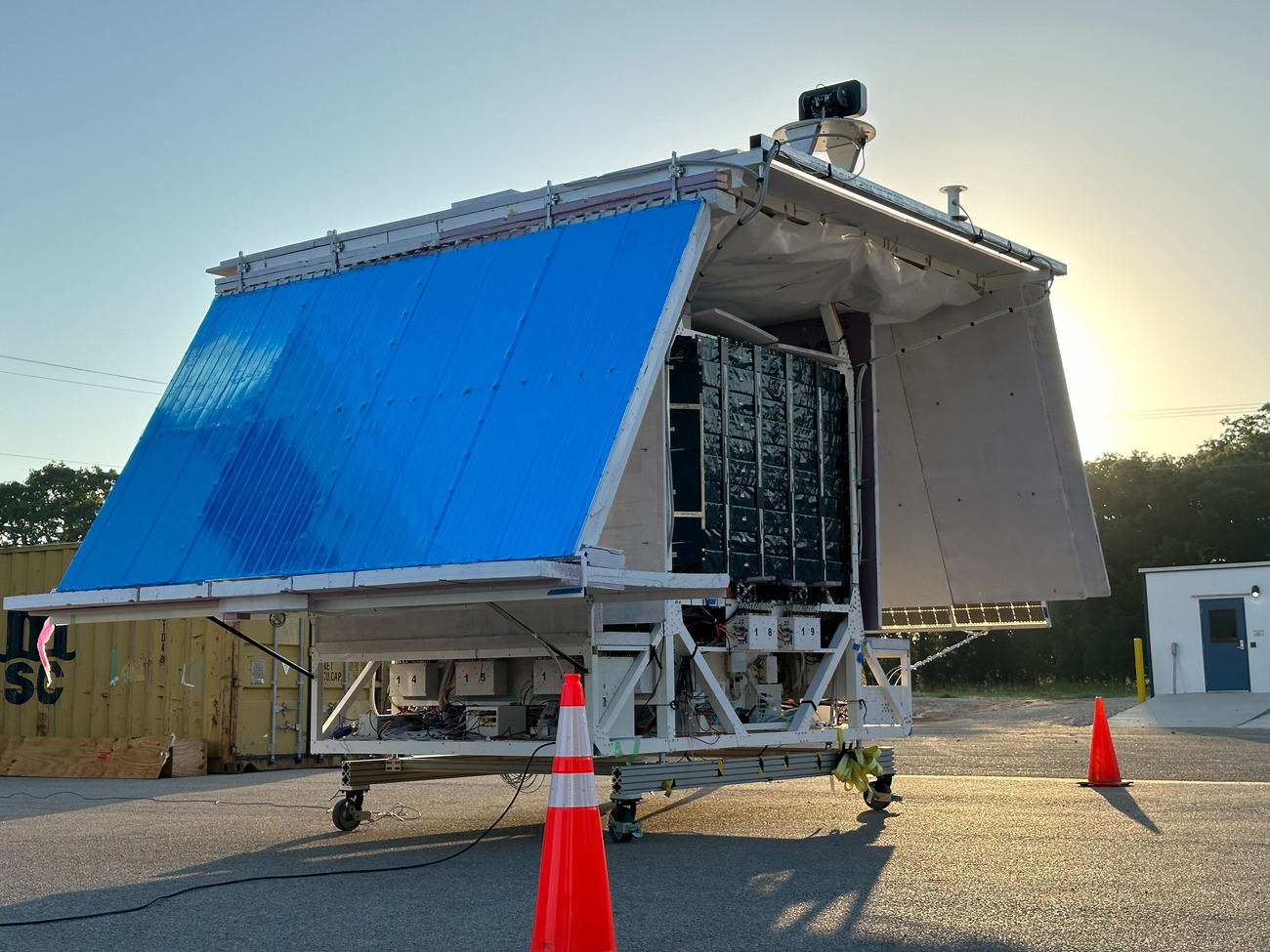}
        \caption{View of the fully integrated GAPS instrument from the radiator side.}
        \label{fig:app-B_gaps_hang_test_rad_side}
    \end{subfigure}
    \hfill
    \begin{subfigure}{0.49\textwidth}
        \centering
        \includegraphics[width=\linewidth]{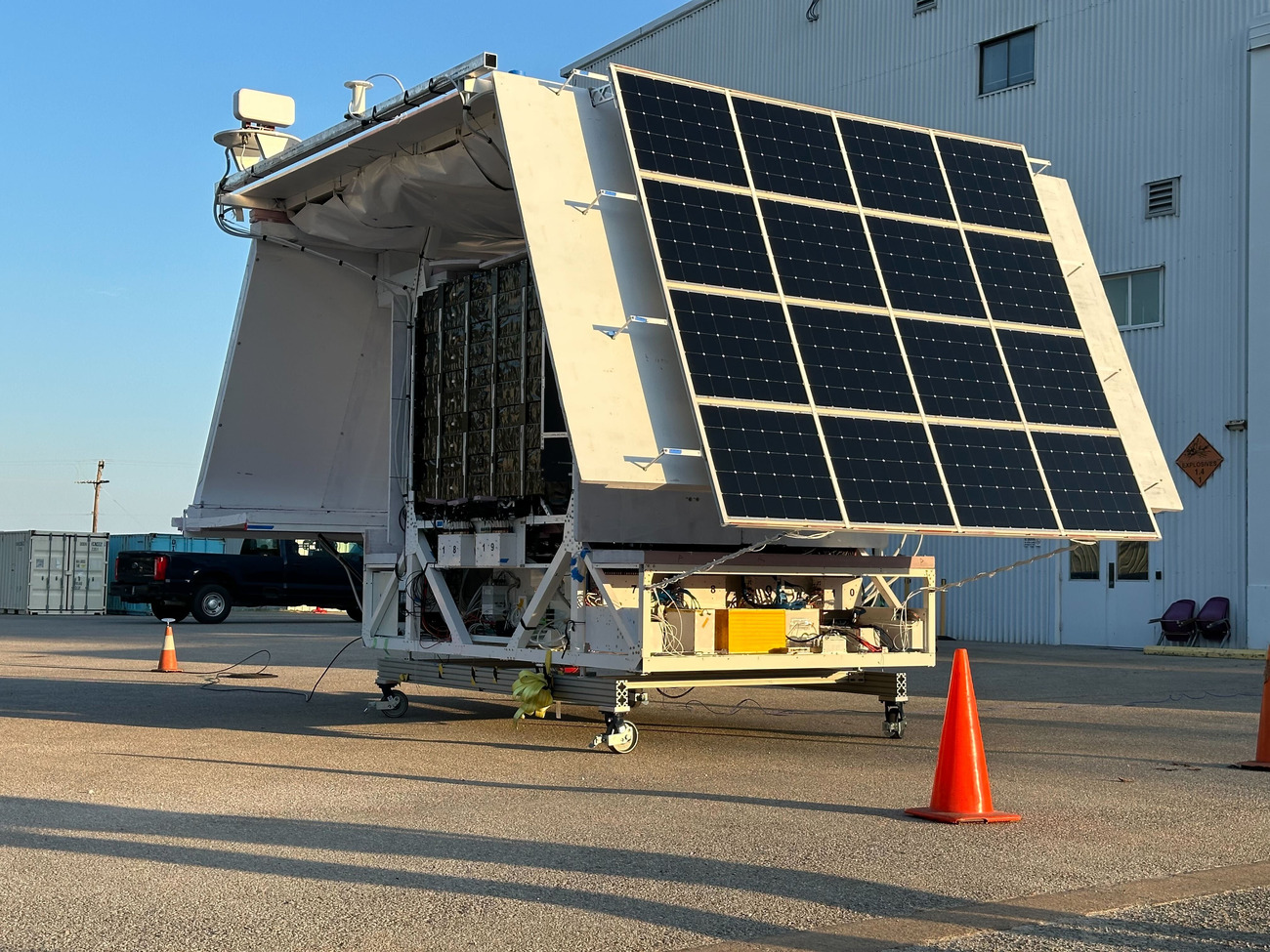}
        \caption{View of the instrument from the side featuring the solar arrays.}
        \label{fig:app-B_gaps_hang_test_solar_side}
    \end{subfigure}
    \caption{The GAPS payload in its final flight configuration prior to the hang test at the Columbia Scientific Balloon Facility. The hang test represented the final system-level validation before shipment to Antarctica.}
    \label{fig:app-B_gaps_hang_test}
\end{figure}

\par
After the TVAC campaign, the payload was transported to the Nevis Laboratories of Columbia University in Irvington, New York, where the final integration steps were carried out, including the installation of the ToF scintillators and their readout electronics.

\par
In May 2024, the instrument was shipped to the NASA Columbia Scientific Balloon Facility (CSBF) in Palestine, Texas, for the final system-level test campaign.
At CSBF, I participated in the final assembly of the payload in its flight configuration, followed by a hang test during which the entire instrument was suspended from a crane.
This test verified the mechanical integrity, weight, and flight readiness of the payload.
As shown in \Cref{fig:app-B_gaps_hang_test_rad_side,fig:app-B_gaps_hang_test_solar_side}, the instrument was fully assembled with radiators and solar panels, and tested with the NASA telemetry and command system.
This system enables real-time transmission of science and housekeeping data during flight, and provides control of the payload orientation via a motorized rotator to align the solar arrays with the Sun.
It also manages altitude control through ballast release and operates the parachute system for recovery.

\par
\begin{figure}[t!]
    \centering
    \captionsetup[subfigure]{justification=justified,singlelinecheck=false}
    \begin{subfigure}{0.49\textwidth}
        \centering
        \includegraphics[width=\linewidth]{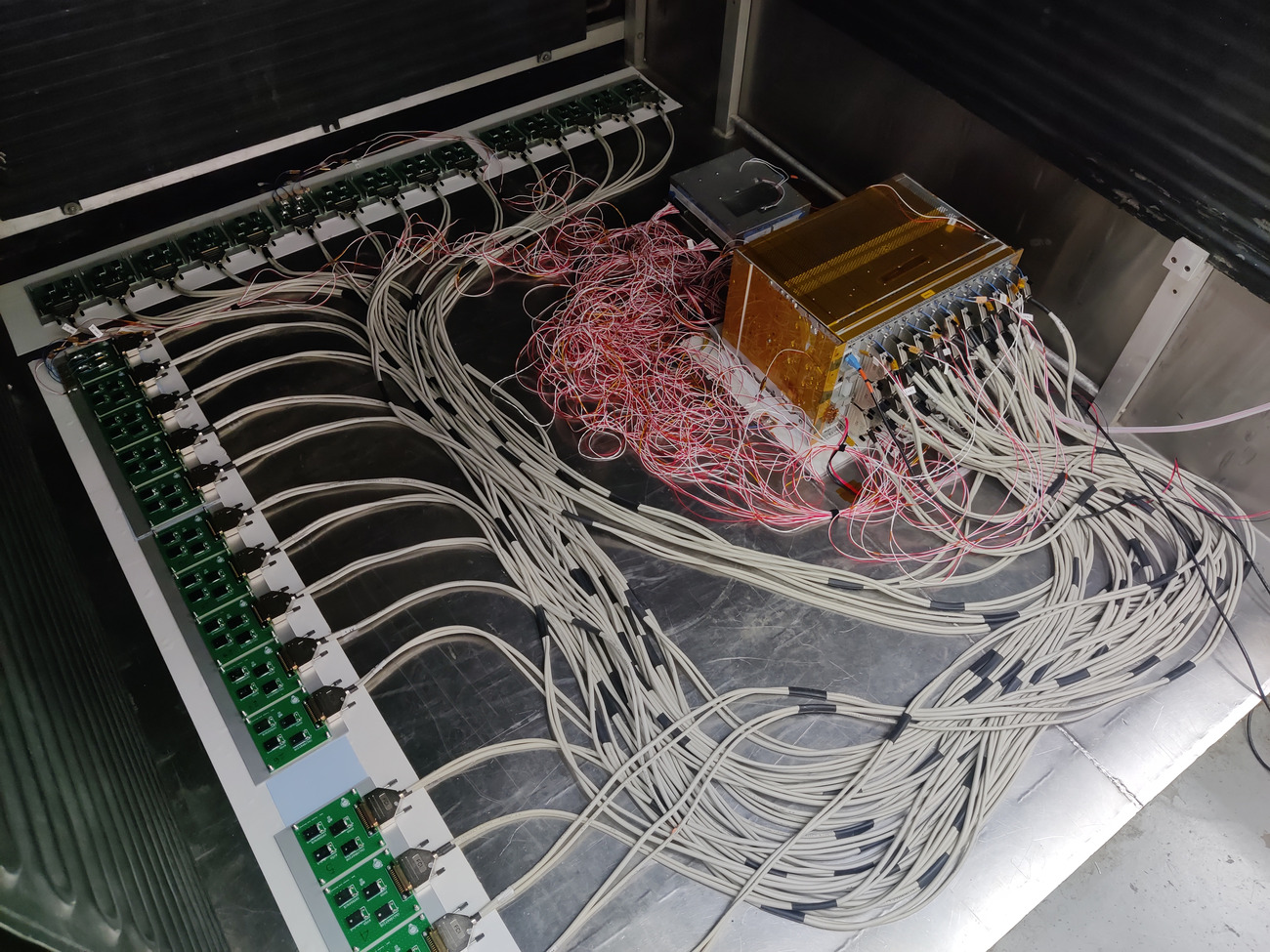}
        \caption{TVAC test setup for the flight-qualified HVLVPS, featuring custom load-emulating boards simulating the tracker rows.}
        \label{fig:app-B_hvlvps_tvac_test}
    \end{subfigure}
    \hfill
    \begin{subfigure}{0.49\textwidth}
        \centering
        \includegraphics[width=\linewidth]{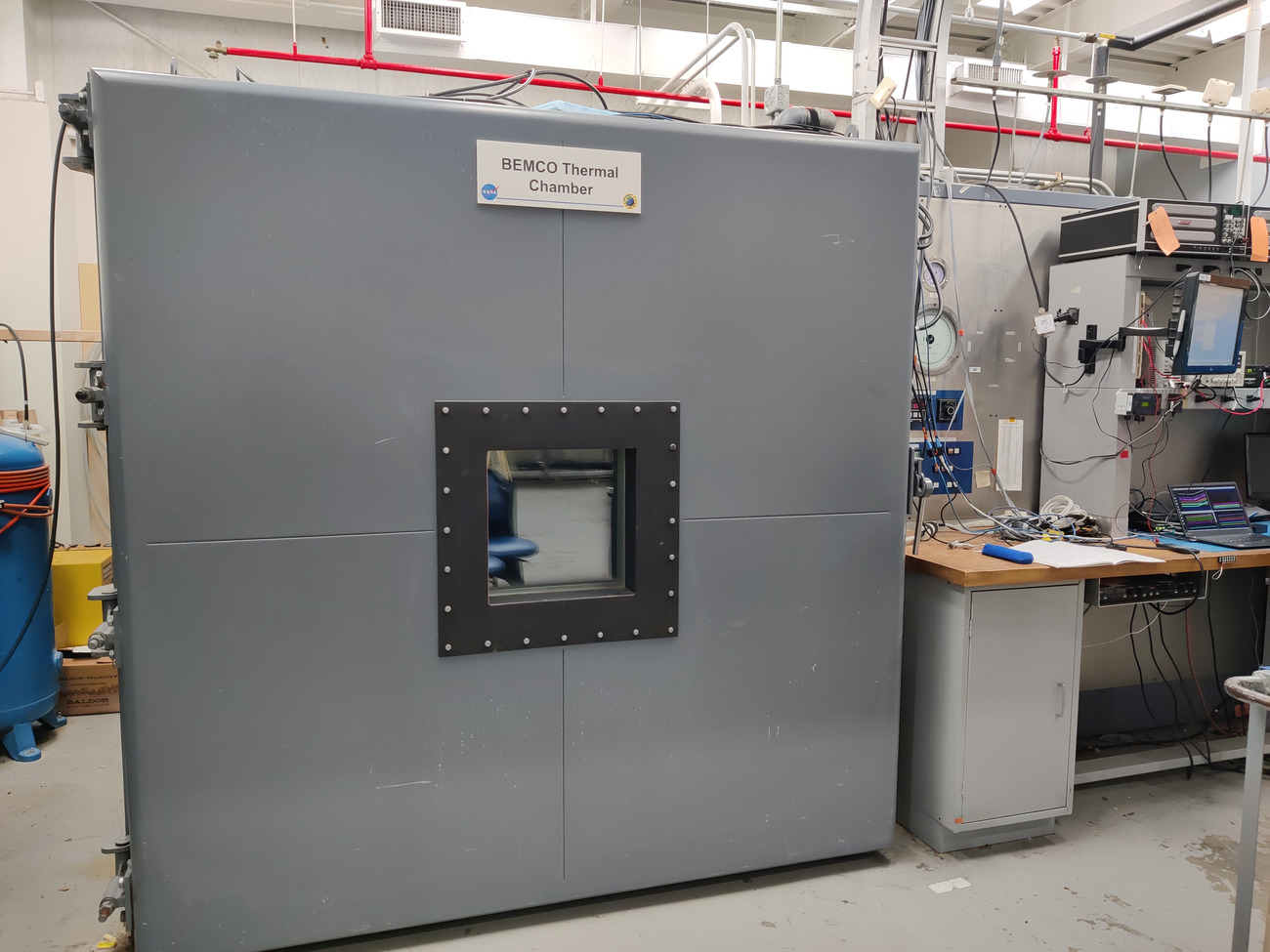}
        \caption{BEMCO thermal vacuum chamber at the NASA Columbia Scientific Balloon Facility in Palestine, Texas.}
        \label{fig:app-B_bemco_tvac_chamber}
    \end{subfigure}
    \caption{Thermal vacuum testing of the HVLVPS at the NASA Columbia Scientific Balloon Facility. The test, conducted prior to payload integration, evaluated the thermal performance of the power supply under flight-like conditions.}
    \label{fig:app-B_nasa_columbia_test}
\end{figure}

During this campaign, I also conducted thermal vacuum tests of the \hbox{high-voltage/low-voltage} power supply (HVLVPS), as shown in \Cref{fig:app-B_hvlvps_tvac_test}.
The tests were performed inside a BEMCO TVAC chamber at CSBF (\Cref{fig:app-B_bemco_tvac_chamber}) to validate the thermal behavior of the power supply under nominal operating conditions at \SI{-40}{\celsius}.
Custom load-emulating boards were used to simulate the tracker rows connected to the HVLVPS.
A dedicated acquisition system logged voltage, current, and board temperature data during operation under varying loads.
This test revealed several thermal issues in the original design that were addressed prior to the Antarctic deployment.

%-------------------------------------------------------------------------------
%   First Antarctic campaign
%-------------------------------------------------------------------------------
\section{First Antarctic campaign}

\begin{figure}[t!]
    \centering
    \captionsetup[subfigure]{justification=justified,singlelinecheck=false}
    \begin{subfigure}{0.49\textwidth}
        \centering
        \includegraphics[width=\linewidth]{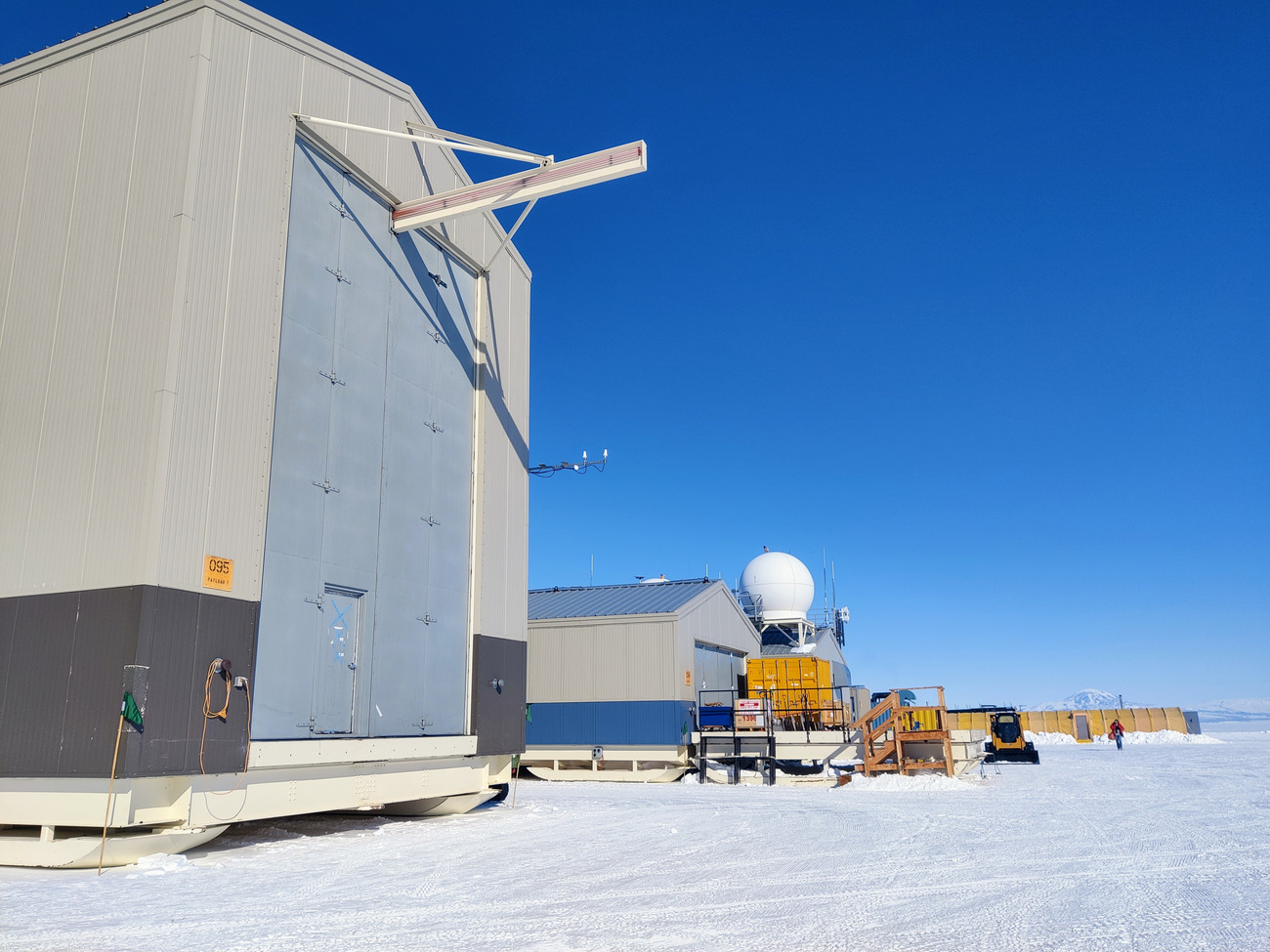}
        \caption{The NASA Long Duration Balloon (LDB) facility near McMurdo Station, Antarctica.}
        \label{fig:app-B_gaps_antarctica_ldb_facility}
    \end{subfigure}
    \hfill
    \begin{subfigure}{0.49\textwidth}
        \centering
        \includegraphics[width=\linewidth]{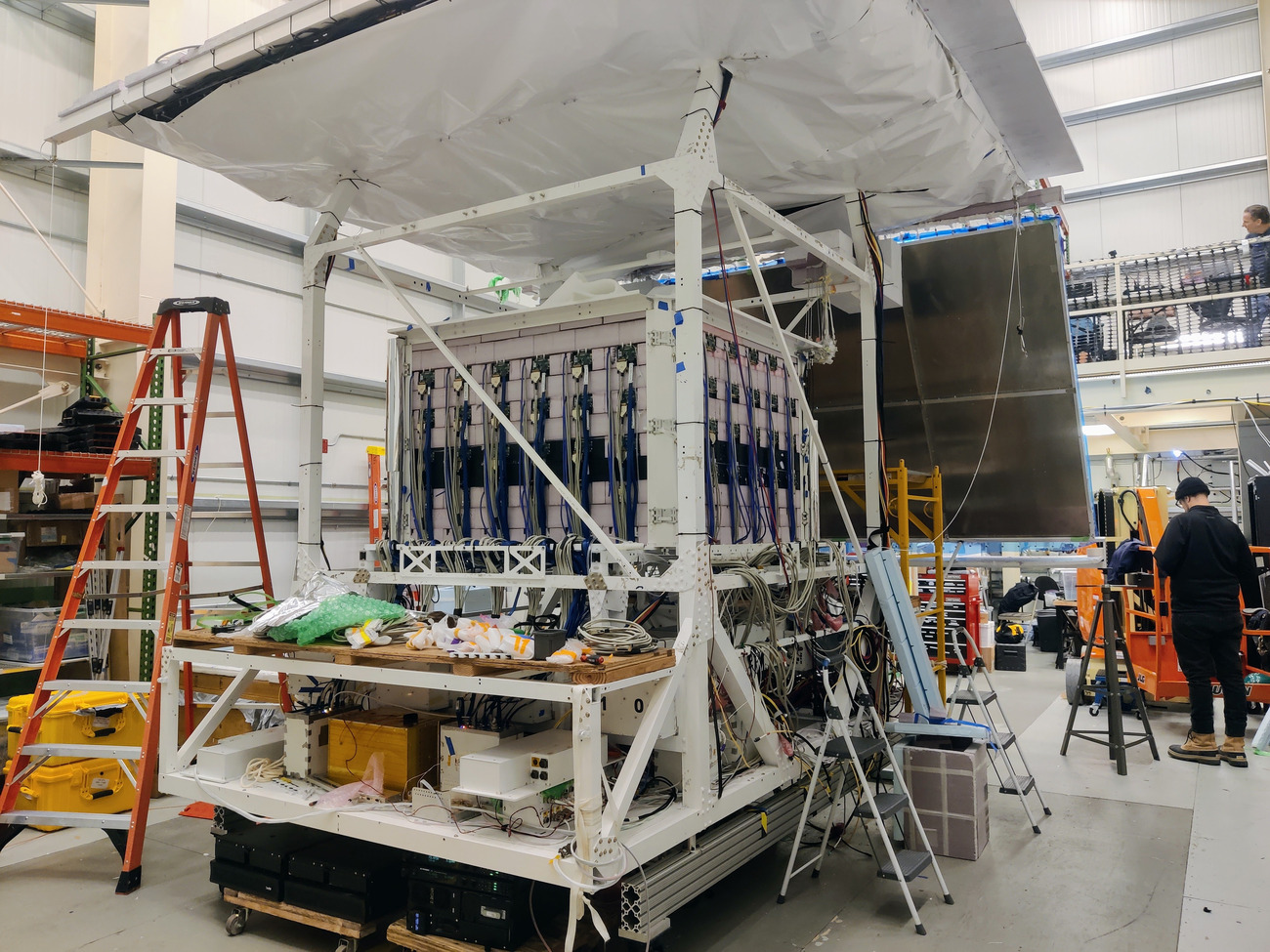}
        \caption{Assembly of the GAPS instrument inside Payload Building 1 during the first Antarctic flight campaign.}
        \label{fig:app-B_gaps_antarctica_tracker_assembly}
    \end{subfigure}
    \caption{GAPS Antarctic deployment activities: (\subref{fig:app-B_gaps_antarctica_ldb_facility}) exterior view of the NASA LDB facility and (\subref{fig:app-B_gaps_antarctica_tracker_assembly}) final integration of the instrument inside the facility during preparations for the first science flight.}
    \label{fig:app-B_gaps_antarctica}
\end{figure}

\begin{figure}[t!]
    \centering
    \captionsetup[subfigure]{justification=justified,singlelinecheck=false}
    \begin{subfigure}{0.49\textwidth}
        \centering
        \includegraphics[width=\linewidth]{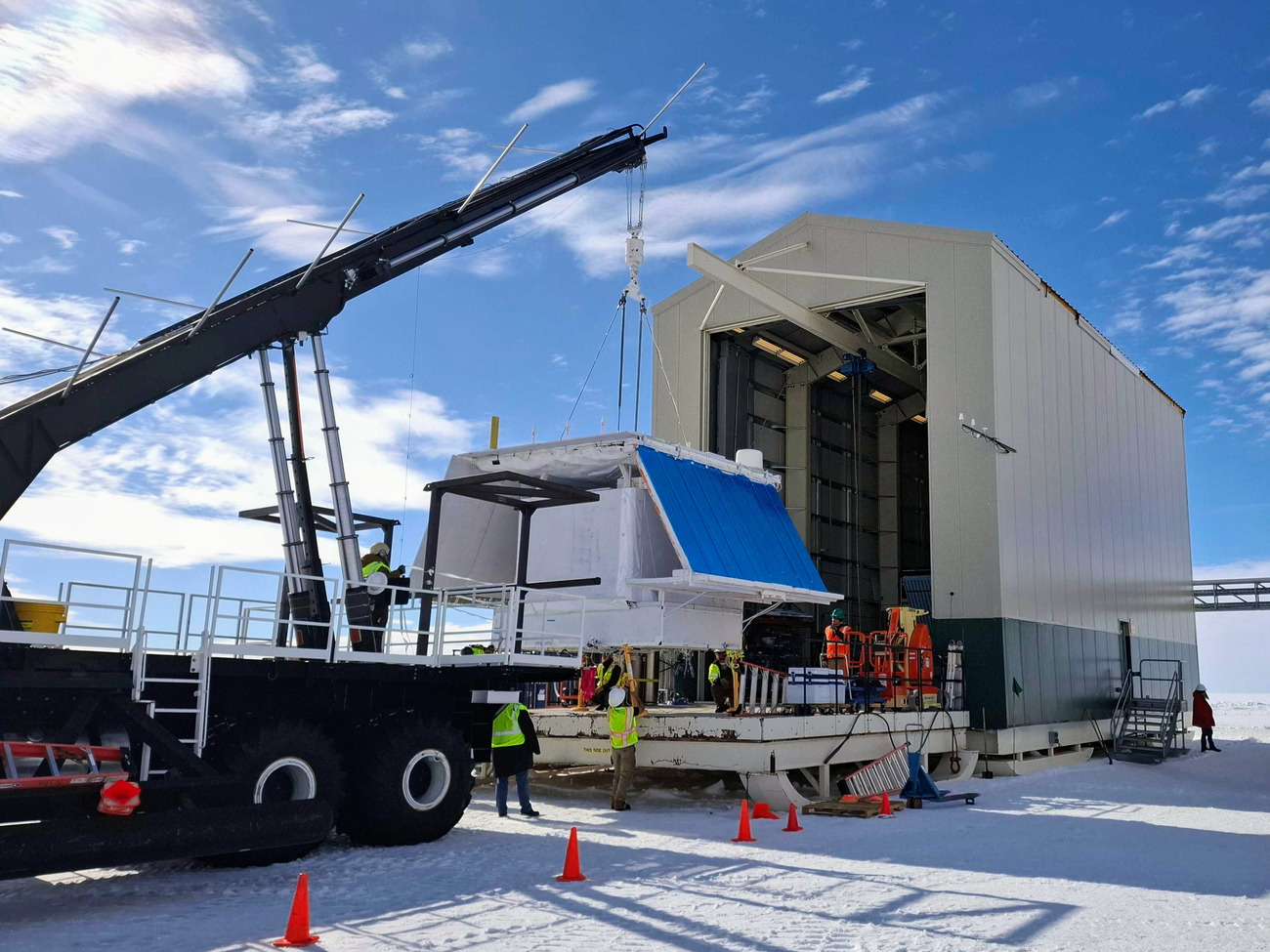}
        \caption{Crane operations moving the fully assembled GAPS payload from the high bay at the LDB facility to the launch site.}
        \label{fig:app-B_gaps_antarctica_ldb_crane}
    \end{subfigure}
    \hfill
    \begin{subfigure}{0.49\textwidth}
        \centering
        \includegraphics[width=\linewidth]{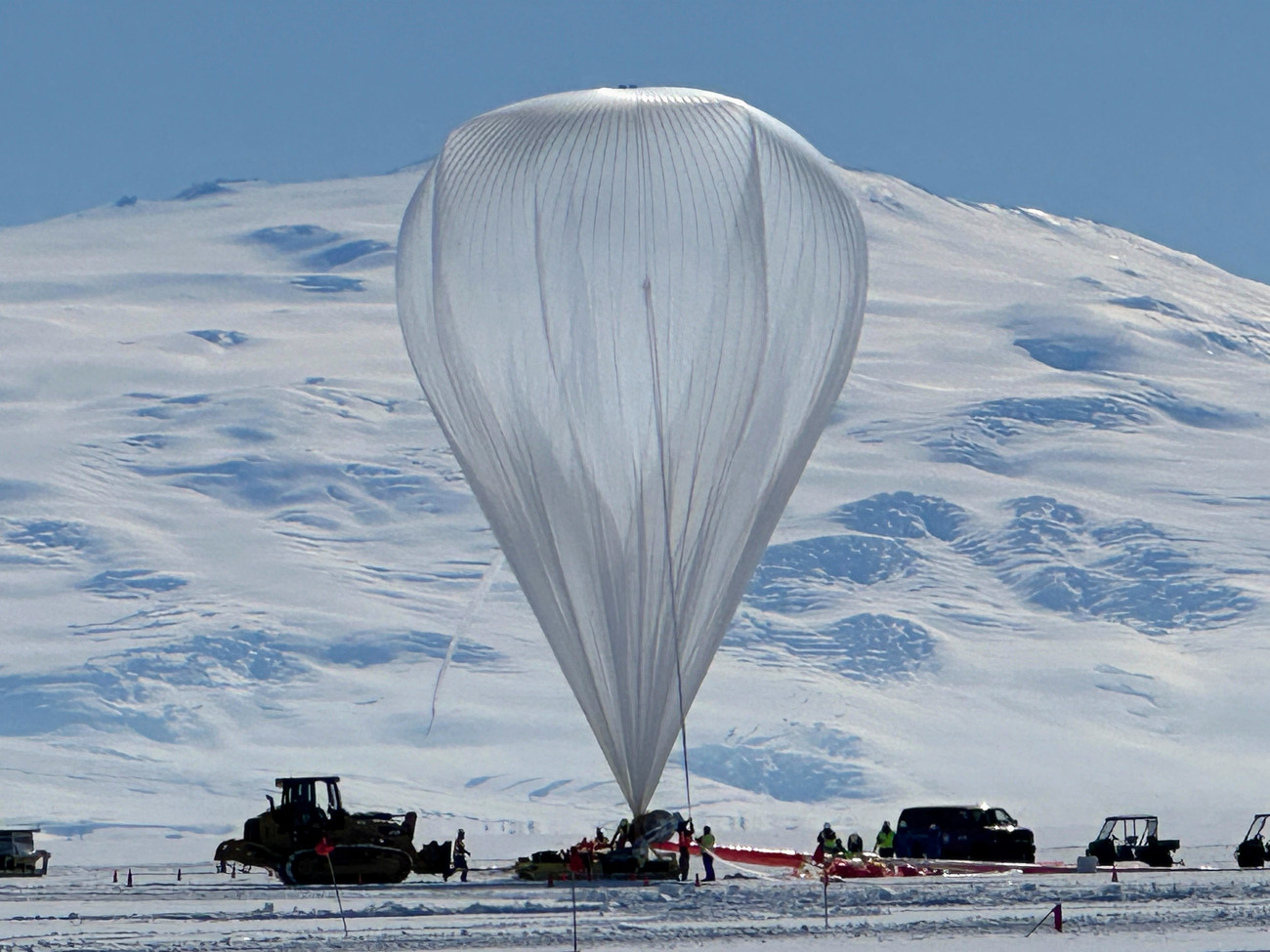}
        \caption{Inflation of the scientific balloon at the NASA LDB launch site prior to the first GAPS flight campaign.}
        \label{fig:app-B_gaps_antarctica_inflation}
    \end{subfigure}
    \caption{Final operations before launch at the NASA LDB facility in Antarctica: (\subref{fig:app-B_gaps_antarctica_ldb_crane}) payload transport and (\subref{fig:app-B_gaps_antarctica_inflation}) balloon inflation.}
    \label{fig:app-B_gaps_antarctica_balloon}
\end{figure}

\par
In November 2024, I was selected and qualified to participate in the first Antarctic campaign of the GAPS experiment, which took place from October 2024 to January 2025 at the NASA Long Duration Balloon (LDB) facility near McMurdo Station, Antarctica, under the United States Antarctic Program (USAP).
The payload was shipped in October 2024, with the main instrument sent by sea and the more sensitive tracker assembly flown separately on a dedicated cargo aircraft.

\par
At the LDB facility, the final integration and validation were performed inside Payload Building 1, as shown in \Cref{fig:app-B_gaps_antarctica_ldb_facility,fig:app-B_gaps_antarctica_tracker_assembly}.
I was responsible for the validation of the tracker readout electronics and power supply during this stage.
Multiple acquisition runs with cosmic muons were conducted to ensure that the tracker system performed within specification prior to launch.

\par
Over a span of ten days from late December 2024 to early January 2025, six launch attempts were conducted.
Adverse weather conditions repeatedly delayed the flight.
During each attempt, the payload was moved to the launch pad using a dedicated balloon launch vehicle, as shown in \Cref{fig:app-B_gaps_antarctica_ldb_crane}.
The first successful balloon inflation took place on January 1, 2025 (\Cref{fig:app-B_gaps_antarctica_inflation}), but the launch campaign was ultimately aborted on January 6 due to persistent unfavorable weather conditions.

\par
At the conclusion of the campaign, the fully integrated GAPS payload was secured at the LDB facility in Antarctica, awaiting its next launch opportunity in the 2025--2026 austral summer.